\documentclass[letterpaper]{article} 
\usepackage[draft]{aaai2026}  
\usepackage{times}  
\usepackage{helvet}  
\usepackage{courier}  
\usepackage[hyphens]{url}  
\usepackage{graphicx} 
\usepackage{natbib}  
\usepackage{caption} 
\usepackage{algorithm}
\usepackage{algorithmicx}
\usepackage{algpseudocode}
\usepackage{amsfonts}
\usepackage{booktabs}
\usepackage{longtable}
\usepackage{array}
\usepackage{multirow}
\usepackage{xcolor}
\usepackage{colortbl}
\usepackage{adjustbox}
\usepackage{mathtools}
\usepackage{makecell}
\usepackage{siunitx}
\usepackage{subcaption}
\usepackage{cleveref}
\usepackage{url}
\usepackage{bm}

\usepackage{newfloat}
\usepackage{listings}
\usepackage{amsmath}
\DeclareCaptionStyle{ruled}{labelfont=normalfont,labelsep=colon,strut=off} 
\floatstyle{ruled}
\newfloat{listing}{tb}{lst}{}
\floatname{listing}{Listing}
\title{\includegraphics[height=0.6cm]{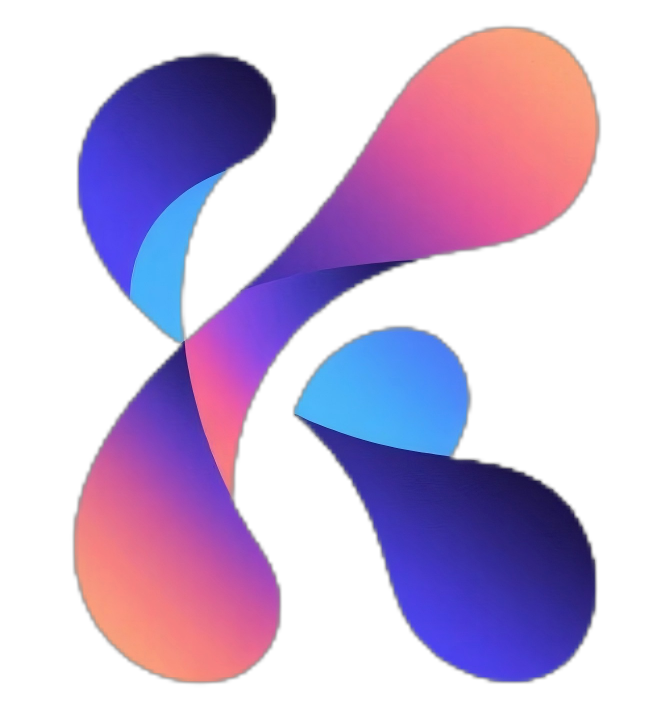}\hspace{0.3em}Kronos: A Foundation Model for the Language of Financial Markets}

\author{
  Yu Shi\textsuperscript{\rm 1,\dag},
  Zongliang Fu\textsuperscript{\rm 2,\dag},
  Shuo Chen\textsuperscript{\rm 1},
  Bohan Zhao\textsuperscript{\rm 1},
  Wei Xu\textsuperscript{\rm 1},
  Changshui Zhang\textsuperscript{\rm 2},
  Jian Li\textsuperscript{\rm 1}
}
\affiliations{
  \textsuperscript{\rm 1}Institute for Interdisciplinary Information Sciences, \textsuperscript{\rm 2}Department of Automation\\
  Tsinghua University\\
  \{shi-y23, fzl22, zhaobh23\}@mails.tsinghua.edu.cn, ChenSh2003@outlook.com, weixu@tsinghua.edu.cn, zcs@mail.tsinghua.edu.cn, lapordge@gmail.com
}

\definecolor{AvgGray}{gray}{0.92}
\definecolor{CntGray}{gray}{0.96}
\newcommand{\best}[1]{\underline{\color{red}{#1}}}
\newcommand{\second}[1]{\underline{\color{blue}{#1}}}
\newcommand{\cnt}[1]{\multicolumn{1}{c}{#1}}
\renewcommand{\footnoterule}{\kern-3pt \hrule width 0.3\columnwidth \kern 2.6pt}

\begin{document}

\maketitle

\let\thefootnote\relax
\footnote{\textsuperscript{\dag}Equal contribution}

\begin{abstract}
 The success of large-scale pre-training paradigm, exemplified by Large Language Models (LLMs), has inspired the development of Time Series Foundation Models (TSFMs). However, their application to financial candlestick (K-line) data remains limited, often underperforming non-pre-trained architectures. Moreover, existing TSFMs often overlook crucial downstream tasks such as volatility prediction and synthetic data generation. To address these limitations, we propose \textbf{Kronos, a unified, scalable pre-training framework tailored to financial K-line modeling}. Kronos introduces a specialized tokenizer that discretizes continuous market information into token sequences, preserving both price dynamics and trade activity patterns. We pre-train Kronos using an autoregressive objective on a massive, multi-market corpus of over 12 billion K-line records from 45 global exchanges, enabling it to learn nuanced temporal and cross-asset representations. Kronos excels in a zero-shot setting across a diverse set of financial tasks. On benchmark datasets, Kronos boosts price series forecasting RankIC by 93\% over the leading TSFM and 87\% over the best non-pre-trained baseline. It also achieves a 9\% lower MAE in volatility forecasting and a 22\% improvement in generative fidelity for synthetic K-line sequences. These results establish Kronos as a robust, versatile foundation model for end-to-end financial time series analysis. Our pre-trained model is publicly available at \url{https://github.com/shiyu-coder/Kronos}.
\end{abstract}

\section{Introduction}

The emergence of Foundation Models (FMs) has initiated a paradigm shift across artificial intelligence, reshaping the methodologies of representation learning and downstream task adaptation. This shift is exemplified by the success of Large Language Models (LLMs) for natural language processing~\cite{brown2020language,achiam2023gpt}, with parallel breakthroughs in computer vision~\cite{radford2021learning,kirillov2023segment}. 

Inspired by these advances, the FM paradigm has recently been extended to temporal data, giving rise to Time Series Foundation Models (TSFMs)~\cite{garza2023timegpt,woo2024unified,xiaoming2025time}. The central aim is to build pre-trained, task-agnostic architectures that serve as universal backbones for diverse time series analytical tasks—from forecasting and anomaly detection to causal inference—thereby substantially reducing the need for bespoke model design in each application domain.

Within this expanding research landscape, financial markets stand out as a critical and challenging application area for TSFMs, given their inherent data richness, high-frequency observations, and complex, non-stationary temporal dynamics. At the core of this domain are K-line sequences, multivariate time series derived from candlestick charts that record \textbf{O}pen, \textbf{H}igh, \textbf{L}ow, and \textbf{C}lose prices, along with trading \textbf{V}olume and \textbf{A}mount (Turnover) over fixed intervals (\textbf{OHLCVA}). These sequences constitute a highly compact, information-dense “language” through which market participants interpret price movements, volatility regimes, liquidity shifts, and collective sentiment~\cite{nison2001japanese}. Consequently, K-line data forms the bedrock of numerous algorithmic trading strategies, portfolio optimization schemes, and risk management systems.

\begin{figure}[t!]
    \centering
    \includegraphics[width=1.0\columnwidth]{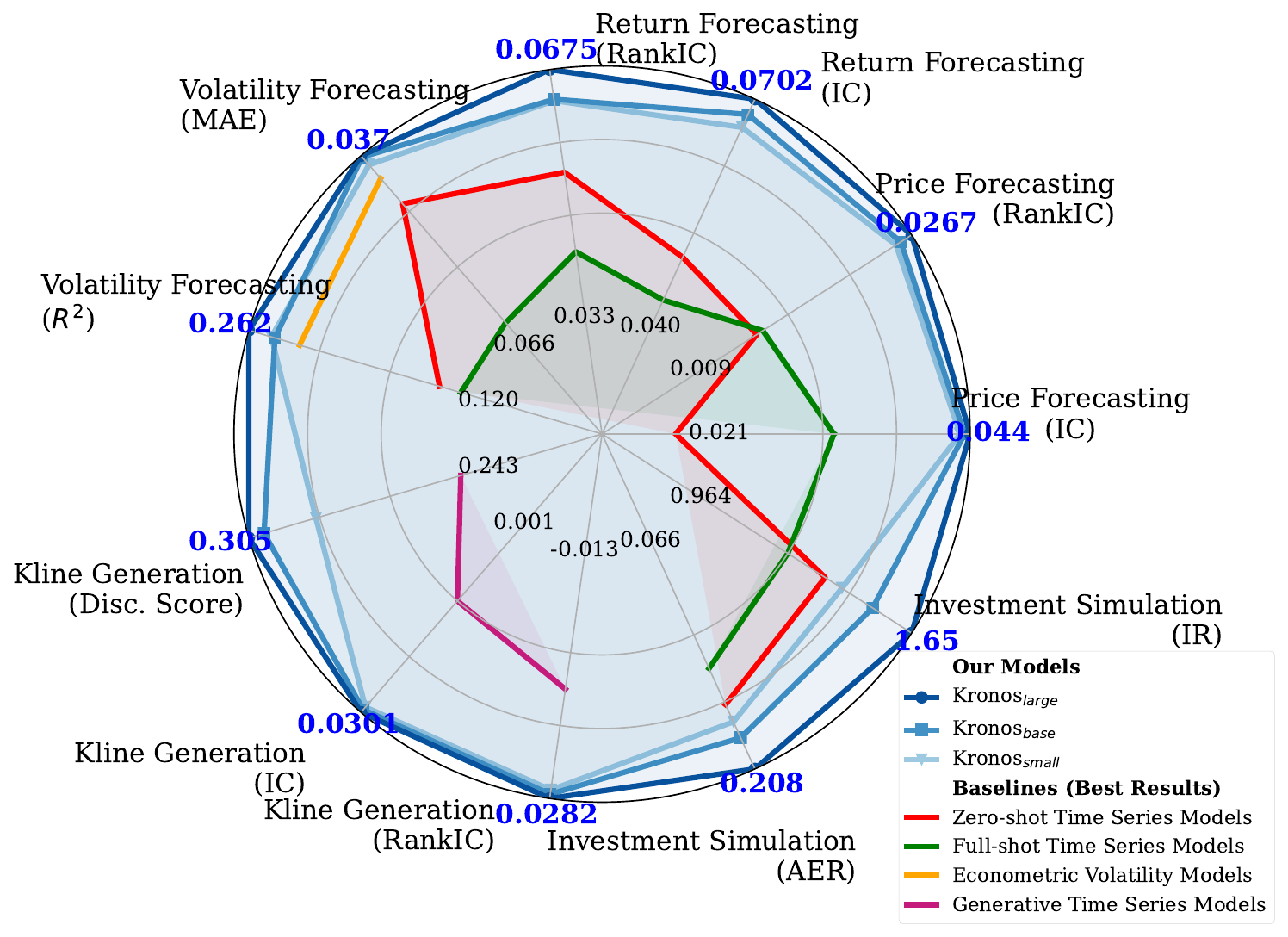}
    \caption{Comprehensive performance of Kronos across several quantitative finance tasks. The chart benchmarks our Kronos models (blue family) against several categories of specialized baselines. 
    A greater distance from the center signifies superior performance.
    }
    \label{fig:performance_overview}
\end{figure}

However, applying general-purpose TSFMs to financial K-line data presents significant challenges, due to two principal factors.
First, K-line sequences exhibit unique statistical properties—such as low signal-to-noise ratios, strong non-stationarities, and intricate, high-order dependencies among OHLCVA attributes~\cite{zhang2025major,baidya2024addressing}—that are often misaligned with the inductive biases of generic TSFMs. 
Second, the financial domain has largely been underserved by mainstream TSFM research; financial sequences constitute a minor fraction of pre-training corpora for most existing TSFMs~\cite{das2024decoder,gao2024units,xiaoming2025time}
, and the spectrum of downstream tasks critical to quantitative finance—spanning volatility estimation, synthetic sequence generation, and risk management—remains largely unaddressed. These factors lead to an important observation, which we empirically validate in this work: general-purpose TSFMs often underperform specialized, non-pre-trained models (e.g., iTransformer~\cite{liuitransformer}) on financial tasks and fail to generalize across the broader landscape of quantitative finance.

To address these shortcomings, we introduce \textbf{Kronos, a unified, scalable pre-training framework designed specifically for financial K-line data}. Kronos employs a specialized tokenizer to discretize continuous, multivariate K-line inputs into a sequence of compact tokens, preserving critical price–volume interactions. It then undergoes autoregressive pre-training on an expansive, heterogeneous corpus of over 12 billion K-line records drawn from over 45 global markets and 7 temporal granularities. 

We validate the efficacy of Kronos through comprehensive experiments across a range of quantitative finance tasks, with a high-level summary presented in Figure \ref{fig:performance_overview}. On the core task of price series forecasting, Kronos establishes a new state-of-the-art, boosting the RankIC by 93\% over the leading TSFM and by 87\% over the best-performing non-pre-trained baseline. Furthermore, it demonstrates strong versatility by achieving a 9\% lower MAE in volatility forecasting and a 22\% improvement in generative fidelity for synthetic K-line generation. These findings highlight the broad effectiveness of our approach and underscore Kronos's potential as a robust foundation model for interpreting the complex ``language'' of financial markets.

Our main contributions can be summarized as follows:

\begin{itemize}
    \item We propose a novel modeling framework for financial K-line data that learns hierarchical representations. It features a specialized tokenizer that quantizes each multivariate K-line record into structured, dual-component (coarse and fine) tokens, coupled with a tailored autoregressive objective that predicts these subtokens sequentially. This coarse-to-fine prediction scheme allows Kronos to explicitly model multi-scale market dynamics.

    \item We conduct large-scale pre-training for a family of Kronos models with varying capacities. This is performed on a massive, diverse financial corpus of over 12 billion K-line records from over 45 global exchanges, which is fundamental to learning the robust and generalizable market representations that underpin the models' effectiveness.

    \item We conduct comprehensive empirical evaluations across a set of quantitative finance tasks. Our results show that Kronos establishes a new state-of-the-art in price series forecasting, significantly outperforming both TSFMs and specialized baselines. The model's versatility is further demonstrated by its strong performance across a broader spectrum of quantitative tasks, including volatility forecasting and synthetic K-line generation.
\end{itemize}

\begin{figure*}[t!]
    \centering
    \includegraphics[width=1.0\textwidth]{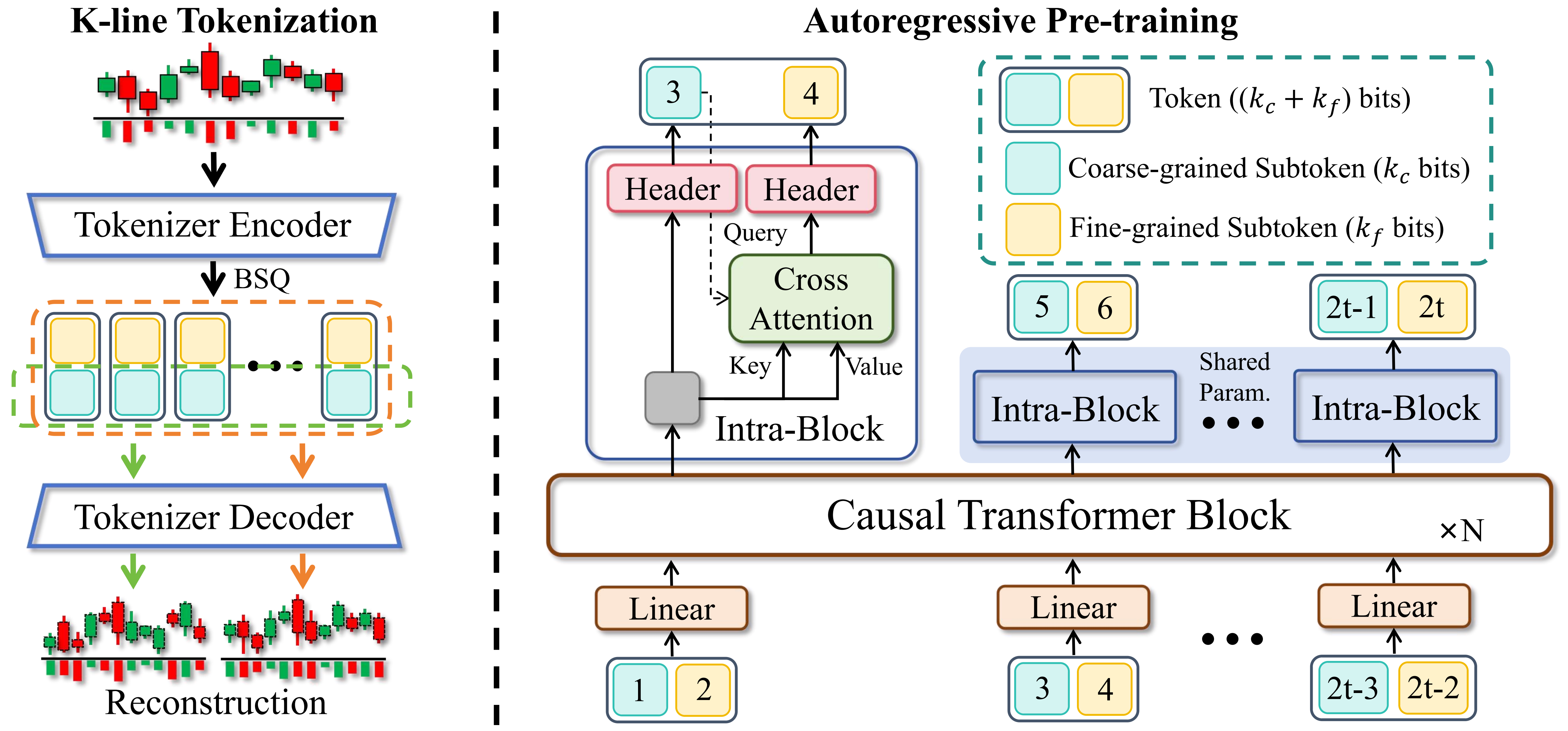}
    \caption{The two-stage framework of Kronos. \textbf{(1) Instance-based K-line Tokenization:} A Transformer-based autoencoder with a dual reconstruction objective quantizes continuous K-line data into a vocabulary of hierarchical discrete tokens, each comprising a coarse and a fine subtoken. \textbf{(2) Autoregressive Pre-training:} A decoder-only Transformer is pre-trained to model the temporal dynamics by sequentially predicting the hierarchical subtokens for the next time step, conditioned on the past.}
    \label{fig:overview}
\end{figure*}

\section{Preliminary}

Let \(D\)-dimensional vector \(\mathbf{x}_t \in \mathbb{R}^D\) denote the K-line observation at discrete time \(t\), comprising \(D\) key financial indicators. In this work, we fix the dimension \(D=6\) to represent OHLCVA attributes (Open, High, Low, Close prices, trading Volume, and Amount). The rationale for this input choice is detailed in Appendix~\ref{sec:discussion} (Q1). Given a historical sequence
\(
\mathbf{x}_{1:T} = (\mathbf{x}_1, \mathbf{x}_2, \ldots, \mathbf{x}_T),
\)
our objective is to predict the following \(H\) observations
\(
\hat{\mathbf{x}}_{T+1:T+H} = (\hat{\mathbf{x}}_{T+1}, \hat{\mathbf{x}}_{T+2}, \ldots, \hat{\mathbf{x}}_{T+H}).
\)

Rather than operating on raw continuous inputs, Kronos first quantizes each multivariate observation \(\mathbf{x}_t\) into a discrete token \(b_t\) via a learnable codebook \(\mathcal{C}\). 
Consequently, the original sequence
\(
\mathbf{x}_{1:T} = (\mathbf{x}_1, \dots, \mathbf{x}_T)
\)
is mapped to
\(
\mathbf{b}_{1:T} = (b_1, \dots, b_T).
\)
The forecasting task then reduces to an autoregressive token-sequence modeling problem:
\begin{equation}
\label{eq:ts_ar}
p(\mathbf{b}_{T+1:T+H} \mid \mathbf{b}_{1:T})
= \prod_{h=1}^{H} p\bigl(b_{T+h} \mid \mathbf{b}_{1:T+h-1}\bigr).
\end{equation}

Such a discrete formulation is inherently scalable and naturally extends to other tasks that can be framed generatively, such as synthetic data generation and volatility forecasting.

\section{Methodology}

Kronos abstracts financial K-line sequences as a discrete language and implements this via a two-phase framework illustrated in Figure~\ref{fig:overview}: \textbf{(1) K-line Tokenization} and \textbf{(2) Autoregressive Pre-training}. In the first phase, we design a specialized Transformer-based tokenizer to quantize a continuous, multivariate K-line sequence into a corresponding sequence of discrete tokens, via a learnable codebook. Each K-line item (OHLCVA) is treated as an individual instance and quantized into a discrete token. Each token is composed of a coarse-grained subtoken and a fine-grained subtoken. 
This property is enforced via a hierarchical reconstruction loss, which explicitly compels the subtokens to model distinct levels of information, thereby creating a coarse-to-fine informational hierarchy.
In the second phase, an autoregressive decoder-only Transformer is pre-trained on these tokenized sequences, using the standard next-token prediction objective to sequentially forecast both subtoken levels at each future time step conditioned on the given historical context. This unified \emph{discretize-and-generate} paradigm enables Kronos to construct a high-fidelity, hierarchical representation of market dynamics, providing a robust foundation for downstream quantitative analysis.

\subsection{K-line Tokenization}

\begin{figure}[t]
    \centering
    \includegraphics[width=1.0\columnwidth]{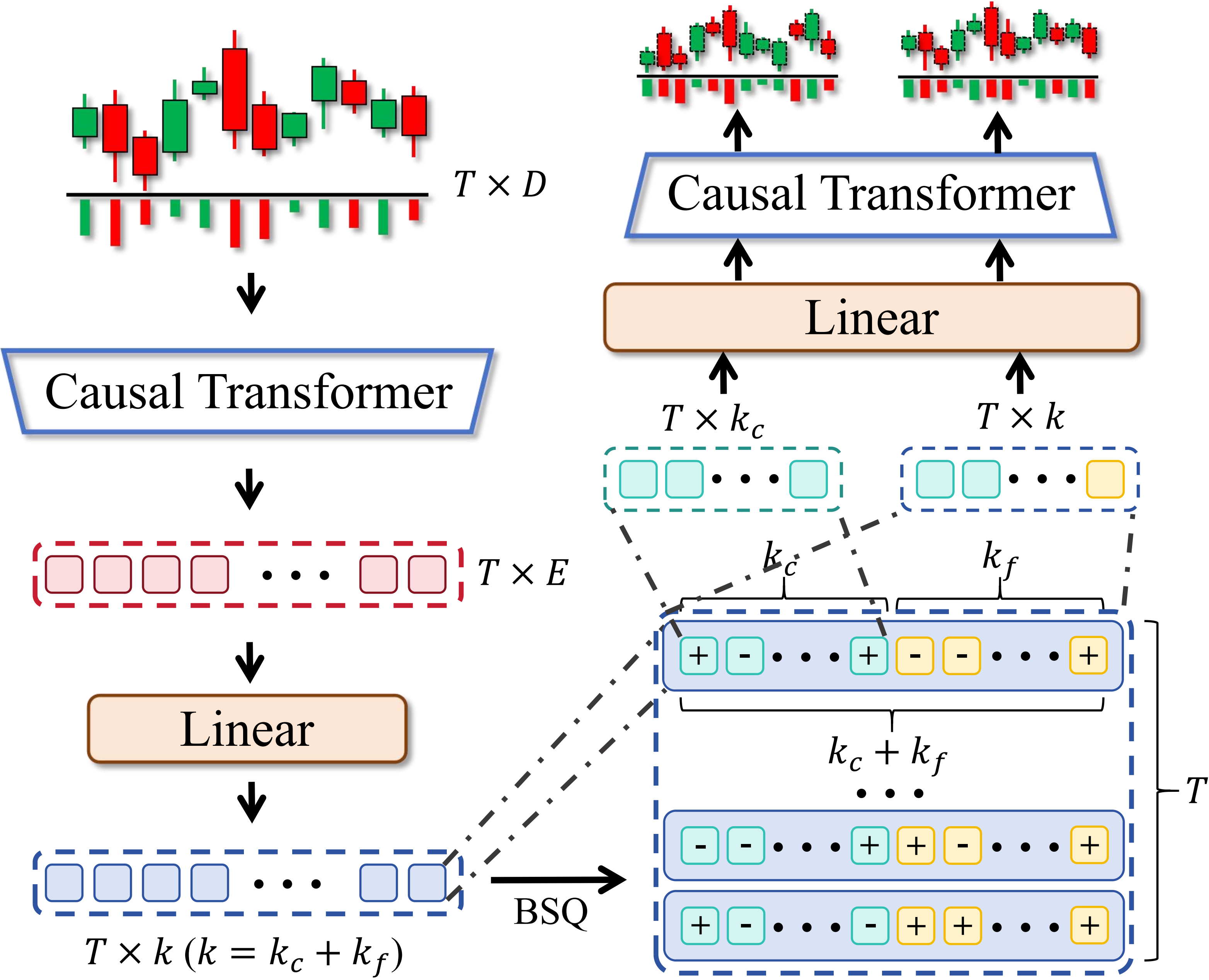}
    \caption{Architecture of the K-line Tokenizer. It employs a Transformer-based autoencoder with a Binary Spherical Quantization (BSQ) layer. 
    }
    \label{fig:tokenizer}
\end{figure}

The first stage of Kronos transforms a continuous, $D$-dimensional K-line sequence $\mathbf{x} = (\mathbf{x}_{1}, \ldots, \mathbf{x}_{T})$, where $\mathbf{x}_{t}\in\mathbb{R}^{D}$ encodes OHLCVA indicators, into a corresponding series of discrete tokens. This is achieved using a Transformer-based autoencoder (Figure~\ref{fig:tokenizer}) composed of an encoder $E_{\text{enc}}$, a quantizer $Q$, and a decoder $E_{\text{dec}}$. Drawing inspiration from video quantization methods in generative modeling~\cite{van2017neural,yu2023language}, we adapt Binary Spherical Quantization (BSQ)~\cite{zhao2024image}, a variant of Look-up Free Quantization (LFQ)~\cite{yu2023language}, for this task. 
We discuss the rationale for this choice in Appendix~\ref{sec:discussion} (Q2). 
BSQ quantizes a continuous latent vector $\bm{\xi}_t$ into a $k$-bit binary code $b_t \in \{-1,1\}^k$ by projecting it onto a set of learnable hyperplanes. While a large 
number of bits $k$ (e.g., $k=20$) is desirable for capturing rich financial patterns, it results in an exponentially large vocabulary of size $2^k$, which introduces significant challenges for the subsequent autoregressive model in terms of computational cost and parameter size. To mitigate this, we follow recent work in video quantization and generation~\cite{yu2023language,wang2025bridging} and factorize the $k$-bit code into $n$ subspaces. Motivated by the trade-off between parameter savings and latency costs detailed in Appendix~\ref{sec:discussion} (Q3), we set $n=2$. We partition the code into a coarse subtoken $b_{t}^{c}$ and a fine subtoken $b_{t}^{f}$ of equal bit length, $k_c = k_f = k/2$, where $k=k_c+k_f$. The resulting code $b_t$ is a concatenation of these two subtokens:
\(
  b_{t} = \bigl[b_{t}^{c},\,b_{t}^{f}\bigr],
\)
with $b_{t}^{c}$, $b_{t}^{f} \in\{-1,1\}^{k/2}$. This decomposition transforms a single prediction over a large vocabulary of size $2^k$ into two sequential predictions over $2^{k/2}$ entries, substantially reducing both computational and parameter complexity.

To enforce a coarse-to-fine structure within each token, we train the tokenizer with a composite objective that combines a hierarchical reconstruction loss and a commitment loss for BSQ:
\begin{equation}
  \mathcal{L}_{\text{tokenizer}}
  = \mathcal{L}_{\text{coarse}} + \mathcal{L}_{\text{fine}} + \lambda \mathcal{L}_{\text{quant}},
\end{equation}
where $\lambda$ is a balancing hyperparameter. The components are defined as:
\begin{itemize}
    \item \(\mathcal{L}_{\text{coarse}} = \mathbb{E}\bigl[\|\mathbf{x} - E_{\text{dec}}(\mathbf{b}^{c})\|^{2}\bigr]\), which trains the coarse subtoken \(\mathbf{b}^{c}\) to form a 
    low-fidelity reconstruction.
    \item \(\mathcal{L}_{\text{fine}} = \mathbb{E}\bigl[\|\mathbf{x} - E_{\text{dec}}(\mathbf{b})\|^{2}\bigr]\), which evaluates the high-fidelity reconstruction using the complete token \(\mathbf{b}\).
    \item \(\mathcal{L}_{\text{quant}}\) is the quantization loss from BSQ~\cite{zhao2024image} that regularizes the learning process. 
    It penalizes the L2 distance between continuous latent vectors $\bm{\xi}$ and their binary codes $\mathbf{b}$, aligning the encoder's outputs with the learned codebook to ensure stable training.
\end{itemize}

This hierarchical reconstruction objective is central to our design. By optimizing \(\mathcal{L}_{\text{coarse}}\), the coarse subtoken \(\mathbf{b}^{c}\) learns to capture the principal structure of the input. Consequently, during the optimization of \(\mathcal{L}_{\text{fine}}\), the fine-grained subtoken \(\mathbf{b}^{f}\) is guided to encode the residual information required to refine the coarse approximation. Prior work has shown that a coarse-to-fine decoding order improves generation quality~\cite{wang2025bridging}. Instead of identifying and prioritizing the decoding of tokens that inherently contain coarse information, our approach is designed to explicitly impose this hierarchy into the tokens during quantization. This ensures that the first subtoken consistently represents coarse-grained information, establishing the desired conditional dependency for the subsequent autoregressive modeling stage.

\subsection{Hierarchical Autoregressive Modeling}
\label{sec:autoregressive}

Following the tokenization stage, the resulting discrete sequences are modeled using a decoder-only Transformer, denoted as $E_{\text{ar}}$, which employs causal-attention to ensure that predictions at each time step depend exclusively on historical context. The primary objective is to estimate the joint distribution over the token sequence $\mathbf{b} = \{b_1, \dots, b_T\}$. A simplified form of Equation~\ref{eq:ts_ar} can be derived as:
\begin{equation}
    p(\mathbf{b}) = \prod_{t=1}^{T} p(b_t | \mathbf{b}_{<t}),
\end{equation}
where $\mathbf{b}_{<t}$ denotes all preceding tokens up to time $t-1$.

Given the hierarchical token design, in which each token is structured as $b_t = [b_t^{c}, b_t^{f}]$, we further decompose the conditional probability using the chain rule to explicitly capture the inherent coarse-to-fine dependency:
\begin{equation}
\label{eq:chain_rule}
p(b_t | \mathbf{b}_{<t}) = p(b_t^{c} | \mathbf{b}_{<t}) \cdot p(b_t^{f} | \mathbf{b}_{<t}, b_t^{c}).
\end{equation}
This formulation allows the model to first predict the coarse-grained subtoken, which serves as a scaffold for subsequently generating the fine-grained residual subtoken. Consequently, the pre-training objective reduces to maximizing the log-likelihood of the observed sequence under this hierarchical factorization.

As depicted in Figure~\ref{fig:overview} (Right), the autoregressive process begins by constructing a unified input vector for each time step. Specifically, at time $i$, the subtokens $b_i^{c}$ and $b_i^{f}$ are independently projected into vector representations using two distinct embedding layers, resulting in representations $e_c(b_i^{c})$ and $e_f(b_i^{f})$, respectively. These embeddings are then concatenated and linearly projected to produce a fused input vector:
\begin{equation}
\label{eq:embedding}
    \mathbf{v}_i = W_{\text{fuse}}([e_c(b_i^{c}); e_f(b_i^{f})]),
\end{equation}
where $[\cdot;\cdot]$ denotes concatenation, and $W_{\text{fuse}}$ is a learnable weight matrix responsible for projecting the combined representation into the model’s latent space.

The sequence of fused inputs $\{\mathbf{v}_1, \dots, \mathbf{v}_{t-1}\}$ is then processed by the Transformer $E_{\text{ar}}$, which outputs contextualized hidden states. The final hidden state from processing $\mathbf{b}_{< t}$, denoted as $\mathbf{h}_t$, is then used to predict the token $b_t$. This hidden state subsequently informs the autoregressive predictions of both coarse and fine subtokens at the next step, thereby enabling the model to effectively capture multi-scale temporal dependencies inherent in the data.

\textbf{Coarse Subtoken Prediction.} The history vector $\mathbf{h}_t$ is projected by a linear head $W_c$ to produce logits for the first subtoken's distribution:
\begin{equation}
    p(b_{t}^{c} | \mathbf{b}_{< t}) = \text{softmax}(W_c \mathbf{h}_t)
\end{equation}

\textbf{Fine Subtoken Prediction.} To model the conditional dependency in Equation~(\ref{eq:chain_rule}), the context needs to be updated with the predicted coarse subtoken, $\hat{b}_{t}^{c}$. During training, we use the model's own prediction from the previous step, $\hat{b}_{t}^{c}$, which is sampled from the predicted distribution $p(b_{t}^{c} | \mathbf{b}{< t})$, rather than using the ground-truth subtoken (i.e., teacher-forcing). We find that this sampling strategy enhances model robustness by mitigating exposure bias, better aligning the training distribution with the auto-regressive nature of multi-step inference where ground-truth tokens are unavailable. We use a cross-attention mechanism where the embedding of $\hat{b}_{t}^{c}$ acts as the query, and the history $\mathbf{h}_t$ provides the key and value. The result is projected by the second head $W_f$:
\begin{equation}
\begin{aligned}
    \mathbf{h}^{\text{update}}_t = \text{CrossAttn}&(q=e_c(\hat{b}_{t}^{c}), k=v=\mathbf{h}_t) \\
    p(b_{t}^{f} | \mathbf{b}_{< t}, b_{t}^{c}) &= \text{softmax}(W_f \mathbf{h}^{\text{update}}_t)
\end{aligned}
\end{equation}
The overall training objective $\mathcal{L}_{\text{ar}}$ is the negative log-likelihood of the data, summed over both prediction steps:
\begin{equation}
    \mathcal{L}_{\text{ar}} = - \mathbb{E}_{\mathbf{b} \sim \mathcal{D}} \sum_{t=1}^{T} \left[ \log p(b_t^{c} | \mathbf{b}_{<t}) + \log p(b_t^{f} | \mathbf{b}_{<t}, b_t^{c}) \right]
\end{equation}
where $\mathcal{D}$ represents the data distribution.

\begin{table}[t]
\centering
\setlength{\tabcolsep}{4pt}

\resizebox{\columnwidth}{!}{%
  \begin{tabular}{l ccccc c}
  \toprule
  & Layers & $\mathbf{d}_{\text{model}}$ & $\mathbf{d}_{\text{ff}}$ & Heads & Vocab. ($2^k$) & Params \\
  \midrule
\text{Kronos}$_{small}$ & 8 & 512  & 1024 & 8 & 20 & 24.7M \\
\text{Kronos}$_{base}$  & 12 & 832 & 2048 & 16 & 20 & 102.3M \\
\text{Kronos}$_{large}$ & 18 & 1664 & 3072 & 32 & 20 & 499.2M \\
  \bottomrule
  \end{tabular}%
}
\caption{Model configurations for the Kronos family. We detail the number of Transformer layers, model dimension ($\mathbf{d}_{\text{model}}$), feed-forward dimension ($\mathbf{d}_{\text{ff}}$), number of attention heads, vocabulary size, and the total number of parameters.}
\label{tab:kronos_configs_resized}
\end{table}

\subsection{Model Pre-training}

\begin{figure*}[t!]
    \centering
    \includegraphics[width=1.0\textwidth]{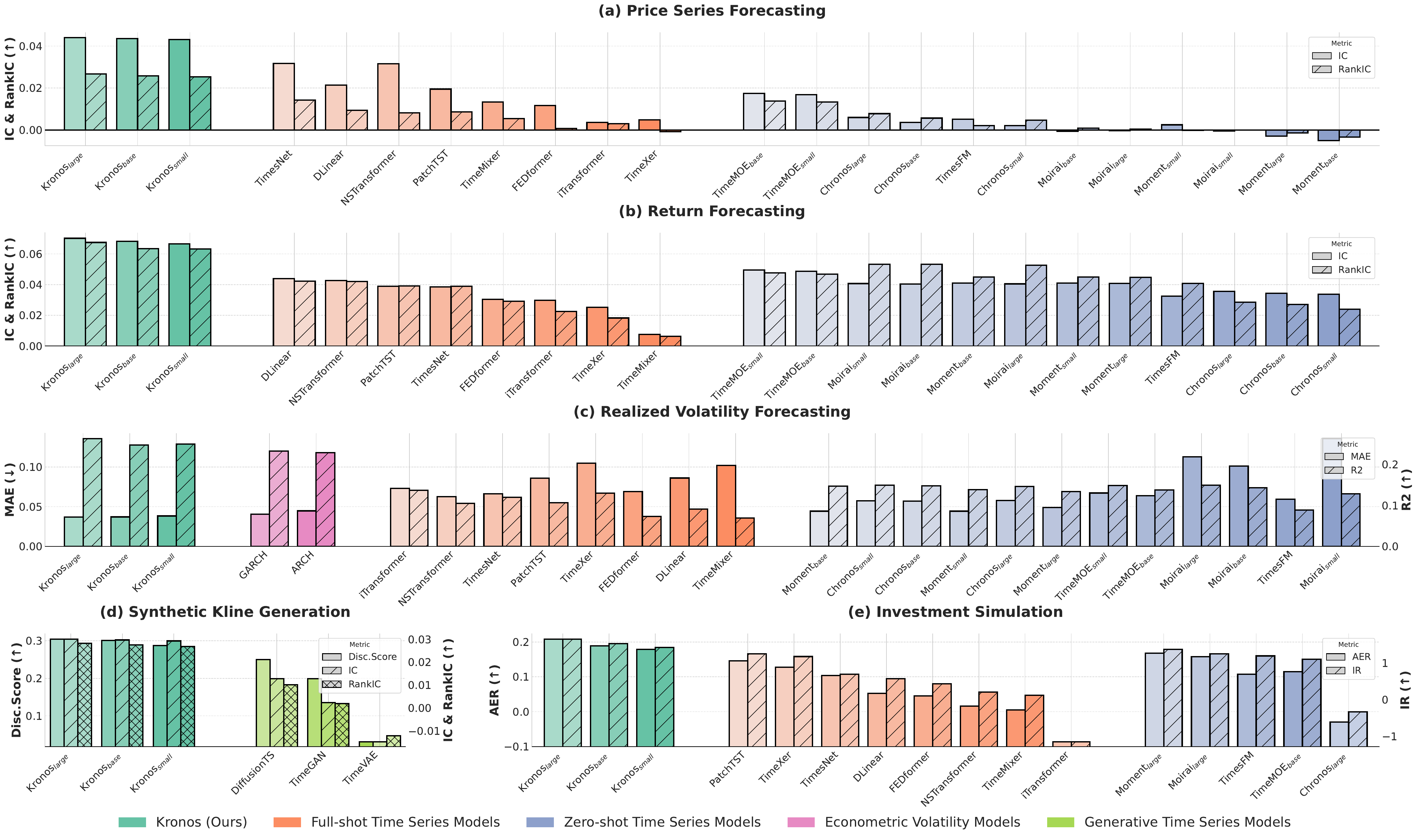}
    \caption{Main experimental results across five representative financial tasks. Subfigures (a-c) show forecasting performance on price series, returns, and realized volatility. Subfigure (d) displays generative model performance in terms of fidelity and usefulness. Subfigure (e) presents the investment simulation backtesting results.}
    \label{fig:main_result}
\end{figure*}

\paragraph{Dataset}
To ensure the quality of pre-training, we curate a large-scale, high-quality financial K-line dataset from the ground up. In contrast to foundation-model research on generic time series—where well-curated public datasets are readily pooled—comprehensive, high-quality financial data remain limited. Our dataset spans over 12 billion observations across 7 sampling frequencies, encompassing a broad spectrum of asset classes drawn from 45 global exchanges. To guarantee data quality, we develop a streamlined data-cleaning pipeline tailored to the unique characteristics of financial K-line data, which identifies and filters out low-quality segments such as those with abnormal price spikes or prolonged periods of inactivity. Further details on the cleaning pipeline are available in Appendix~\ref{sec:dataset_details}. 

\paragraph{Model Training} Informed by the scaling laws observed in LLMs ~\cite{kaplan2020scaling}, we trained three variants of Kronos with increasing parameter counts, up to nearly 0.5 billion, to provide a trade-off between performance and inference budget. The detailed model configurations are presented in Table~\ref{tab:kronos_configs_resized}. 
Considering resource constraints and practical deployment scenarios, we limit the maximum context length to 512 tokens. Nevertheless, this design remains fully compatible with arbitrary forecasting horizons by leveraging K-line data at varying frequencies; for instance, using 1-minute data for short-term forecasting and daily data for weekly or monthly predictions. Complete training details are provided in Appendix~\ref{sec:implementation_details}.

\paragraph{Inference} At inference time, we generate future token sequences autoregressively, analogous to text generation. The stochasticity of this process is controlled via standard techniques like temperature scaling and top-$p$ (nucleus) sampling~\cite{holtzman2019curious}. The probability of sampling token $i$ from logits $\mathbf{z}$ is given by $p_i \propto \exp(z_i / T)$, where $T$ is the temperature. For tasks requiring high precision, prediction accuracy can be enhanced by generating multiple future trajectories (i.e., Monte Carlo rollouts) and averaging the decoded continuous values to produce a more stable forecast. As demonstrated in our experiments, this approach consistently improves forecast quality.

\section{Experiments}
\label{sec:experiments}

\begin{table*}[t!]

\resizebox{\textwidth}{!}{
\begin{tabular}{@{} l l l c c c c c c c c @{}}
\toprule

\multirow{2.5}{*}{\textbf{Model}} & \multirow{2.5}{*}{\textbf{Prediction Space}} & \multirow{2.5}{*}{\textbf{Training Objective}} & \multicolumn{2}{c}{\textbf{Price Series Forecasting}} && \multicolumn{2}{c}{\textbf{Return Forecasting}} && \multicolumn{2}{c}{\textbf{Volatility Forecasting}} \\
\cmidrule(lr){4-5} \cmidrule(lr){7-8} \cmidrule(lr){10-11}

& & & IC ($\uparrow$) & RankIC ($\uparrow$) && IC ($\uparrow$) & RankIC ($\uparrow$) && MAE ($\downarrow$) & R$^2$ ($\uparrow$) \\
\midrule

Direct-AR       & Continuous & Mean Squared Error (MSE)  & 0.0212 & 0.0149 && 0.0416 & 0.0399 && 0.0565 & 0.1608 \\
Prob-AR         & Continuous & Negative Log-Likelihood (NLL) & 0.0179 & 0.0102 && 0.0356 & 0.0329 && 0.0464 & 0.1383 \\
\midrule 

Kronos-Parallel & Discrete   & Cross-Entropy             & 0.0345 & 0.0226 && 0.0529 & 0.0505 && 0.0461 & 0.1784 \\
\textbf{$\text{Kronos}_{small}$} & \textbf{Discrete}   & \textbf{Cross-Entropy}    & \textbf{0.0431} & \textbf{0.0254} && \textbf{0.0665} & \textbf{0.0622} && \textbf{0.0384} & \textbf{0.2490} \\
\bottomrule
\end{tabular}
}
\centering
\caption{
    Ablation study dissecting the architectural choices of Kronos.
    We compare our model against variants targeting different \textbf{Prediction Spaces} (continuous vs. discrete) with corresponding \textbf{Training Objectives}.
    \textit{Direct-AR} serves as a standard regression baseline.
    \textit{Prob-AR} evaluates the benefit of probabilistic modeling in the continuous space.
    \textit{Kronos-Parallel} ablates our sequential subtoken design by predicting subtokens concurrently.
    Best results are in \textbf{bold}.
}
\label{tab:ablation_study_refined}
\end{table*}

\begin{figure*}[htbp]
    \centering
    \includegraphics[width=1.0\textwidth]{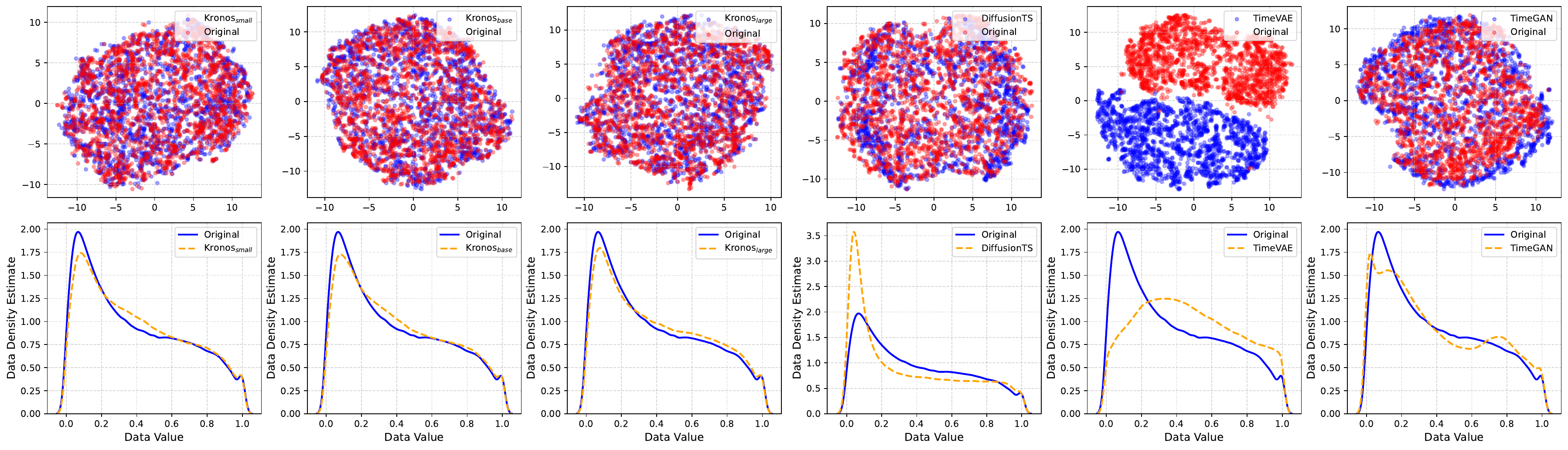}
    \caption{Visual comparison of generative models on the dataset of Shanghai Stock Exchange, 15-minute frequency. 
    \textbf{Top row:} t-SNE embeddings of original (\textcolor{red}{red}) versus synthetic (\textcolor{blue}{blue}) data. 
    \textbf{Bottom row:} Kernel Density Estimates (KDE) of original versus synthetic data.}
    \label{fig:gen_diversity_XSHG_15min}
\end{figure*}

To comprehensively evaluate the capabilities of Kronos as a foundation model for financial K-line data, we design a suite of experiments spanning 5 representative tasks. These tasks are selected to evaluate Kronos's performance in both predictive and generative applications, thereby demonstrating its versatility in practical quantitative finance scenarios.

\subsection{Experimental Setup}

The experimental tasks span predictive applications (price series, return and realized volatility forecasting), generative capabilities (synthetic K-line generation), and an investment simulation to gauge real-world applicability.

For a rigorous comparison, we benchmark Kronos against a comprehensive suite of 25 baseline models. These baselines are carefully selected to represent the state-of-the-art across four distinct paradigms: non-pre-trained full-shot models (e.g., iTransformer~\cite{liuitransformer}), zero-shot time series foundation models (e.g., TimeMOE~\cite{xiaoming2025time}), econometric volatility models (e.g., GARCH~\cite{bollerslev1986generalized}, classical approaches for volatility prediction from econometrics), and generative time series models (e.g., DiffusionTS~\cite{yuan2024diffusion}).
Task details and baselines are in Appendix~\ref{sec:exp_setup_details}. An overview of our main experimental results is presented in Figure~\ref{fig:main_result}, with a complete results breakdown in Appendix~\ref{sec:full_exp_results}.

\subsection{Main Results}

\subsubsection{Prediction Tasks}
Figure~\ref{fig:main_result}(a-c) presents the results for the three forecasting tasks. Kronos achieves consistent state-of-the-art performance across all of them. In particular, for price series forecasting, Kronos achieves a remarkable 93\% improvement in RankIC compared to the strongest TSFM baseline, and an 87\% gain over the best non-pre-trained model. Furthermore, as the model size scales up, performance on these tasks consistently improves, empirically validating the scaling laws for time series foundation models~\cite{yao2024towards}.

\subsubsection{Generative Tasks}

Following established practices~\cite{yoon2019time}, we evaluate the quality of synthetic data from three perspectives: \textit{diversity}, \textit{fidelity}, and \textit{usefulness}.
To assess \textit{diversity}—how well generated samples cover the real data's distribution—we use two visual methods: projecting original and synthetic data into a 2D space using t-SNE, and comparing their distributions via kernel density estimation (KDE). As shown in Figure~\ref{fig:gen_diversity_XSHG_15min} and Appendix~\ref{sec:full_exp_results}, the t-SNE plots show that Kronos's synthetic data better overlaps the original data space, and the KDE plots confirm a higher similarity in distributions.

For quantitative evaluation, we assess \textit{fidelity} (i.e., data realism) using the discriminative score, which measures how difficult it is for a classifier to distinguish between original and synthetic samples. We also evaluate \textit{usefulness} (the synthetic data's effectiveness for training downstream models) via the Train-on-Synthetic, Test-on-Real (TSTR) protocol, where a forecasting model is trained on synthetic data and its resulting IC and RankIC are evaluated on a test set composed of real data. As shown in Figure~\ref{fig:main_result}(d), Kronos achieves the best performance in both \textit{fidelity} and \textit{usefulness}. This superiority is also enhanced as the model size scales.

\subsubsection{Investment Simulation}
To validate Kronos's performance in a realistic investment scenario, we simulate a long-only investment strategy on the Chinese A-shares market by constructing portfolios with the top-$k$ stocks ranked by each model's predictive signals. 
As shown in Figure~\ref{fig:main_result}(e), Kronos outperforms all other baselines, achieving the highest Annualized Excess Return (AER) and Information Ratio (IR). This demonstrates that the model can effectively translate its superior predictive accuracy into tangible investment gains.

\subsection{Ablation Study}

We conduct ablation studies to validate our core design choices, focusing on two questions: (Q1) the effectiveness of our modeling paradigm compared to other alternatives, and (Q2) the impact of vocabulary size. An additional ablation on the tokenizer is provided in Appendix~\ref{sec:tokenizer_ablation}.

\noindent\textbf{Analysis of Modeling Paradigms.} 
To address Q1, we compare Kronos against variants that differ in their prediction spaces and objectives, while maintaining comparable parameter counts. (Table~\ref{tab:ablation_study_refined}).
Detailed descriptions of these architectural variants are provided in Appendix~\ref{sec:ablation_baselines}. We test two continuous-space models: \textit{Direct-AR} (a regression baseline with MSE) and \textit{Prob-AR}. Following established work~\cite{yao2024towards}, \textit{Prob-AR} uses a Student-t mixture distribution to better model heavy-tailed data distributions. The results show that our discrete-space models markedly outperform these continuous alternatives.We also find that \textit{Kronos-Parallel}, a variant that predicts subtokens concurrently, performs worse than our sequential approach, demonstrating the importance of modeling subtoken dependencies.
These findings validate our discrete, sequential modeling framework as a more effective approach for this domain.

\begin{figure}[t!] 
    \centering 
    \includegraphics[width=1.0\columnwidth]{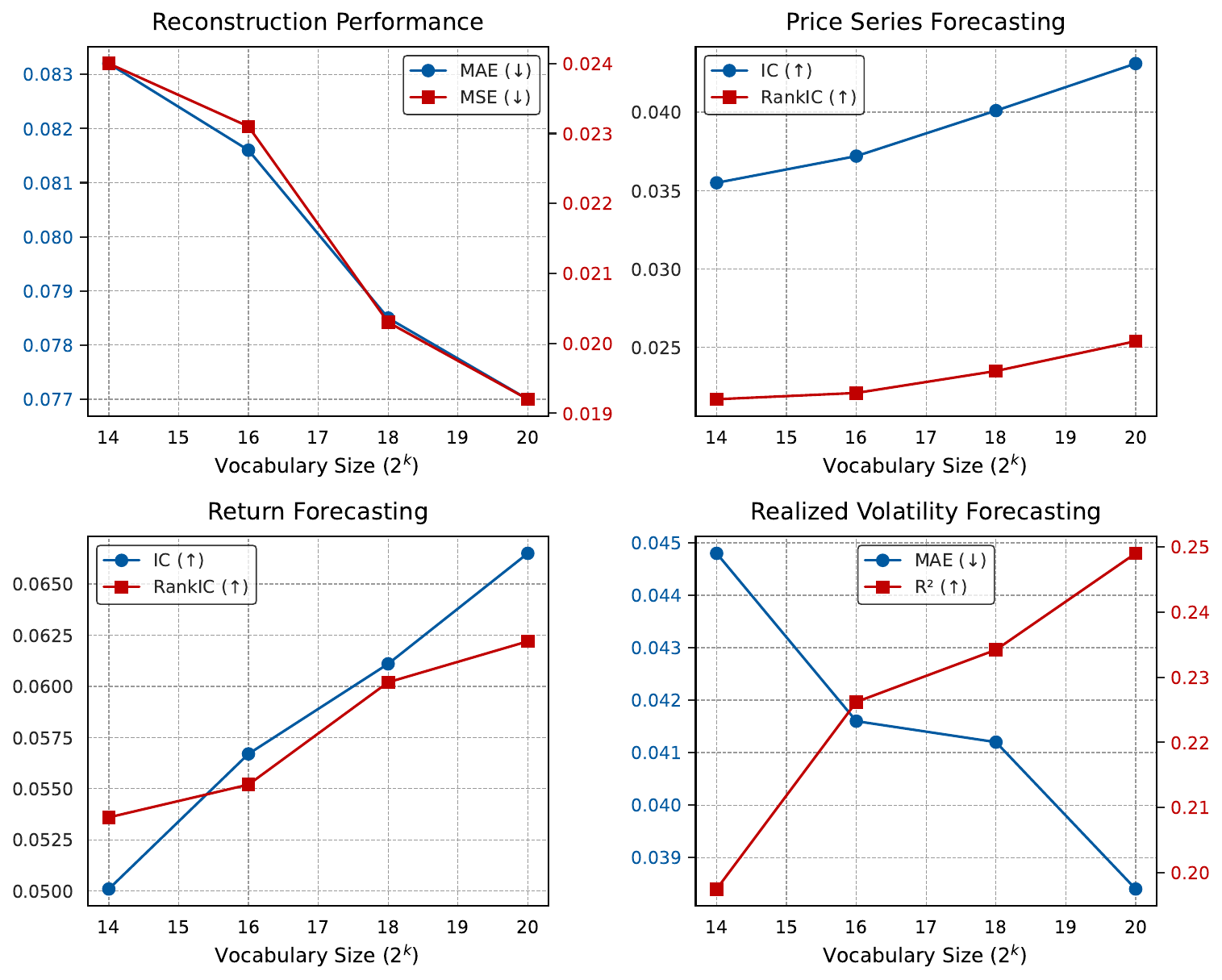} 
    \caption{Impact of vocabulary size on model performance. We plot reconstruction quality and downstream forecasting performance as vocabulary size increases.
    } 
    \label{fig:vocab_size} 
\end{figure}

\noindent\textbf{Impact of Vocabulary Size.} 
To answer Q2, we investigate how vocabulary size affects model performance. As shown in Figure~\ref{fig:vocab_size}, increasing the vocabulary size improves both reconstruction quality and forecasting accuracy. A larger vocabulary provides a finer-grained representation, reducing quantization error. Crucially, this enhanced representational precision translates to better predictive outcomes. This finding aligns with observations in video generation, where for quantization techniques like LFQ and BSQ, larger vocabularies have been shown to lead to improved generation quality~\cite{zhao2024image,yu2023language}.

\begin{figure}[t!]
    \centering
    \includegraphics[width=1.0\columnwidth]{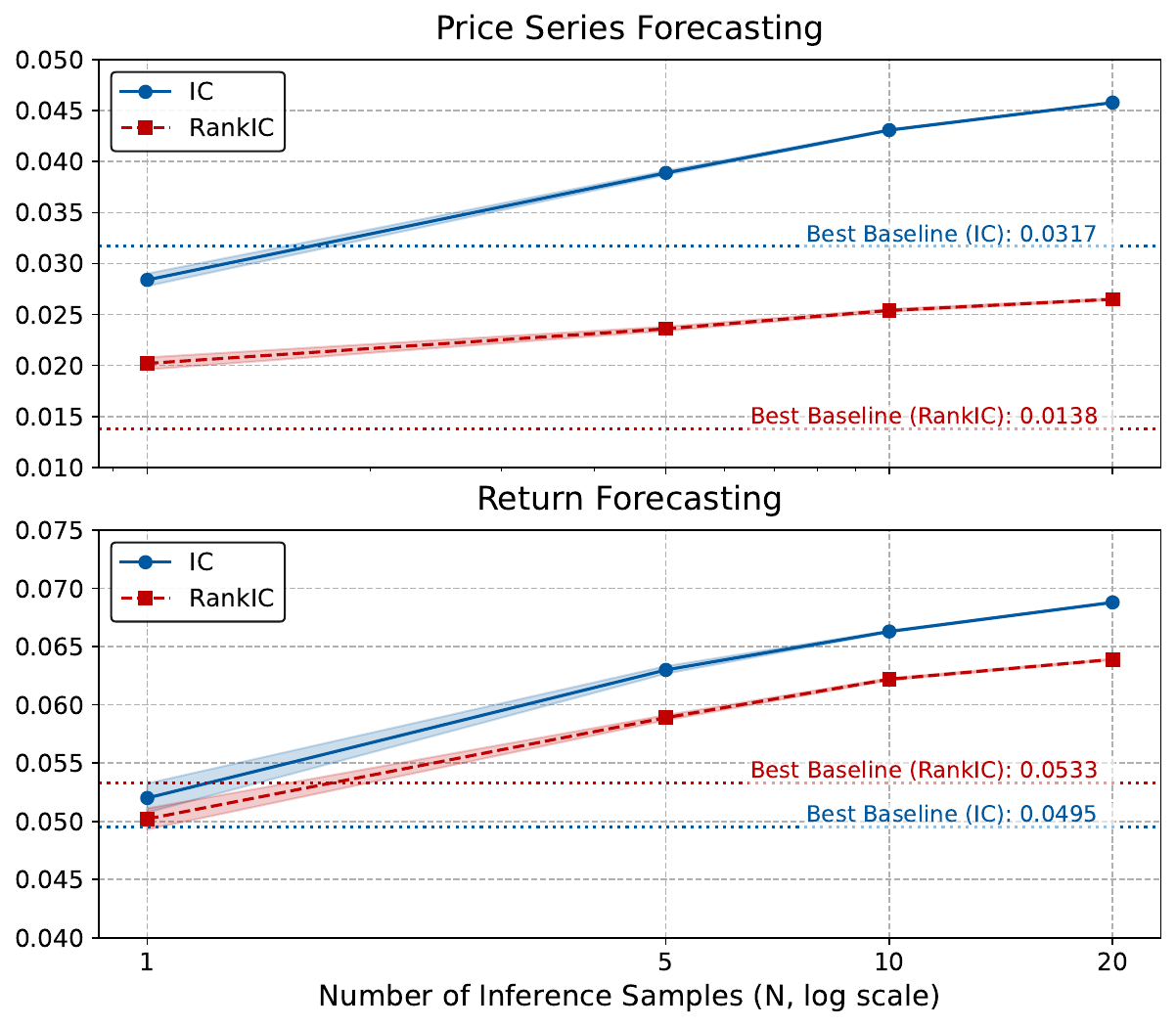}
    \caption{Impact of the number of inference samples (N) on forecasting performance.The lines represent the mean performance over 5 runs with different random seeds, while the shaded areas indicate the standard deviation.
    }
    \label{fig:inference_sample}
\end{figure}

\subsection{Test-Time Scaling} 
\label{sec:test_time_scaling}

A notable advantage of our probabilistic, generative framework is the ability to enhance predictive accuracy at inference time without retraining the model. By leveraging stochastic sampling, Kronos can generate multiple distinct future trajectories from the same context. We investigate the effect of ensembling these predictions by averaging the outcomes from an increasing number of sampled paths. Figure~\ref{fig:inference_sample} presents the performance on forecasting tasks as a function of the number of samples. The results demonstrate a consistent improvement in both IC and RankIC as more samples are included in the ensemble. Averaging across multiple paths mitigates the stochasticity inherent in the generation process and reduces prediction variance, yielding a more robust and stable estimate. This capability offers a trade-off, allowing practitioners to balance computational cost at inference with desired levels of predictive accuracy.

\section{Conclusion}

In this work, we introduce Kronos, a foundation model specifically designed for financial K-line sequences. Kronos employs a novel two-stage framework, where an instance-based tokenizer first discretizes continuous market data into hierarchical coarse-to-fine tokens, which are then modeled by a large autoregressive Transformer. Comprehensive empirical evaluations demonstrate that Kronos establishes new state-of-the-art benchmarks in 
price series forecasting, as well as in other relevant applications such as synthetic K-line generation and volatility forecasting, significantly outperforming existing TSFMs and other baselines. These results position Kronos as a robust and versatile foundation for a range of applications in quantitative finance.


\clearpage
\appendix

\setcounter{secnumdepth}{1}

\section*{Overview of Appendix}

This appendix provides supplementary materials to support the main paper. We detail our data preprocessing pipeline, model and training configurations, experimental setups for all tasks, and present additional results including hyperparameter sensitivity analyses, full result tables, and forecasting showcases.

\section{Related Work}

\subsection{Time Series Tokenization}

The recent success of large, token-based models has spurred a growing interest in discretizing continuous time series. This tokenization process is pivotal for adapting such architectures for time series analysis, yet dedicated research in this area remains sparse. Early efforts like Chronos~\cite{ansari2024chronos} employ scaling and uniform quantization, while TOTEM~\cite{talukder2024totem} utilizes a Vector Quantized Variational Autoencoder (VQ-VAE)~\cite{van2017neural}—a seminal approach that maps encoder outputs to learned discrete latent codes—for codebook-based tokenization. Given this nascent landscape, we draw inspiration from the more mature field of visual tokenization. Beyond the foundational VQ-VAE, innovations include Lookup-Free Quantization (LFQ)~\cite{yu2023language}, achieving high-fidelity reconstruction via an implicit codebook without explicit lookups. Binary Spherical Quantization (BSQ)~\cite{zhao2024image} advances implicit codebooks using spherical projection for an exponentially growing vocabulary, offering bounded quantization error and improved trainability over LFQ. Further, Index Backpropagation Quantization (IBQ)~\cite{shi2025scalableimagetokenizationindex} tackles codebook collapse by making all code entries differentiable, enabling stable joint optimization of large-scale codebooks and the visual encoder. While primarily designed for visual data, these methods can also be applied to discretize general multivariate time series.

\subsection{General-Purpose Time Series Foundation Models}

The paradigm of time series analysis has recently been reshaped by Time Series Foundation Models (TSFMs), drawing inspiration from the success of Large Language Models in leveraging massive pre-trained Transformers. These models are trained on vast, multi-domain corpora—some with over a hundred billion data points—to achieve remarkable zero-shot or few-shot performance on general forecasting benchmarks. This versatility is enabled by diverse architectures, including decoder-only models like Lag-Llama~\cite{rasul2023lag}, TimesFM~\cite{das2024decoder}, Timer~\cite{liu2024timer}, Time-MoE~\cite{xiaoming2025time}, and Sumdial~\cite{liu2025sundial}; encoder-only frameworks like MOMENT~\cite{goswami2024moment} and Moirai~\cite{woo2024unified}; encoder-decoder structures such as TimeGPT~\cite{garza2023timegpt}; and models with modified Transformer blocks for multi-task learning like UniTS~\cite{gao2024units}. At the input level, they employ generic representations such as direct value patching (e.g., TimesFM~\cite{das2024decoder}, MOMENT~\cite{goswami2024moment}), value quantization into a fixed vocabulary (e.g., Chronos~\cite{ansari2024chronos}), or treating consecutive time points as tokens (e.g., Timer~\cite{liu2024timer}). Several of these models also extend to probabilistic forecasting (e.g., Lag-Llama~\cite{rasul2023lag}, Moirai~\cite{woo2024unified}, Chronos~\cite{ansari2024chronos} and Sumdial~\cite{liu2025sundial}). 

However, the very generality that drives their success on broad benchmarks becomes a limitation in specialized domains. To provide a concrete comparison, we summarize key attributes of prominent TSFMs in Table~\ref{tab:tsfm_comparison}. A important observation from the table is the minuscule proportion of financial data within the pre-training corpora of these general-purpose models, with most dedicating less than 1\% of their data to this domain. This data imbalance means that the unique structural properties, non-stationarity, and complex dynamics of financial K-line sequences are largely overlooked or averaged out during pre-training, often resulting in suboptimal performance for financial tasks. To address this fundamental gap in pre-training, we introduce Kronos, a foundation model built from the ground up on a massive corpus composed exclusively of financial K-line data.

\subsection{Financial Time Series Foundation Models}

The development of foundation models specifically for finance time series is a nascent but rapidly growing field. These efforts can be divided into two main streams. The first focuses on general financial time series, including K-line data. For instance, PLUTUS~\cite{xu2024plutus} introduces an invertible embedding and multi-scale temporal attention, pre-trained on massive datasets to uncover market regularities. DELPHYNE~\cite{ding2025delphyne} is designed explicitly to counteract the negative transfer from non-financial data. While promising, neither of these works has released their code or models, precluding direct empirical comparison. The second stream targets order flow data, where models like MarketGPT~\cite{wheeler2024marketgpt} and MarS~\cite{li2024mars} act as generative engines for realistic market simulation. These pioneering efforts validate the value of domain-specific pre-training. However, K-line data possesses broader applicability than order flow, as it is readily available across all markets and suitable for diverse time horizons where order flow data is often inaccessible. Despite its central importance, a versatile and open-source foundation model for K-line analysis remains a notable gap. We introduce Kronos to fill this void, offering a unified, scalable framework designed specifically for financial K-line data.

\begin{table*}[t]
\centering

\resizebox{\textwidth}{!}{%
\begin{tabular}{@{}lllccl@{}}
\toprule
\textbf{Model} & \textbf{Architecture} & \textbf{Tokenization} & \textbf{Probabilistic} & \textbf{Financial Data Ratio (Est.)} & \textbf{Primary Domain} \\ \midrule
\textbf{Kronos (Ours)} & \textbf{Decoder-only} & \textbf{Discrete (BSQ)} & \textbf{Yes} & \textbf{100\%} & \textbf{Financial K-lines} \\
\midrule
Sundial~\cite{liu2025sundial} & Decoder-only & Continuous & Yes & 1.02\% & General \\
Time-MoE~\cite{xiaoming2025time} & Decoder-only & Continuous & No & $<$0.01\% & General \\
Moirai~\cite{woo2024unified} & Encoder-only & Continuous & Yes & 0.10\% & General \\
MOMENT~\cite{goswami2024moment} & Encoder-only & Continuous & No & 1.60\% & General \\
Chronos~\cite{ansari2024chronos} & Encoder-Decoder & Discrete (Quantization) & Yes & 0.45\% & General \\
Timer~\cite{liu2024timer} & Decoder-only & Continuous & No & 0.03\% & General \\
TimesFM~\cite{das2024decoder} & Decoder-only & Continuous & No & $<$0.01\% & General \\
UniTS~\cite{gao2024units} & Encoder-only & Continuous & No & Unknown & General \\
Lag-Llama~\cite{rasul2023lag} & Decoder-only & Continuous & Yes & 0.01\% & General \\
\bottomrule
\end{tabular}%
}
\caption{Comparison of time series foundation models. The table highlights architectural choices, tokenization methods, probabilistic forecasting capabilities, and the estimated proportion of financial data in their pre-training corpora.}
\label{tab:tsfm_comparison}
\end{table*}

\section{Dataset Details}
\label{sec:dataset_details}

\begin{table}[t]
\centering
\small

\setlength{\tabcolsep}{4pt} 
\resizebox{\columnwidth}{!}{%
\begin{tabular}{l c c cc}
\toprule
& & & \multicolumn{2}{c}{\textbf{Max. Consecutive Bars}} \\
\cmidrule(lr){4-5}
\textbf{Frequency} & \textbf{\begin{tabular}[c]{@{}c@{}}Min. Length \\ (bars)\end{tabular}} & \textbf{\begin{tabular}[c]{@{}c@{}}Price Jump \\ Threshold\end{tabular}} & \textbf{Illiquid} & \textbf{Stagnant} \\
\midrule
1min    & 2048 & 0.10 & 15 & 45 \\
5min    & 1024 & 0.15 & 3 & 10 \\
10min   & 512  & 0.15 & 3 & 6  \\
15min   & 512  & 0.15 & 2 & 5  \\
20min   & 512  & 0.15 & 2 & 5  \\
30min   & 512  & 0.20 & 2 & 3  \\
40min   & 256  & 0.20 & 1 & 3  \\
60min   & 256  & 0.20 & 1 & 3  \\
2H      & 128  & 0.25 & 1 & 3  \\
4H      & 128  & 0.25 & 1 & 3  \\
Daily   & 128  & 0.30 & 1 & 3  \\
Weekly  & 16   & 0.50 & 0 & 2  \\
\bottomrule
\end{tabular}}
\caption{Frequency-specific parameters for the low-quality data filtering pipeline. Thresholds are adjusted to reflect the distinct dynamics of different time frequencies.}
\label{tab:cleaning_parameters}
\end{table}

\subsection{Data Preprocessing and Cleaning}
\label{appendix:data_cleaning}

This section details the preprocessing and cleaning pipeline applied to the large-scale K-line dataset used for pre-training. The dataset is aggregated from over 40 exchanges across more than 30 countries, comprising a diverse range of asset classes at multiple temporal frequencies (1-minute to weekly). A statistical overview is provided in Table~\ref{tab:dataset_overview}. The integrity of large-scale pre-training is contingent upon high-quality input data. Raw K-line series, however, are frequently contaminated by artifacts stemming from low liquidity, price limits, or data feed errors. To mitigate the impact of such issues, we implement a rigorous, two-stage pipeline designed to process missing values and filter out low-quality data segments.

\begin{algorithm}[t]
\caption{Low-Quality Segment Filtering Pipeline}
\label{alg:segment_filtering}
\begin{algorithmic}[1]
\Statex \textbf{Input:} Raw K-line series $S_{raw}$, Parameter set $\Theta$ for a given frequency (from Table~\ref{tab:cleaning_parameters})
\Statex \textbf{Output:} A set of clean K-line segments $\mathcal{C}$

\Function{FilterLowQualitySegments}{$S_{raw}, \Theta$}
    \State $\mathcal{C} \leftarrow \emptyset$ \Comment{Initialize the set of final clean segments}
    \State $\mathcal{S}_{initial} \leftarrow \text{PartitionByPriceJumps}(S_{raw}, \Theta_{\text{jump}})$ \Comment{Split by structural breaks}
    
    \ForAll{segment $S$ in $\mathcal{S}_{initial}$}
        \State $M_{illiquid} \leftarrow \text{FlagConsecutiveIlliquid}(S, \Theta_{\text{illiquid}})$ \Comment{Identify illiquid periods}
        \State $M_{stagnant} \leftarrow \text{FlagConsecutiveStagnant}(S, \Theta_{\text{stagnant}})$ \Comment{Identify stagnant periods}
        \State $M_{invalid} \leftarrow M_{illiquid} \lor M_{stagnant}$ \Comment{Combine masks for all invalid points}
        
        \State $\mathcal{S}_{clean} \leftarrow \text{ExtractValidSubsequences}(S, M_{invalid})$ \Comment{Split segment on invalid boundaries}
        
        \ForAll{subsequence $S_{sub}$ in $\mathcal{S}_{clean}$}
            \If{$\text{Length}(S_{sub}) \ge \Theta_{\text{min\_len}}$}
                \State $\mathcal{C} \leftarrow \mathcal{C} \cup \{S_{sub}\}$ \Comment{Add valid, sufficiently long segment}
            \EndIf
        \EndFor
    \EndFor
    
    \State \textbf{return} $\mathcal{C}$
\EndFunction
\end{algorithmic}
\end{algorithm}

\subsubsection{Missing Value Processing}
We employ a field-specific strategy to handle missing values, which are typically represented as `NaN' (Not a Number) or `Inf' (Infinity).

\begin{itemize}
    \item \textbf{Price Fields (Open, High, Low, Close):} For price-related fields, we treat missing values as hard boundaries. Inspired by TimeMOE~\cite{xiaoming2025time}, we partition the time series into contiguous, valid sub-sequences at each occurrence of a missing price value. This approach ensures that each resulting segment maintains its internal temporal integrity without unwarranted imputation.
    \item \textbf{Volume and Amount Fields:} In contrast, for volume and amount fields, which primarily serve as auxiliary covariates, we impute missing values with zero. To enhance model robustness to sparse or unavailable volumetric data, we introduce a regularization technique: during training, both volume and amount are randomly set to zero for 5\% of the input samples. This encourages the model to learn to make effective predictions from price information alone.
\end{itemize}

\subsubsection{Low-Quality Segment Filtering}
Beyond addressing discrete missing values, our pipeline systematically identifies and removes entire segments of low-quality data. This is achieved through a multi-stage filtering process where tolerance thresholds are dynamically adjusted according to the data's temporal frequency, as detailed in Table~\ref{tab:cleaning_parameters}. The procedure, formalized in Algorithm~\ref{alg:segment_filtering}, consists of the following steps:

\begin{itemize}
    \item \textbf{Structural Break Segmentation.} The initial filtering stage partitions the series based on significant price discontinuities. We identify these breaks by calculating the relative price jump between the previous bar's close and the current bar's open ($| \text{open}_t / \text{close}_{t-1} - 1 |$). If this jump exceeds a frequency-specific threshold, the sequence is split. This step effectively isolates artifacts arising from events such as contract rollovers, stock splits, or dividend distributions.
    \item \textbf{Filtering of Illiquid Periods.} Within each segment from the previous step, we screen for periods of sustained illiquidity. A bar is deemed illiquid if its trading volume is zero or near-zero. If the number of consecutive illiquid bars exceeds a frequency-dependent threshold, the corresponding period is flagged as invalid.
    \item \textbf{Filtering of Price Stagnation.} We apply a similar method to filter periods of price stagnation, where the closing price remains constant over an extended duration. This often indicates potential data feed issues or market inactivity. If the length of a stagnant streak surpasses its frequency-specific tolerance, it is also flagged as an invalid period.
    \item \textbf{Final Segment Validation.} After flagging all illiquid and stagnant periods, the initial segments are further split at the boundaries of these flagged regions. Finally, only the resulting sub-segments that meet the frequency-specific minimum length requirement ($\Theta_{\text{min\_len}}$ in Table~\ref{tab:cleaning_parameters}) are retained for the final pre-training dataset. This ensures each sample is sufficiently long to support meaningful model learning.
\end{itemize}

\section{Implementation Details}
\label{sec:implementation_details}

In this section, we provide further details on the implementation of Kronos, covering data preprocessing, model architecture, and configurations for training and inference.

\subsection{Input Preprocessing}
\label{sec:input_processing}
Each input K-line sequence $\mathbf{x} = (\mathbf{x}_1, \mathbf{x}_2, \dots, \mathbf{x}_T)$, where $\mathbf{x}_t \in \mathbb{R}^D$, is normalized in a two-step procedure before being passed to the tokenizer. First, we apply z-score normalization independently to each of the $D$ feature dimensions (e.g., Open, High, Low, Close, Volume and Amount). Second, to mitigate the potential impact of extreme outliers on training stability, the normalized values are clipped to the range $[-5, 5]$. This process ensures that all input features have a consistent scale while preserving the model's robustness against anomalous data points.

\begin{table*}[t!]
  \centering
  
  \resizebox{0.9\textwidth}{!}{%
  \begin{tabular}{@{}lcccccc@{}}
    \toprule
    \textbf{Model} & \textbf{FFN Dropout} & \textbf{Residual Dropout} & \textbf{Attention Dropout} & \textbf{Token Dropout} & \textbf{Learning Rate} & \textbf{Weight Decay} \\
    \midrule
    $\text{Kronos}_{small}$ & 0.25 & 0.25 & 0.1 & 0.1 & \num{1e-3} & 0.01 \\
    $\text{Kronos}_{base}$  & 0.20 & 0.20 & 0.0 & 0.0 & \num{5e-4} & 0.05 \\
    $\text{Kronos}_{large}$ & 0.00 & 0.00 & 0.0 & 0.0 & \num{2e-4} & 0.10 \\
    \bottomrule
  \end{tabular}%
  }
  \caption{Hyperparameter configurations for the Kronos model series. All models are trained with the AdamW optimizer.}
  \label{tab:training_params}
\end{table*}

\subsection{Model Architecture}

\paragraph{Temporal Embeddings.} To capture cyclical patterns inherent in financial markets, such as intraday, weekly, and monthly seasonality~\cite{ozenbas2008intra,kohli1992week}, we incorporate learnable temporal embeddings. We extract five time-related features for each K-line entry: minute-of-day, hour-of-day, day-of-week, day-of-month, and month-of-year. Each feature is mapped to a dense vector via a dedicated embedding layer. These temporal embeddings are summed and then added to the input representation of each corresponding token, providing the model with explicit temporal context.

\paragraph{K-line Tokenization.} The tokenizer's autoencoder is designed to be lightweight. The encoder and decoder each consist of 3 Transformer layers, with a model dimension of 256, a feed-forward network dimension of 512, and 4 attention heads. Following the official open-source implementation of BSQ\footnote{\url{https://github.com/zhaoyue-zephyrus/bsq-vit}}, we configure the key quantization hyperparameters as follows: a commitment weight $\beta=0.05$, entropy penalty weights $\gamma_0=1.0$ and $\gamma=1.1$, and an overall entropy scale $\zeta=0.05$. The balancing hyperparameter $\lambda$ for the quantization loss in our objective is set to 1. The quantization group size is set to 5 for tractable entropy computation. 

\paragraph{Transformer Block Architecture.} To encode the sequential nature of the data, we employ causal self-attention with Rotary Position Embeddings (RoPE)~\cite{su2024roformer}, which injects relative positional information. The attention operation is formulated as follows:
\begin{equation}
    \text{Attention}(Q, K, V) = \text{CausalMask}\left(\frac{Q' (K')^T}{\sqrt{d_k}}\right)V
\end{equation}
where $d_k$ is the dimension of the key vectors, and $\text{CausalMask}$ prevents attending to future positions. The matrices $Q'$ and $K'$ represent the original query and key matrices with RoPE transformations applied. Furthermore, we adopt the Pre-Layer Normalization (Pre-LN)~\cite{xiong2020layer} to improve training stability, specifically utilizing Root Mean Square Layer Normalization (RMSNorm)~\cite{zhang2019root} for its computational efficiency and performance.

\subsection{Training Configuration}
The training hyperparameters are carefully selected for each model size to ensure a stable pre-training process. As model scale increases, we decrease the peak learning rate and dropout probability while increasing the weight decay. We employ the AdamW optimizer~\cite{loshchilov2017decoupled} and a cosine learning rate schedule with a linear warm-up phase. The learning rate warms up from 10\% of its peak value over the first 15,000 training steps. Table~\ref{tab:training_params} details the specific hyperparameter settings for each model variant.

\begin{table*}[t!]
\centering
\setlength{\tabcolsep}{4pt}

\begin{tabular}{lccc}
\toprule 
\textbf{Task} & \textbf{Temperature (T)} & \textbf{Top-p} & \textbf{Number of Inference Samples (N)} \\
\midrule 
Price Series Forecasting      & 0.6 & 0.90 & 10 \\
Return Forecasting            & 0.6 & 0.90 & 10 \\
Realized Volatility Forecasting & 0.9 & 0.90 & 1 \\
Synthetic K-line Generation      & 1.0 & 0.95 & 1 \\
Investment  Simulation        & 0.6 & 0.90 & 10 \\
\bottomrule 
\end{tabular}
\caption{Inference hyperparameters for downstream tasks. T denotes the temperature for sampling, Top-p controls nucleus sampling, and N is the number of inference samples generated for each test instance.}
\label{tab:inference_hyperparams}
\end{table*}

\subsection{Inference Hyperparameters}
The generation process at inference time is controlled by temperature scaling ($T$) and nucleus (top-$p$) sampling. The optimal choice of these hyperparameters is task-dependent. For example, forecasting tasks generally benefit from lower temperatures to reduce randomness, whereas generative tasks may require higher temperatures to increase diversity. A detailed analysis of hyperparameter sensitivity is available in Appendix~\ref{sec:sampling_hyperparams}. The inference hyperparameters used for each task are detailed in Table~\ref{tab:inference_hyperparams}.

\subsection{Pre-training Data Rebalancing}
The raw pre-training corpus exhibits a natural imbalance across asset classes, with equities being more prevalent than cryptocurrencies, futures, and foreign exchange (forex) assets. To prevent potential underfitting on these less-represented classes, we apply strategic resampling to the training data. Specifically, we increase the sampling weights for data from crypto, futures, and forex markets. This rebalancing ensures the model gains more balanced exposure to the diverse dynamics across different financial instruments.

\section{Experimental Design and Implementation}
\label{sec:exp_setup_details}

In this section, we present the comprehensive experimental design and implementation for the evaluation of Kronos. We begin by outlining the core evaluation tasks and their corresponding metrics. Next, we introduce the suite of baseline models used for comparison and detail their specific configurations. Finally, we provide a detailed account of the implementation for each experimental task, covering the datasets, parameters, and specific protocols used in our evaluation.

\subsection{Tasks and Evaluation Metrics}
We evaluate Kronos on a diverse set of tasks that are central to quantitative finance. The tasks and their respective evaluation metrics are as follows:
\begin{itemize}
    \item \textbf{Price Series Forecasting:} We assess the model's ability to predict future price series. Performance is measured by the Information Coefficient (IC) and Rank Information Coefficient (RankIC) between the predicted and actual values.
    \item \textbf{Return Forecasting:} Similarly, we evaluate the model's proficiency in forecasting asset returns, also using IC and RankIC as the metrics to gauge predictive accuracy.
    \item \textbf{Realized Volatility Forecasting:} We use the model's high-frequency forecasts to estimate realized volatility. The accuracy of these estimations is evaluated using Mean Absolute Error (MAE) and the Coefficient of Determination ($R^2$).
    \item \textbf{Synthetic K-line Generation:} Following established practices in time series generation~\cite{yoon2019time}, we assess the quality of synthetic K-line sequences from three perspectives: \textit{diversity}, assessing how well the generated samples cover the distribution of the real data; \textit{fidelity}, assessing whether synthetic samples are indistinguishable from real data; and \textit{usefulness}, evaluating if synthetic data is as effective as real data for downstream predictive tasks (i.e., the Train-on-Synthetic, Test-on-Real paradigm).
    \item \textbf{Investment Simulation:} To measure the practical applicability of the model's forecasts, we perform backtesting simulations. The performance is reported using Annualized Excess Return (AER) and Information Ratio (IR).
\end{itemize}

\subsection{Baselines and Configurations}
For a rigorous evaluation, we benchmark Kronos against a comprehensive suite of 25 baseline models. These models are selected from prior works (e.g.,~\cite{xiaoming2025time,wang2024deep,yuan2024diffusion}) to represent a diverse range of established and state-of-the-art approaches across different paradigms. They are organized into four distinct groups:
\begin{itemize}
    \item \textbf{Full-shot Time Series Models:} This category consists of modern, non-pre-trained time series models that are trained from scratch on the specific downstream task. It includes TimeXer~\cite{wang2024timexer}, TimesNet~\cite{wu2022timesnet}, TimeMixer~\cite{wang2024timemixer}, PatchTST~\cite{nie2022time}, Non-stationary Transformer (NSTransformer)~\cite{liu2022non}, DLinear~\cite{zeng2023transformers}, FEDformer~\cite{zhou2022fedformer}, and iTransformer~\cite{liuitransformer}.
    \item \textbf{Zero-shot Time Series Models:} This group comprises large-scale, pre-trained foundation models designed for general time series analysis. The baselines are TimeMOE~\cite{xiaoming2025time}, Moirai~\cite{woo2024unified}, TimesFM~\cite{das2024decoder}, Moment~\cite{goswami2024moment}, and Chronos~\cite{ansari2024chronos}, which we evaluate in a zero-shot setting.
    \item \textbf{Econometric Volatility Models:} For the volatility forecasting task, we include established econometric models as specialized baselines, namely ARCH~\cite{engle1982autoregressive} and GARCH~\cite{bollerslev1986generalized}.
    \item \textbf{Generative Time Series Models:} For the K-line generation task, we compare Kronos against models representing three mainstream generative architectures: DiffusionTS (diffusion-based)~\cite{yuan2024diffusion}, TimeVAE (VAE-based)~\cite{desai2021timevae}, and TimeGAN (GAN-based)~\cite{yoon2019time}.
\end{itemize}

\noindent\textbf{Full-shot Time Series Models.}
For all non-pre-trained deep learning models, we employ a composite loss function that combines Mean Squared Error (MSE) with an Information Coefficient (IC) term. We find this objective empirically improves predictive performance on financial tasks compared to using MSE alone, as it directly rewards the model for capturing the directional accuracy of price movements. The loss function is defined as:
\begin{equation}
\mathcal{L} = \frac{1}{M \cdot H} \sum_{i=1}^{M} \sum_{j=1}^{H} (y_{i,j} - \hat{y}_{i,j})^2 - \lambda \cdot \frac{1}{M} \sum_{i=1}^{M} \text{IC}(y_i, \hat{y}_i)
\end{equation}
where $y_i$ and $\hat{y}_i$ are the true and predicted sequences for the $i$-th feature, respectively, $M$ is the number of features, $H$ is the prediction horizon, and $\lambda$ is a balancing hyperparameter, set to 4 in our experiments.

All models are trained with a batch size of 256 and an Adam optimizer with a learning rate of $5 \times 10^{-4}$. We train for a maximum of 12 epochs, employing an early stopping mechanism with a patience of 3 epochs based on the validation loss. For each model, we test two sets of hyperparameters corresponding to smaller and larger model sizes to ensure a fair and robust comparison. The configuration that yields the best performance on the validation set is selected for final evaluation. For DLinear, instead of varying model dimensions, we evaluate two configurations based on its `individual' parameter: one where a single linear layer is shared across all variates (`individual=False') and another where a separate linear layer is trained for each variate (`individual=True'). The specific hyperparameter configurations are detailed in Table~\ref{tab:full_shot_hyperparams}.

\begin{table}[ht]
\centering

\resizebox{\columnwidth}{!}{%
\begin{tabular}{l cccc}
\toprule
\textbf{Model} & \textbf{Layers} & \textbf{$\mathbf{d}_{\text{model}}$} & \textbf{$\mathbf{d}_{\text{ff}}$} & \textbf{Heads} \\
\midrule
TimeXer       & 3 / 5 & 128 / 256 & 256 / 512 & 4 / 8 \\
TimesNet      & 3 / 5 & 128 / 256 & 256 / 512 & ---   \\
TimeMixer     & 3 / 5 & 128 / 256 & 256 / 512 & 4 / 8 \\
PatchTST      & 3 / 5 & 128 / 256 & 256 / 512 & 4 / 8 \\
NSTransformer & 2 / 3 & 128 / 256 & 256 / 512 & 4 / 8 \\
FEDformer     & 2 / 3 & 128 / 256 & 256 / 512 & 4 / 8 \\
iTransformer  & 3 / 5 & 128 / 256 & 256 / 512 & 4 / 8 \\
\bottomrule
\end{tabular}}
\caption{Hyperparameter configurations for the baseline models. Values for the two evaluated sets are separated by a slash (/). We detail the number of layers, model dimension ($\mathbf{d}_{\text{model}}$), feed-forward dimension ($\mathbf{d}_{\text{ff}}$), and the number of attention heads.}
\label{tab:full_shot_hyperparams}
\end{table}

\noindent\textbf{Econometric Volatility Models.}
For the specialized volatility forecasting baselines, we follow standard econometric practices for model selection.
\begin{itemize}
    \item \textbf{ARCH}: For each time series, we fit ARCH models with lag orders $p \in \{1, 2, 3\}$. The model with the lowest Bayesian Information Criterion (BIC) is selected for forecasting. The BIC penalizes model complexity, helping to prevent overfitting.
    \item \textbf{GARCH}: We perform a grid search over the lag orders for both the autoregressive term ($p$) and the moving average term ($q$), with $p, q \in \{1, 2, 3\}$. Similar to ARCH, the GARCH(p,q) model with the minimum BIC is chosen as the final model for that series.
\end{itemize}

\subsection{Task Implementation Details}
Below, we describe the specific setups for each of our evaluation tasks.

\paragraph{Forecasting Task Setup}

The pre-training data for Kronos extends up to June 2024. Consequently, our test period for all tasks begins in July 2024 to ensure a strict temporal separation between training and evaluation. We select a diverse set of assets and K-line frequencies to rigorously test model generalization.

\subparagraph{Assets} We evaluate on three major asset classes:
\begin{itemize}
    \item \textbf{Stocks:} To test both in-distribution and out-of-distribution generalization, we use data from nine global stock exchanges.
    \begin{itemize}
        \item \textit{In-distribution exchanges:} Shanghai (XSHG), NASDAQ (XNAS), Japan (XJPX), India (XNSE), Korea (XKRX), and Hong Kong (XHKG).
        \item \textit{Out-of-distribution exchanges:} Indonesia (XIDX), Malaysia (XKLS), and Taiwan (XTAI).
    \end{itemize}
    \item \textbf{Cryptocurrency:} All spot trading pairs available on the Binance exchange.
    \item \textbf{Forex:} A comprehensive dataset of over 1,000 foreign exchange pairs.
\end{itemize}
For cryptocurrency and forex assets, we intentionally exclude volume and amount fields, providing only the OHLC price series. This setup tests the models' ability to make predictions based solely on price dynamics, a common scenario where reliable volume data is unavailable.

\subparagraph{Frequencies and Horizons} We test on a range of K-line frequencies, again including both in-distribution and out-of-distribution settings. For each frequency, we define look-back and forecast horizons that are relevant to practical applications in quantitative finance. These settings are detailed in Table~\ref{tab:forecast_horizons}.

\begin{table}[th]
\centering

\resizebox{\columnwidth}{!}{%
\begin{tabular*}{\columnwidth}{@{\extracolsep{\fill}} ccc}
\toprule
\textbf{Frequency} & \textbf{Look-back Window} & \textbf{Forecast Horizon} \\
\midrule
5min    & 480 & 96 \\
10min   & 240 & 48 \\
15min   & 160 & 32 \\
20min   & 120 & 24 \\
40min   &  90 & 24 \\
\midrule
1-hour & 80 & 12 \\
2-hour  &  60 & 12 \\
4-hour  &  90 & 18 \\
Daily   &  40 & 12 \\
\bottomrule
\end{tabular*}}
\caption{Look-back and forecast horizon settings for each K-line frequency in the forecasting tasks.}
\label{tab:forecast_horizons}
\end{table}

\subparagraph{Metric Calculation Details}
\begin{itemize}
    \item \textbf{Price Series Forecasting:} For each sample, the IC and RankIC are calculated between the predicted and true series for each of the four price channels (Open, High, Low, Close). The final reported metrics are the average across these four channels.
    \item \textbf{Return Forecasting:} We define the predicted return $\hat{r}$ based on the last value of the predicted close price sequence $\hat{p}_{t+H}$ and the last value of the historical close price sequence $p_{t}$:
    \begin{equation}
    \hat{r} = \frac{\hat{p}_{t+H}}{p_t} - 1
    \end{equation}
    The IC and RankIC are then computed between the vector of predicted returns and the vector of actual returns for all samples within a given asset class and frequency.
    \item \textbf{Realized Volatility Forecasting:} We estimate the realized volatility from a high-frequency price series. Using the model's predicted closing prices $\{ \hat{p}_i \}_{i=1}^H$ over the forecast horizon, the realized volatility is calculated as the sum of squared log returns:
    \begin{equation}
    \hat{\sigma}^2 = \sum_{i=1}^{H-1} \left( \log(\hat{p}_{i+1}) - \log(\hat{p}_i) \right)^2
    \end{equation}
    We then compute the Mean Absolute Error (MAE) and Coefficient of Determination ($R^2$) between the predicted and actual realized volatilities across all samples.
\end{itemize}

\paragraph{Synthetic K-line Generation Setup}

\begin{figure*}[t!]
    \centering
    \includegraphics[width=1.0\textwidth]{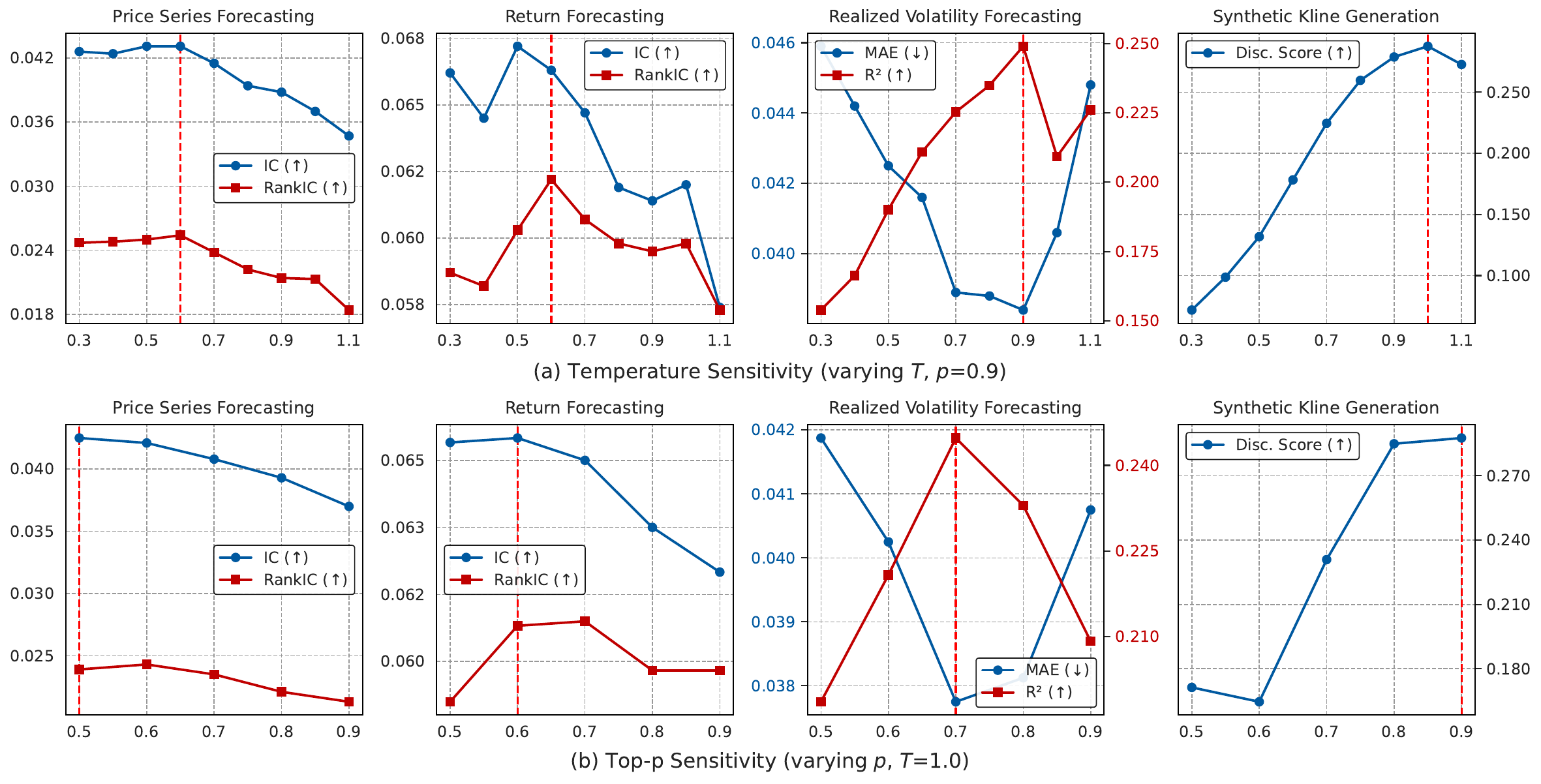}
    \caption{Sensitivity analysis of Kronos's performance on downstream tasks with respect to inference sampling hyperparameters. (a) Varying temperature $T$ while keeping top-$p=0.9$ fixed. (b) Varying top-$p$ while keeping temperature $T=1.0$ fixed. Optimal values, indicated by red dashed lines, are task-dependent, highlighting different requirements for precision versus diversity.}
    \label{fig:T_and_top_p}
\end{figure*}

\subparagraph{Datasets and Generation Parameters} We use data from two stock exchanges (in-distribution XSHG and out-of-distribution XTAI), as well as the cryptocurrency and forex datasets. We evaluate generation on two frequencies: 15-minute and daily. For the 15-minute frequency, we use a look-back window of 120 and generate a future sequence of length 96. For the daily frequency, the look-back is 96 and the generation horizon is 35. For each asset-frequency pair, we generate 6,000 synthetic sequences for evaluation.

\subparagraph{Evaluation Metrics}
\begin{itemize}
    \item \textbf{Discriminative Score:} To assess the fidelity of the generated data, we employ a post-hoc LSTM-based classifier to distinguish between real and synthetic sequences. The classifier consists of a single LSTM layer with a hidden dimension of 32. For training, we construct a balanced dataset of 6,000 samples (3,000 real, 3,000 synthetic) and a held-out test set of the same size and composition. The model is trained for 20 epochs with a batch size of 64, using the Adam optimizer (learning rate = 0.0005) and the binary cross-entropy (BCE) loss function. The Discriminative Score is defined as the classification error on the test set. A score approaching 0.5 indicates higher fidelity, signifying that the classifier struggles to differentiate generated data from real data.

    \item \textbf{Usefulness (TSTR):} To measure the practical usefulness of the synthetic data, we adopt the Train-on-Synthetic, Test-on-Real (TSTR) methodology. We train a post-hoc LSTM prediction model to forecast a future K-timestep window given a historical one. This model comprises two LSTM layers with a hidden dimension of 64. It is trained exclusively on 6,000 generated synthetic sequences for 20 epochs using the Adam optimizer (learning rate = 0.001) and a batch size of 64, with the Mean Squared Error (MSE) loss as the objective function. The look-back and horizon windows are set to (80, 16) for 15-minute data and (30, 5) for daily data, respectively. The trained model is then evaluated on the original, real test data. The final usefulness score is reported as the average Information Coefficient (IC) and Rank Information Coefficient (RankIC) of the predicted price series.
\end{itemize}

\paragraph{Investment Simulation Setup}

To evaluate the practical profitability of Kronos and other baselines in real-world markets, we conduct an investment simulation on the Chinese A-share market. For simplicity, regarding the Zero-shot Time Series Models, we only select the largest-sized model from each family for comparison.

\subparagraph{Data} Our empirical analysis utilizes daily market data for the Chinese A-share market, sourced from the Qlib platform~\cite{yang2020qlib}, an open-source framework for quantitative finance. To promote transparency and reproducibility, we apply no additional filtering or preprocessing to the data, using it in its original, unprocessed state. Furthermore, we conduct all backtesting simulations within the Qlib framework. This approach leverages its integrated backtesting engine to ensure a standardized and consistent evaluation protocol for all models under review.

\subparagraph{Strategy} We employ the top-$k$/drop-$n$ portfolio construction strategy. On each trading day, all stocks in the investment universe are ranked based on their predicted return signal. An equal-weight portfolio is formed by taking long positions in the top $k$ stocks. To manage turnover and trading costs, a maximum of $n$ stocks are bought or sold daily, and a minimum holding period of 5 days is enforced for all positions.

\subparagraph{Signal and Backtest}

The predictive signal is formulated as an expected return derived from a multi-step price forecast over a horizon of $H$ days. This signal generation pipeline is applied uniformly to all models under evaluation, including Kronos and the baselines, to ensure a fair comparison. For any given stock on trading day $t$, a sequence of forecasted closing prices for the subsequent $H$ days, denoted as $\{\hat{p}_{t+i}\}_{i=1}^{H}$, is first generated by the respective model. The signal, which we term the $H$-day average expected return ($R_{t \to t+H}$), is then calculated by comparing the arithmetic mean of these forecasted prices to the current closing price $p_t$:
\begin{equation}
R_{t \to t+H} = \frac{\left(\frac{1}{H}\sum_{i=1}^{H} \hat{p}_{t+i}\right) - p_t}{p_t}
\end{equation}
In our experiments, we set the forecast horizon to $H=10$. All price forecasts are generated using daily K-line data with a 90-day look-back window. This methodology is designed to produce a robust signal by averaging the forecasted price path, thereby mitigating the influence of short-term prediction noise and capturing the underlying trend more effectively.

Backtests are performed on the constituents of the CSI 300 and CSI 800 indices. These indices are chosen as they represent two key segments of the Chinese A-share market: the CSI 300 comprises large-cap, highly liquid stocks, while the CSI 800 provides broader market coverage by including both large- and mid-cap stocks. This allows for a comprehensive assessment of the model's performance across different market segments.

\subparagraph{Parameters and Costs} For the CSI 300 index, we set $k=50$ and $n=5$. For the broader CSI 800 index, we set $k=200$ and $n=10$. The relatively large portfolio sizes are chosen to ensure diversification and produce more stable backtesting results, reducing the influence of idiosyncratic stock movements. To ensure a realistic performance assessment, a conservative transaction cost of 0.15\% is applied to each trade.

\begin{figure*}[t!]
    \centering
    \begin{minipage}[t]{1.0\columnwidth}
        \centering
        \includegraphics[width=\columnwidth]{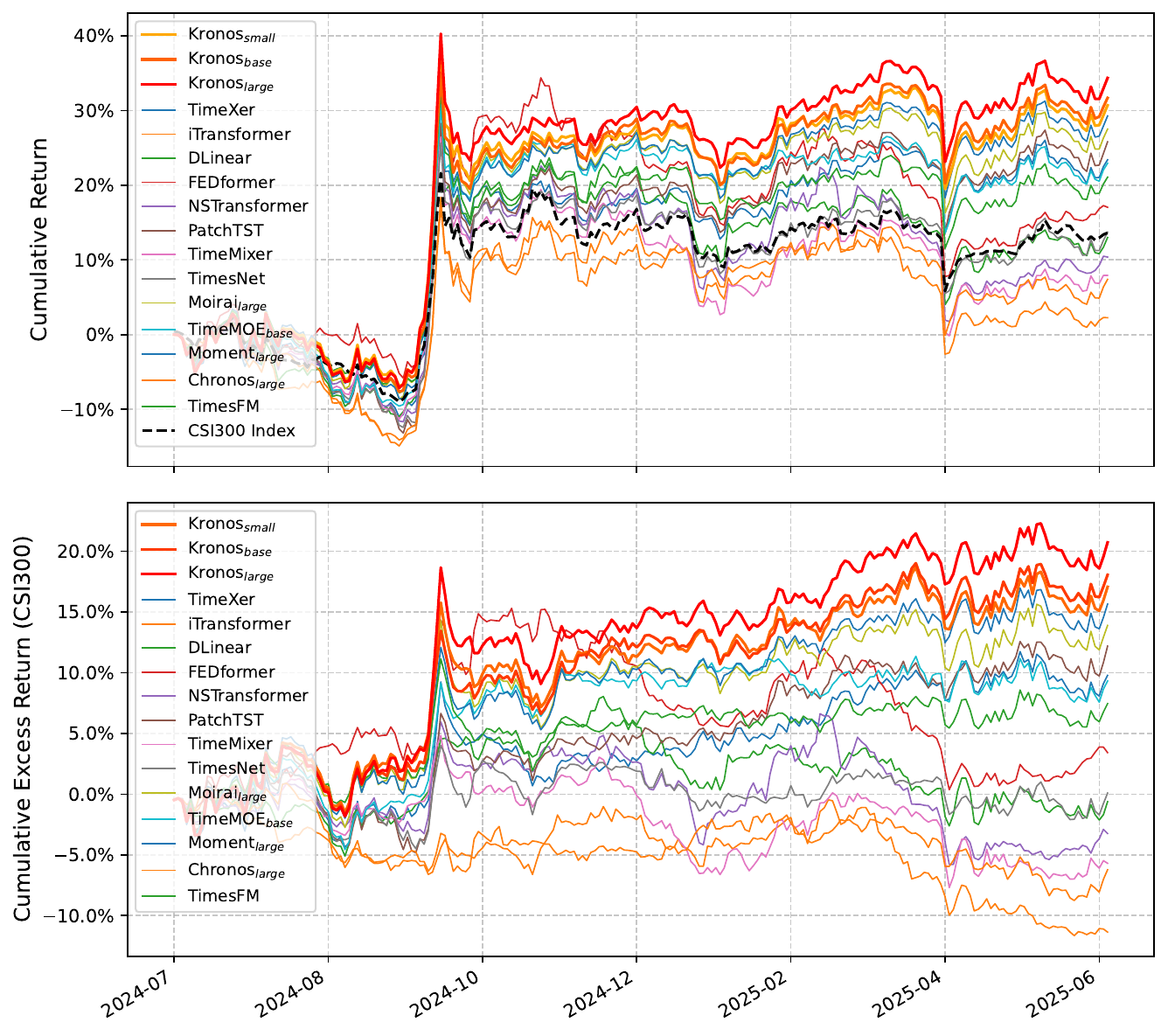}
        \caption*{ (a) CSI300 Index}
    \end{minipage}
    \hspace{0.02\columnwidth}
    \begin{minipage}[t]{1.0\columnwidth}
        \centering
        \includegraphics[width=\columnwidth]{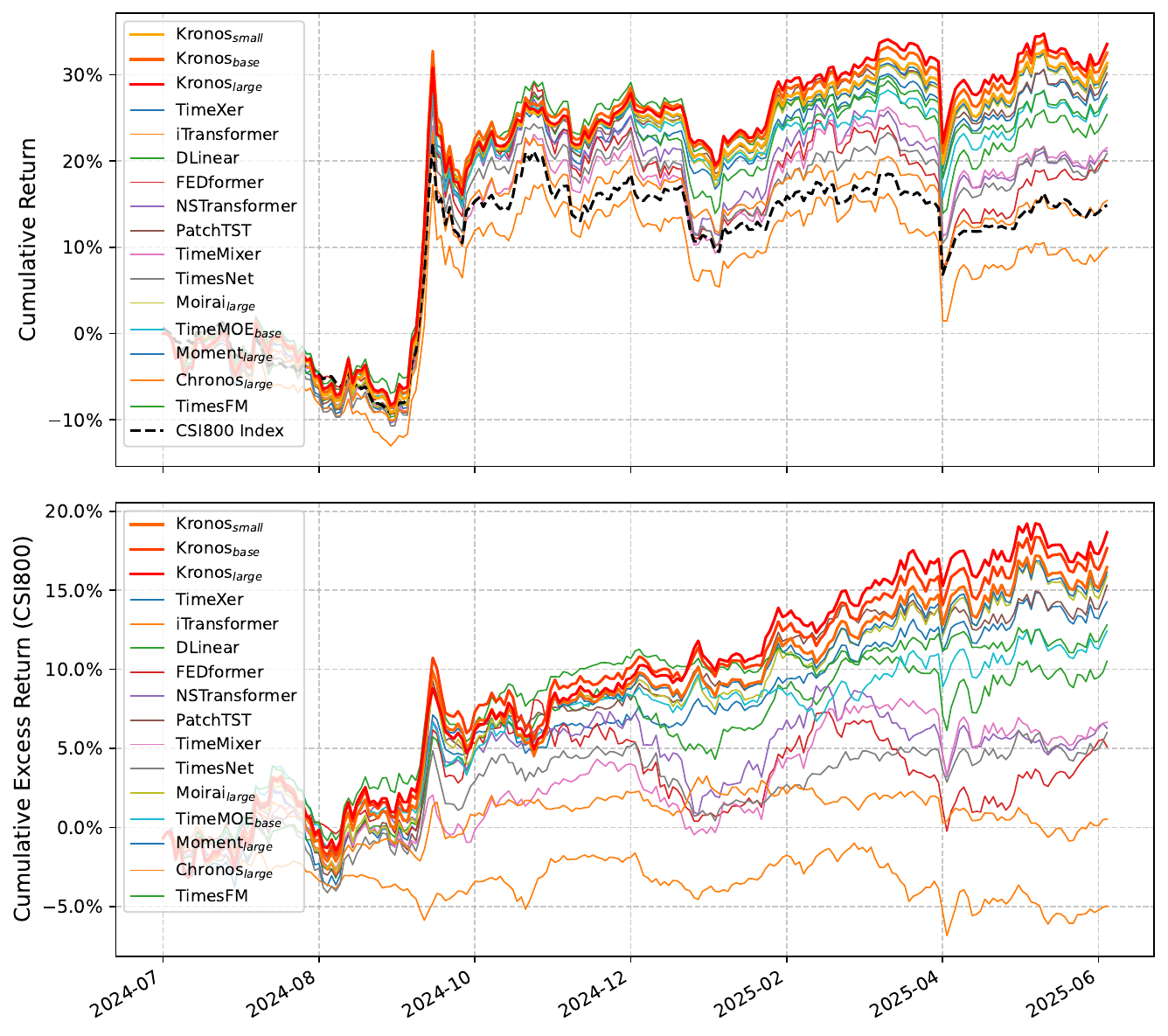}
        \caption*{ (b) CSI800 Index}
    \end{minipage}
    \caption{Cumulative return curves of backtest using signals generated by different models.}
    \label{fig:backtest_profit}
\end{figure*}

\subsection{Details of Ablation Study Baselines}
\label{sec:ablation_baselines}

To investigate the architectural choices of Kronos, we design three baseline variants for our ablation study (Table~\ref{tab:ablation_study_refined}). Each variant targets a different modeling paradigm, allowing us to isolate the benefits of our proposed discrete, sequential framework. Below we provide a detailed description of each model.

\noindent\textbf{Direct-AR.} This model serves as a standard autoregressive forecasting baseline in the continuous space. Given a sequence of input features $\{x_1, \dots, x_T\}$, each feature vector $x_t \in \mathbb{R}^D$ is first mapped to a higher-dimensional embedding via a linear projection. The sequence of embeddings is then processed by a Transformer decoder backbone. The model is trained to directly predict the value of the next time step, $\hat{x}_{T+1}$, from the historical context. The training objective is to minimize the Mean Squared Error (MSE) between the predicted and ground-truth values. This approach represents the most common regression-based formulation for time series forecasting.

\noindent\textbf{Prob-AR.} This is a probabilistic forecasting model operating in the continuous space. Following established practices~\cite{yao2024towards}, instead of a point estimate, Prob-AR predicts the parameters of a probability distribution for the next time step. We use a mixture of four Student-t distributions to model the predicted distribution. The probability density function (PDF) for a random variable $x$ following a single Student-t distribution is:
\begin{equation}
p(x | \nu, \mu, \sigma) = \frac{\Gamma(\frac{\nu+1}{2})}{\Gamma(\frac{\nu}{2})\sqrt{\pi\nu}\sigma} \left(1 + \frac{1}{\nu}\left(\frac{x-\mu}{\sigma}\right)^2\right)^{-\frac{\nu+1}{2}}
\end{equation}
where $\nu > 0$, $\mu \in \mathbb{R}$, and $\sigma > 0$ are the degrees of freedom, location, and scale parameters, respectively, and $\Gamma(\cdot)$ is the gamma function. The model employs independent linear layers to predict the parameters for each of the four components—degrees of freedom ($\nu_k$), location ($\mu_k$), scale ($\sigma_k$), and mixture weights ($w_k$). To ensure parameter validity, a softplus transformation is applied to $\nu_k$ and $\sigma_k$ to enforce positivity, and a softmax function is applied to the weights $w_k$ to ensure they form a valid probability distribution. The model is trained by minimizing the Negative Log-Likelihood (NLL) of the true value under the predicted mixture distribution.

\noindent\textbf{Kronos-Parallel.} This variant is a direct ablation of the sequential subtoken generation mechanism within Kronos. While it shares the same input quantization and discrete prediction space as Kronos, it removes the intra-block module. After the Transformer backbone produces a context vector from the input history, a single prediction head is used to concurrently predict the logits for both subtokens of the next time step. The training objective is the sum of the cross-entropy losses for each subtoken, optimized jointly.

\subsection{Experimental Environment}
\label{sec:exp_environment}

All experiments are conducted within a Kubernetes (k8s) cluster. For all computational tasks, we utilize three identical pods. Each pod is provisioned with a dedicated set of resources comprising 96 CPU cores (Intel Xeon Gold 6330 @ 2.00 GHz), 200 GB of system memory (RAM), and eight NVIDIA GeForce RTX 4090D GPUs. This configuration provides a total of 24 GPUs, which are collectively employed for model training and all subsequent evaluations.

The software environment is containerized and standardized across all pods. The primary components and their versions are detailed below:
\begin{itemize}
    \item \textbf{Operating System:} Ubuntu 24.04.1 LTS
    \item \textbf{Software versions:} Python 3.13.2, PyTorch 2.7.0, NumPy 1.26.2, Pandas 2.2.2, Matplotlib 3.9.3, Hugging Face Hub (`huggingface\_hub') 1.57.4
\end{itemize}

\section{Additional Results}

\subsection{Impact of Inference Sampling Hyperparameters}
\label{sec:sampling_hyperparams}

The autoregressive generation process of Kronos is governed by sampling strategies that introduce controlled stochasticity, namely temperature scaling ($T$) and top-$p$ (nucleus) sampling. The choice of these hyperparameters can significantly influence model performance on different downstream tasks. To provide guidance on their optimal settings, we conduct a sensitivity analysis. Figure~\ref{fig:T_and_top_p} illustrates the performance of Kronos across our four main tasks while varying one hyperparameter and holding the other constant.

As shown in Figure~\ref{fig:T_and_top_p}, the optimal sampling hyperparameters are task-dependent. For forecasting tasks (price series and return), which demand precision, lower temperatures (e.g., $T \approx 0.6$) are preferable. This sharpens the next-token distribution, compelling the model towards more deterministic and high-confidence predictions. Conversely, realized volatility forecasting and synthetic K-line generation benefit from greater stochasticity, achieving optimal performance at temperatures closer to 1.0. A higher temperature encourages the generation of more diverse sequences, which is essential for capturing the probabilistic nature of volatility and for producing realistic, non-repetitive market data.

The analysis of top-$p$ sampling reveals a similar pattern: forecasting tasks favor smaller $p$ values to restrict the sampling pool, whereas generative tasks perform better with a larger nucleus ($p \ge 0.9$) to preserve diversity. When comparing the two techniques, we observe that temperature scaling generally offers more effective and nuanced control, leading to slightly better peak performance across tasks. This suggests that the global probability rescaling of temperature may be a more suitable tuning mechanism than the hard truncation of nucleus sampling.

\subsection{Ablation on Tokenizer Architecture}
\label{sec:tokenizer_ablation}
We perform an ablation study on the tokenizer architecture to justify our design choices. We compare our proposed Transformer-based tokenizer using a hierarchical loss against two alternatives: (1) a Transformer-based tokenizer with a standard, non-hierarchical reconstruction loss and (2) a CNN-based architecture with a comparable parameter count. All models are trained with a vocabulary size of $2^{18}$.

\begin{table}[ht]
\centering

\resizebox{\columnwidth}{!}{%
\begin{tabular}{@{}l S[table-format=1.4] S[table-format=1.4]@{}}
\toprule
\textbf{Tokenizer Architecture} & {\textbf{MAE} ($\downarrow$)} & {\textbf{MSE} ($\downarrow$)} \\
\midrule
\textbf{Transformer w/ Hierarchical Loss (Ours)} & 0.0785 & 0.0203 \\
Transformer w/ Standard Loss                   & 0.0781 & 0.0202 \\
\midrule
CNN-based                                      & 0.0916 & 0.0251 \\
\bottomrule
\end{tabular}%
}
\caption{
    Ablation study on the K-line tokenizer architecture. We compare our proposed Transformer-based tokenizer, which employs a hierarchical reconstruction loss, against two key variants: a Transformer-based tokenizer with a standard reconstruction loss and a CNN-based architecture. All models are trained with a vocabulary size of $2^{18}$. The table reports reconstruction quality measured by MAE and MSE.
}
\label{tab:tokenizer_ablation}
\end{table}

As shown in Table~\ref{tab:tokenizer_ablation}, the results indicate that Transformer-based architectures outperform the CNN-based model in reconstruction quality, highlighting the effectiveness of self-attention for capturing dependencies in K-line data. More importantly, our model with hierarchical loss achieves reconstruction quality nearly identical to that of the standard loss variant. This confirms that our approach successfully engineers a coarse-to-fine structure within the tokens—a property beneficial for the subsequent autoregressive model—without a notable trade-off in representational fidelity.

\subsection{K-line Reconstruction Visualizations}

\begin{figure*}[ht]
    \centering

    \begin{subfigure}[b]{0.49\textwidth}
        \centering
        \includegraphics[width=\linewidth]{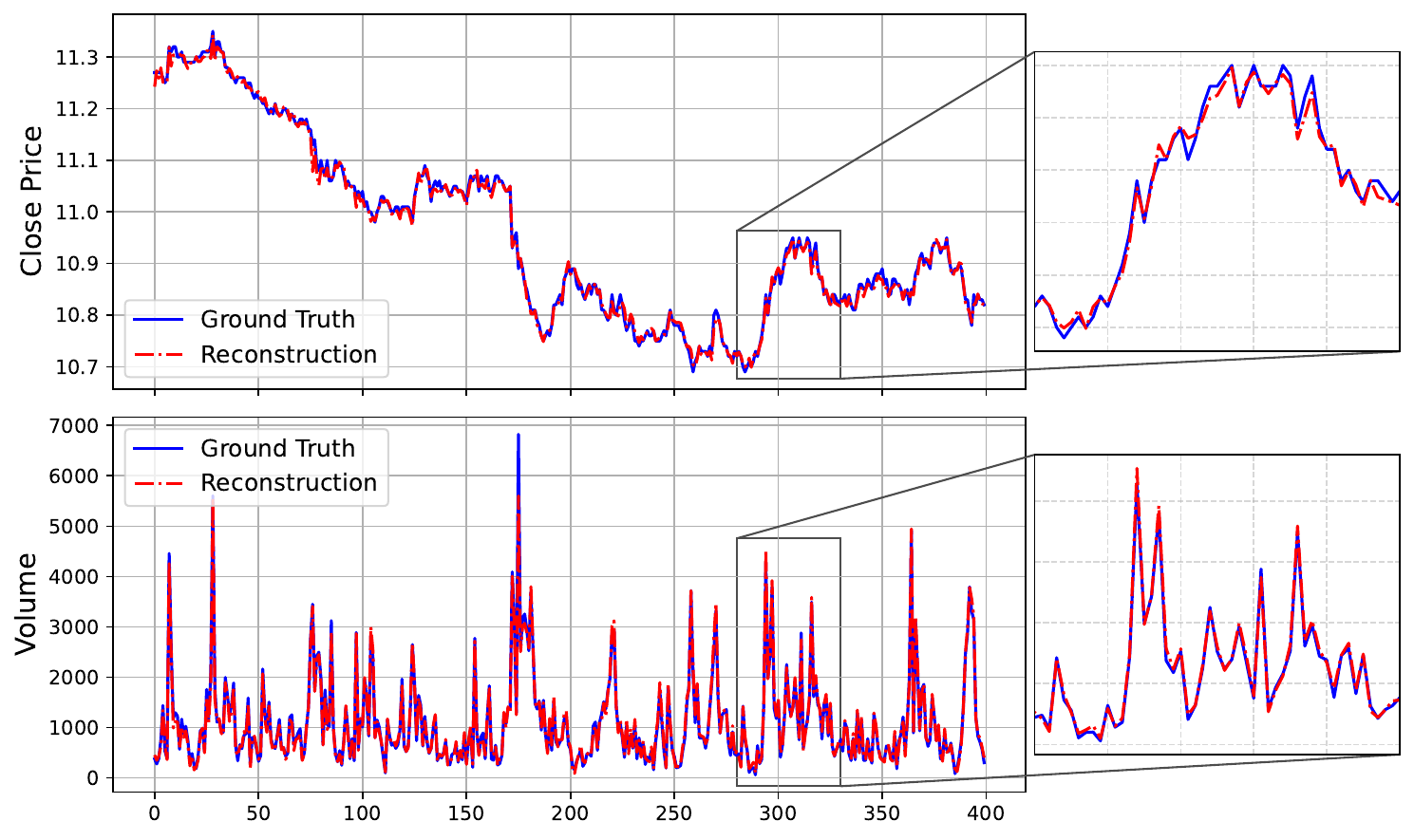}
        \caption{China Film Co.,Ltd. (SSE: 600977)}
    \end{subfigure}
    \hfill
    \begin{subfigure}[b]{0.49\textwidth}
        \centering
        \includegraphics[width=\linewidth]{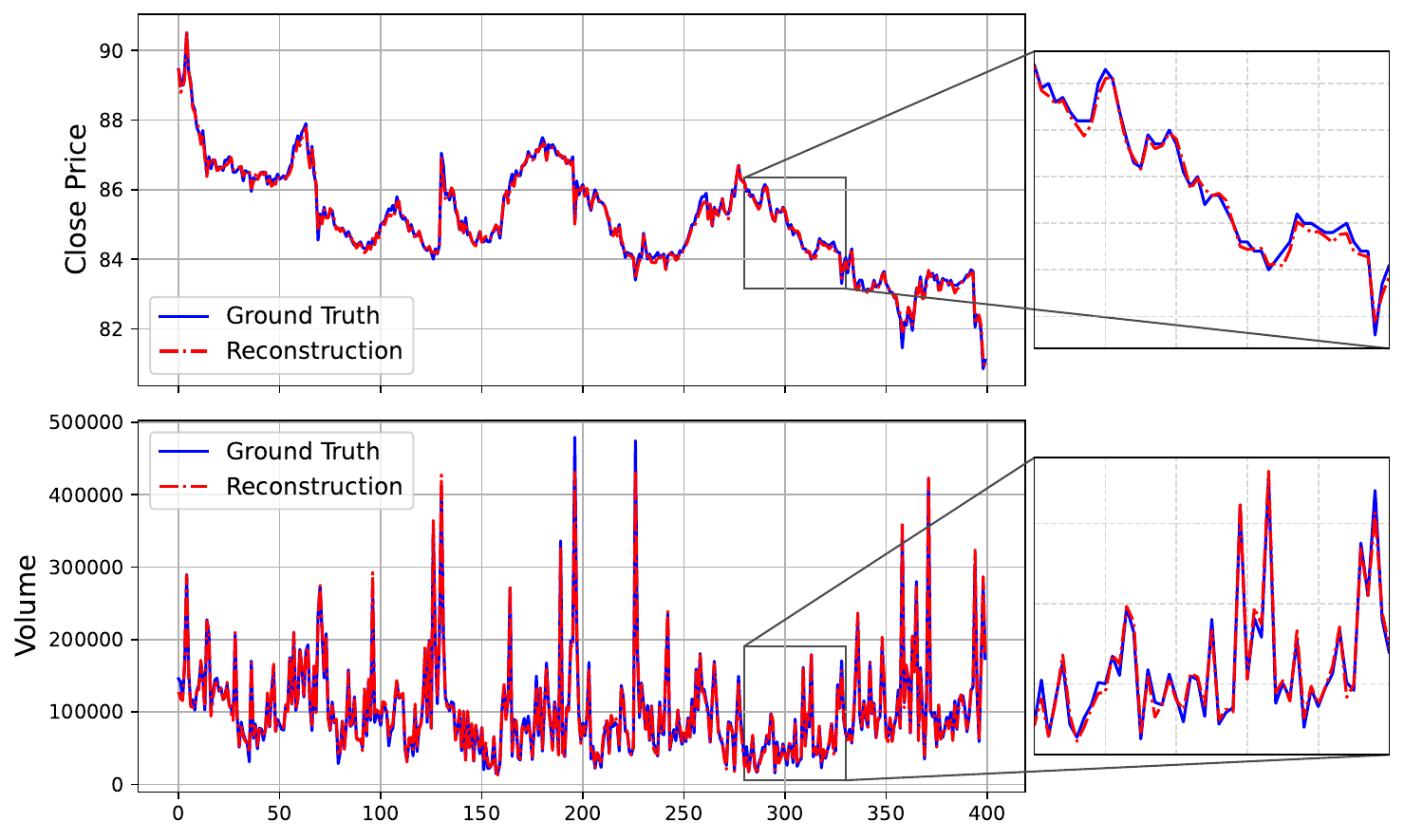}
        \caption{Pop Mart (HKEX: 09992)}
    \end{subfigure}
    
    \vspace{0.1cm}

    \begin{subfigure}[b]{0.49\textwidth}
        \centering
        \includegraphics[width=\linewidth]{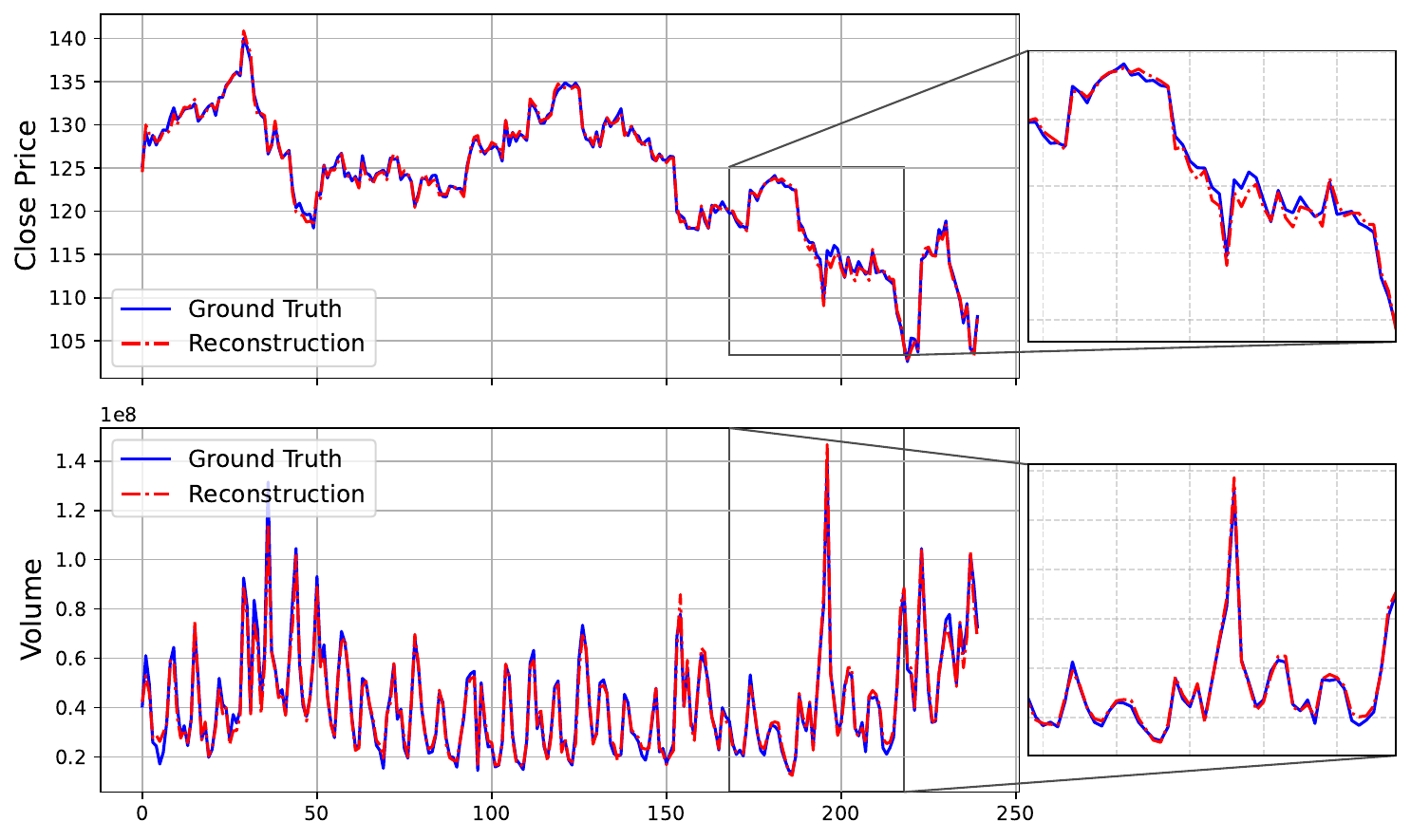}
        \caption{NVIDIA (NASDAQ: NVDA)}
    \end{subfigure}
    \hfill
    \begin{subfigure}[b]{0.49\textwidth}
        \centering
        \includegraphics[width=\linewidth]{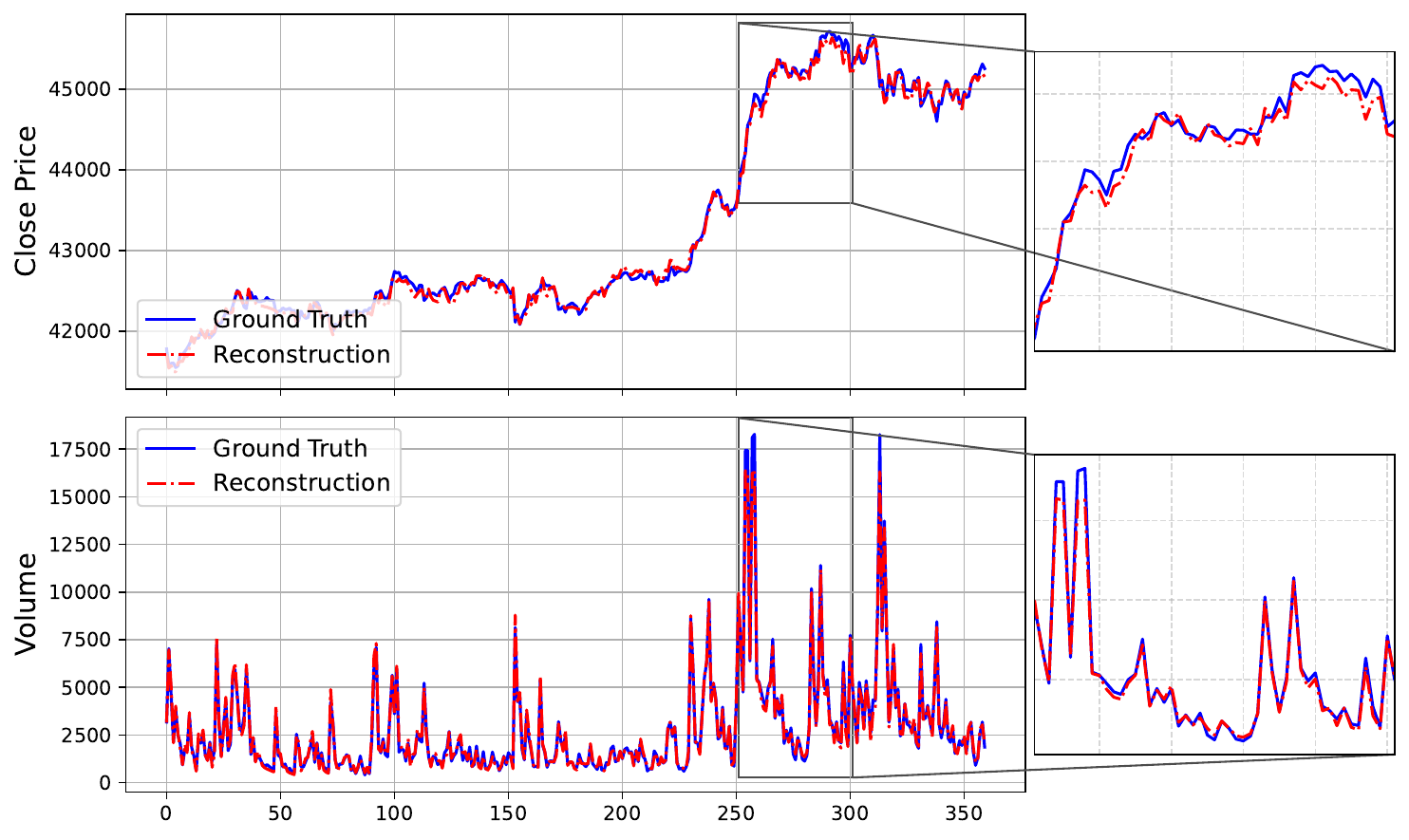}
        \caption{BTC/USDT Perpetual (Binance: BTCUSDT)}
    \end{subfigure}
    
    \vspace{0.1cm}

    \begin{subfigure}[b]{0.49\textwidth}
        \centering
        \includegraphics[width=\linewidth]{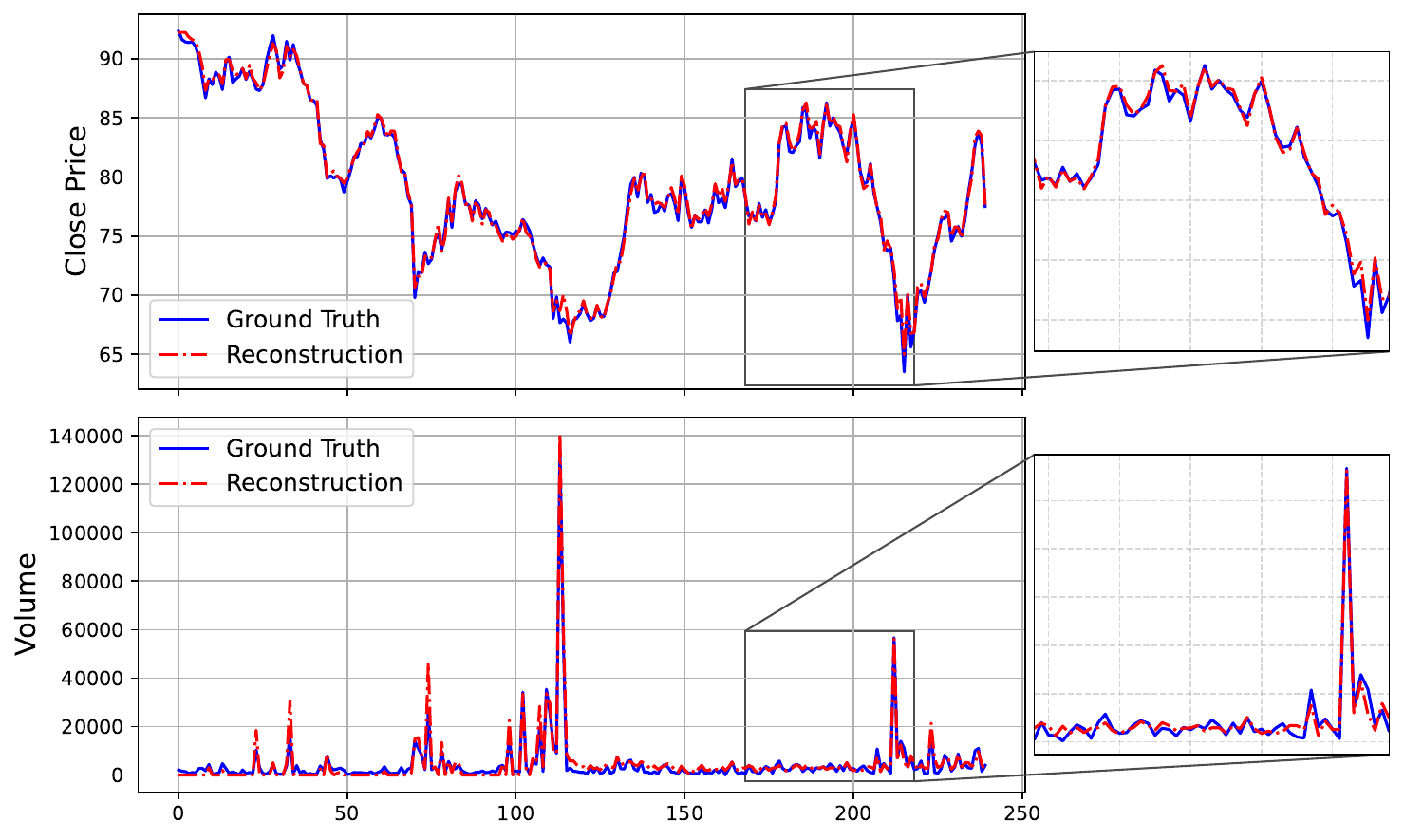}
        \caption{BMW (FWB: BMW)}
    \end{subfigure}
    \hfill
    \begin{subfigure}[b]{0.49\textwidth}
        \centering
        \includegraphics[width=\linewidth]{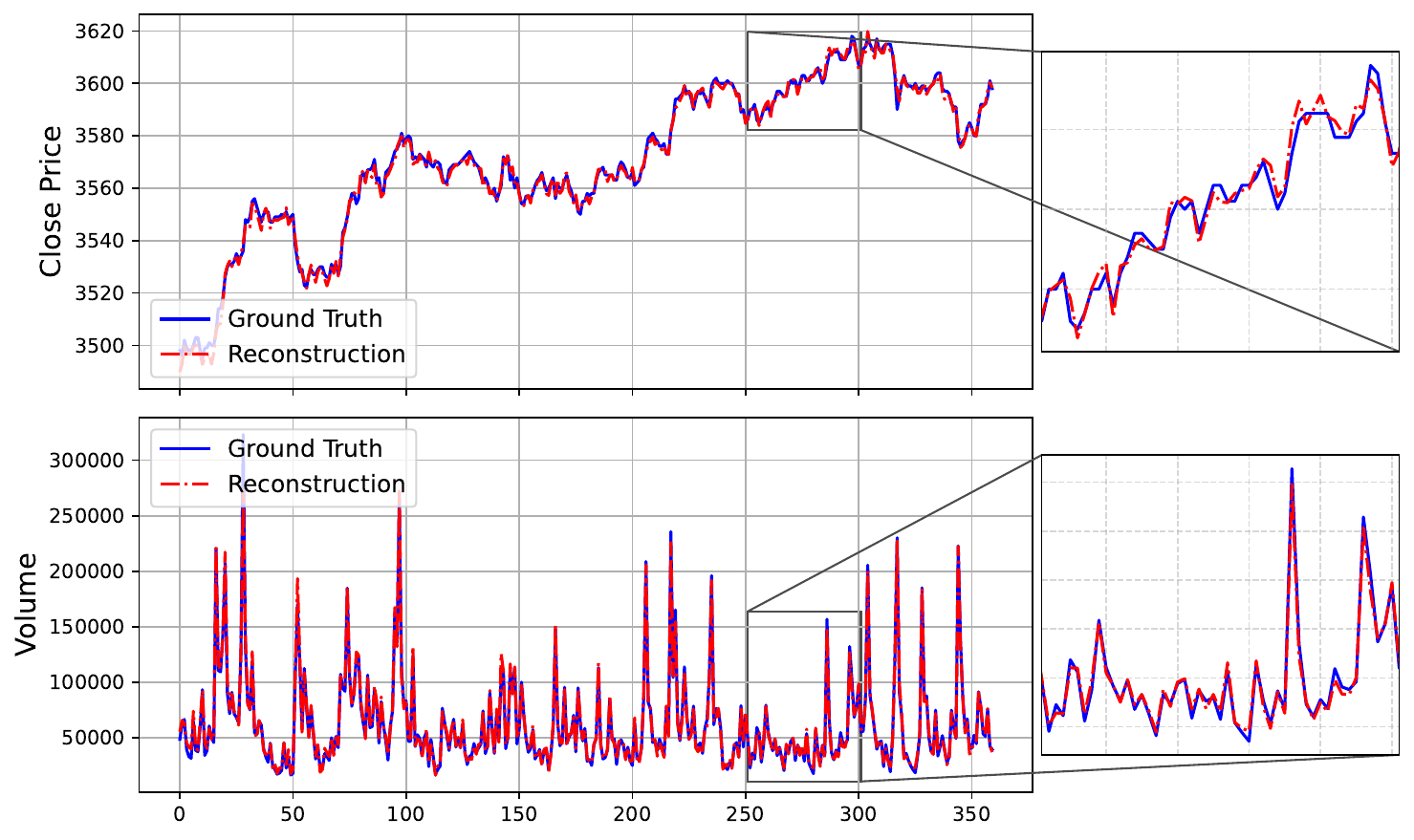}
        \caption{Rebar Steel Futures (SHFE: RB)}
    \end{subfigure}
    
    \vspace{0.1cm}

    \caption{Visualization of reconstruction results for the `Close Price' and `Volume' from our K-line Tokenizer. \textcolor{blue}{Blue} lines denote the ground truth, while \textcolor{red}{red} lines indicate the reconstructions generated by our model.}
    \label{fig:recon_case}
\end{figure*}

Figure~\ref{fig:recon_case} visualizes our tokenizer's reconstruction results on a diverse set of financial instruments. The plots show that the reconstructed `Close Price' and `Volume' series closely track the ground truth, confirming that our tokenizer effectively preserves the essential dynamics of the original continuous data within its discrete token representation. 

\subsection{Cumulative Return Curve Visualizations}

Figure~\ref{fig:backtest_profit} presents the cumulative return curves derived from backtesting using predictive signals by different models. As illustrated, Kronos consistently demonstrates superior performance, achieving the highest cumulative returns among the evaluated models.

\begin{table}[t]
  \centering
  \footnotesize
  \setlength{\tabcolsep}{4pt}

  \adjustbox{max width=\columnwidth}{
  \begin{tabular}{l *{6}{S}}
    \toprule
    \multirow{2}{*}{\textbf{Model}} & \multicolumn{2}{c}{\textbf{CSI300 Index}} & \multicolumn{2}{c}{\textbf{CSI800 Index}} & \multicolumn{2}{c}{\textbf{Average}} \\
    
    \cmidrule(lr){2-3} \cmidrule(lr){4-5} \cmidrule(lr){6-7}
    
    & \textbf{AER} & \textbf{IR} & \textbf{AER} & \textbf{IR} & \textbf{AER} & \textbf{IR} \\
    \midrule

    TimeXer       & 0.1035 & 0.7988 & 0.1509 & 1.5471 & 0.1272 & 1.1730 \\
    TimeMixer       & -0.0600 & -0.5721 & 0.0705 & 0.8113 & 0.0053 & 0.1196 \\
    iTransformer       & -0.1202 & -1.4441 & -0.0525 & -0.8558 & -0.0864 & -1.1500 \\
    PatchTST       & 0.1289 & 0.9895 & 0.1620 & 1.5033 & 0.1455 & 1.2464 \\
    TimesNet       & 0.1441 & 0.6558 & 0.0634 & 0.7225 & 0.1038 & 0.6892 \\
    DLinear       & -0.0066 & -0.0605 & 0.1112 & 1.2003 & 0.0523 & 0.5699 \\
    FEDformer       & 0.0362 & 0.2943 & 0.0539 & 0.5602 & 0.0451 & 0.4273 \\
    NSTransformer       & -0.0343 & -0.2889 & 0.0664 & 0.6979 & 0.0161 & 0.2045 \\
    
    \midrule
    
    $\text{Time-MOE}_{base}$       & 0.0985 & 0.8230 & 0.1315 & 1.3726 & 0.1150 & 1.0978 \\
    $\text{Moirai}_{large}$       & 0.1470 & 0.9747 & 0.1683 & 1.5215 & 0.1577 & 1.2481 \\
    TimesFM       & 0.0788 & 0.7357 & 0.1355 & 1.6427 & 0.1072 & 1.1892 \\
    $\text{Moment}_{large}$       & 0.1655 & 1.1993 & 0.1707 & 1.5361 & 0.1681 & 1.3677 \\
    $\text{Chronos}_{large}$       & -0.0659 & -0.7670 & 0.0056 & 0.0902 & -0.0302 & -0.3384 \\

    \midrule
    \textbf{$\text{Kronos}_{small}$} & 0.1805 & 1.2394 & 0.1772 & 1.6050 & 0.1789 & 1.4222 \\
    \textbf{$\text{Kronos}_{base}$} & \second{0.1911} & \second{1.3782} & \second{0.1867} & \second{1.6652} & \second{0.1889} & \second{1.5217} \\
    \textbf{$\text{Kronos}_{large}$} & \best{0.2193} & \best{1.4177} & \best{0.1974} & \best{1.8805} & \best{0.2084} & \best{1.6491}\\
    \bottomrule
  \end{tabular}}
  \caption{Full results of investment simulation. We report Annualized Excess Return (AER) and Information Ratio (IR). Best and second best results are marked with \best{red underline} and \second{blue underline}, respectively.
  }
  \label{tab:full_backtest_result}
\end{table}

\section{Full Experiment Results}
\label{sec:full_exp_results}

\begin{figure*}[htbp]
    \centering

    \begin{subfigure}{\textwidth}
        \centering
        \includegraphics[width=1.0\textwidth]{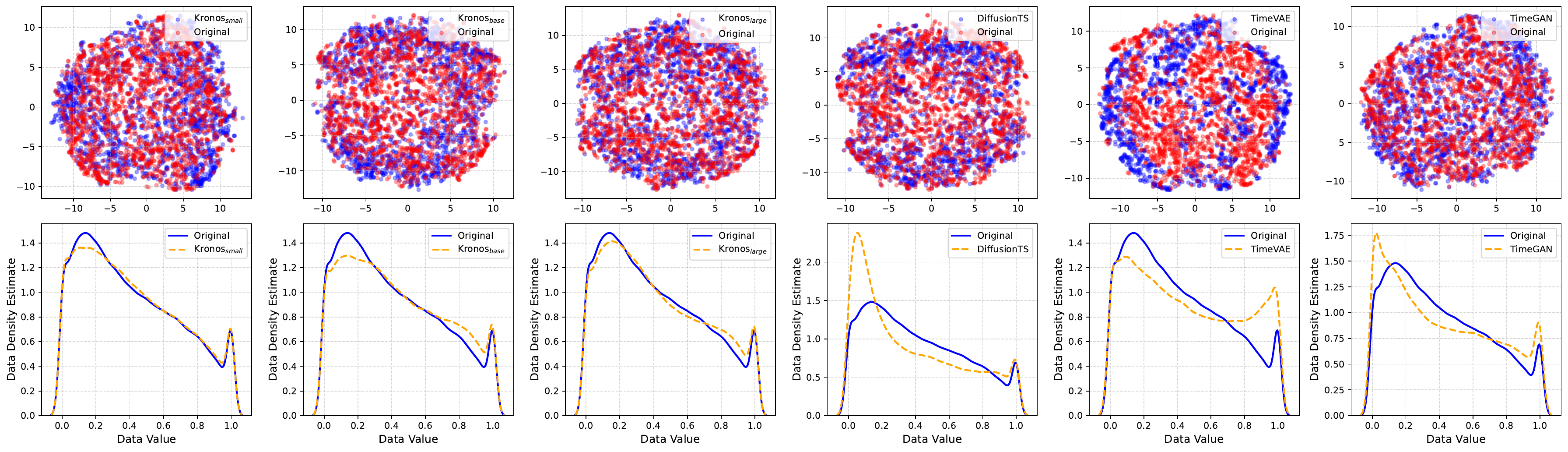}
        \caption{Shanghai Stock Exchange (XSHG), Daily frequency}
    \end{subfigure}

    \begin{subfigure}{\textwidth}
        \centering
        \includegraphics[width=1.0\textwidth]{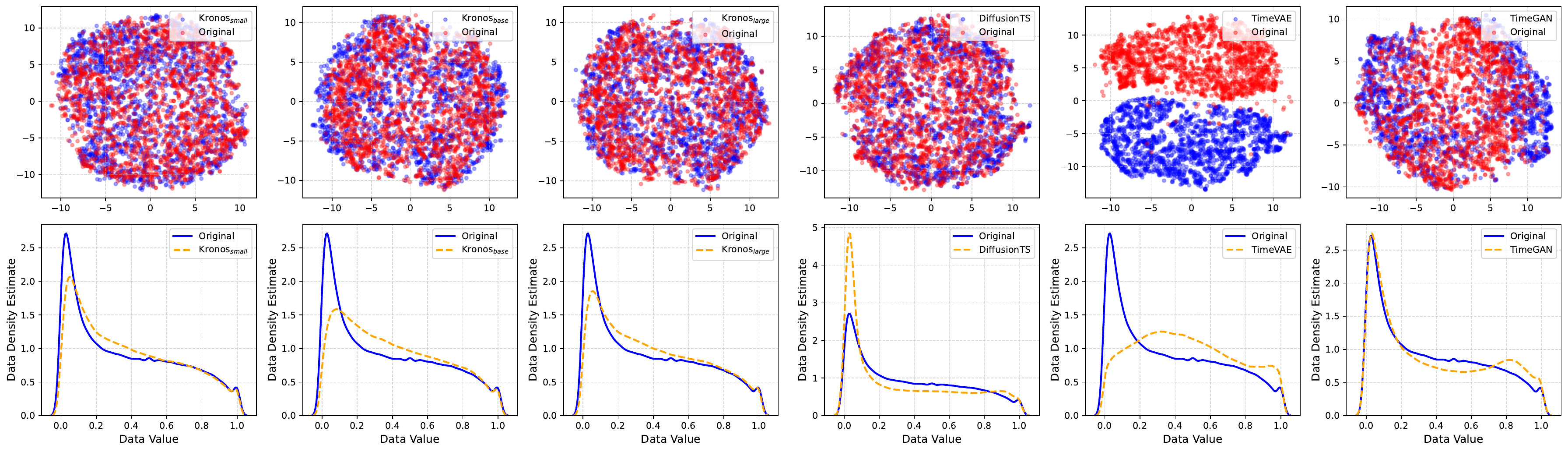}
        \caption{Taiwan Stock Exchange (XTAI), 15-minute frequency}
    \end{subfigure}

    \begin{subfigure}{\textwidth}
        \centering
        \includegraphics[width=1.0\textwidth]{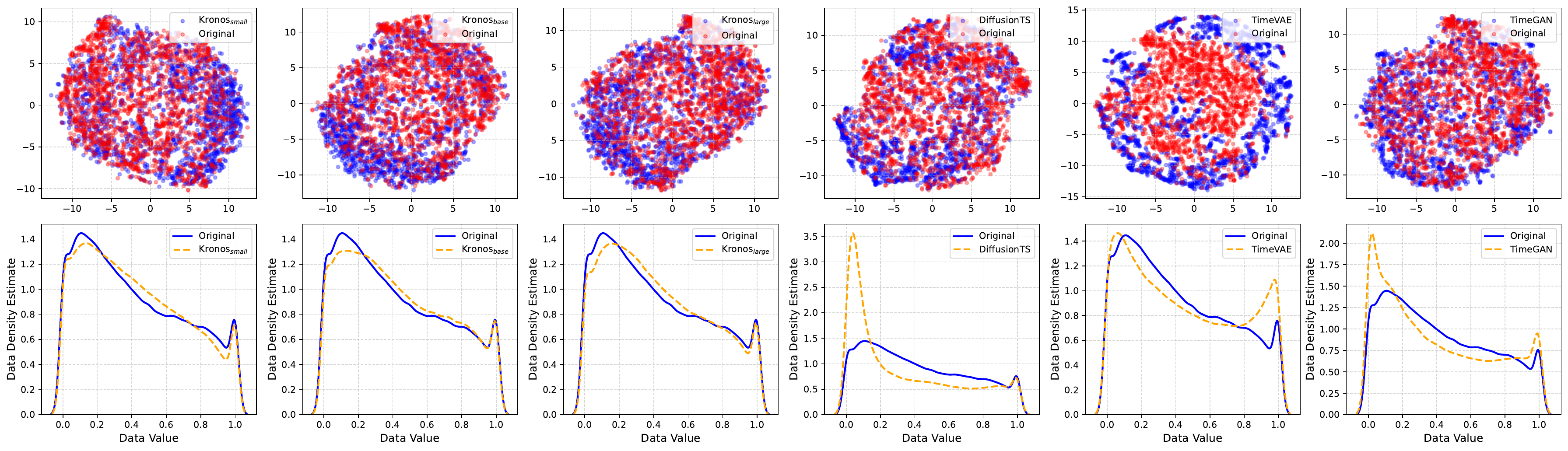}
        \caption{Taiwan Stock Exchange (XTAI), Daily frequency}
    \end{subfigure}
    
    \begin{subfigure}{\textwidth}
        \centering
        \includegraphics[width=1.0\textwidth]{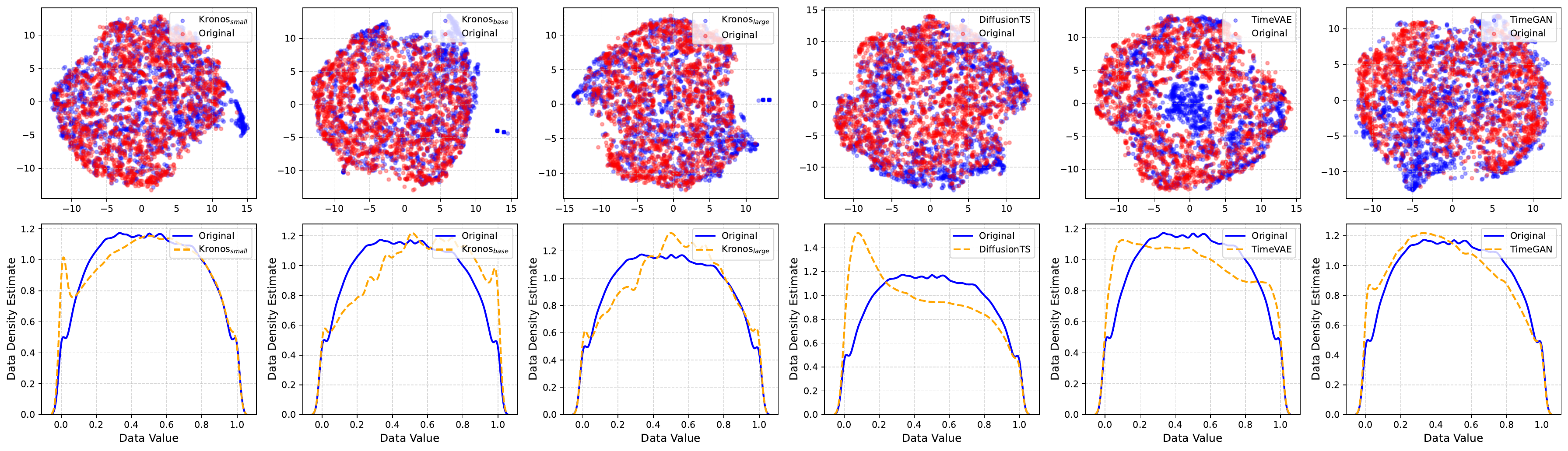}
        \caption{Cryptocurrency (Crypto), 15-minute frequency}
    \end{subfigure}

    \caption{
    Visual comparison of generative models on different datasets. 
    \textbf{Top row in each subfigure:} t-SNE embeddings of original (\textcolor{red}{red}) versus synthetic (\textcolor{blue}{blue}) data. 
    \textbf{Bottom row in each subfigure:} Kernel Density Estimates (KDE) of original versus synthetic data.}
    \label{fig:gen_visual_1}
\end{figure*}

\begin{figure*}[htbp]
    \centering


    \begin{subfigure}{\textwidth} 
        \centering
        \includegraphics[width=1.0\textwidth]{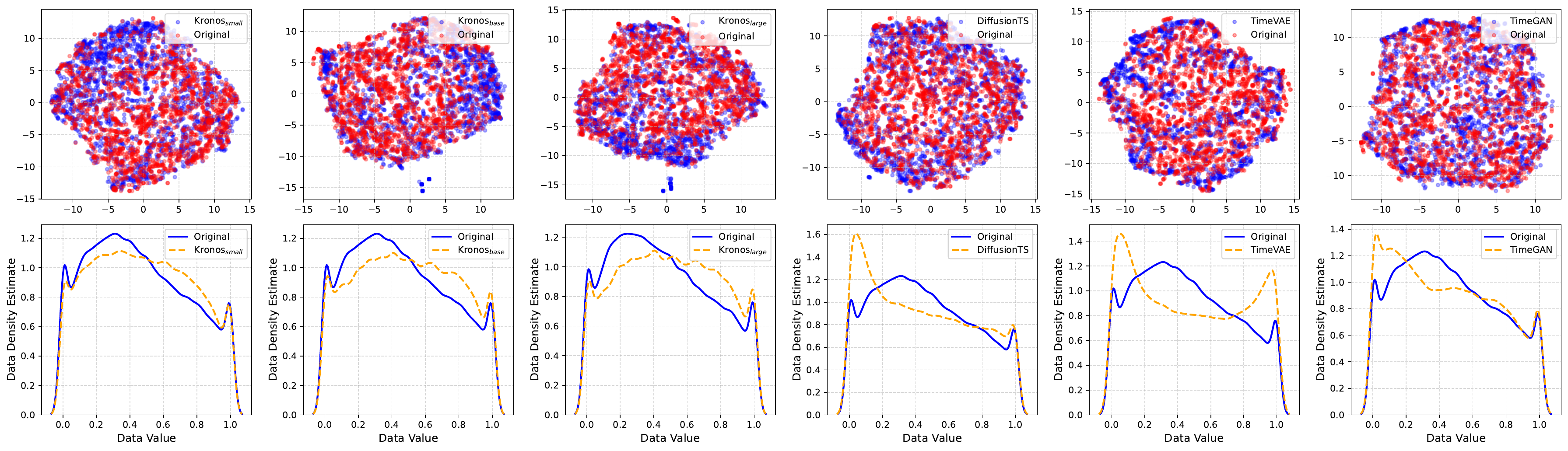}
        \caption{Cryptocurrency (Crypto), Daily frequency}
    \end{subfigure} 

    \begin{subfigure}{\textwidth}
        \centering
        \includegraphics[width=1.0\textwidth]{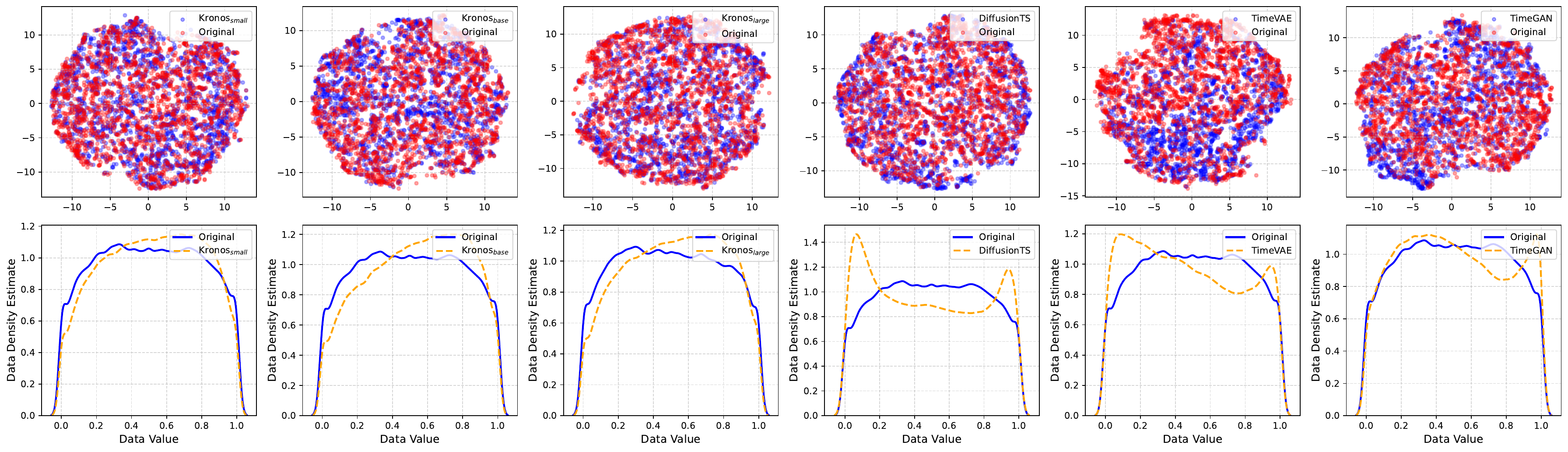}
        \caption{Foreign Exchange (Forex), 15-minute frequency}
    \end{subfigure}

    \begin{subfigure}{\textwidth}
        \centering
        \includegraphics[width=1.0\textwidth]{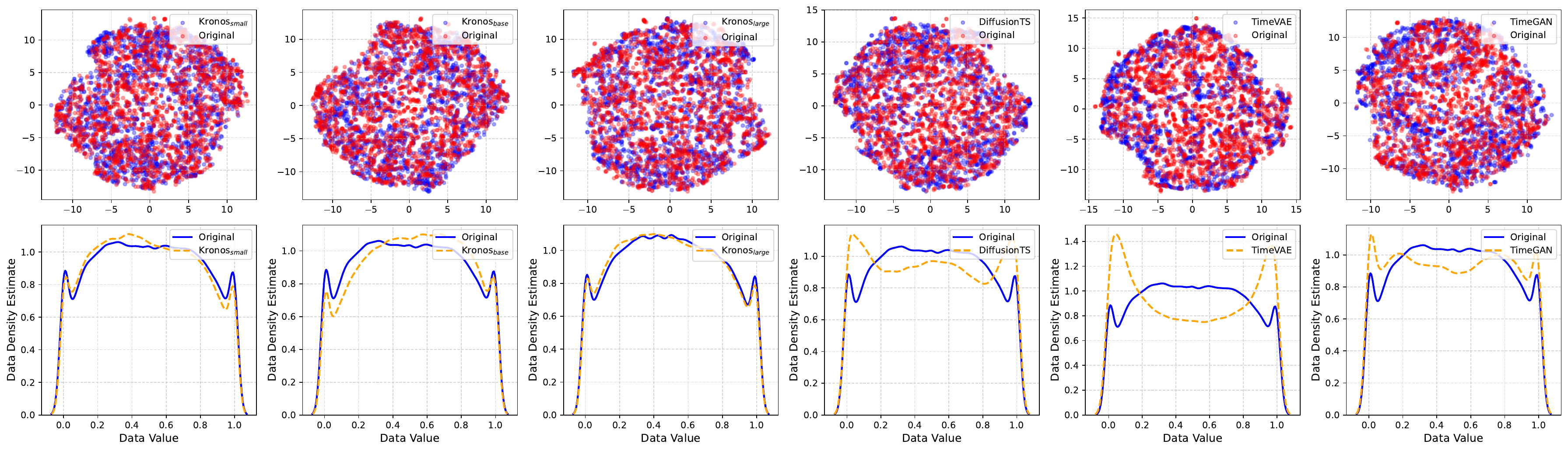}
        \caption{Foreign Exchange (Forex), Daily frequency}
    \end{subfigure}

    \caption{Visual comparison of generative models on different datasets. 
    \textbf{Top row in each subfigure:} t-SNE embeddings of original (\textcolor{red}{red}) versus synthetic (\textcolor{blue}{blue}) data. 
    \textbf{Bottom row in each subfigure:} Kernel Density Estimates (KDE) of original versus synthetic data.}
    \label{fig:gen_visual_2}
\end{figure*}

In this section, we present the complete experimental results for three forecasting tasks and the synthetic K-line generation task. For the forecasting tasks, we report the results for each asset, averaged over all tested frequencies. 
Tables~\ref{tab:full_price_forecasting_1} and~\ref{tab:full_price_forecasting_2} show the results of the price series forecasting experiments. 
The outcomes for return forecasting are presented in Tables~\ref{tab:full_return_forecasting_1} and~\ref{tab:full_return_forecasting_2}, while those for realized volatility forecasting are in Tables~\ref{tab:full_volatility_forecasting_1} and~\ref{tab:full_volatility_forecasting_2}.
Furthermore, for the synthetic K-line generation task, Figures~\ref{fig:gen_visual_1} and~\ref{fig:gen_visual_2} provide visualizations of the \textit{diversity} of the generated sequences by different models. The results for the discriminative score and predictive usefulness are presented in Table~\ref{tab:generate_discriminative} and Table~\ref{tab:generate_usefulness}, respectively. Finally, the results of the investment simulation experiment are presented in Table~\ref{tab:full_backtest_result}.

\section{Forecast Showcases}

\begin{figure*}[ht]
    \centering

    \begin{subfigure}[b]{0.33\textwidth}
        \centering
        \includegraphics[width=\linewidth]{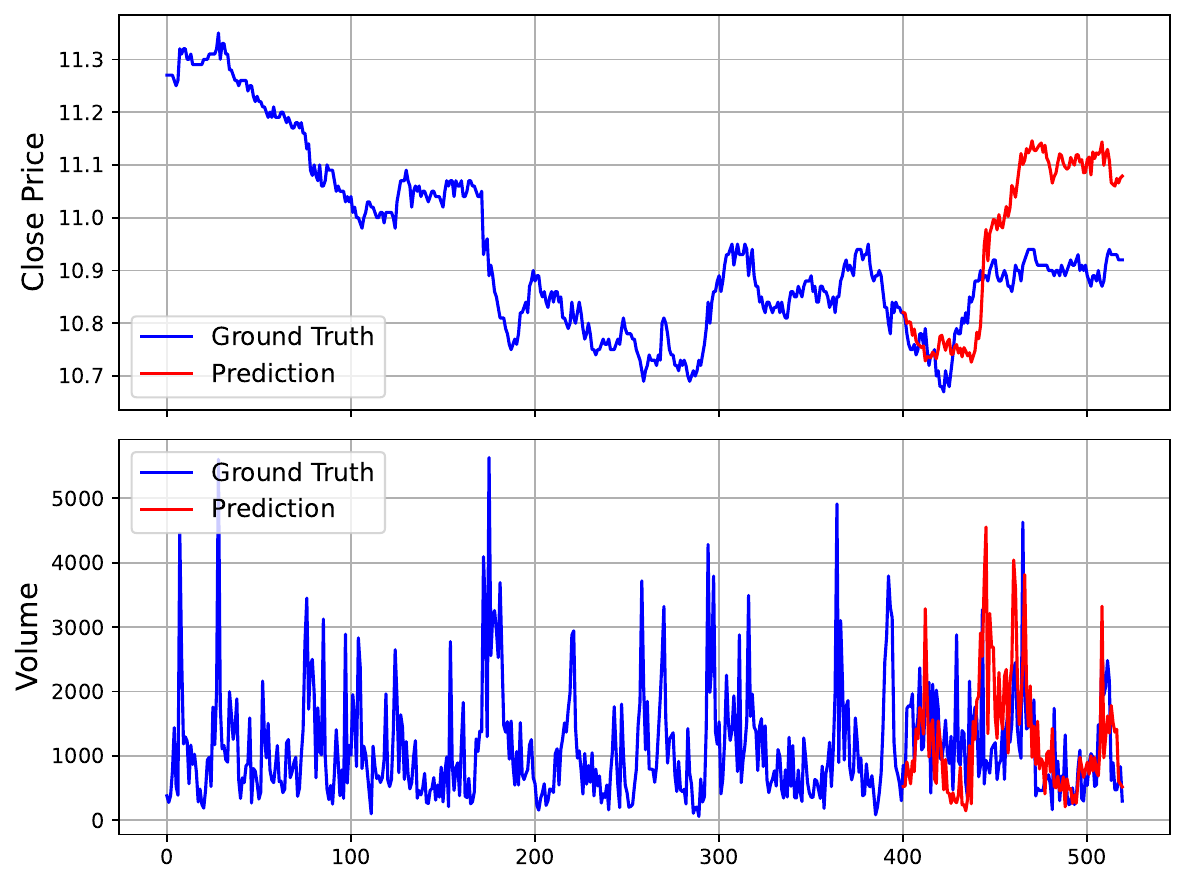}
        \caption{$\text{Kronos}_{small}$}
    \end{subfigure}
    \hfill
    \begin{subfigure}[b]{0.33\textwidth}
        \centering
        \includegraphics[width=\linewidth]{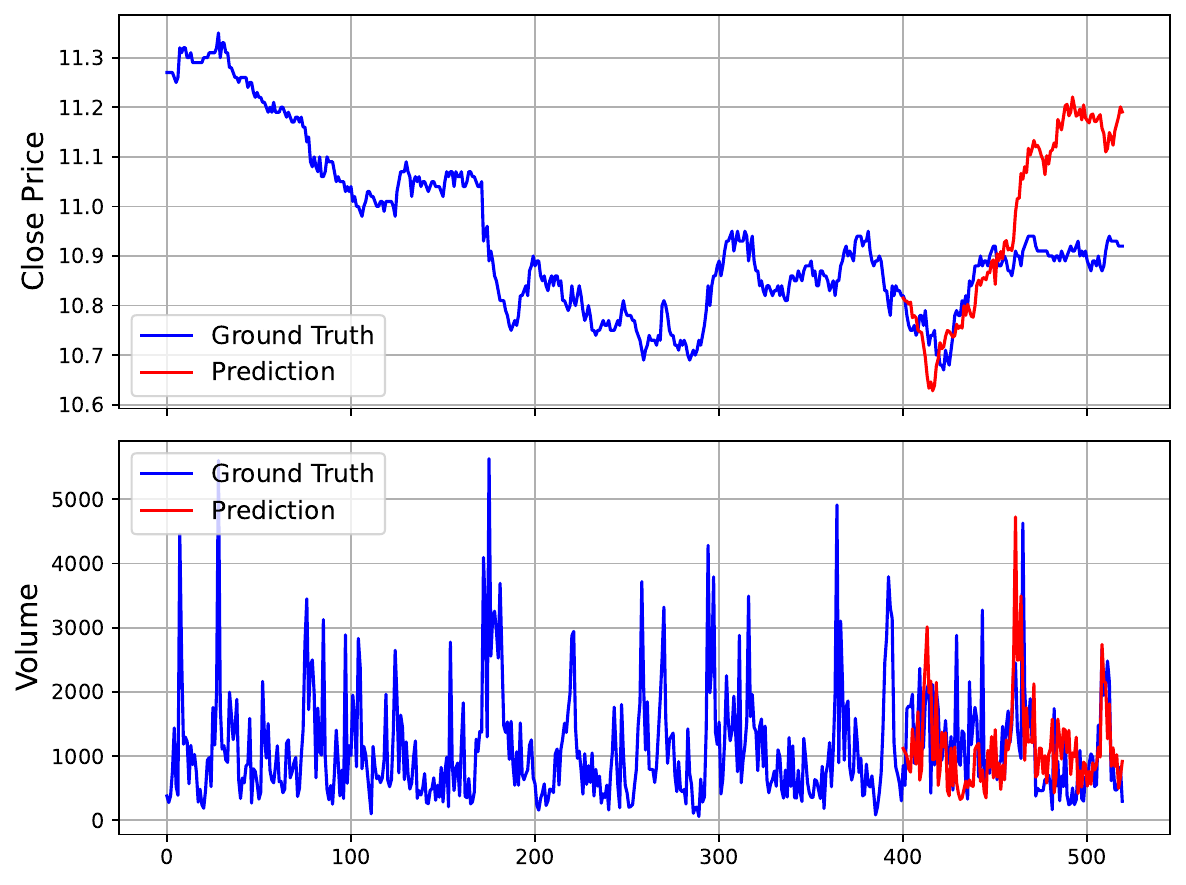}
        \caption{$\text{Kronos}_{base}$}
    \end{subfigure}
    \hfill
    \begin{subfigure}[b]{0.33\textwidth}
        \centering
        \includegraphics[width=\linewidth]{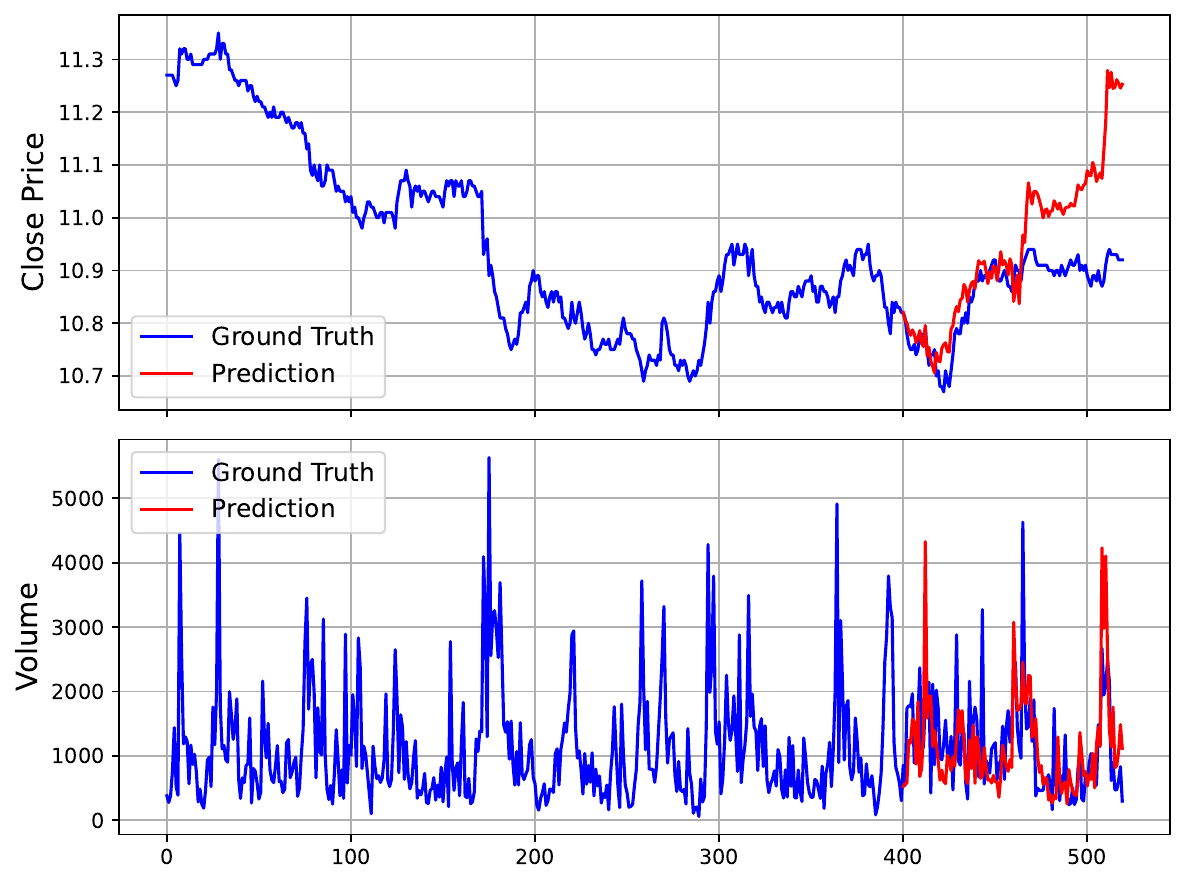}
        \caption{$\text{Kronos}_{large}$}
    \end{subfigure}
    
    \vspace{0.1cm}

    \begin{subfigure}[b]{0.33\textwidth}
        \centering
        \includegraphics[width=\linewidth]{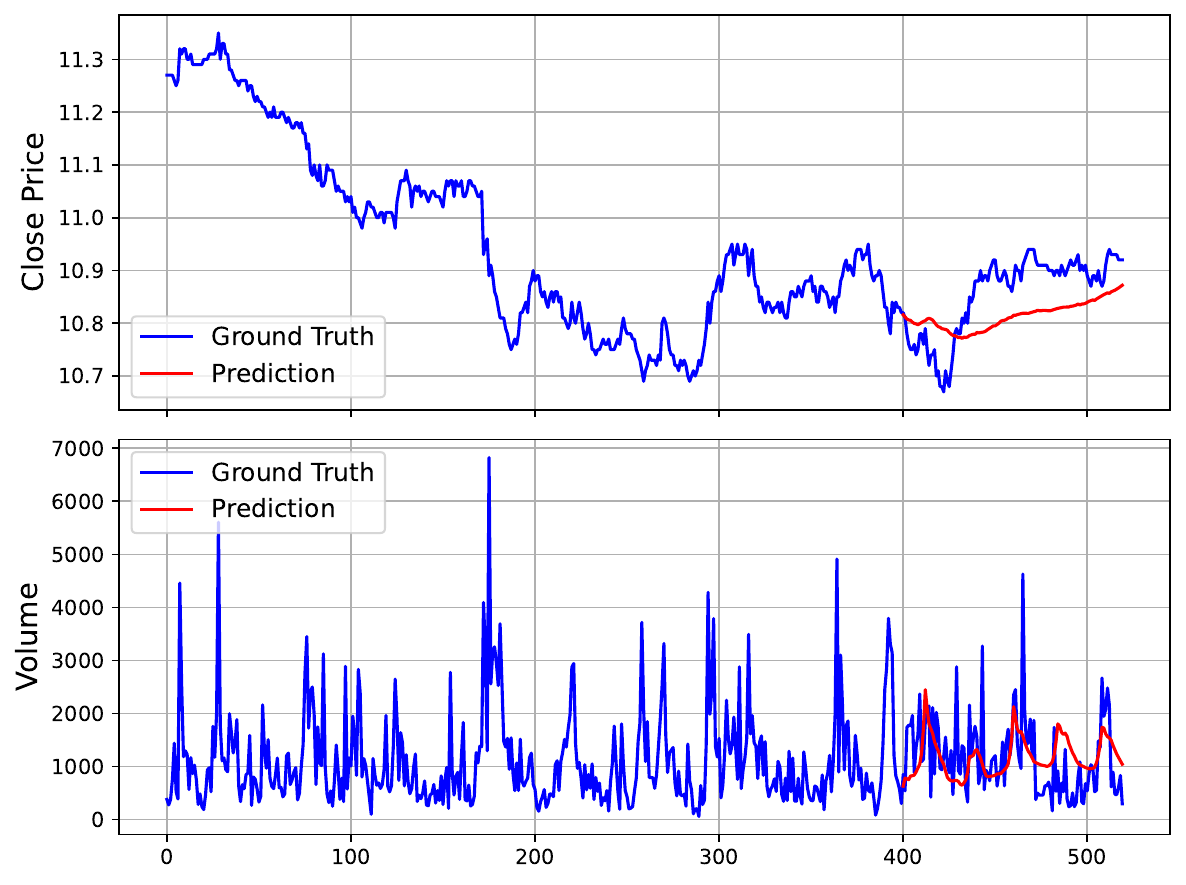}
        \caption{$\text{TimeMOE}_{small}$}
    \end{subfigure}
    \hfill
    \begin{subfigure}[b]{0.33\textwidth}
        \centering
        \includegraphics[width=\linewidth]{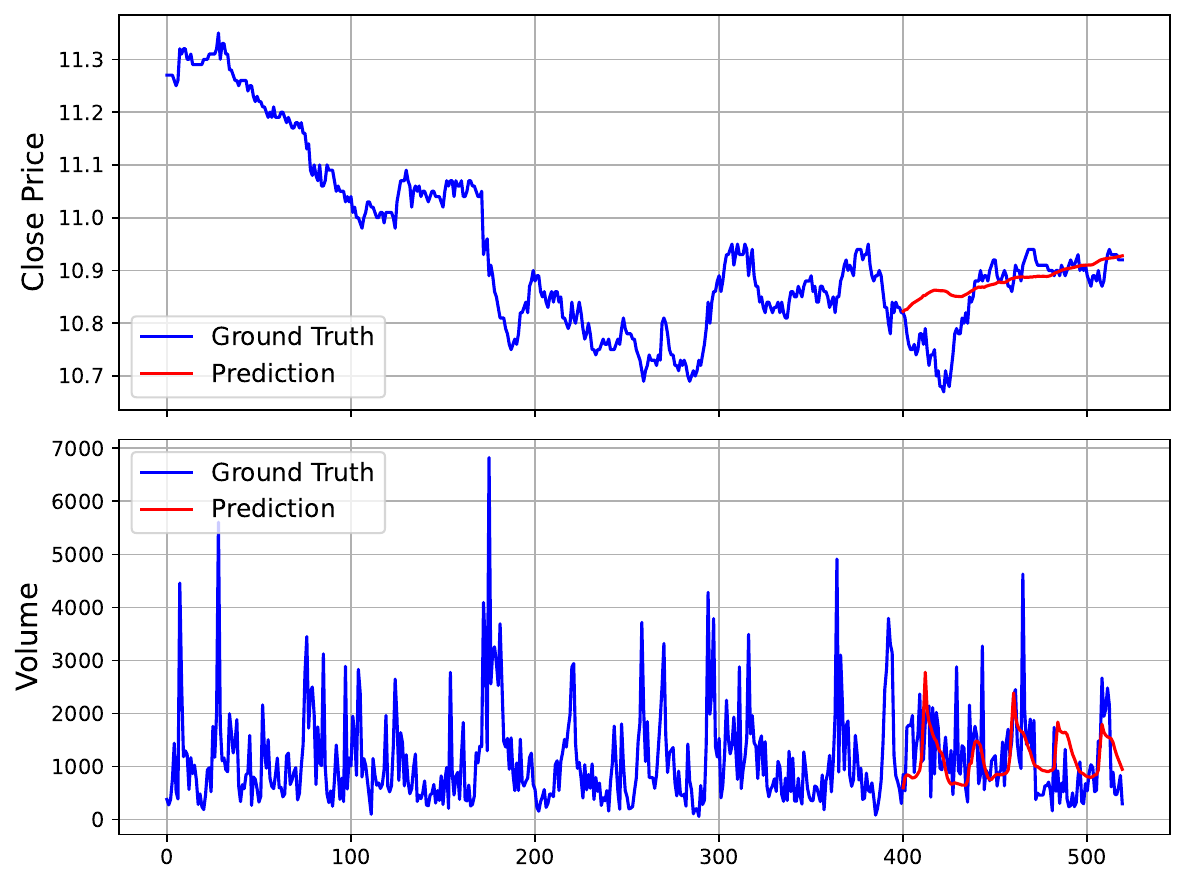}
        \caption{$\text{TimeMOE}_{large}$}
    \end{subfigure}
    \hfill
    \begin{subfigure}[b]{0.33\textwidth}
        \centering
        \includegraphics[width=\linewidth]{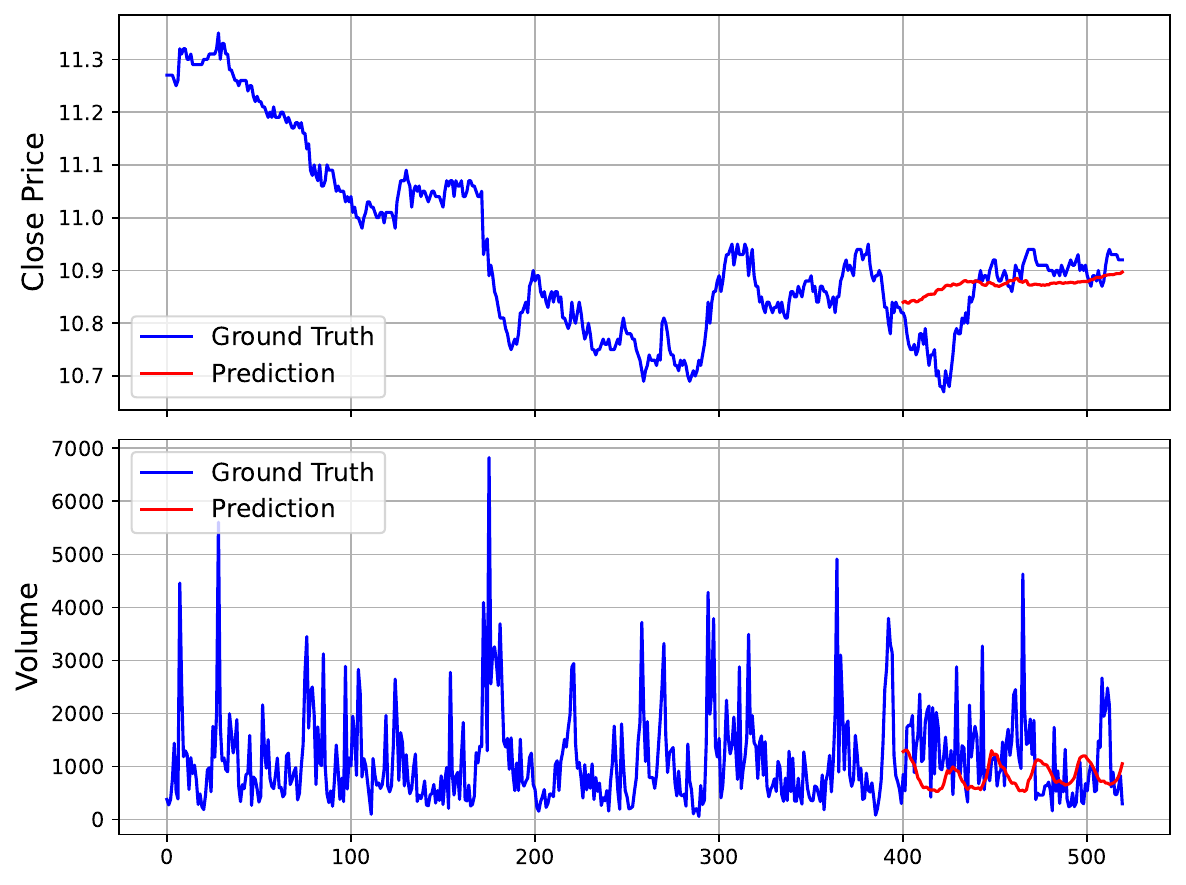}
        \caption{TimesFM}
    \end{subfigure}

    \vspace{0.1cm}

    \begin{subfigure}[b]{0.33\textwidth}
        \centering
        \includegraphics[width=\linewidth]{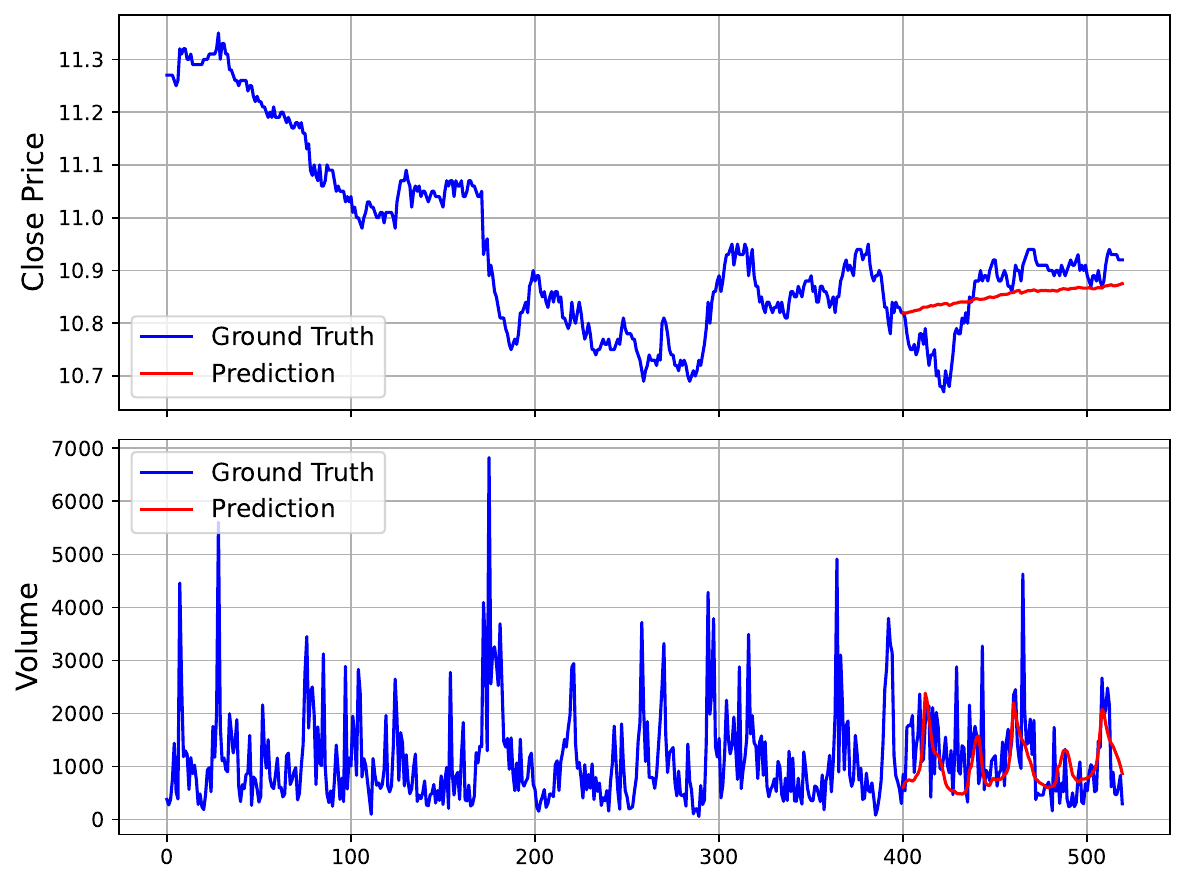}
        \caption{$\text{Chronos}_{small}$}
    \end{subfigure}
    \hfill
    \begin{subfigure}[b]{0.33\textwidth}
        \centering
        \includegraphics[width=\linewidth]{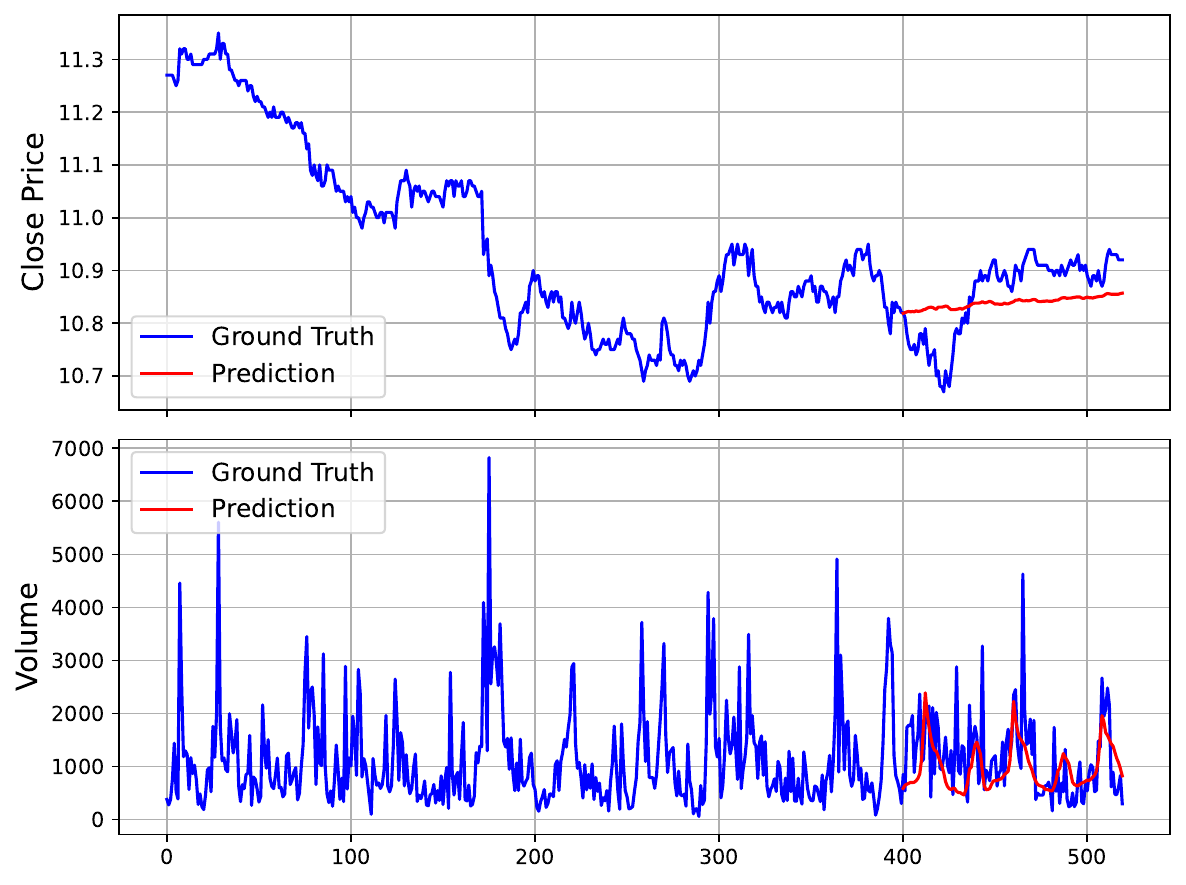}
        \caption{$\text{Chronos}_{base}$}
    \end{subfigure}
    \hfill
    \begin{subfigure}[b]{0.33\textwidth}
        \centering
        \includegraphics[width=\linewidth]{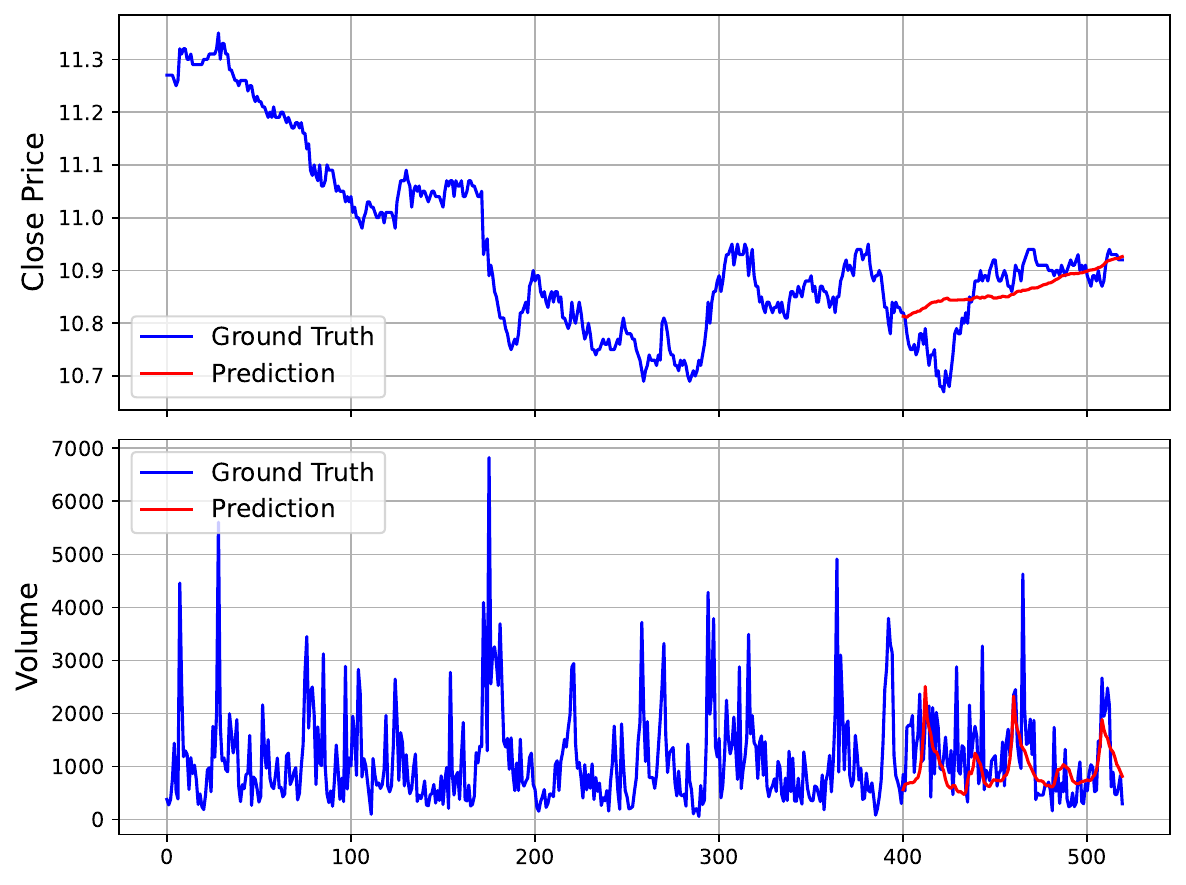}
        \caption{$\text{Chronos}_{large}$}
    \end{subfigure}
    
    \vspace{0.1cm}

    \begin{subfigure}[b]{0.33\textwidth}
        \centering
        \includegraphics[width=\linewidth]{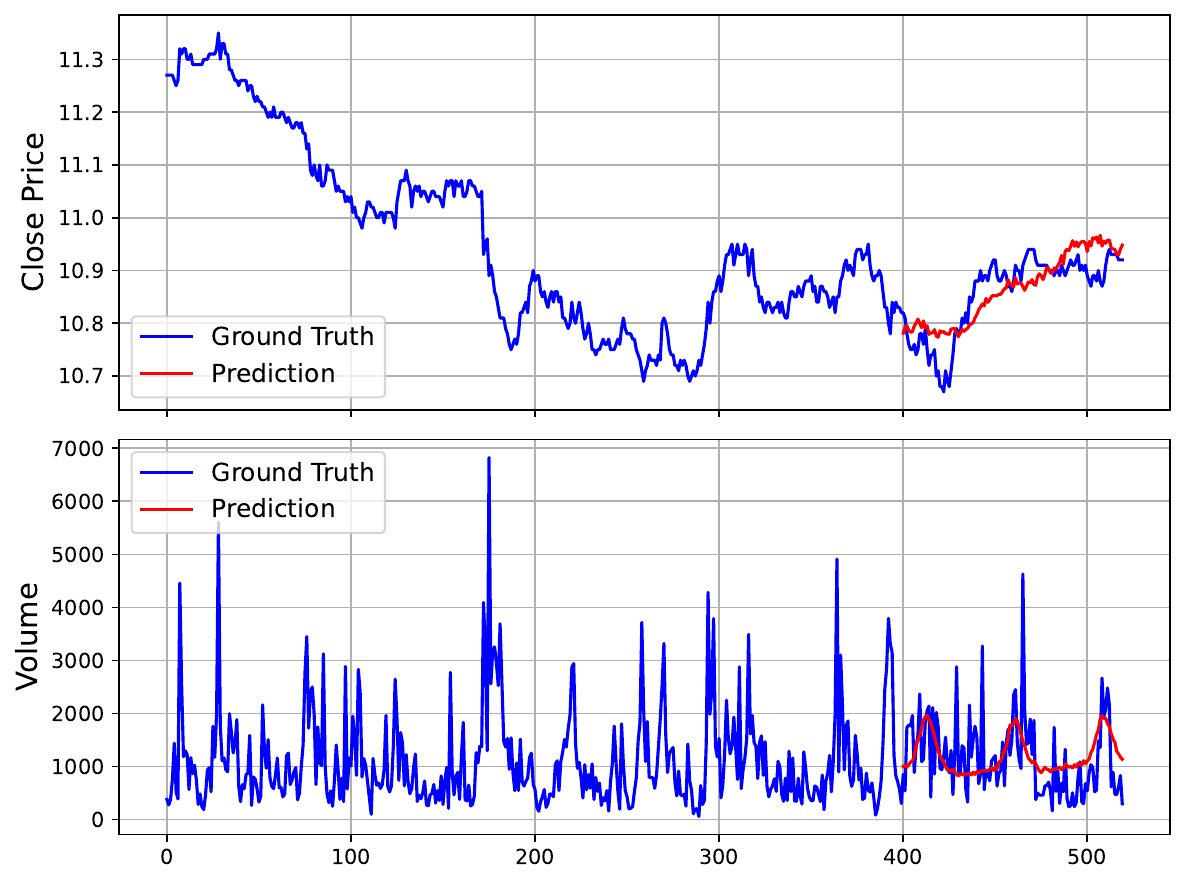}
        \caption{iTransformer}
    \end{subfigure}
    \hfill
    \begin{subfigure}[b]{0.33\textwidth}
        \centering
        \includegraphics[width=\linewidth]{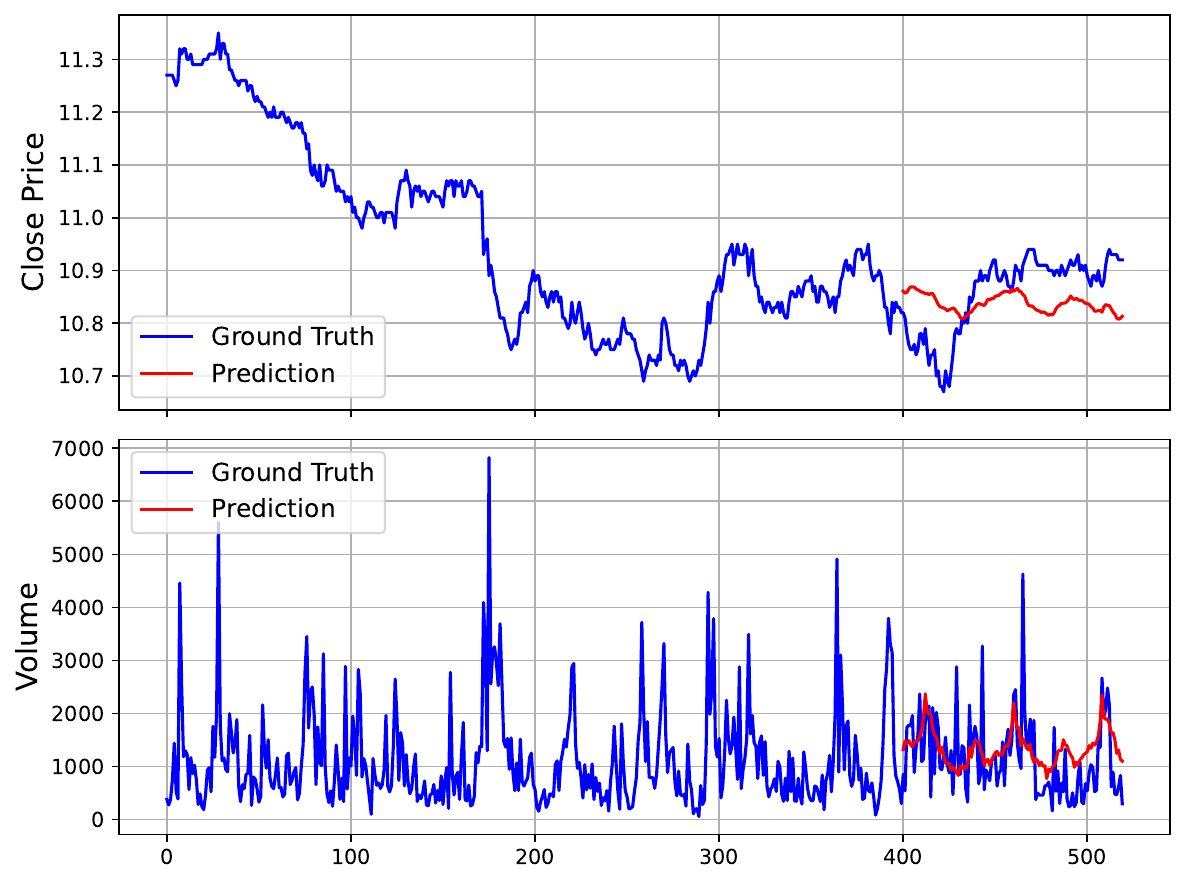}
        \caption{DLinear}
    \end{subfigure}
    \hfill
    \begin{subfigure}[b]{0.33\textwidth}
        \centering
        \includegraphics[width=\linewidth]{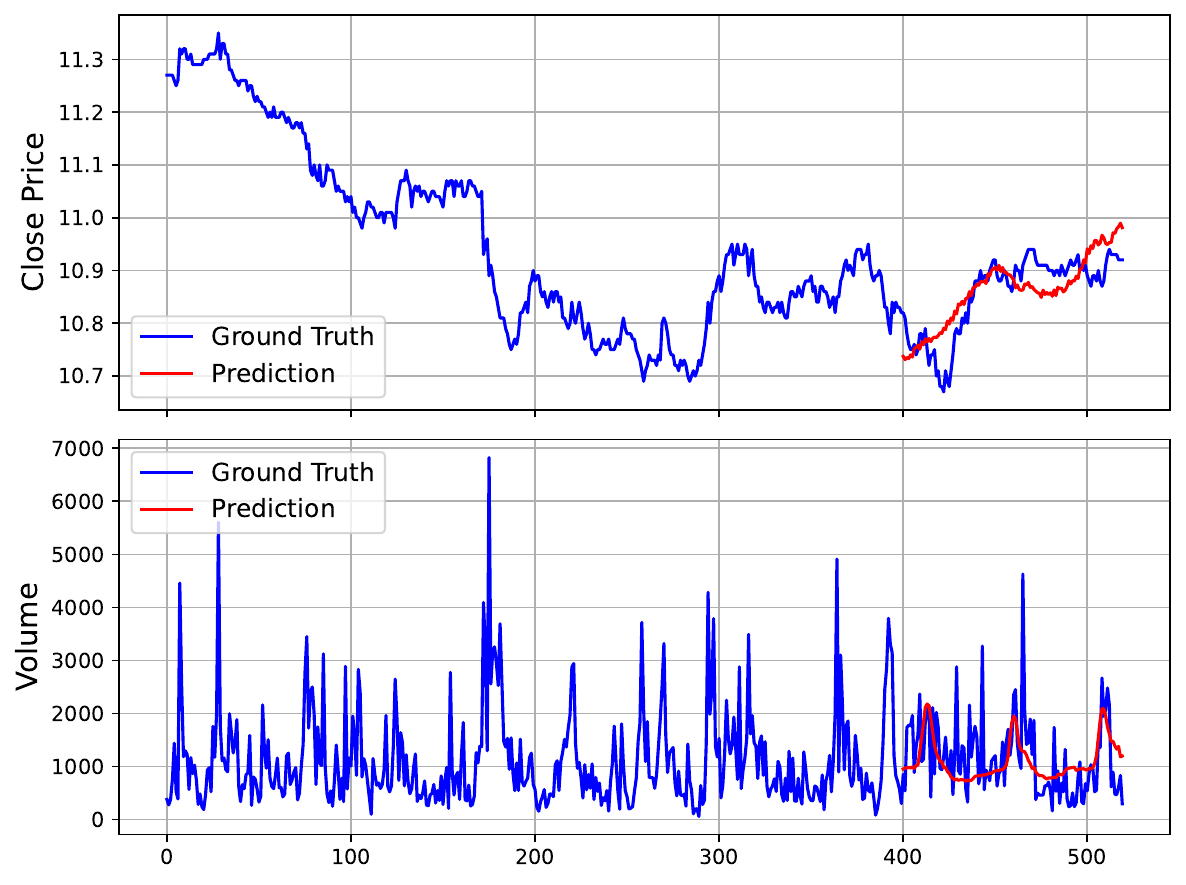}
        \caption{TimesNet}
    \end{subfigure}

    \caption{Forecasting results for the `Close Price' and `Volume' of China Film Co.,Ltd. (SSE: 600977), based on 5-minute K-line data. The model uses a 400-step look-back window to predict a 120-step horizon. \textcolor{blue}{Blue} lines represent the ground truths and \textcolor{red}{red} lines are the model's predictions.}
    \label{fig:pred_case_1}
\end{figure*}

\begin{figure*}[ht]
    \centering

    \begin{subfigure}[b]{0.33\textwidth}
        \centering
        \includegraphics[width=\linewidth]{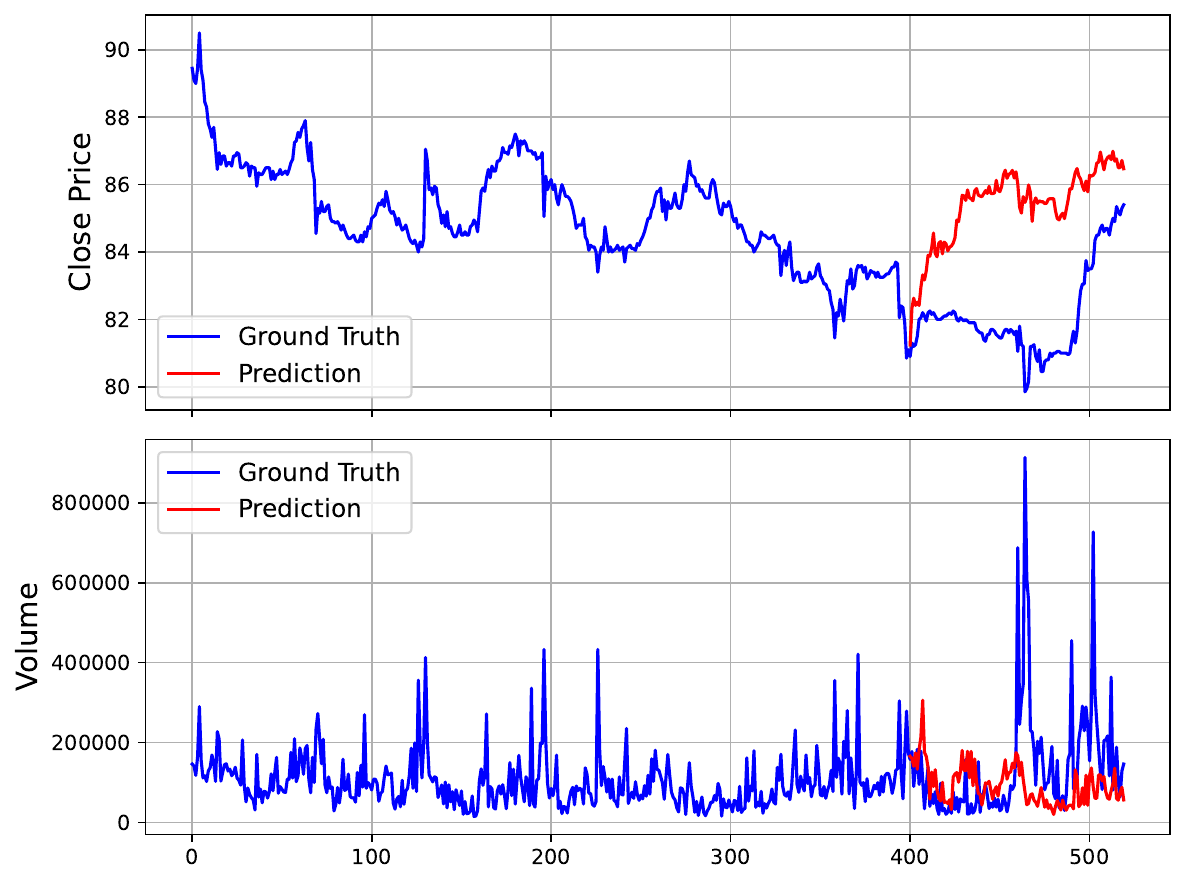}
        \caption{$\text{Kronos}_{small}$}
    \end{subfigure}
    \hfill
    \begin{subfigure}[b]{0.33\textwidth}
        \centering
        \includegraphics[width=\linewidth]{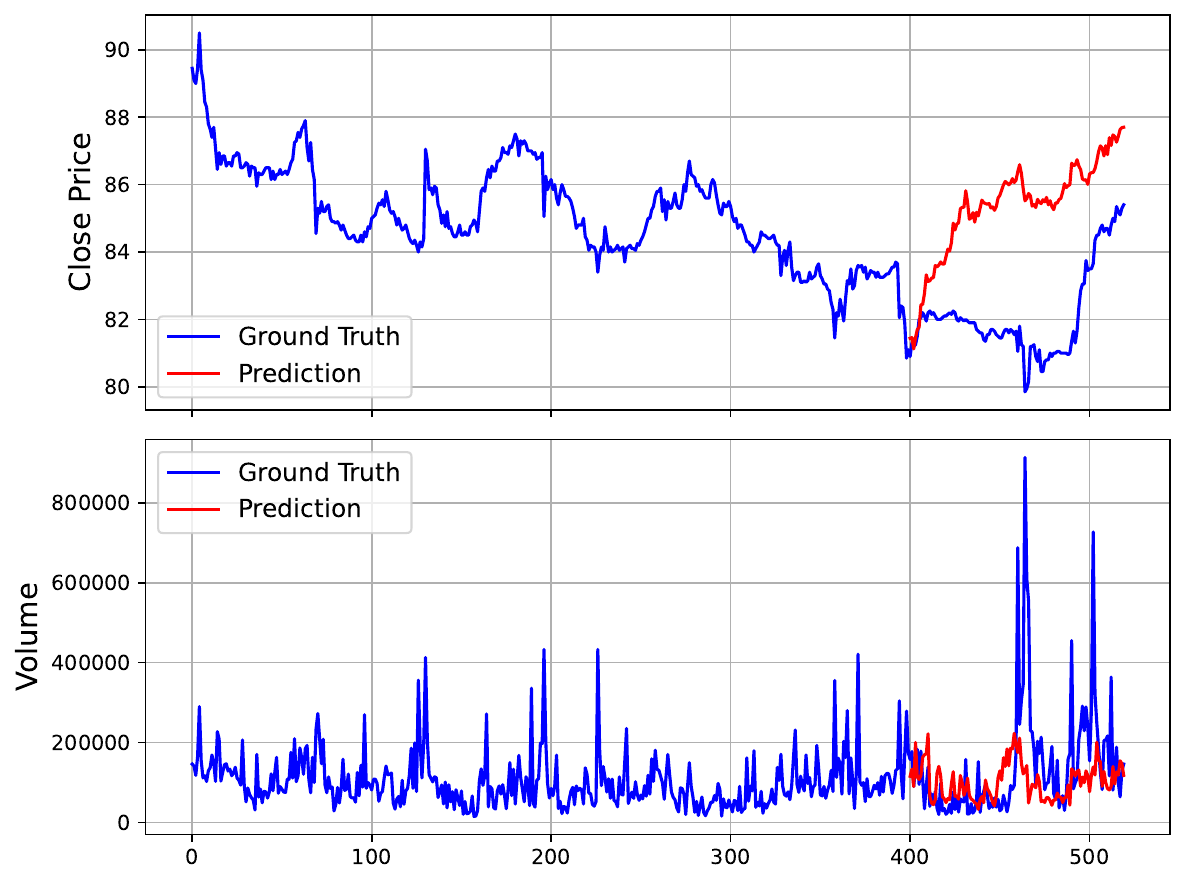}
        \caption{$\text{Kronos}_{base}$}
    \end{subfigure}
    \hfill
    \begin{subfigure}[b]{0.33\textwidth}
        \centering
        \includegraphics[width=\linewidth]{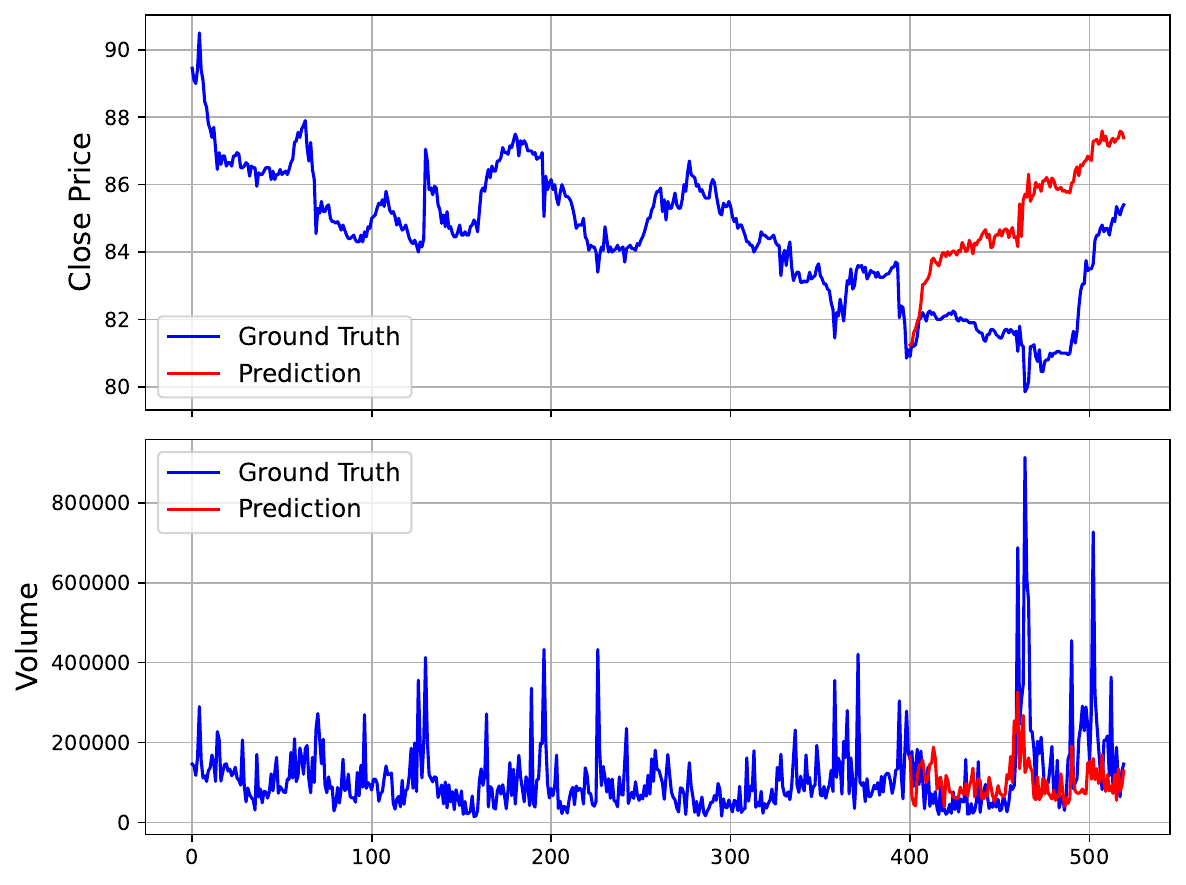}
        \caption{$\text{Kronos}_{large}$}
    \end{subfigure}
    
    \vspace{0.1cm}

    \begin{subfigure}[b]{0.33\textwidth}
        \centering
        \includegraphics[width=\linewidth]{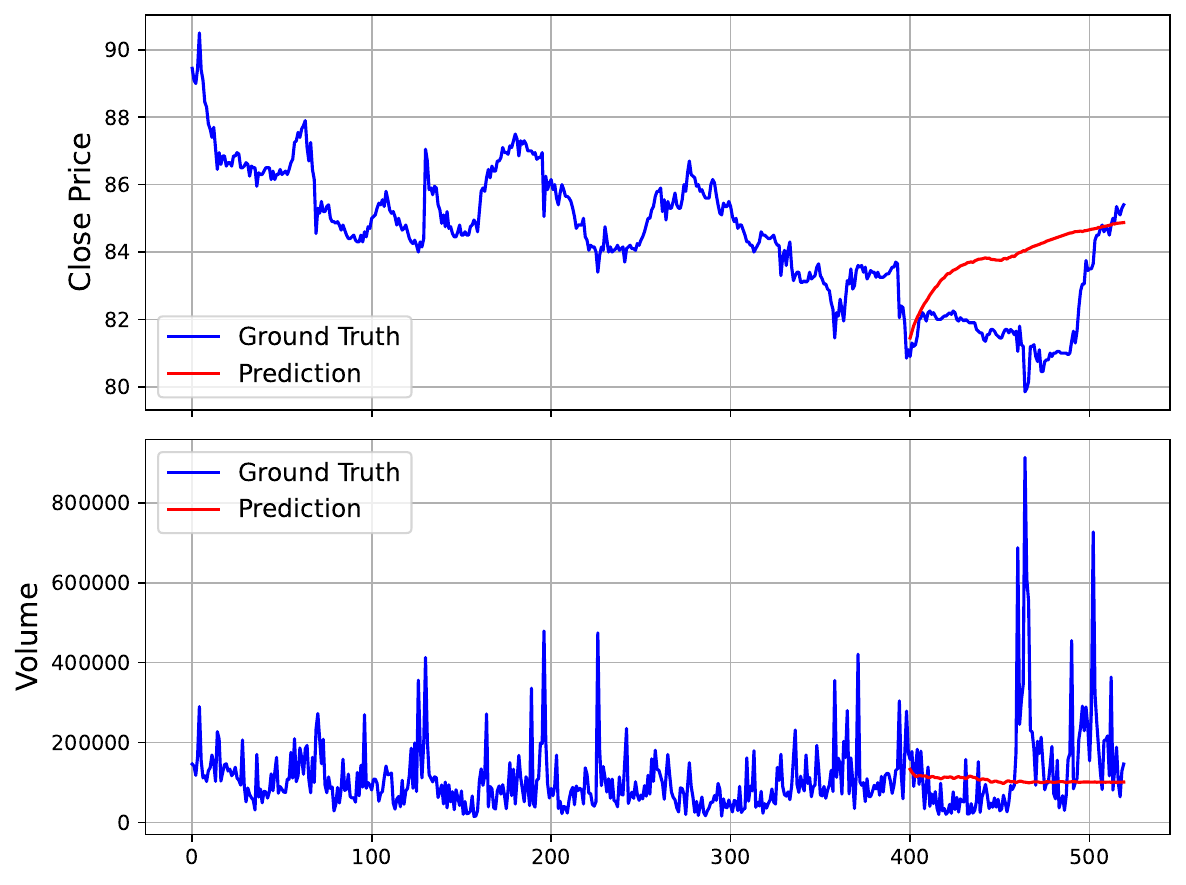}
        \caption{$\text{TimeMOE}_{small}$}
    \end{subfigure}
    \hfill
    \begin{subfigure}[b]{0.33\textwidth}
        \centering
        \includegraphics[width=\linewidth]{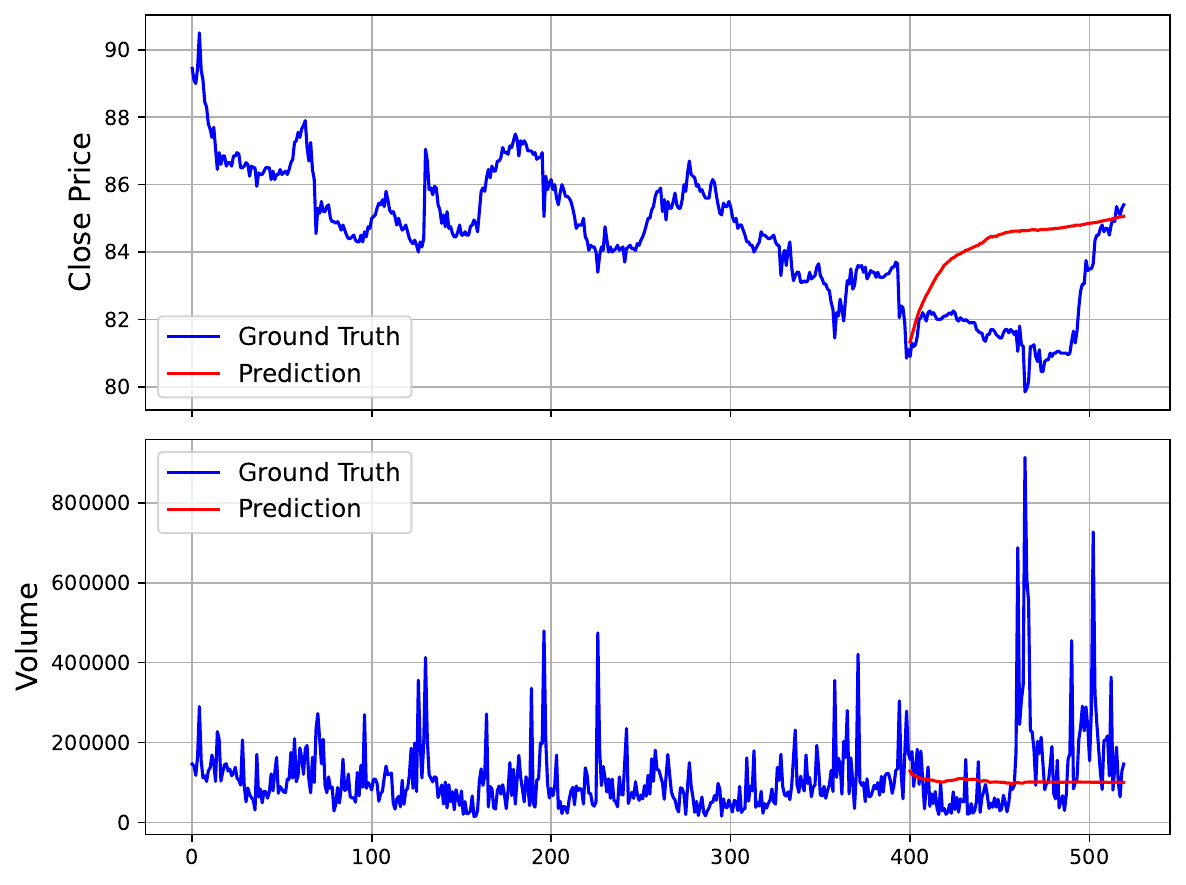}
        \caption{$\text{TimeMOE}_{large}$}
    \end{subfigure}
    \hfill
    \begin{subfigure}[b]{0.33\textwidth}
        \centering
        \includegraphics[width=\linewidth]{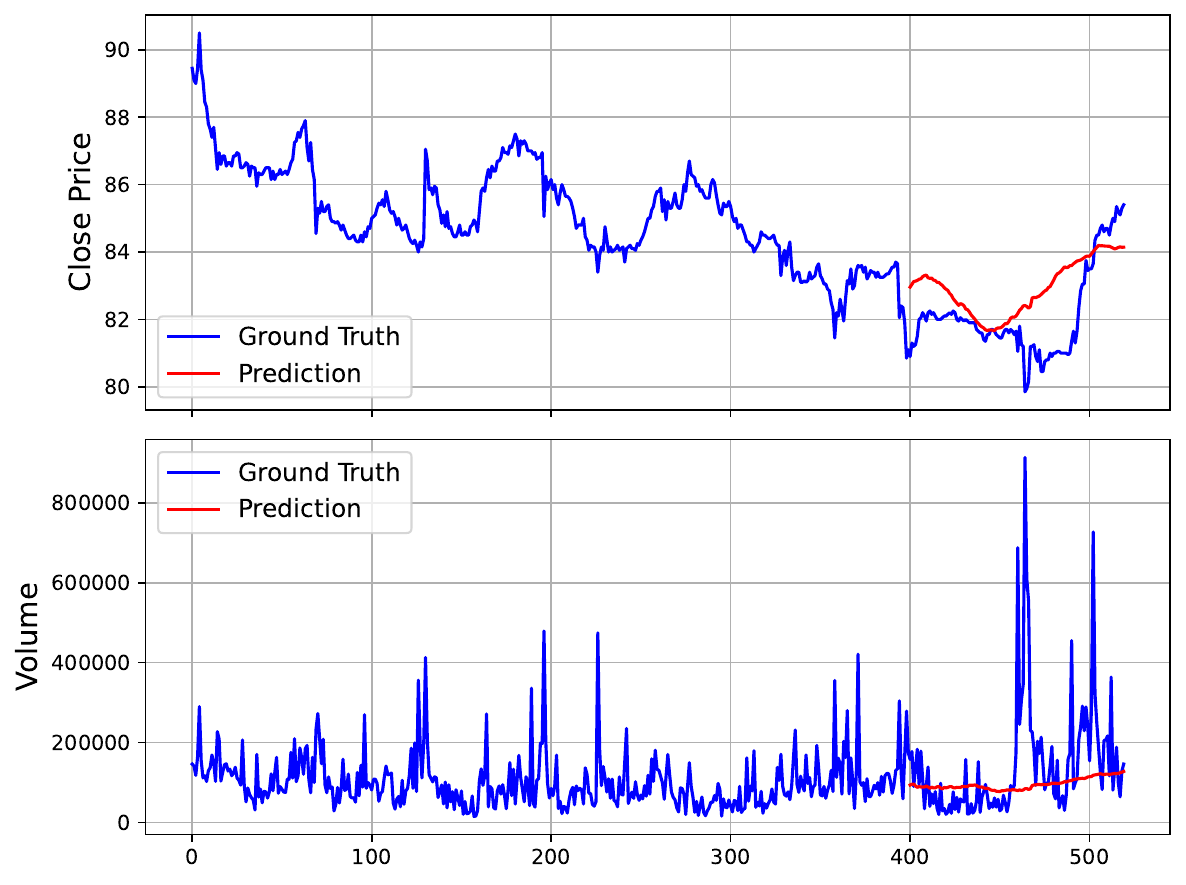}
        \caption{TimesFM}
    \end{subfigure}

    \vspace{0.1cm}

    \begin{subfigure}[b]{0.33\textwidth}
        \centering
        \includegraphics[width=\linewidth]{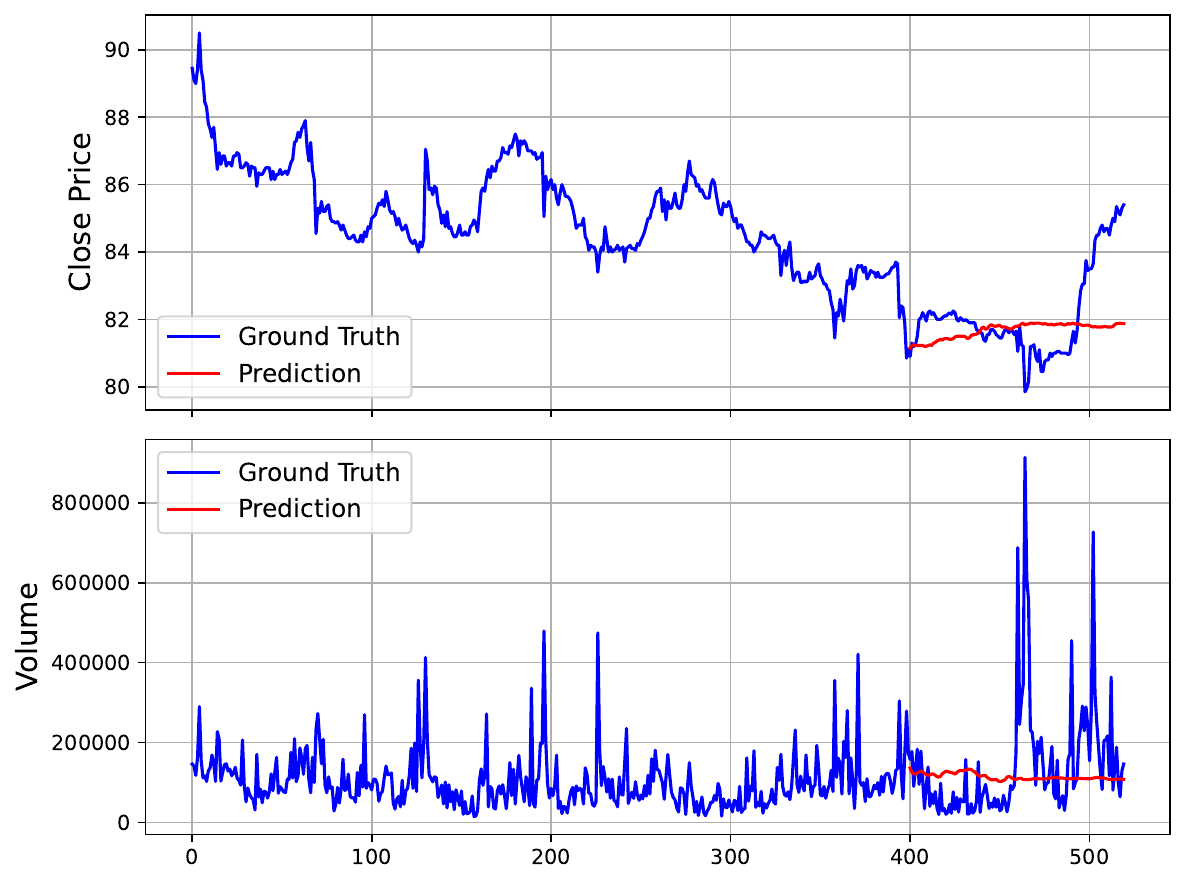}
        \caption{$\text{Chronos}_{small}$}
    \end{subfigure}
    \hfill
    \begin{subfigure}[b]{0.33\textwidth}
        \centering
        \includegraphics[width=\linewidth]{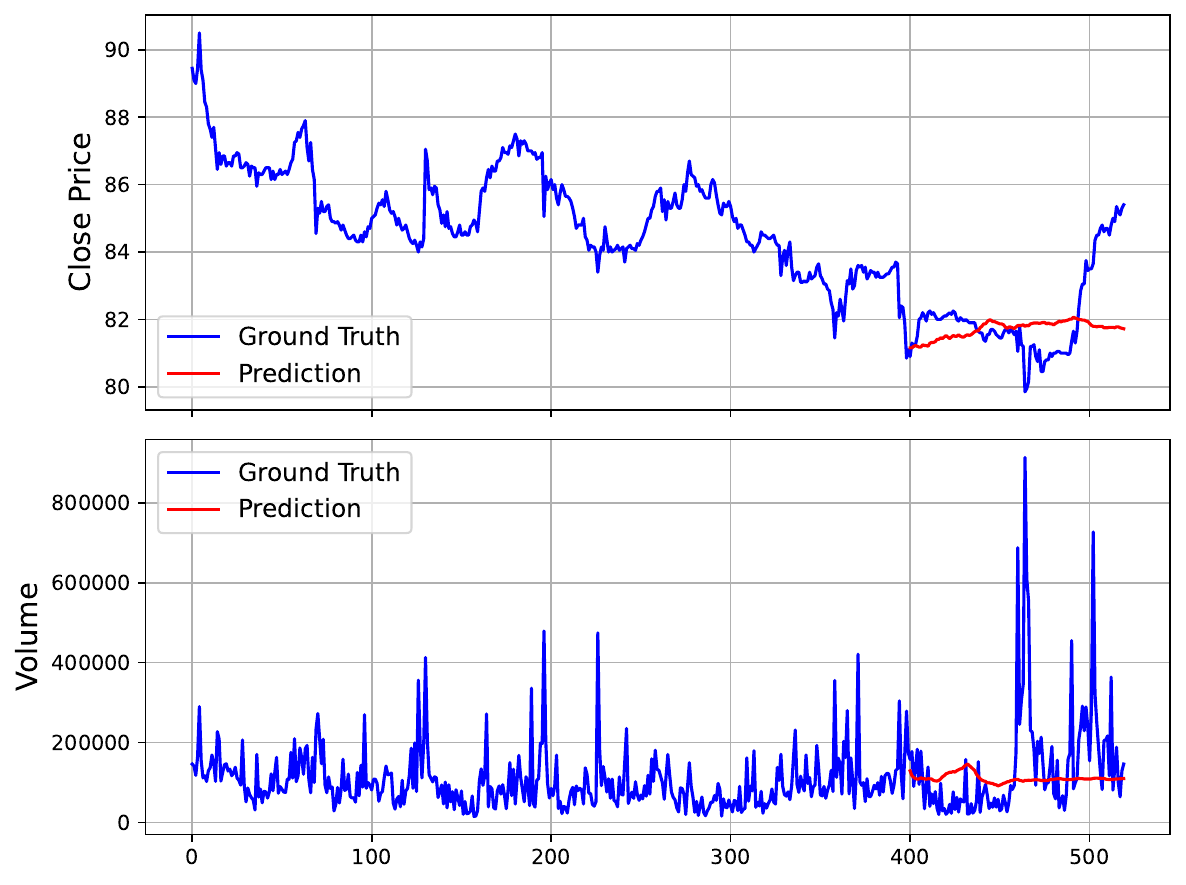}
        \caption{$\text{Chronos}_{base}$}
    \end{subfigure}
    \hfill
    \begin{subfigure}[b]{0.33\textwidth}
        \centering
        \includegraphics[width=\linewidth]{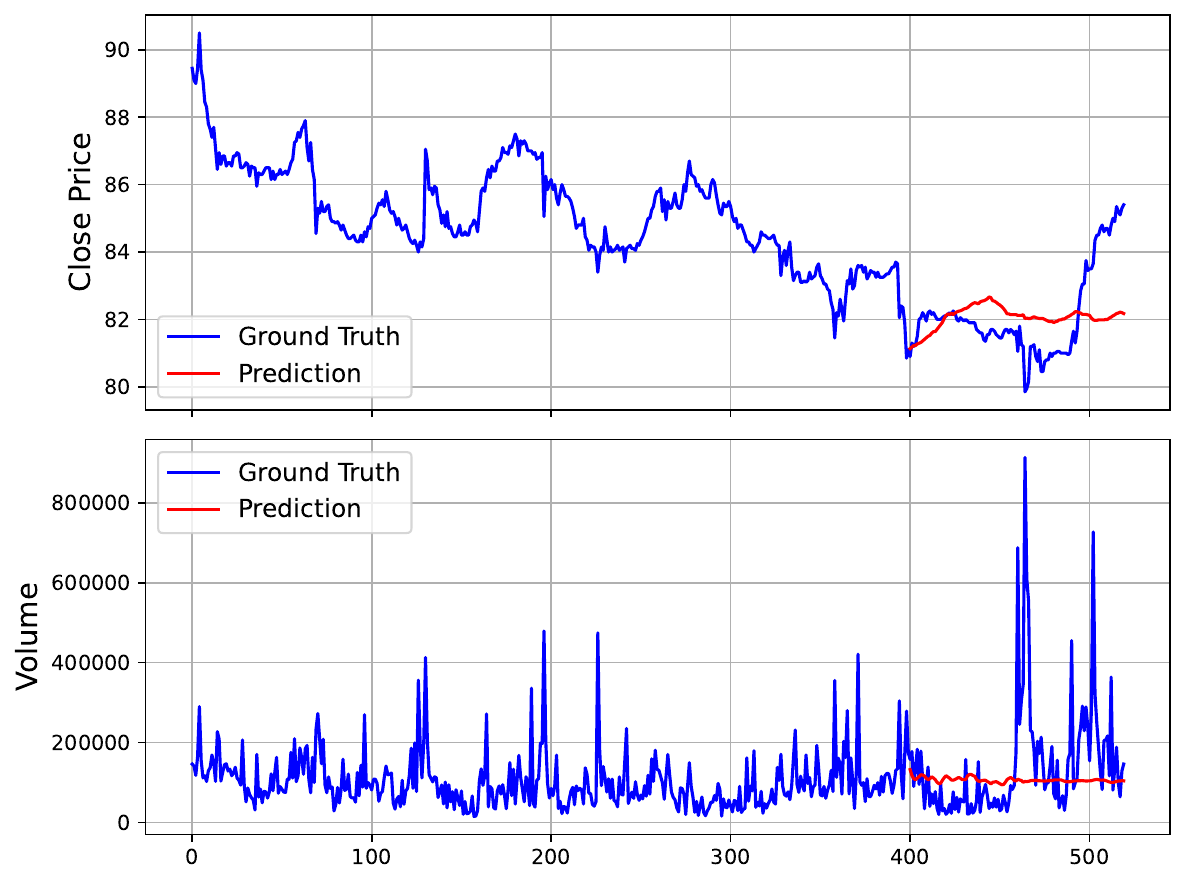}
        \caption{$\text{Chronos}_{large}$}
    \end{subfigure}
    
    \vspace{0.1cm}

    \begin{subfigure}[b]{0.33\textwidth}
        \centering
        \includegraphics[width=\linewidth]{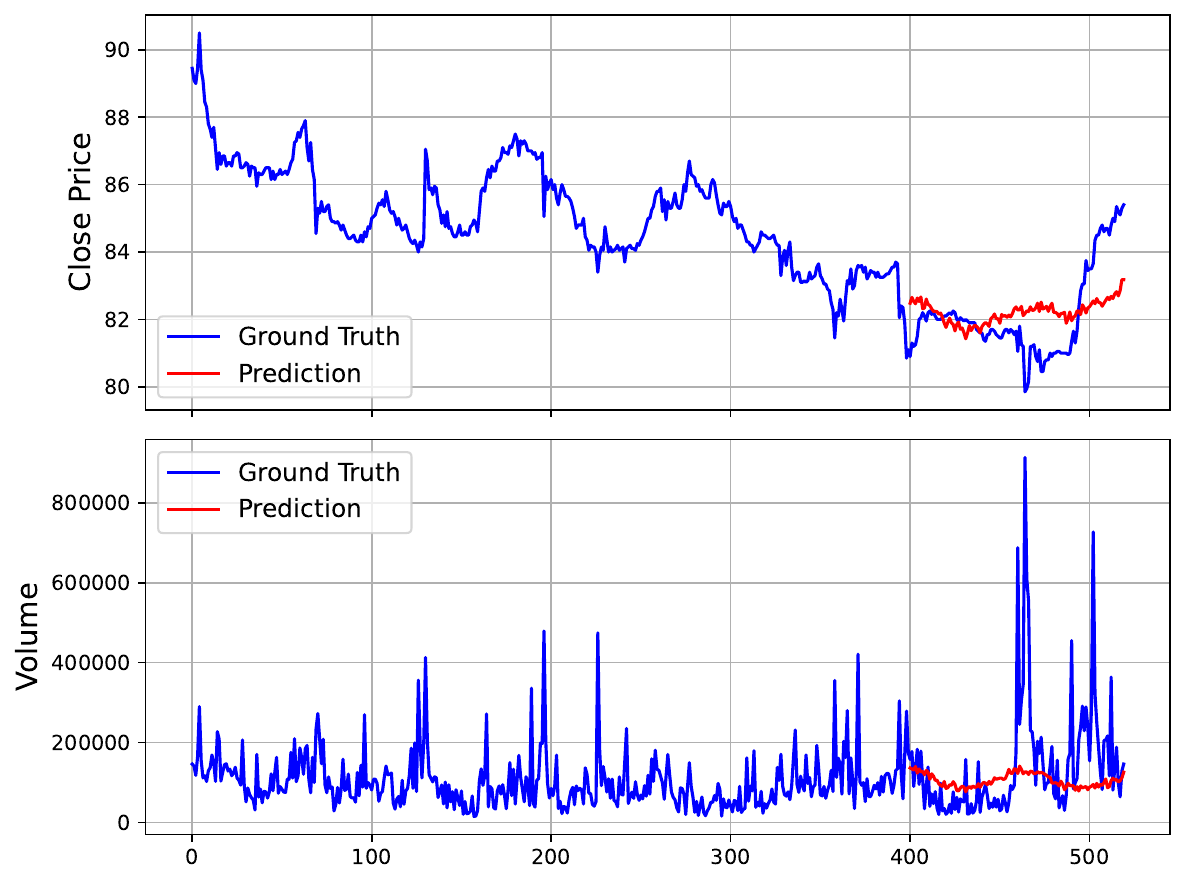}
        \caption{iTransformer}
    \end{subfigure}
    \hfill
    \begin{subfigure}[b]{0.33\textwidth}
        \centering
        \includegraphics[width=\linewidth]{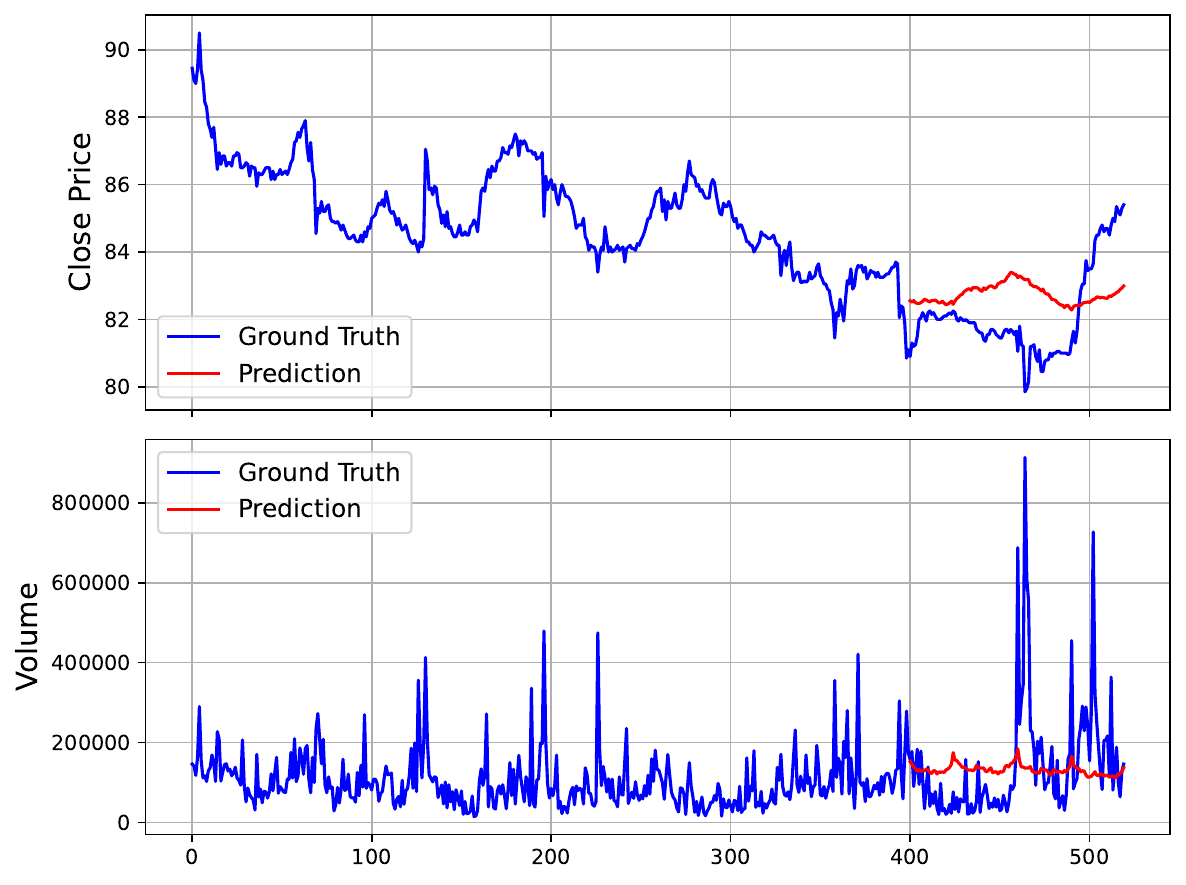}
        \caption{DLinear}
    \end{subfigure}
    \hfill
    \begin{subfigure}[b]{0.33\textwidth}
        \centering
        \includegraphics[width=\linewidth]{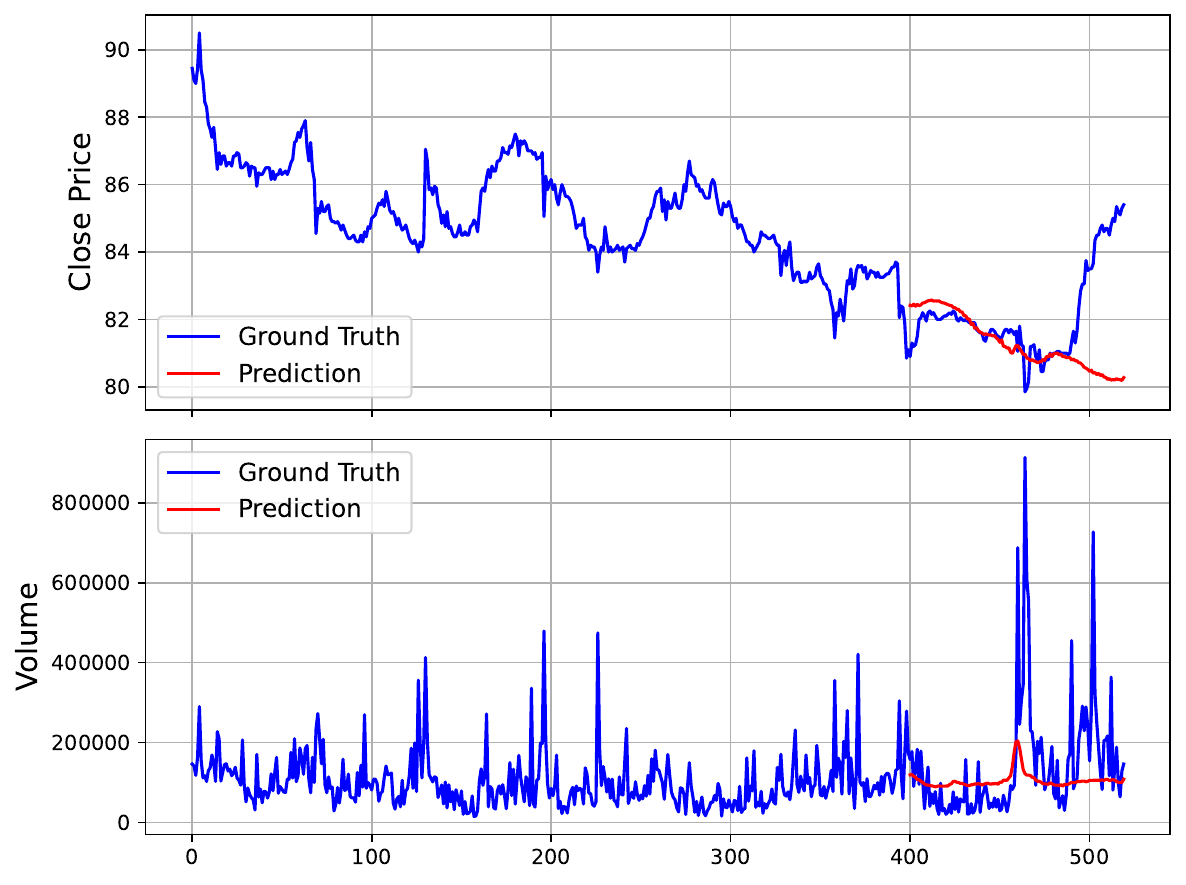}
        \caption{TimesNet}
    \end{subfigure}

    \caption{Forecasting results for the `Close Price' and `Volume' of Pop Mart (HKEX: 09992), based on 5-minute K-line data. The model uses a 400-step look-back window to predict a 120-step horizon. \textcolor{blue}{Blue} lines represent the ground truths and \textcolor{red}{red} lines are the model's predictions.}
    \label{fig:pred_case_2}
\end{figure*}

\begin{figure*}[ht]
    \centering

    \begin{subfigure}[b]{0.33\textwidth}
        \centering
        \includegraphics[width=\linewidth]{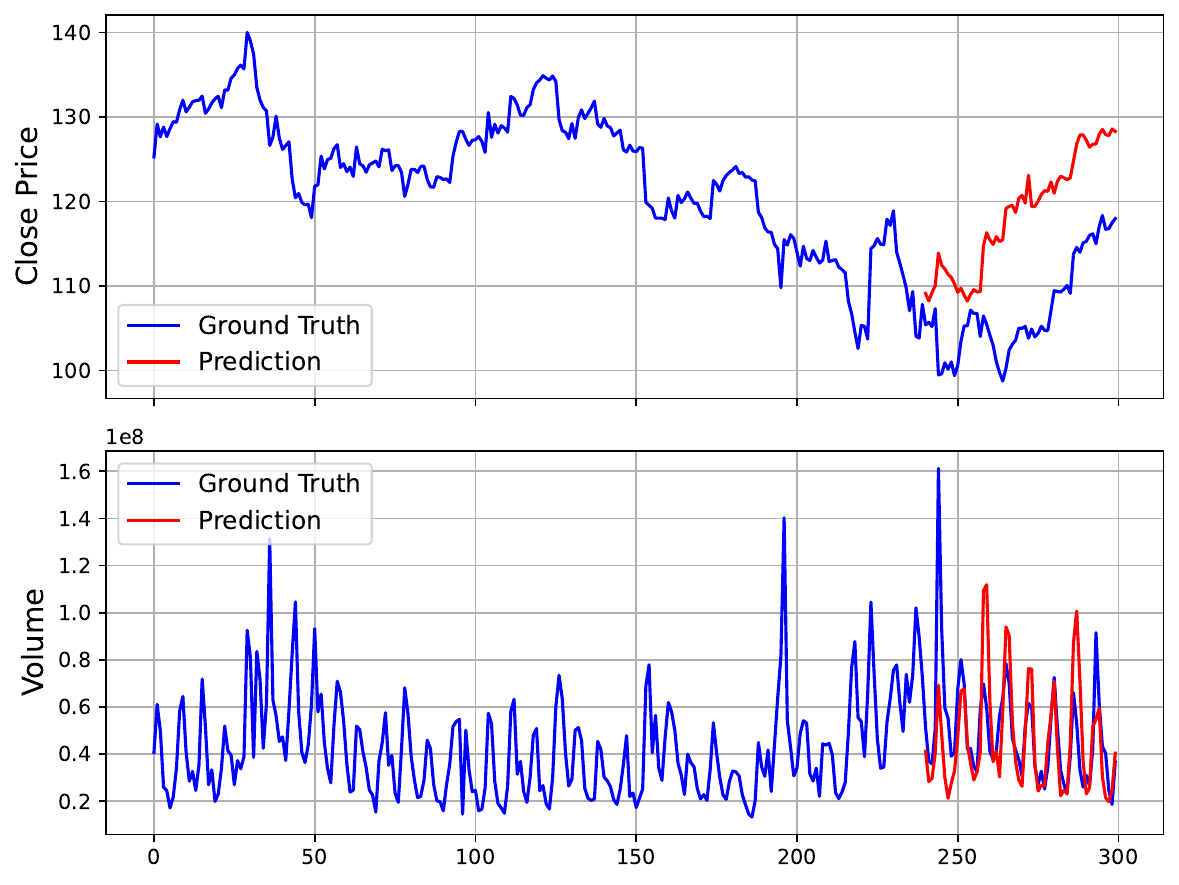}
        \caption{$\text{Kronos}_{small}$}
    \end{subfigure}
    \hfill
    \begin{subfigure}[b]{0.33\textwidth}
        \centering
        \includegraphics[width=\linewidth]{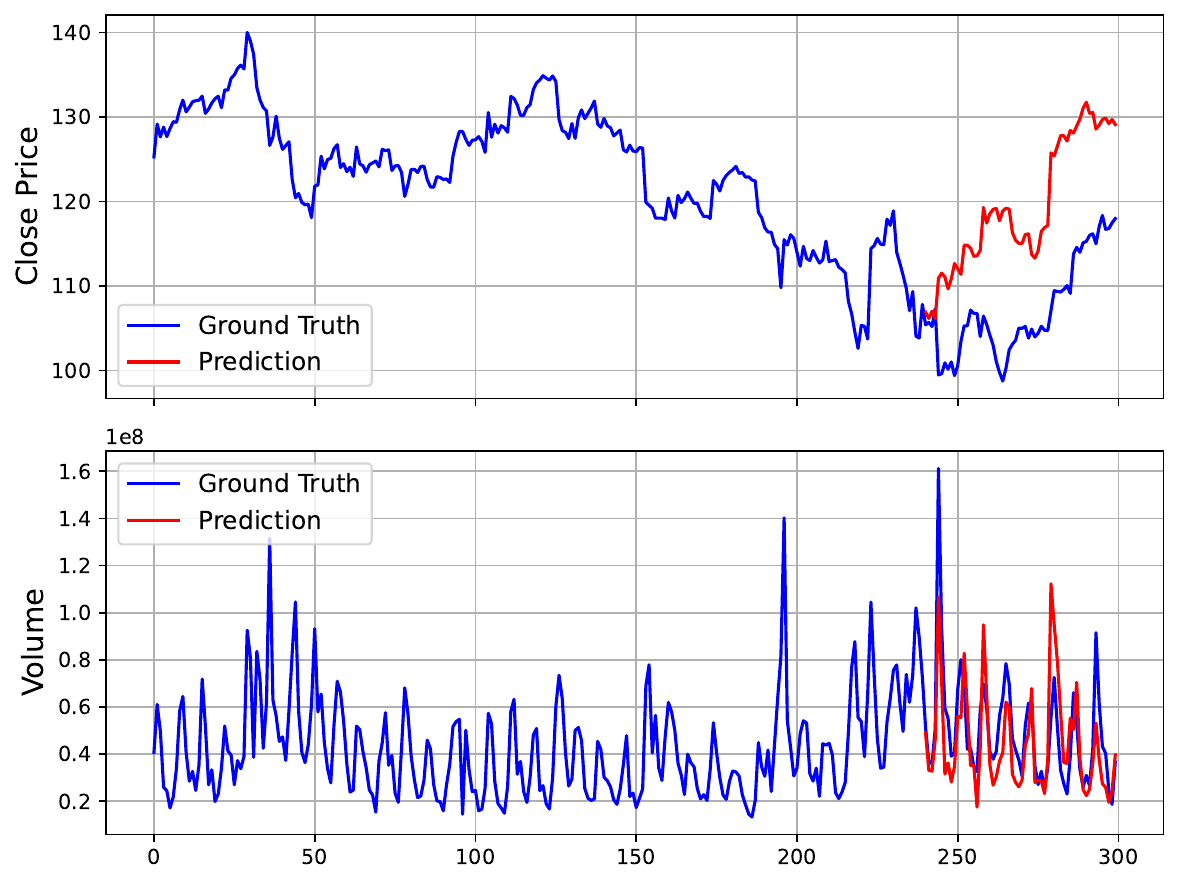}
        \caption{$\text{Kronos}_{base}$}
    \end{subfigure}
    \hfill
    \begin{subfigure}[b]{0.33\textwidth}
        \centering
        \includegraphics[width=\linewidth]{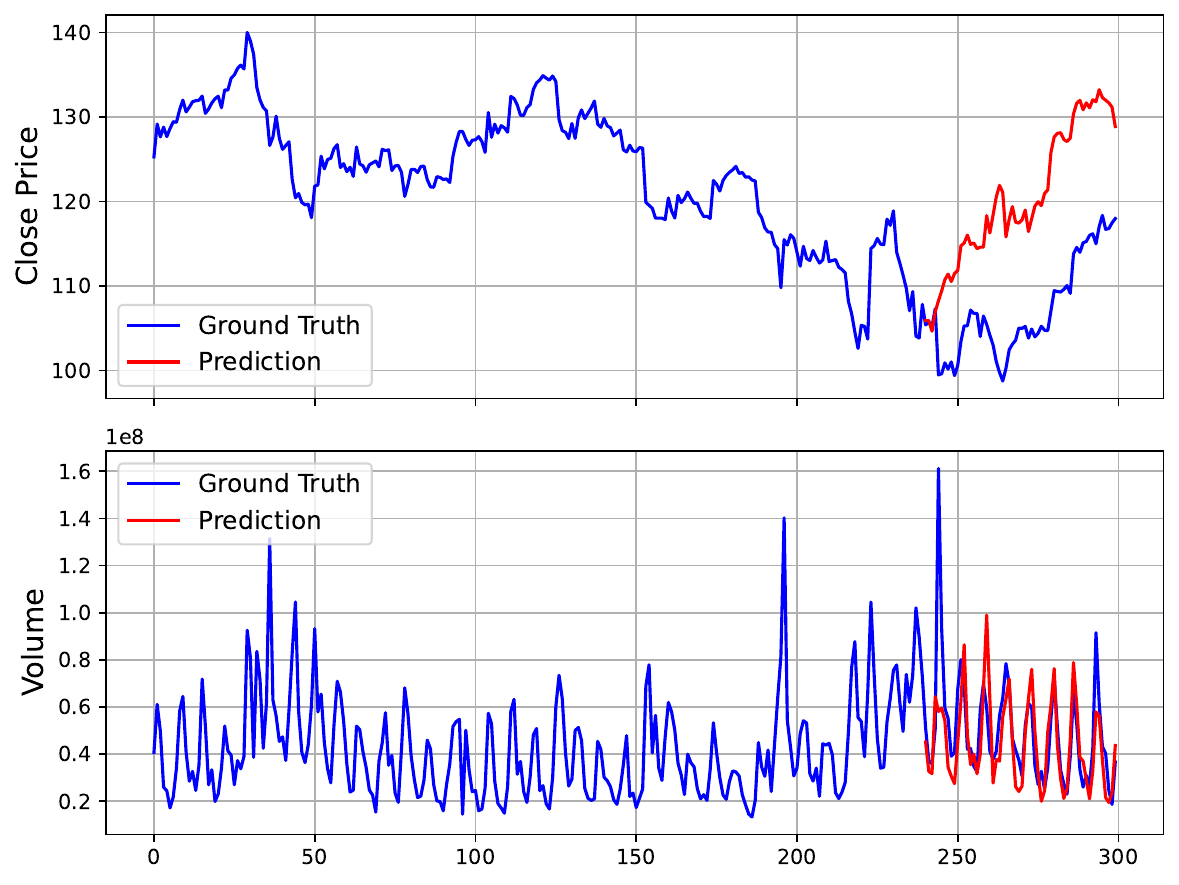}
        \caption{$\text{Kronos}_{large}$}
    \end{subfigure}
    
    \vspace{0.1cm}

    \begin{subfigure}[b]{0.33\textwidth}
        \centering
        \includegraphics[width=\linewidth]{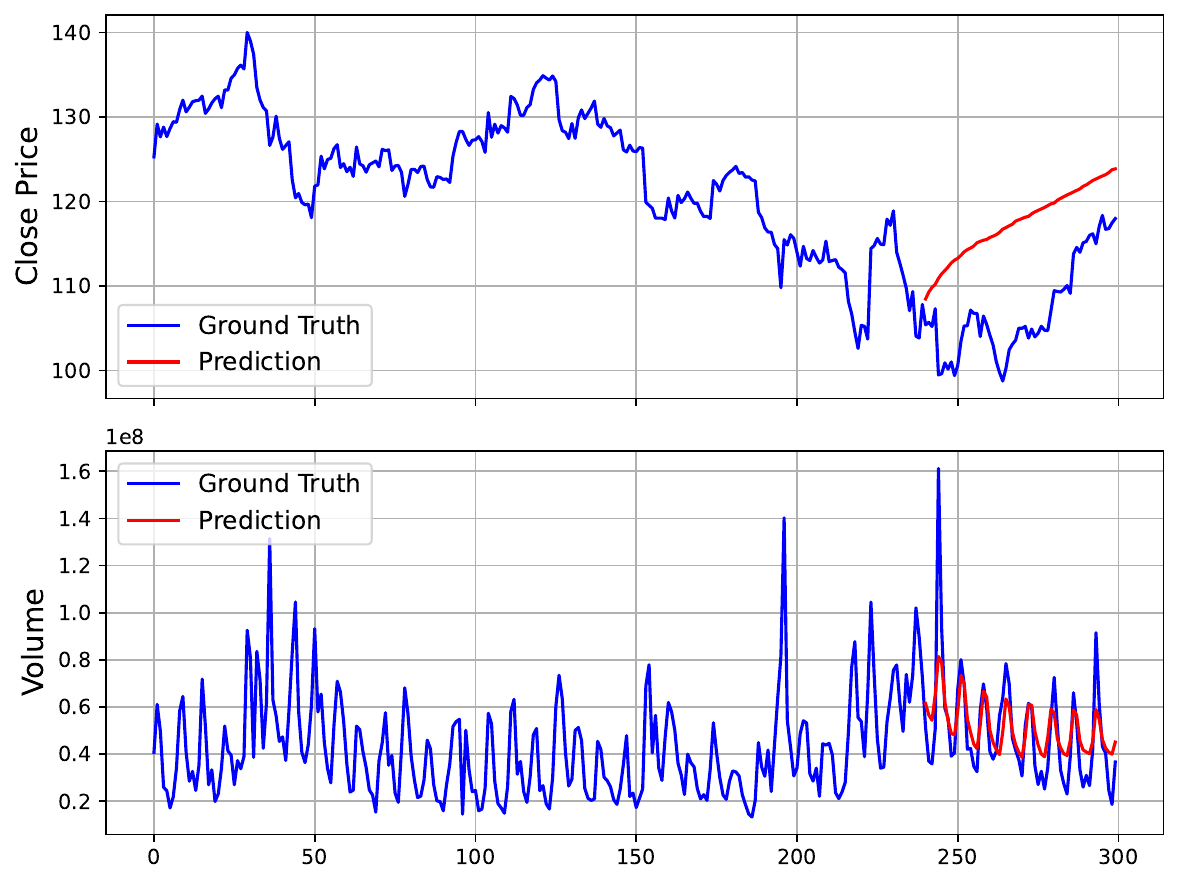}
        \caption{$\text{TimeMOE}_{small}$}
    \end{subfigure}
    \hfill
    \begin{subfigure}[b]{0.33\textwidth}
        \centering
        \includegraphics[width=\linewidth]{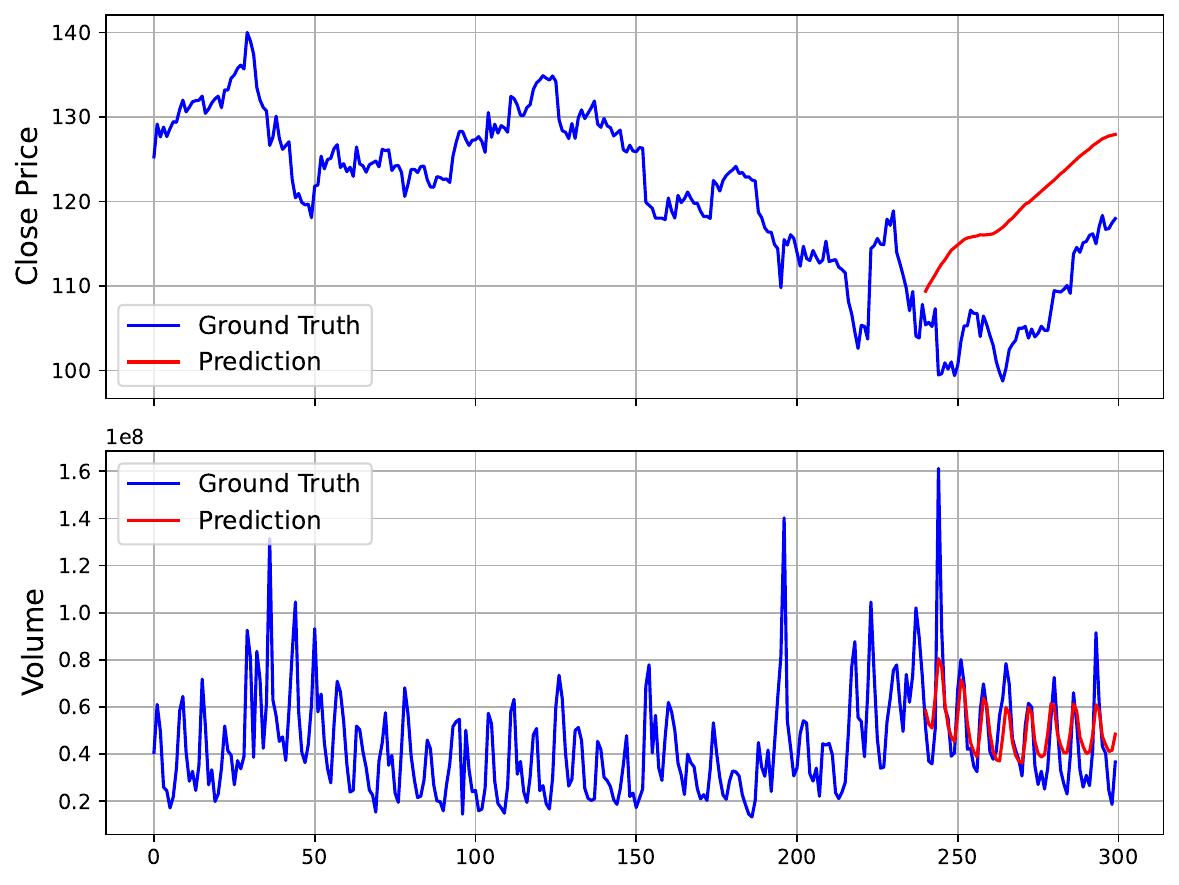}
        \caption{$\text{TimeMOE}_{large}$}
    \end{subfigure}
    \hfill
    \begin{subfigure}[b]{0.33\textwidth}
        \centering
        \includegraphics[width=\linewidth]{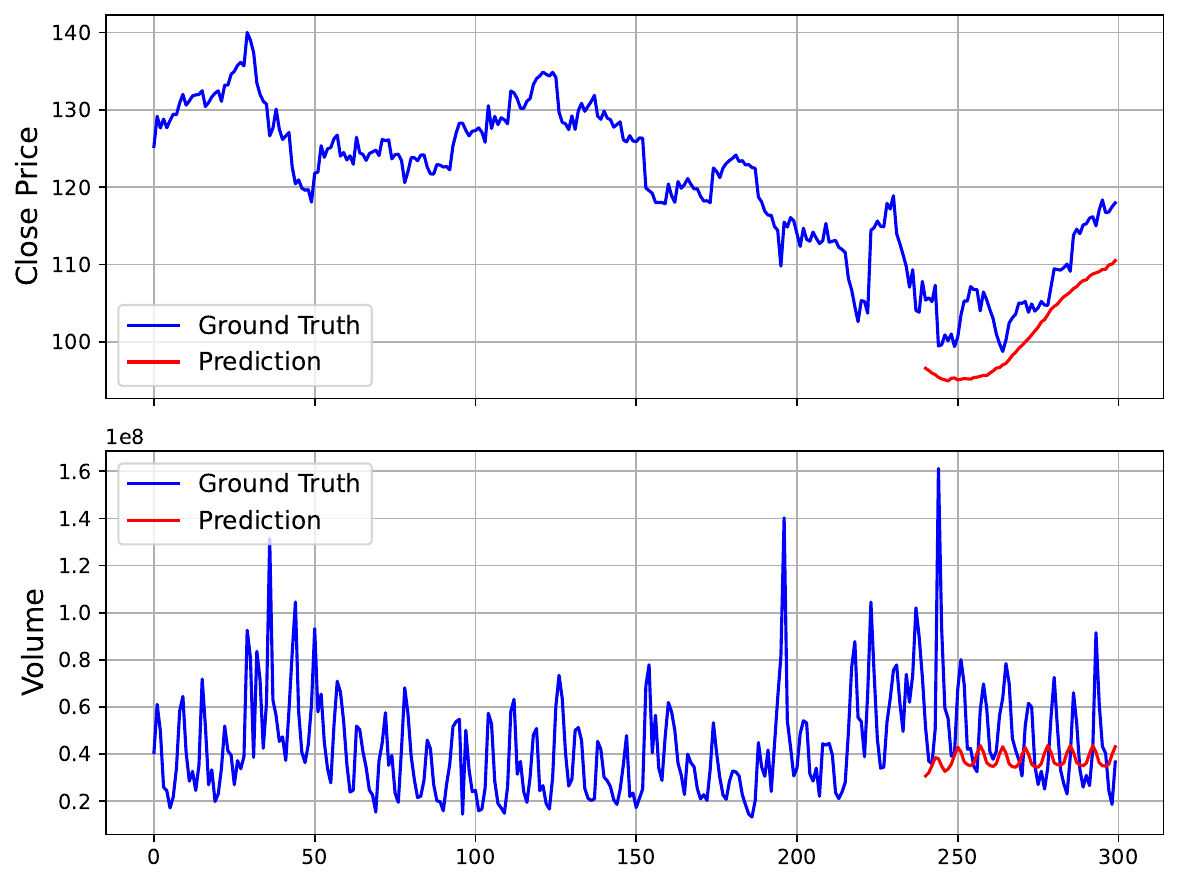}
        \caption{TimesFM}
    \end{subfigure}

    \vspace{0.1cm}

    \begin{subfigure}[b]{0.33\textwidth}
        \centering
        \includegraphics[width=\linewidth]{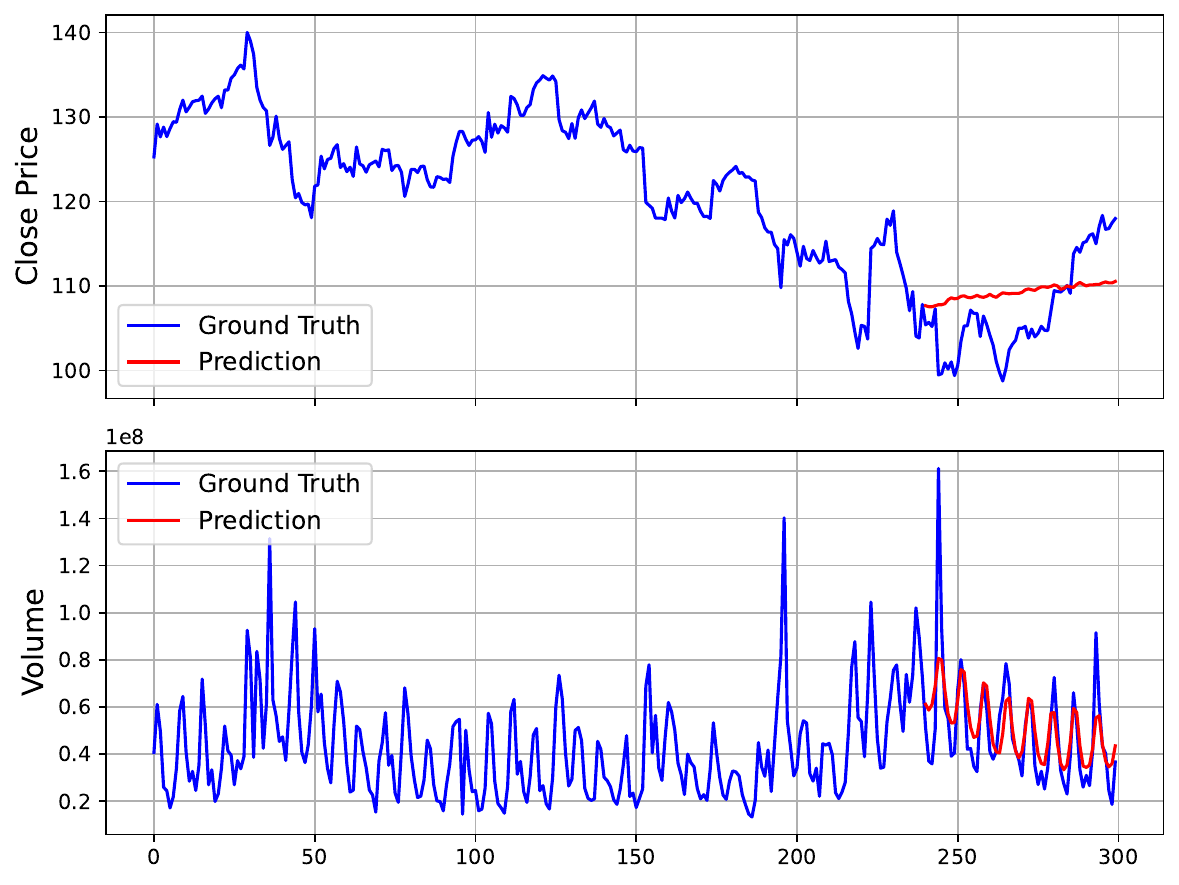}
        \caption{$\text{Chronos}_{small}$}
    \end{subfigure}
    \hfill
    \begin{subfigure}[b]{0.33\textwidth}
        \centering
        \includegraphics[width=\linewidth]{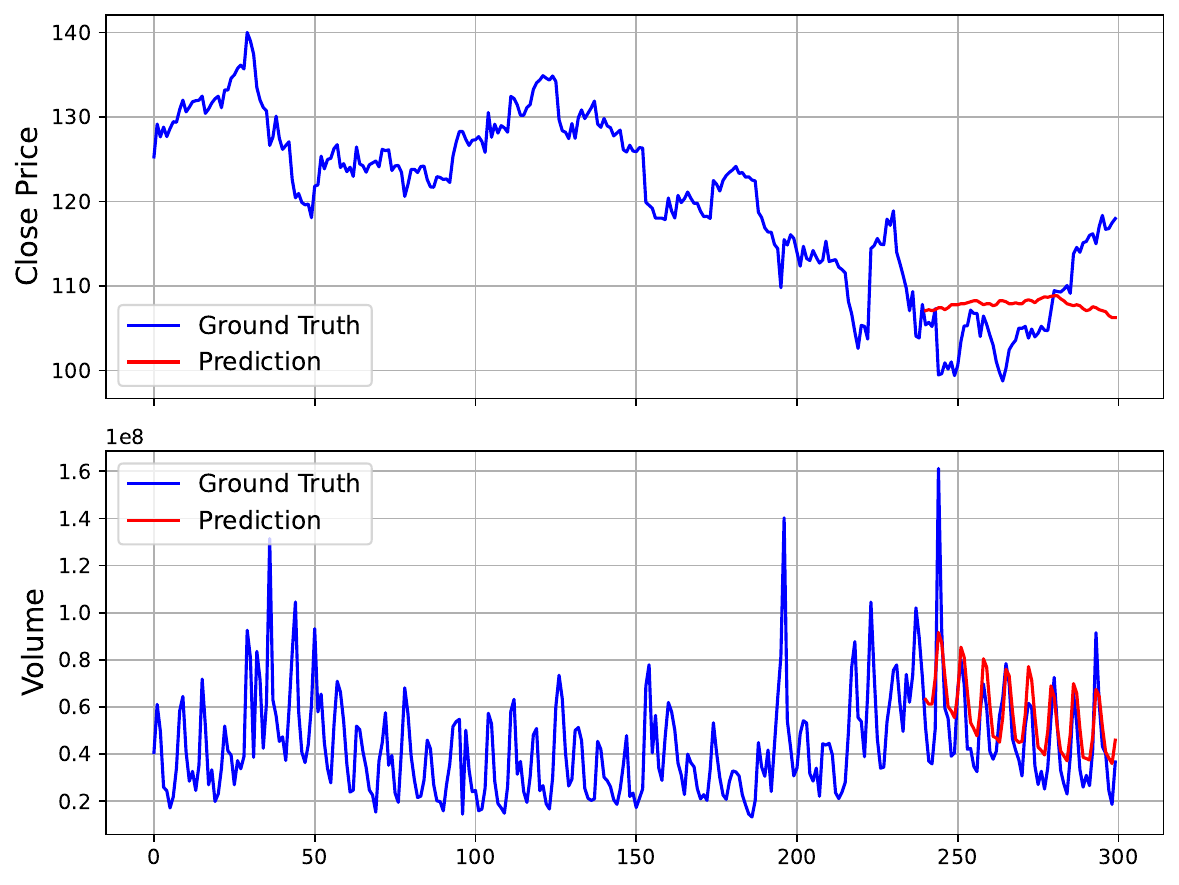}
        \caption{$\text{Chronos}_{base}$}
    \end{subfigure}
    \hfill
    \begin{subfigure}[b]{0.33\textwidth}
        \centering
        \includegraphics[width=\linewidth]{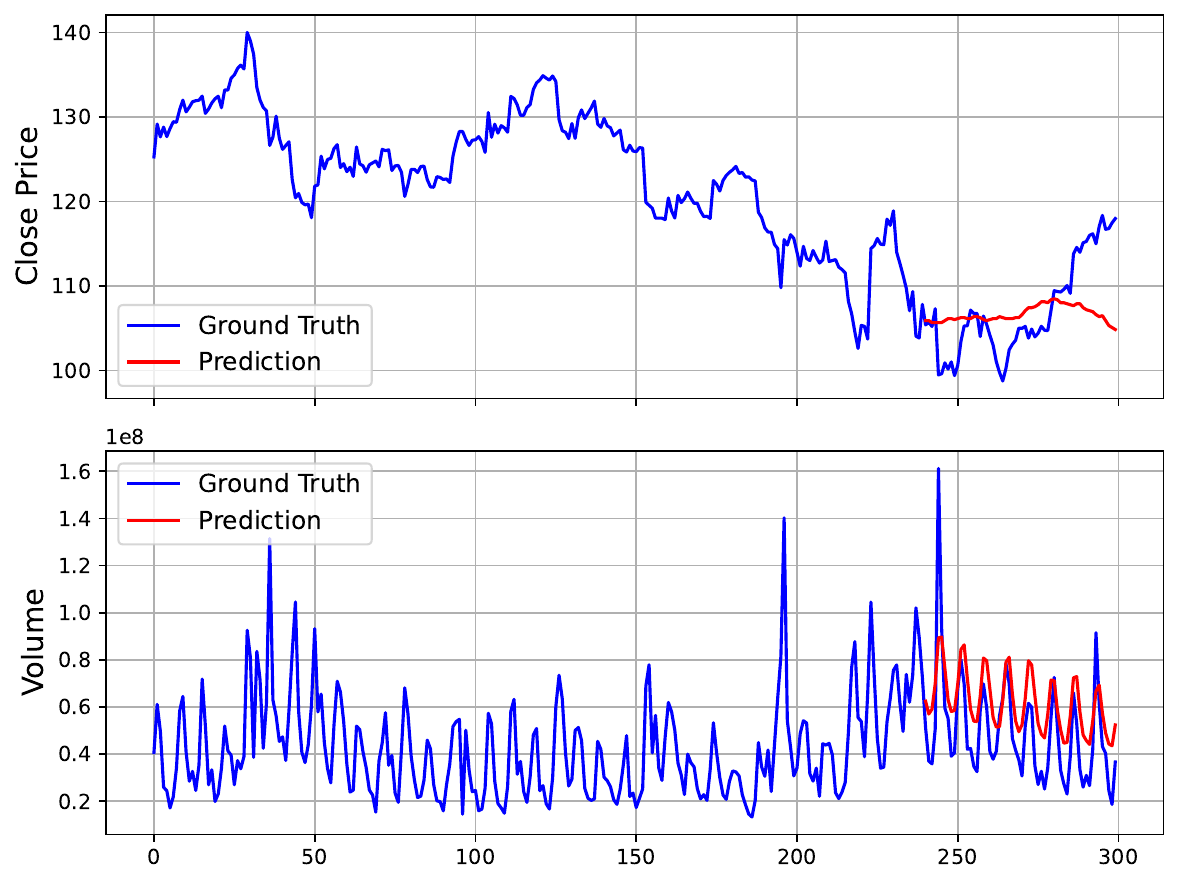}
        \caption{$\text{Chronos}_{large}$}
    \end{subfigure}
    
    \vspace{0.1cm}

    \begin{subfigure}[b]{0.33\textwidth}
        \centering
        \includegraphics[width=\linewidth]{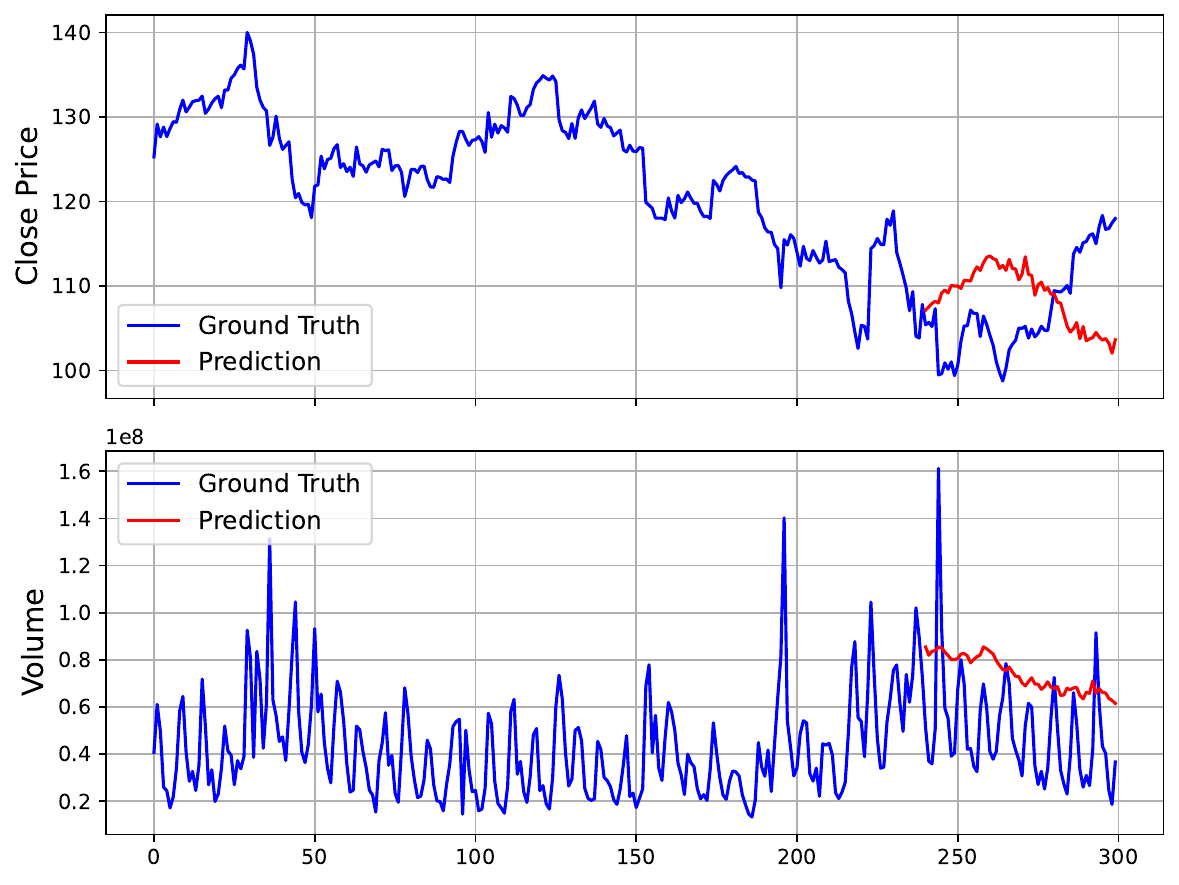}
        \caption{iTransformer}
    \end{subfigure}
    \hfill
    \begin{subfigure}[b]{0.33\textwidth}
        \centering
        \includegraphics[width=\linewidth]{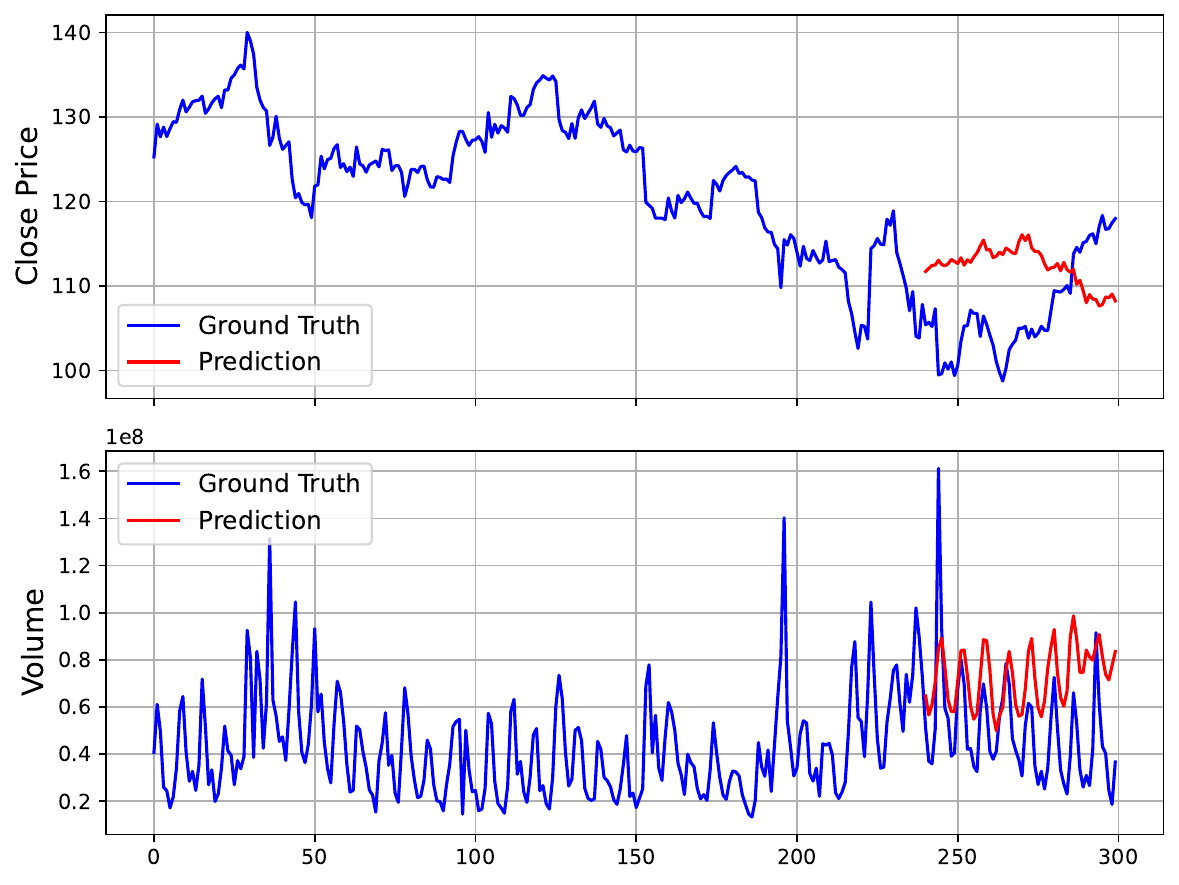}
        \caption{DLinear}
    \end{subfigure}
    \hfill
    \begin{subfigure}[b]{0.33\textwidth}
        \centering
        \includegraphics[width=\linewidth]{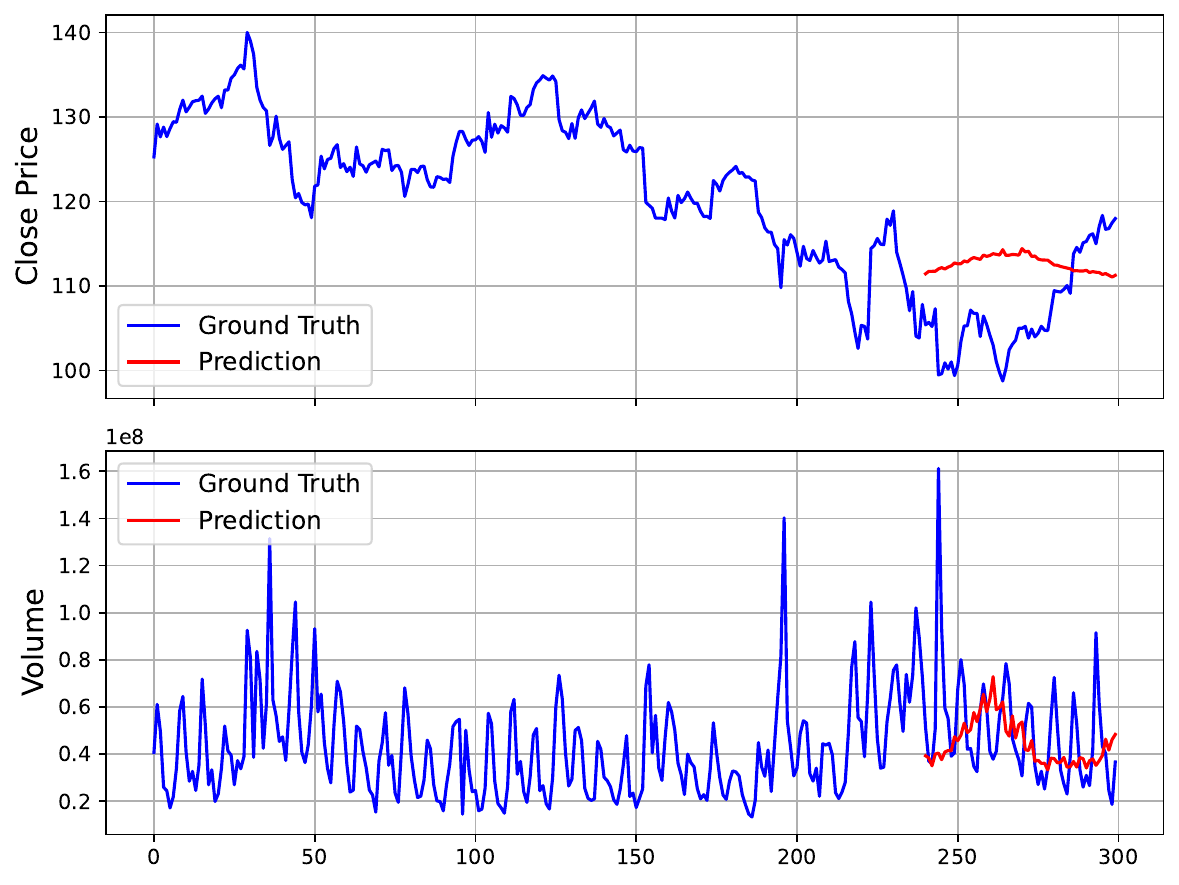}
        \caption{TimesNet}
    \end{subfigure}

    \caption{Forecasting results for the `Close Price' and `Volume' of NVIDIA (NASDAQ: NVDA), based on 1-hour K-line data. The model uses a 240-step look-back window to predict a 60-step horizon. \textcolor{blue}{Blue} lines represent the ground truths and \textcolor{red}{red} lines are the model's predictions.}
    \label{fig:pred_case_3}
\end{figure*}

\begin{figure*}[ht]
    \centering

    \begin{subfigure}[b]{0.33\textwidth}
        \centering
        \includegraphics[width=\linewidth]{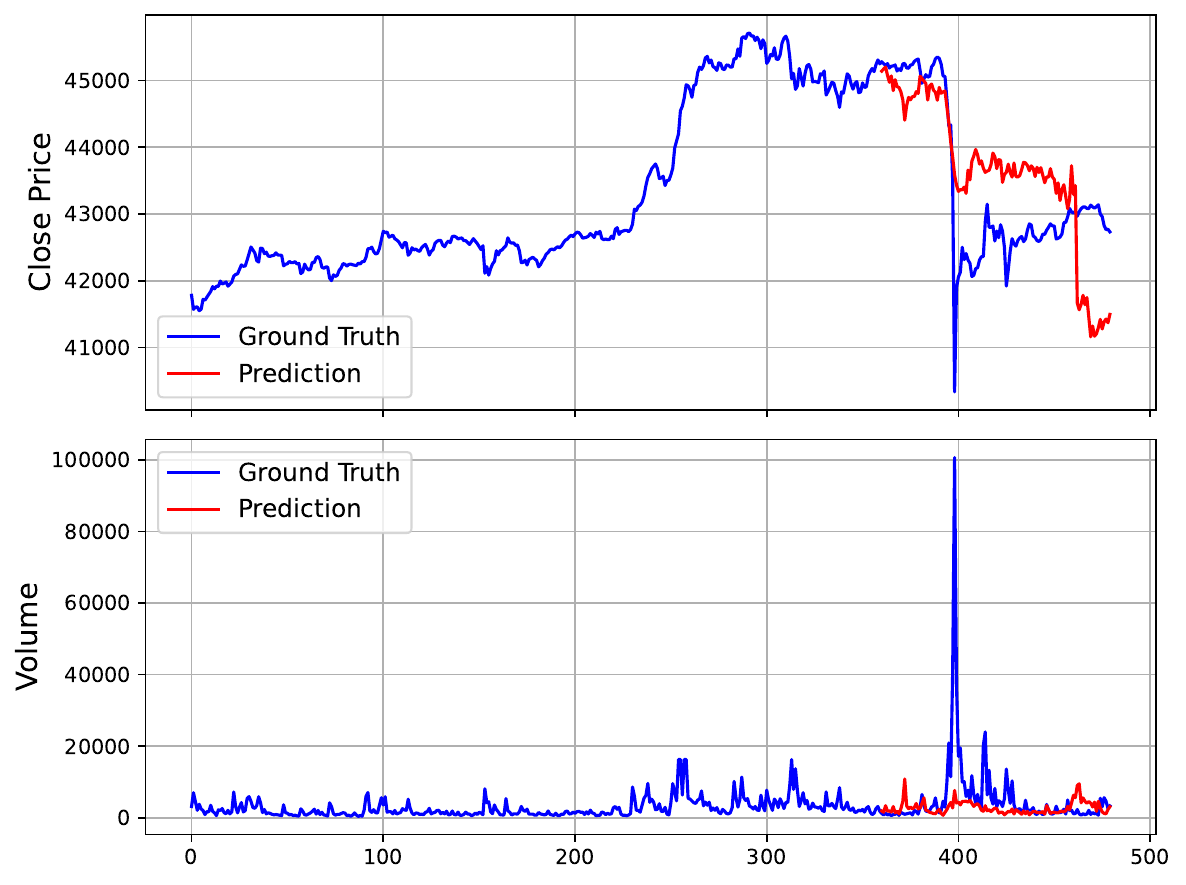}
        \caption{$\text{Kronos}_{small}$}
    \end{subfigure}
    \hfill
    \begin{subfigure}[b]{0.33\textwidth}
        \centering
        \includegraphics[width=\linewidth]{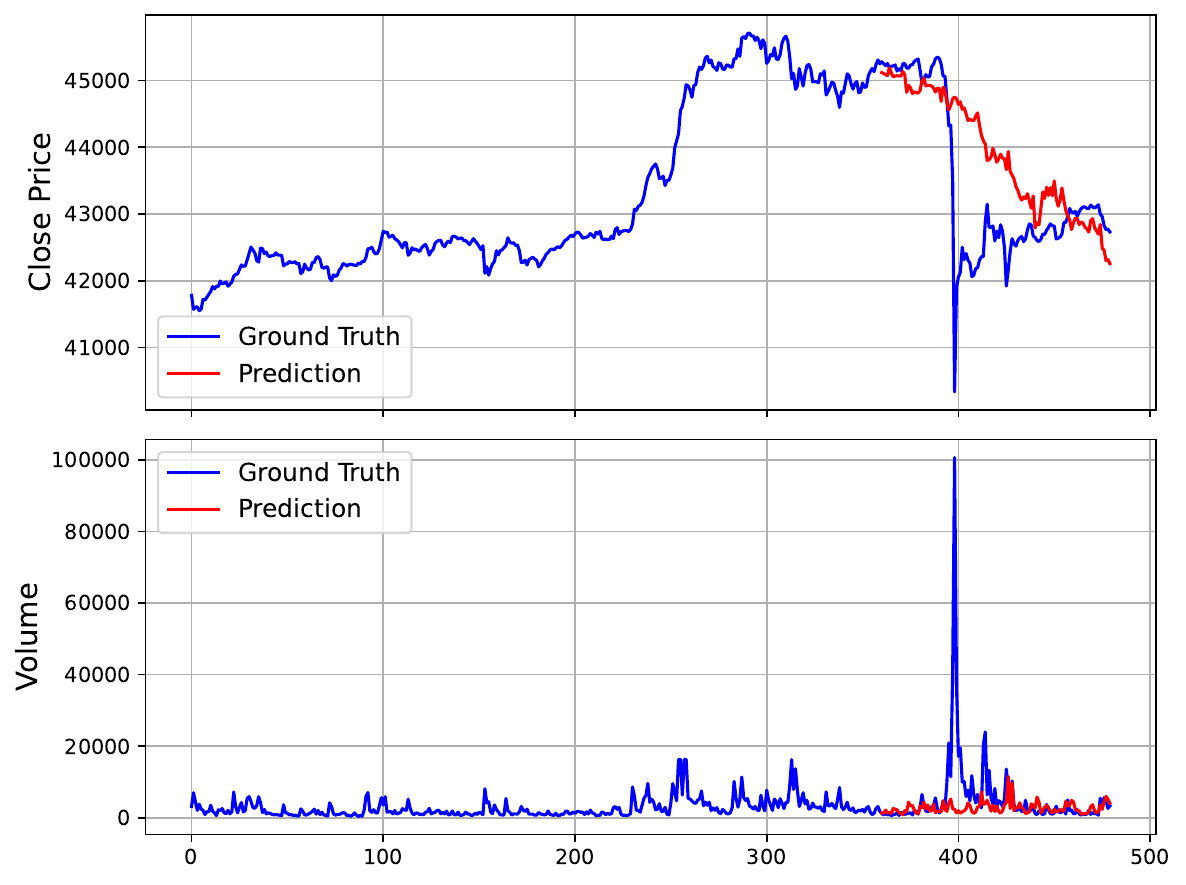}
        \caption{$\text{Kronos}_{base}$}
    \end{subfigure}
    \hfill
    \begin{subfigure}[b]{0.33\textwidth}
        \centering
        \includegraphics[width=\linewidth]{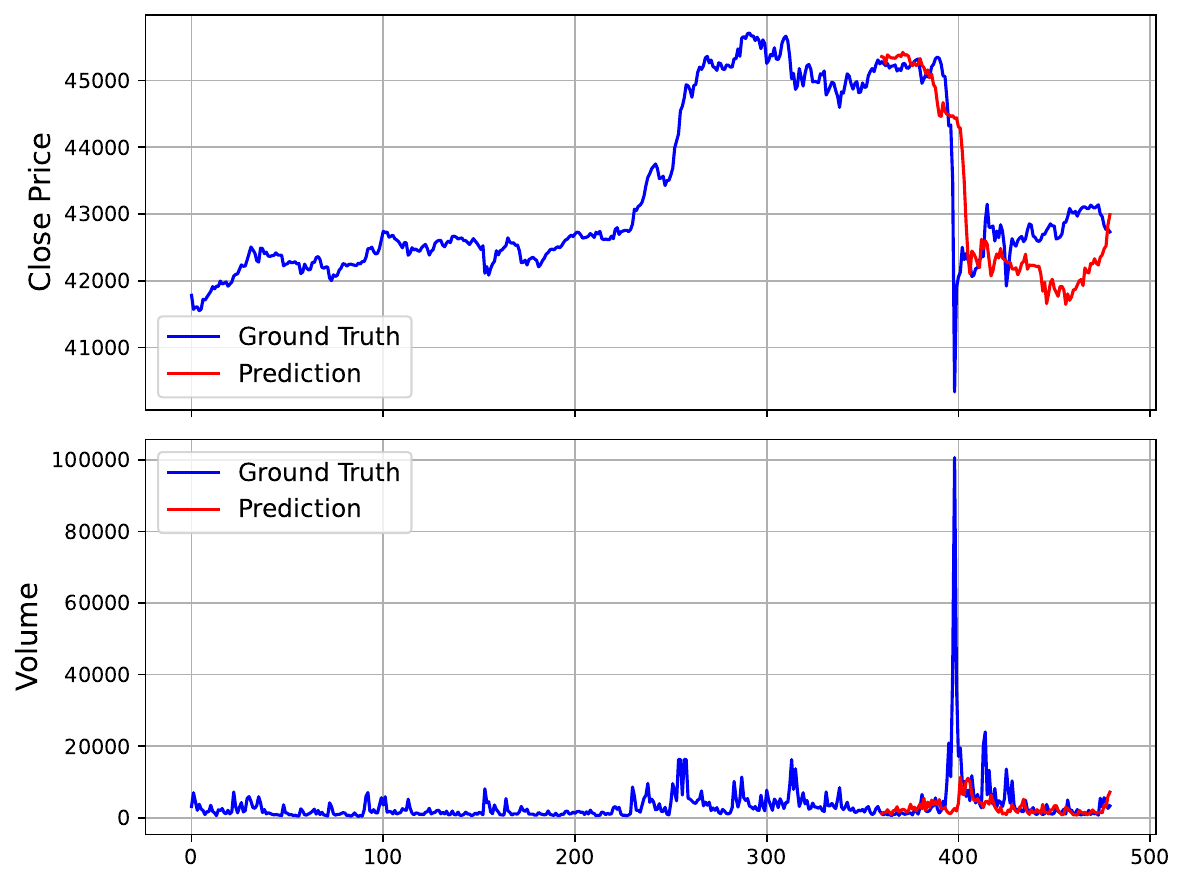}
        \caption{$\text{Kronos}_{large}$}
    \end{subfigure}
    
    \vspace{0.1cm}

    \begin{subfigure}[b]{0.33\textwidth}
        \centering
        \includegraphics[width=\linewidth]{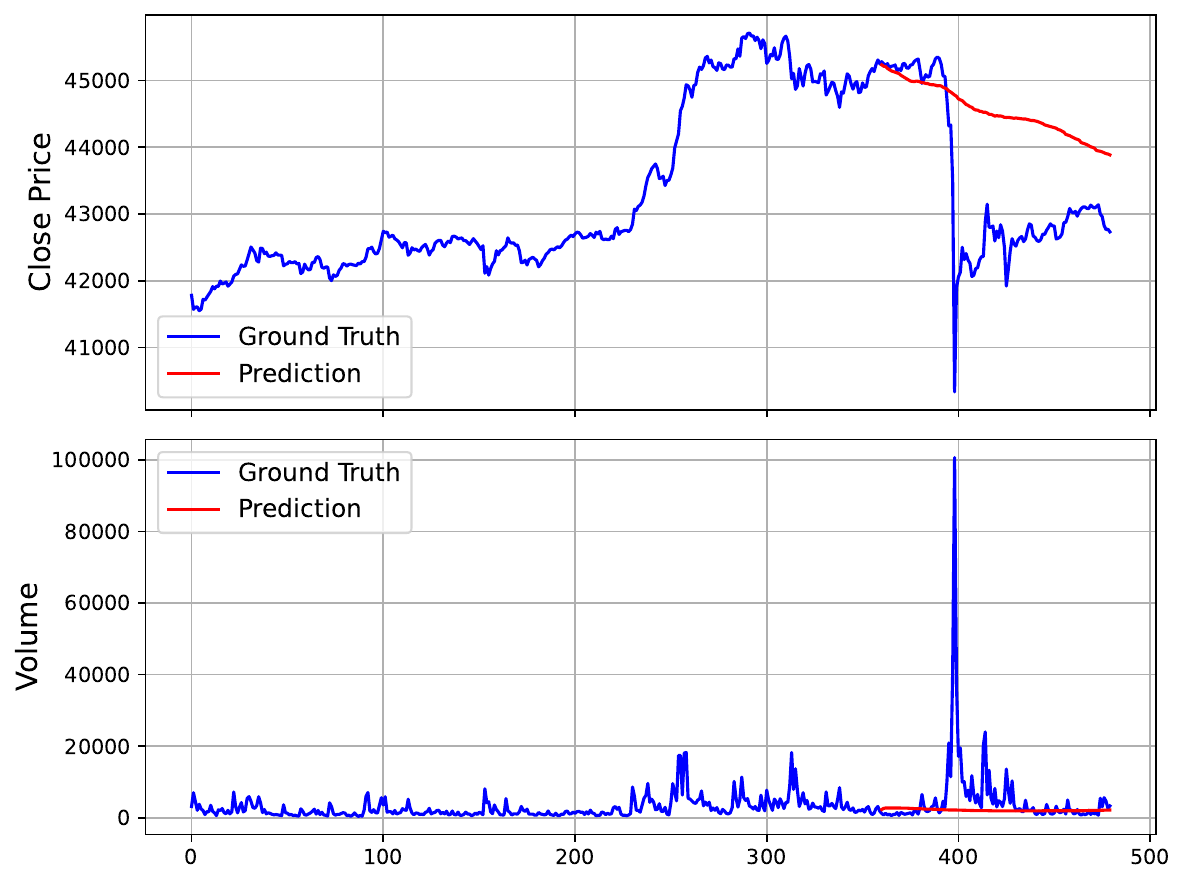}
        \caption{$\text{TimeMOE}_{small}$}
    \end{subfigure}
    \hfill
    \begin{subfigure}[b]{0.33\textwidth}
        \centering
        \includegraphics[width=\linewidth]{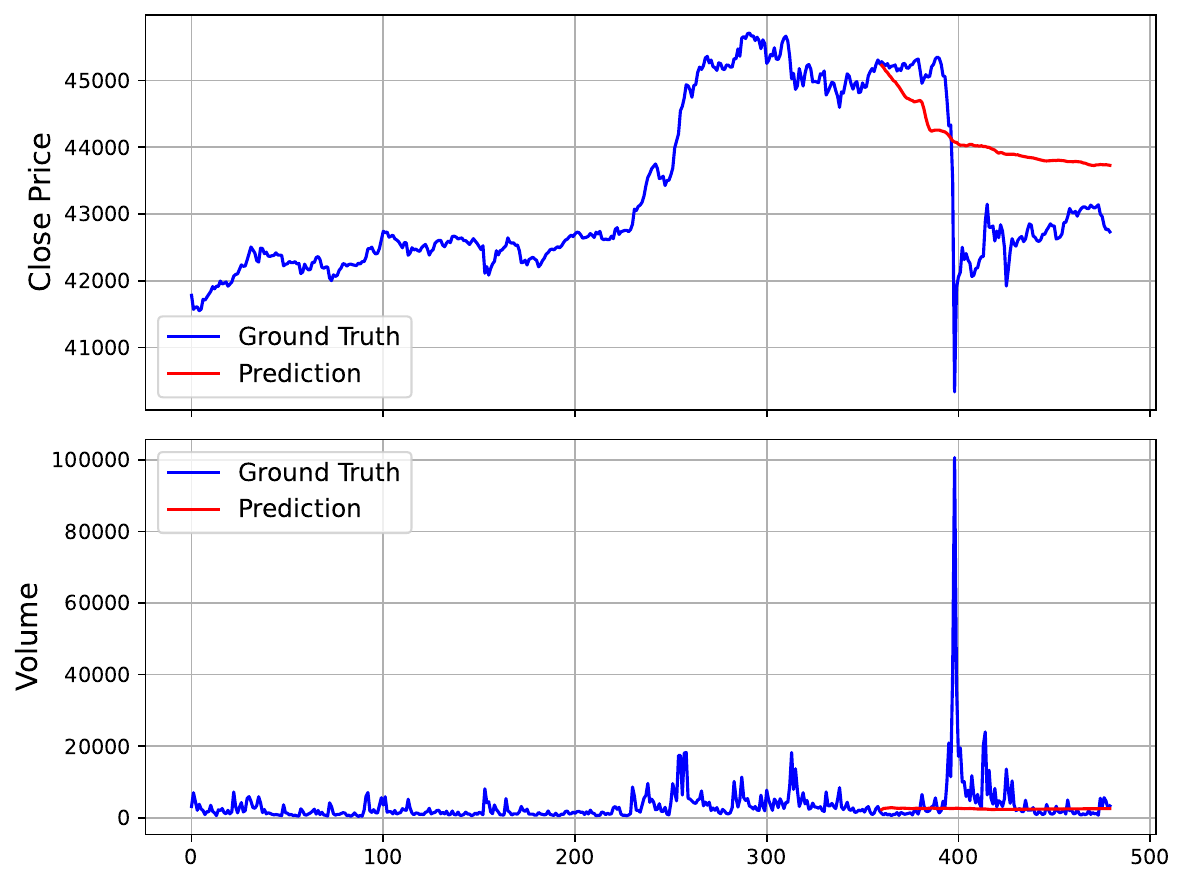}
        \caption{$\text{TimeMOE}_{large}$}
    \end{subfigure}
    \hfill
    \begin{subfigure}[b]{0.33\textwidth}
        \centering
        \includegraphics[width=\linewidth]{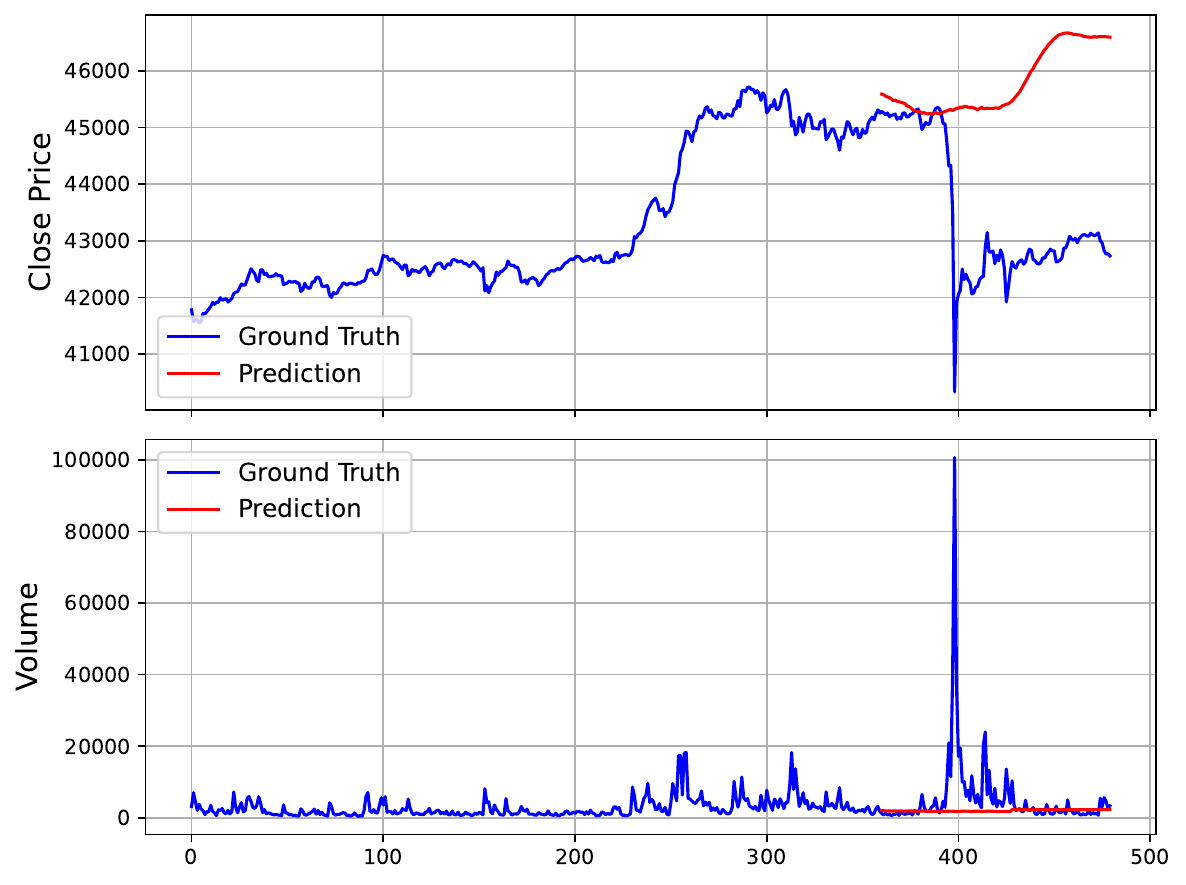}
        \caption{TimesFM}
    \end{subfigure}

    \vspace{0.1cm}

    \begin{subfigure}[b]{0.33\textwidth}
        \centering
        \includegraphics[width=\linewidth]{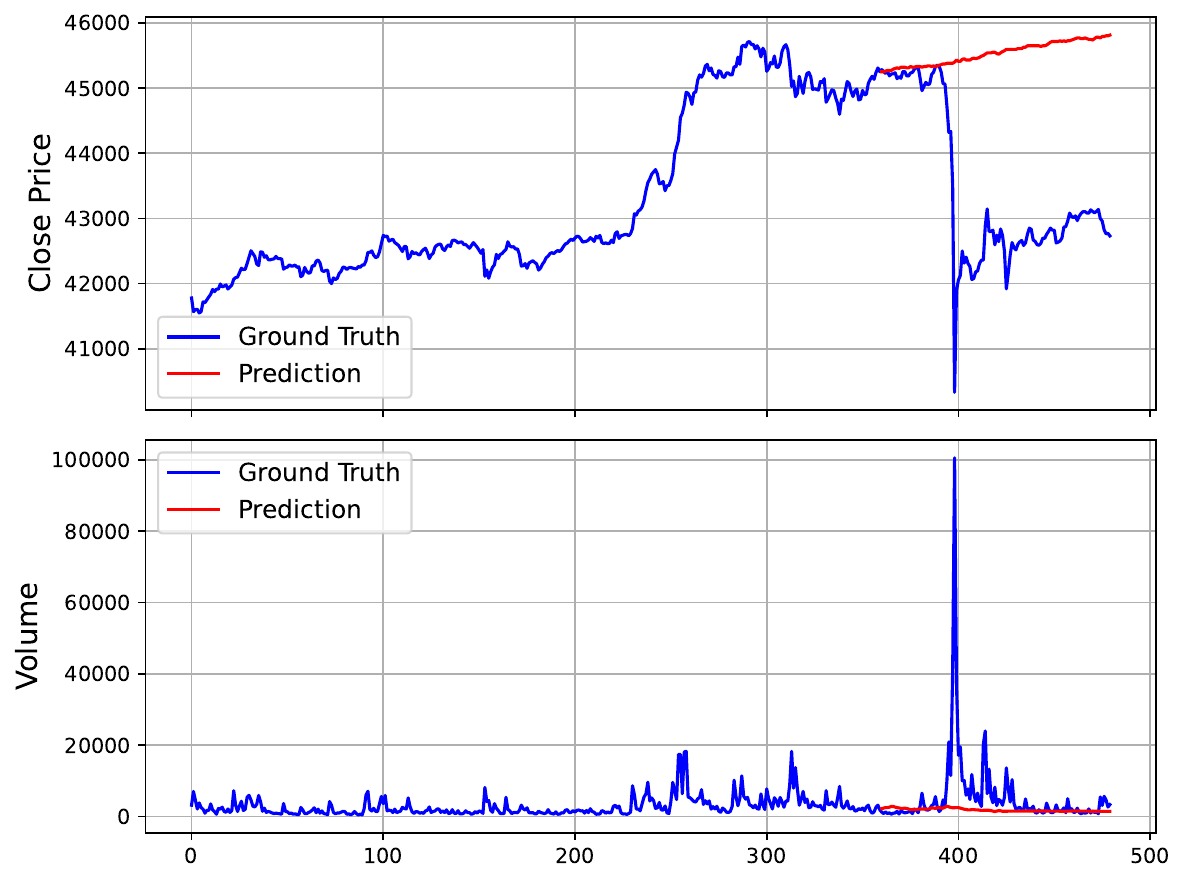}
        \caption{$\text{Chronos}_{small}$}
    \end{subfigure}
    \hfill
    \begin{subfigure}[b]{0.33\textwidth}
        \centering
        \includegraphics[width=\linewidth]{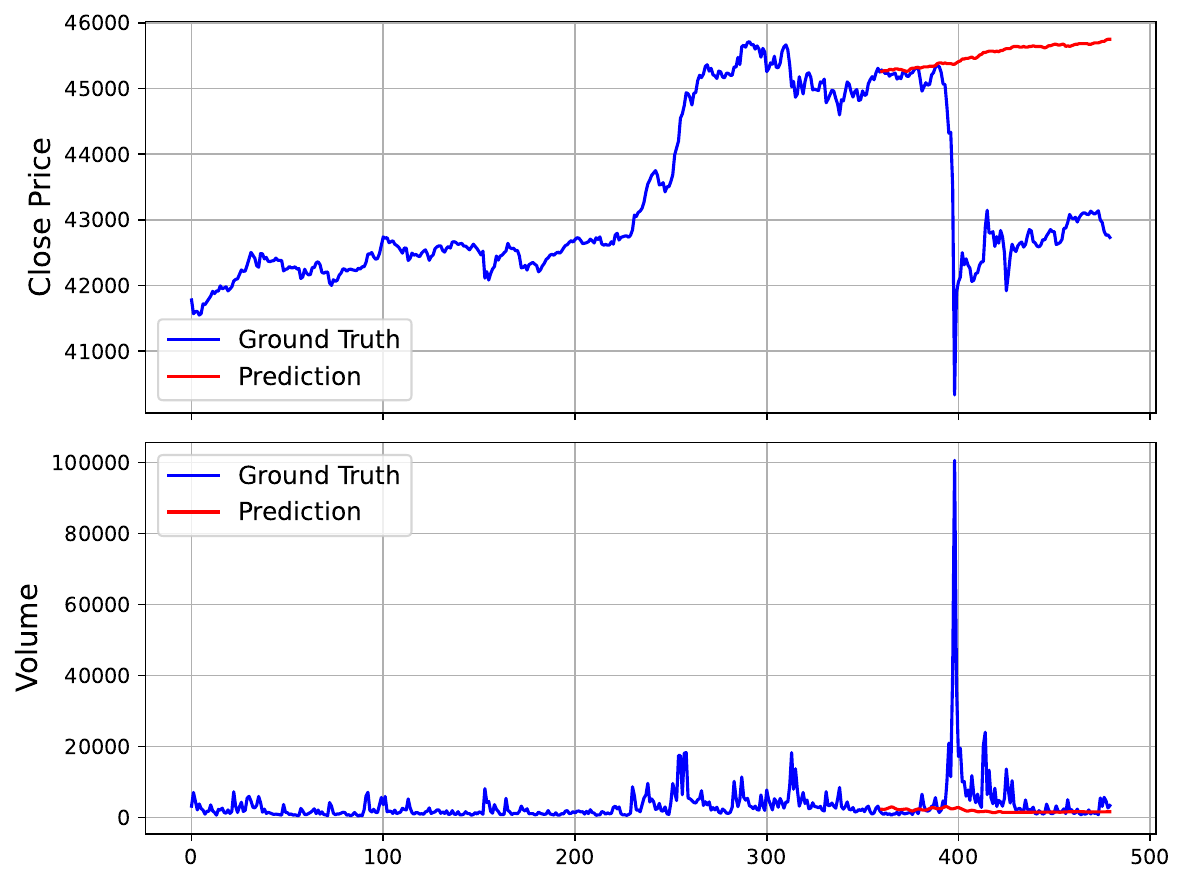}
        \caption{$\text{Chronos}_{base}$}
    \end{subfigure}
    \hfill
    \begin{subfigure}[b]{0.33\textwidth}
        \centering
        \includegraphics[width=\linewidth]{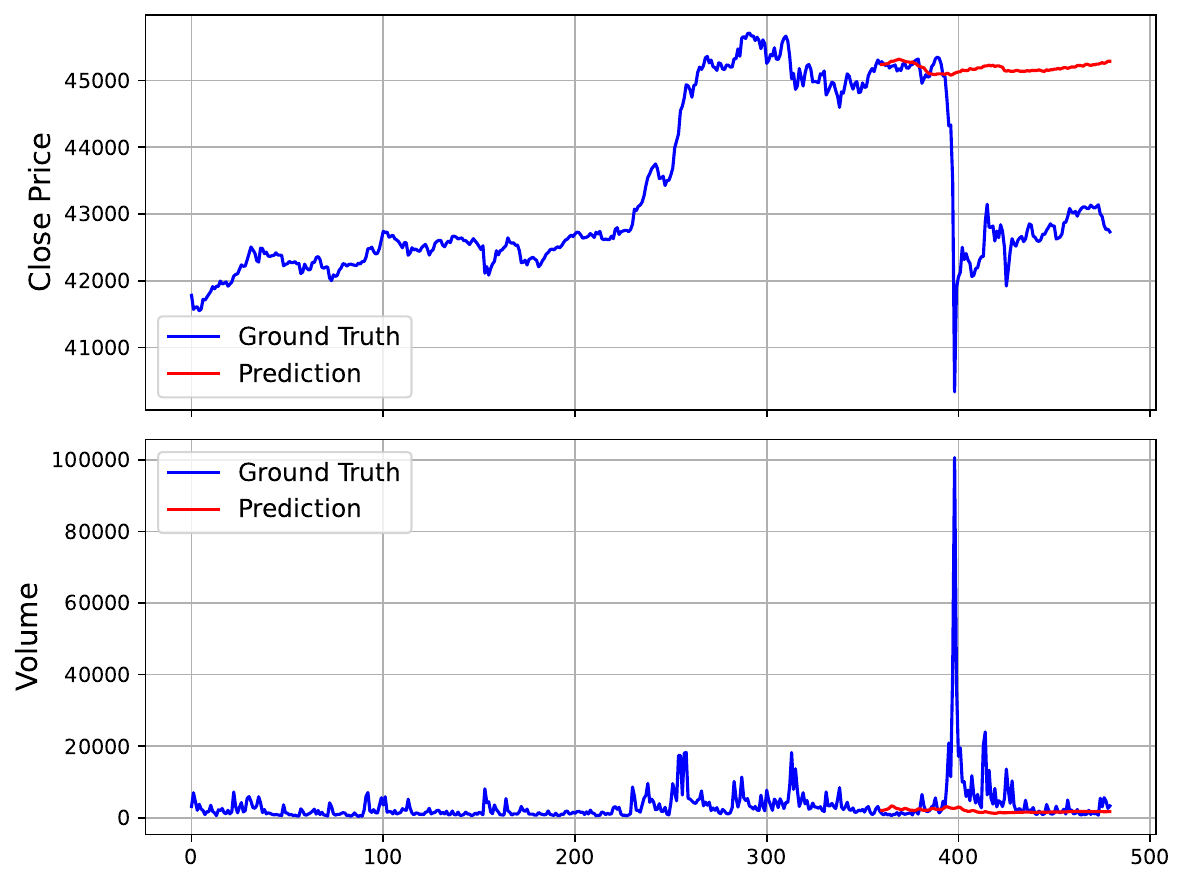}
        \caption{$\text{Chronos}_{large}$}
    \end{subfigure}
    
    \vspace{0.1cm}

    \begin{subfigure}[b]{0.33\textwidth}
        \centering
        \includegraphics[width=\linewidth]{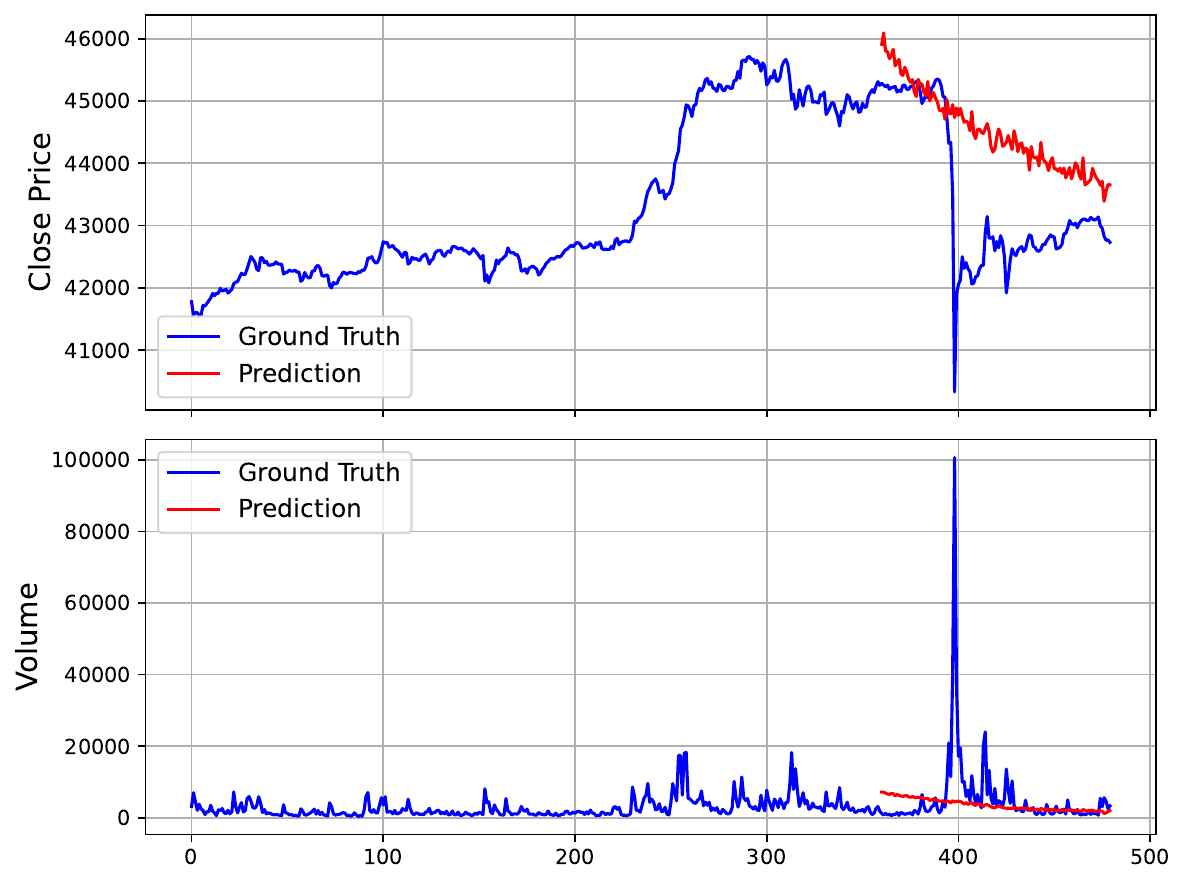}
        \caption{iTransformer}
    \end{subfigure}
    \hfill
    \begin{subfigure}[b]{0.33\textwidth}
        \centering
        \includegraphics[width=\linewidth]{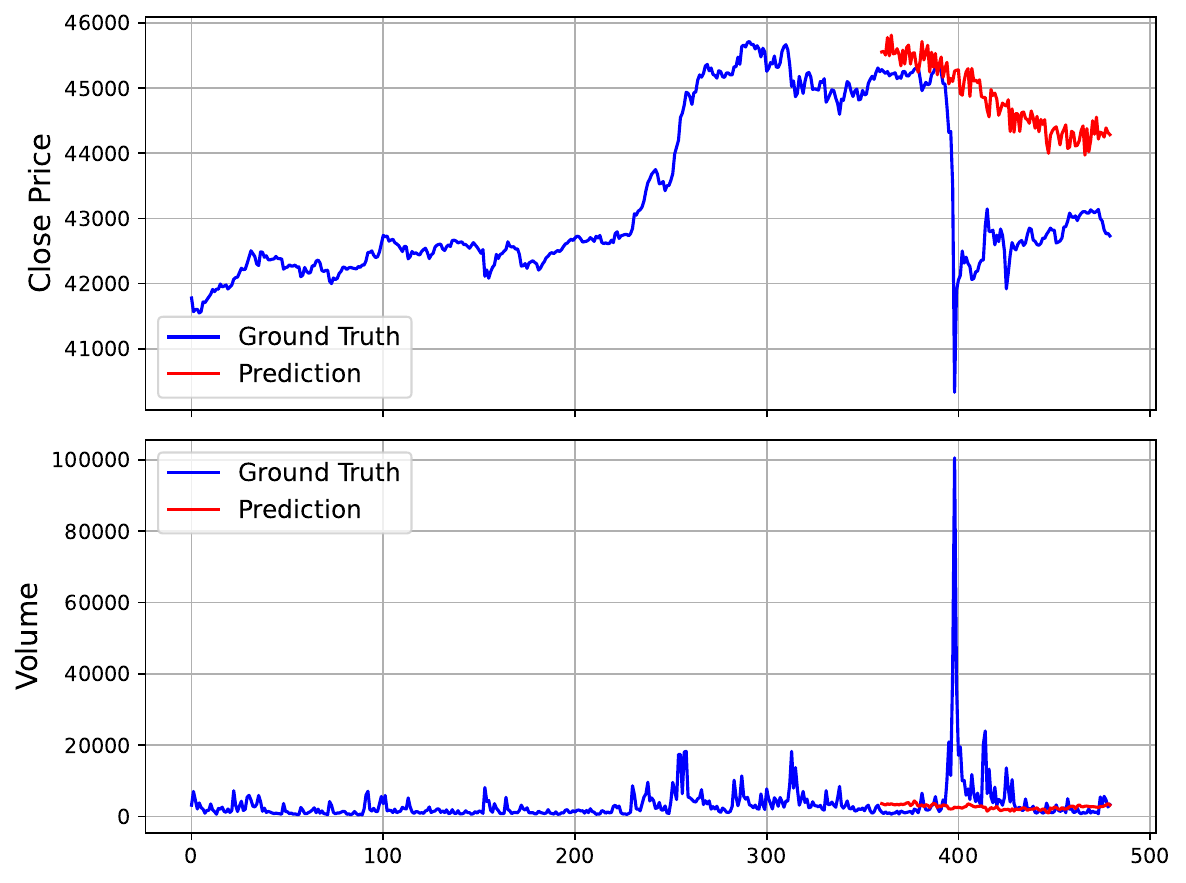}
        \caption{DLinear}
    \end{subfigure}
    \hfill
    \begin{subfigure}[b]{0.33\textwidth}
        \centering
        \includegraphics[width=\linewidth]{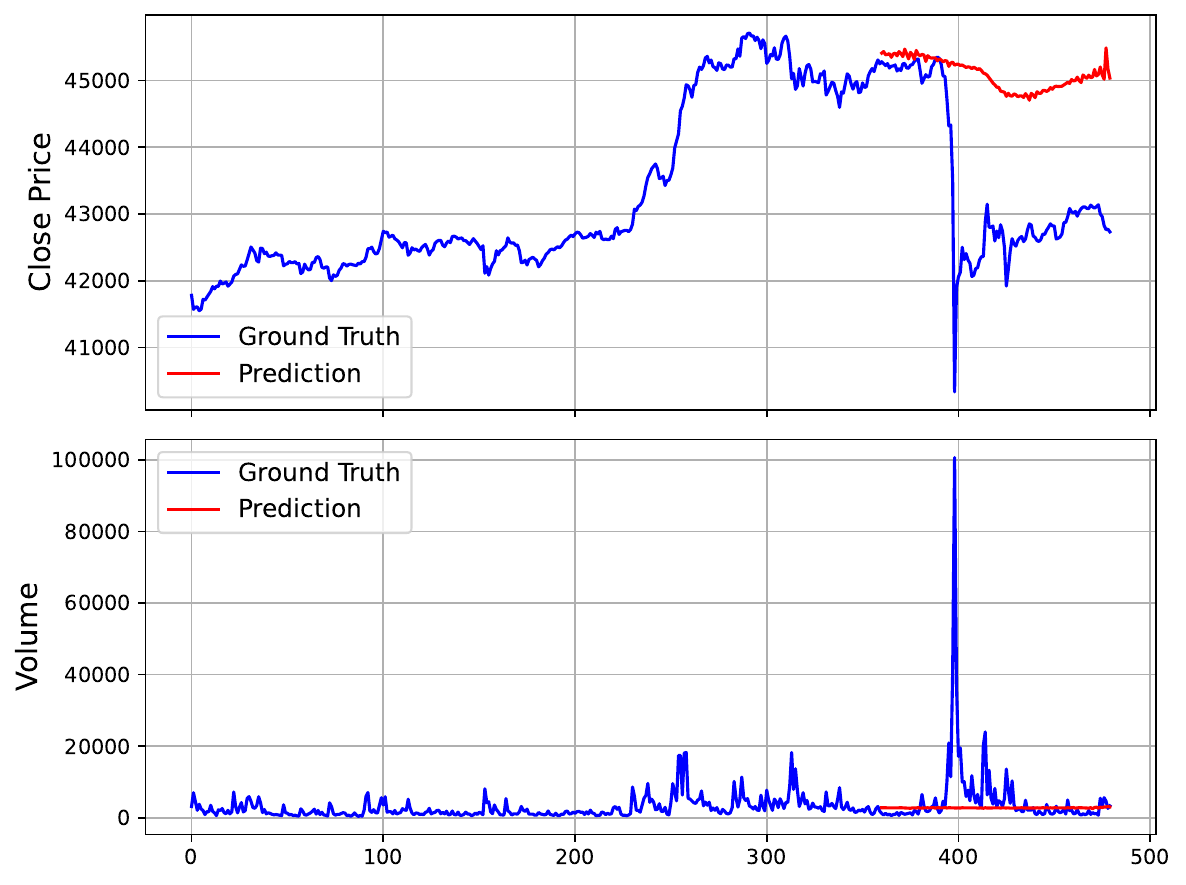}
        \caption{TimesNet}
    \end{subfigure}

    \caption{Forecasting results for the `Close Price' and `Volume' of the BTC/USDT perpetual contract on Binance, based on 15-minute K-line data. The model uses a 360-step look-back window to predict a 120-step horizon. \textcolor{blue}{Blue} lines represent the ground truths and \textcolor{red}{red} lines are the model's predictions.}
    \label{fig:pred_case_4}
\end{figure*}

\begin{figure*}[ht]
    \centering

    \begin{subfigure}[b]{0.33\textwidth}
        \centering
        \includegraphics[width=\linewidth]{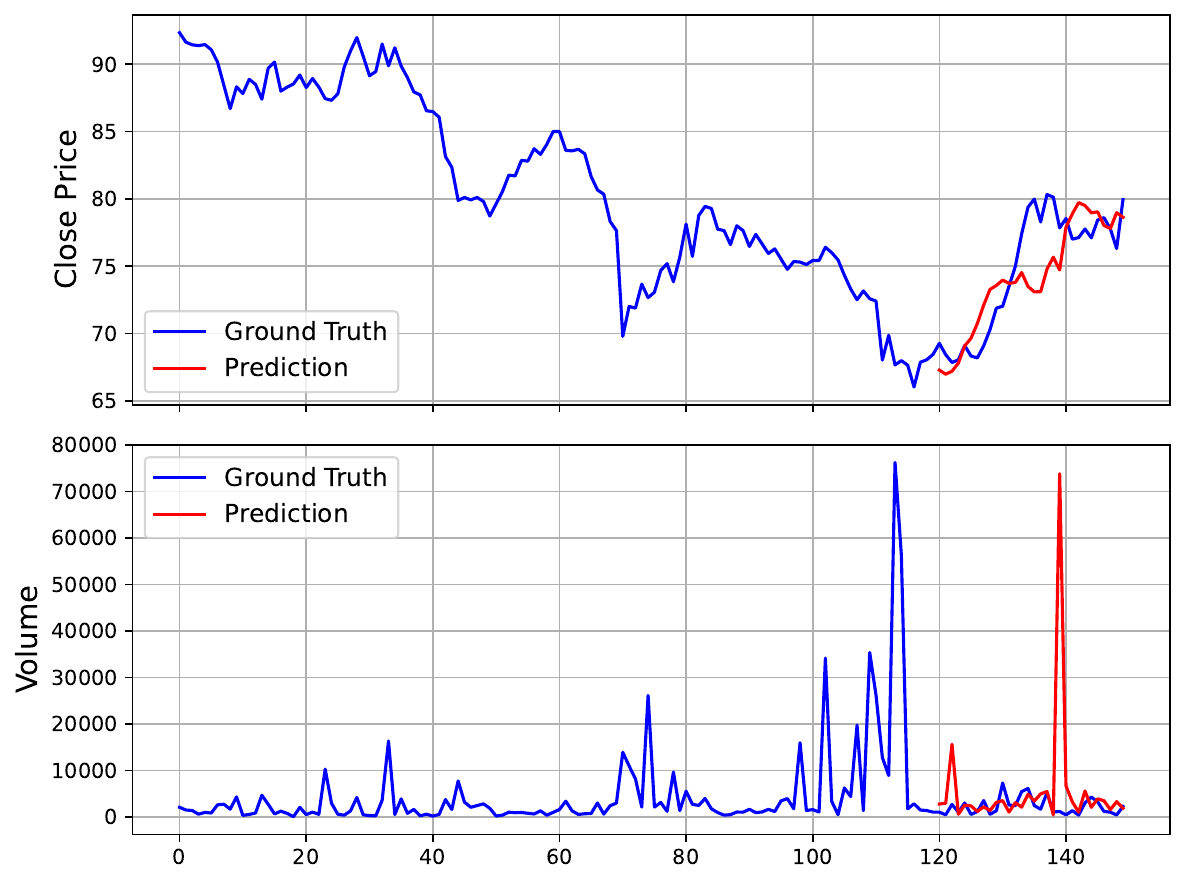}
        \caption{$\text{Kronos}_{small}$}
    \end{subfigure}
    \hfill
    \begin{subfigure}[b]{0.33\textwidth}
        \centering
        \includegraphics[width=\linewidth]{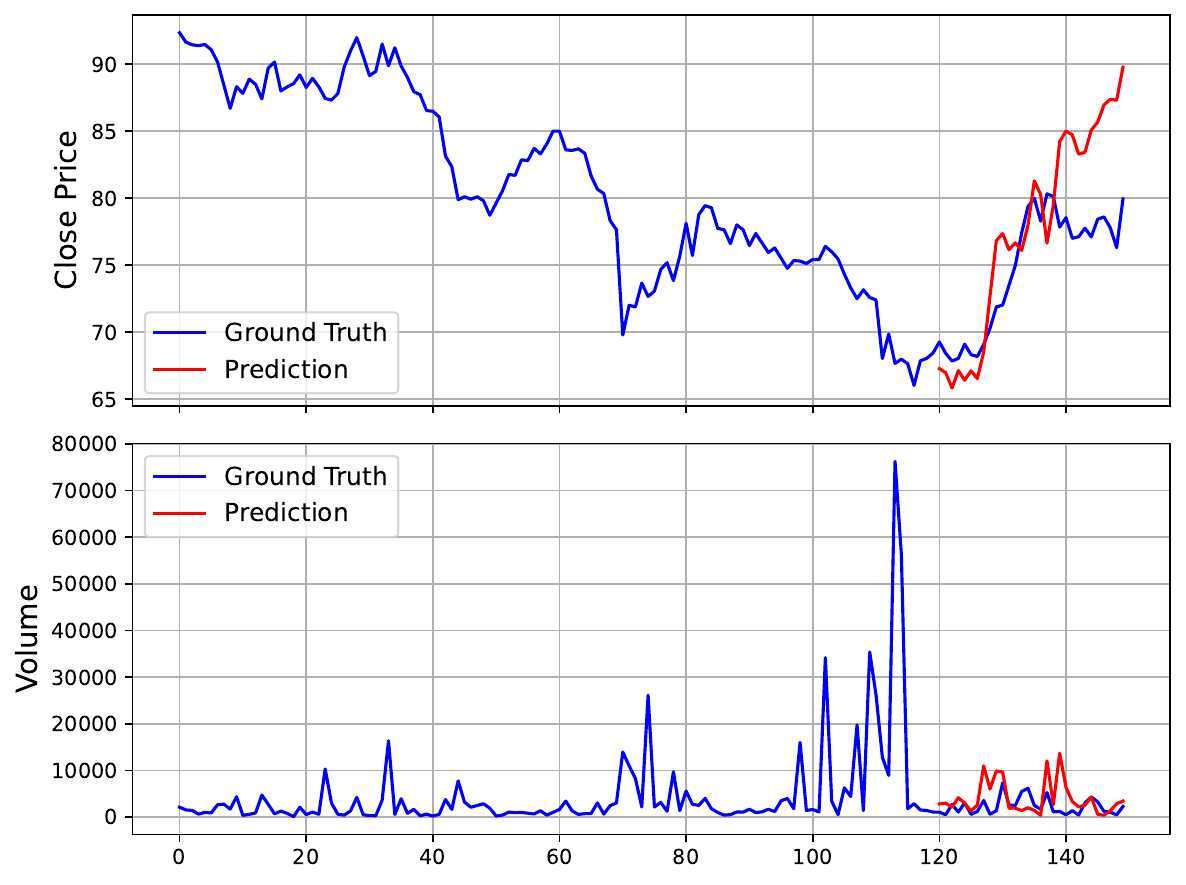}
        \caption{$\text{Kronos}_{base}$}
    \end{subfigure}
    \hfill
    \begin{subfigure}[b]{0.33\textwidth}
        \centering
        \includegraphics[width=\linewidth]{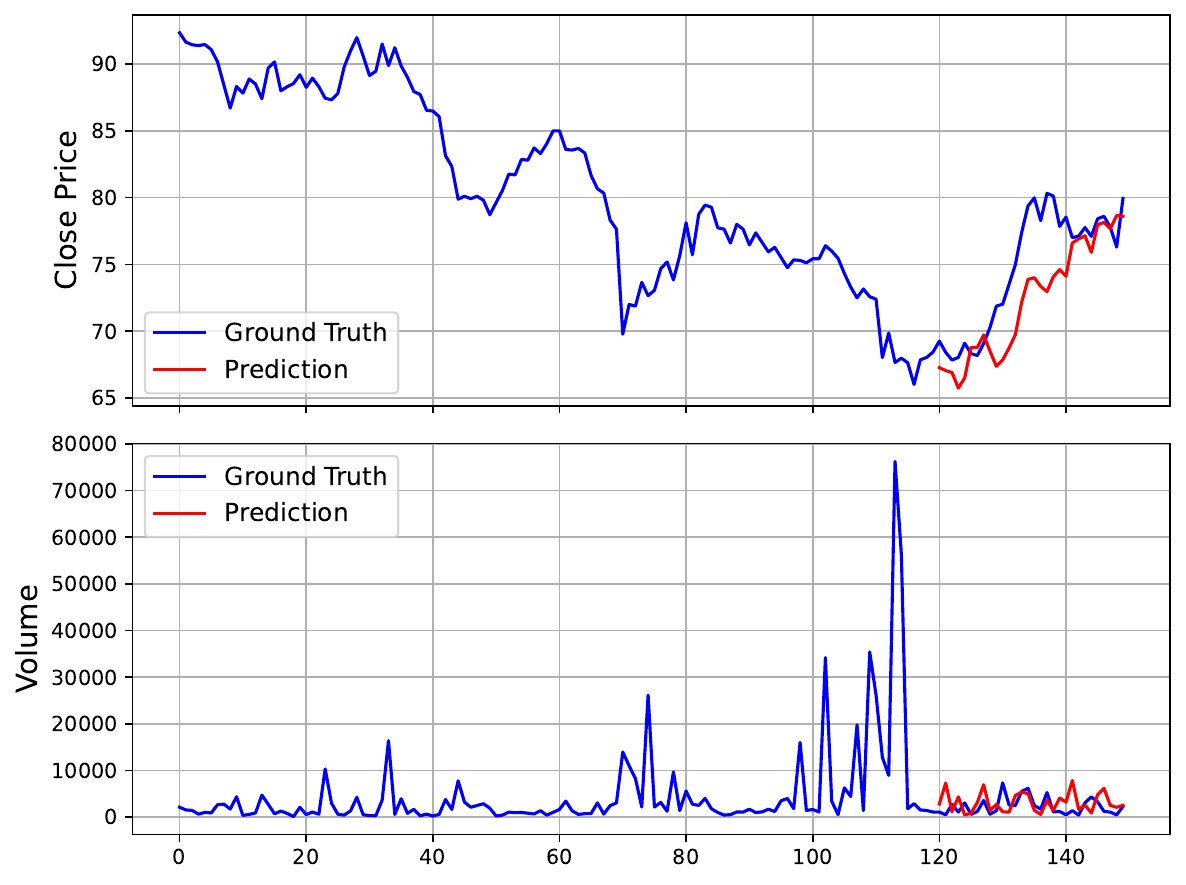}
        \caption{$\text{Kronos}_{large}$}
    \end{subfigure}
    
    \vspace{0.1cm}

    \begin{subfigure}[b]{0.33\textwidth}
        \centering
        \includegraphics[width=\linewidth]{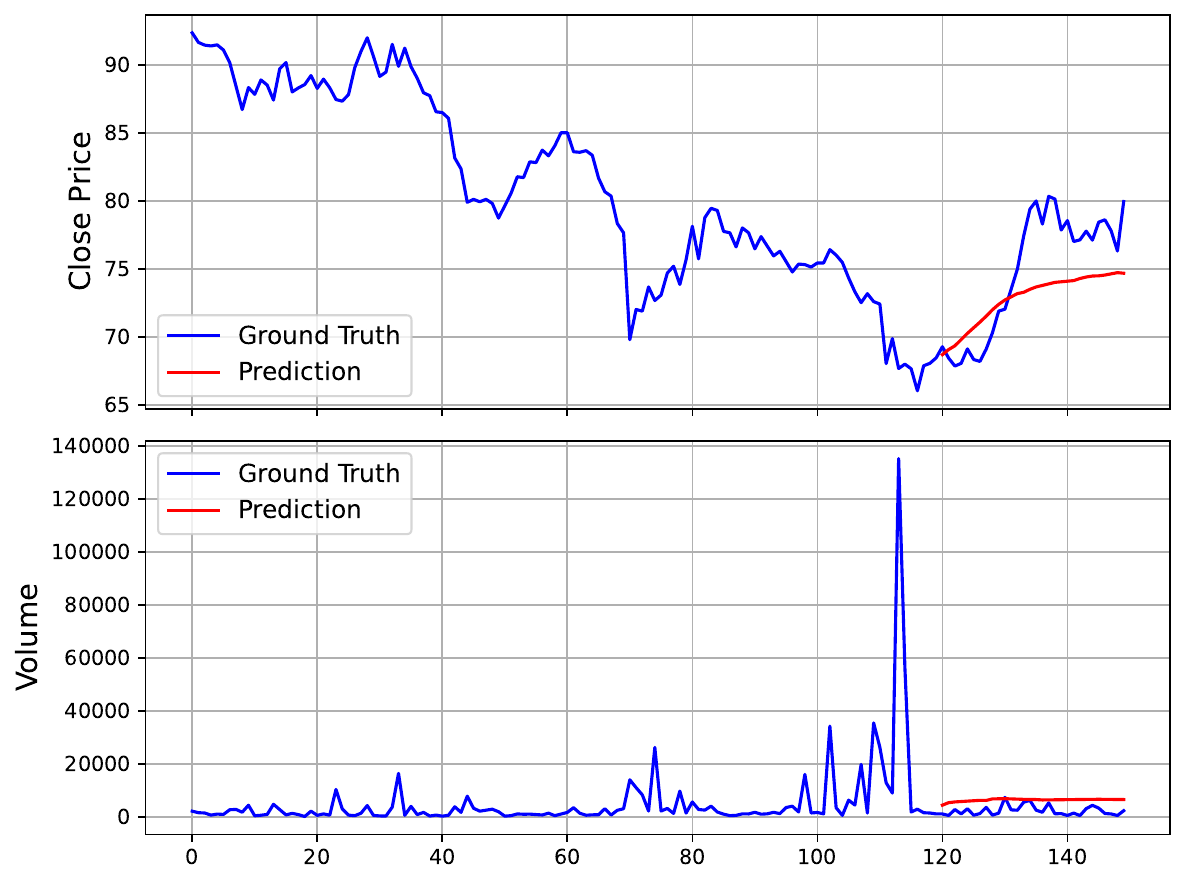}
        \caption{$\text{TimeMOE}_{small}$}
    \end{subfigure}
    \hfill
    \begin{subfigure}[b]{0.33\textwidth}
        \centering
        \includegraphics[width=\linewidth]{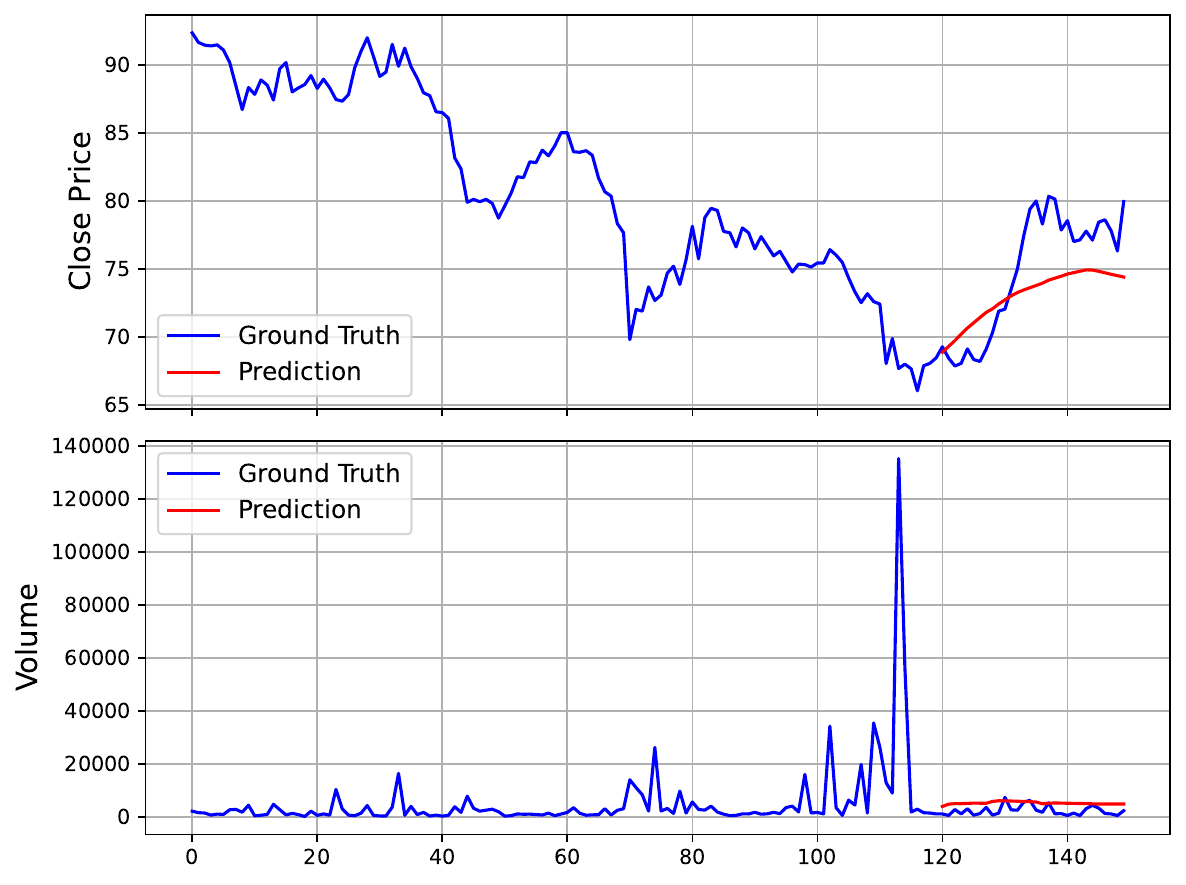}
        \caption{$\text{TimeMOE}_{large}$}
    \end{subfigure}
    \hfill
    \begin{subfigure}[b]{0.33\textwidth}
        \centering
        \includegraphics[width=\linewidth]{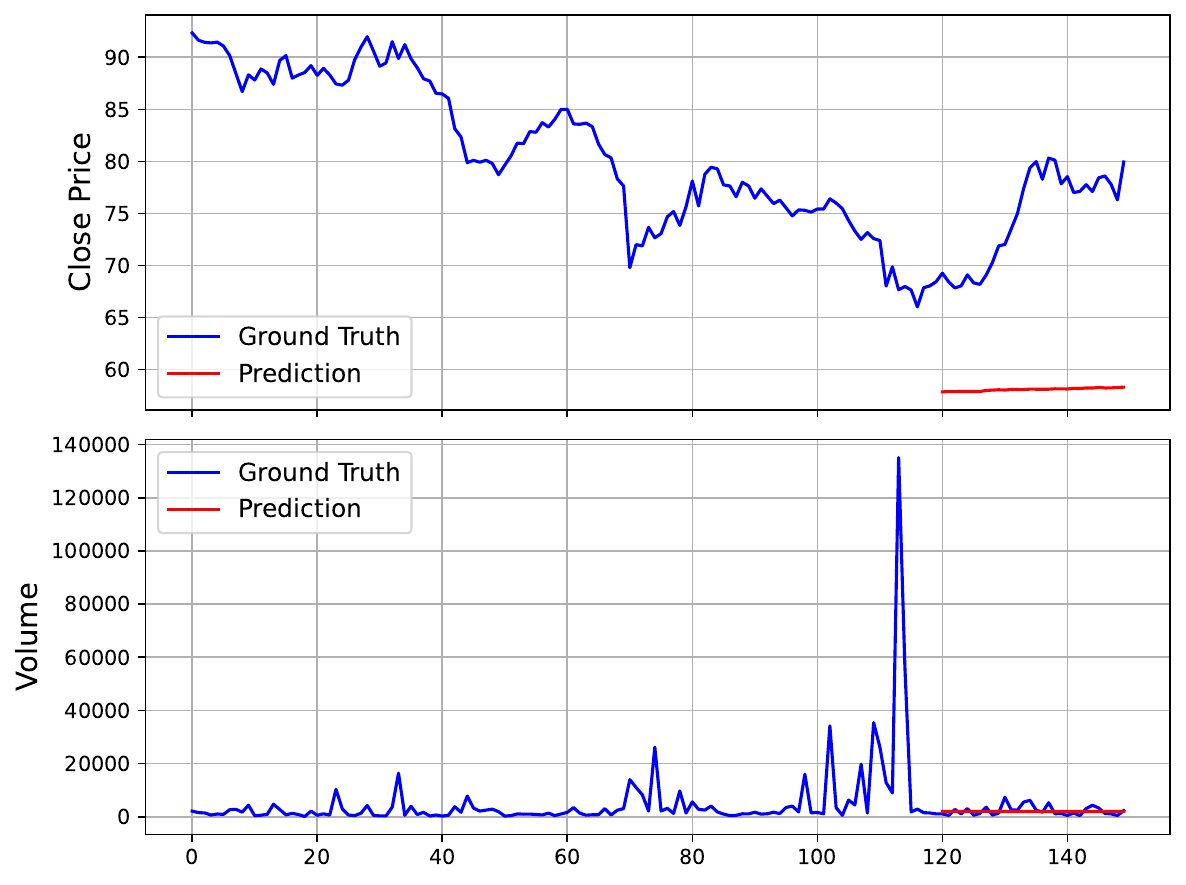}
        \caption{TimesFM}
    \end{subfigure}

    \vspace{0.1cm}

    \begin{subfigure}[b]{0.33\textwidth}
        \centering
        \includegraphics[width=\linewidth]{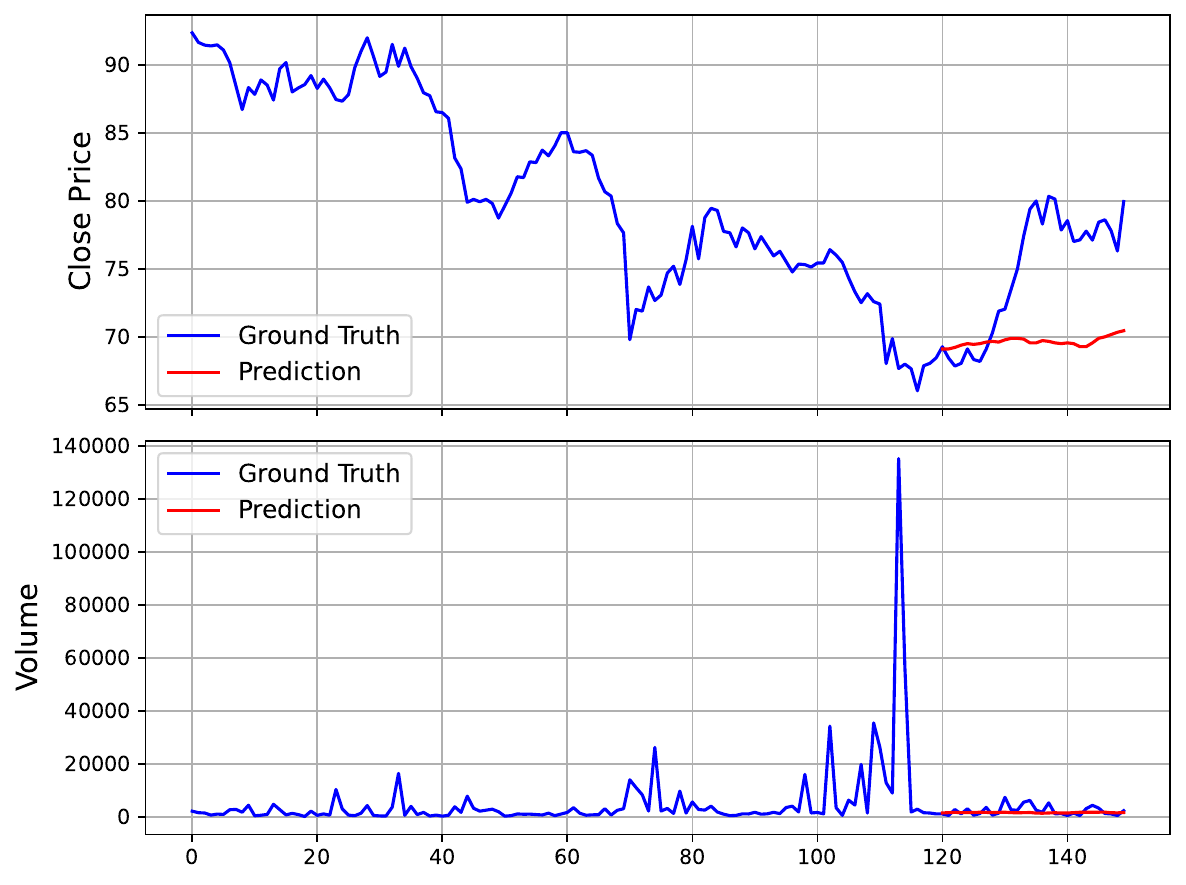}
        \caption{$\text{Chronos}_{small}$}
    \end{subfigure}
    \hfill
    \begin{subfigure}[b]{0.33\textwidth}
        \centering
        \includegraphics[width=\linewidth]{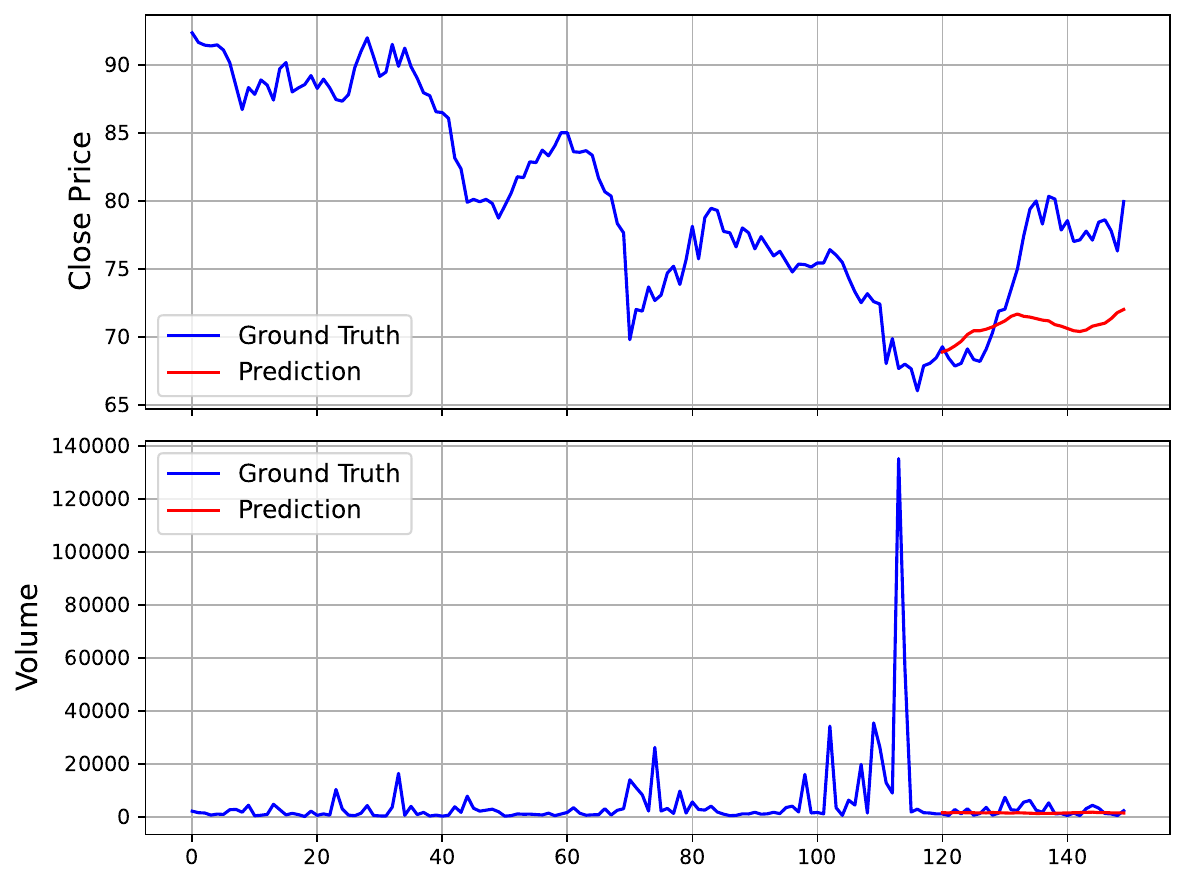}
        \caption{$\text{Chronos}_{base}$}
    \end{subfigure}
    \hfill
    \begin{subfigure}[b]{0.33\textwidth}
        \centering
        \includegraphics[width=\linewidth]{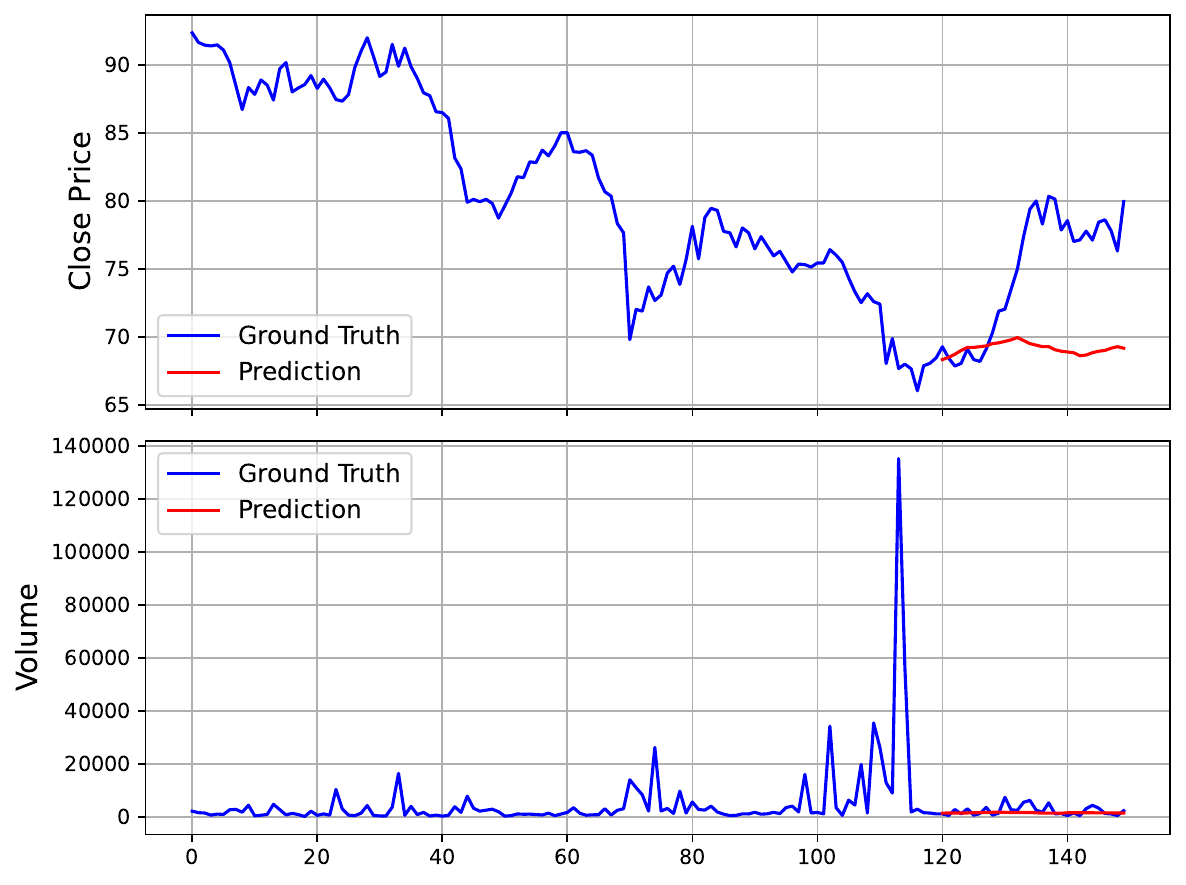}
        \caption{$\text{Chronos}_{large}$}
    \end{subfigure}
    
    \vspace{0.1cm}

    \begin{subfigure}[b]{0.33\textwidth}
        \centering
        \includegraphics[width=\linewidth]{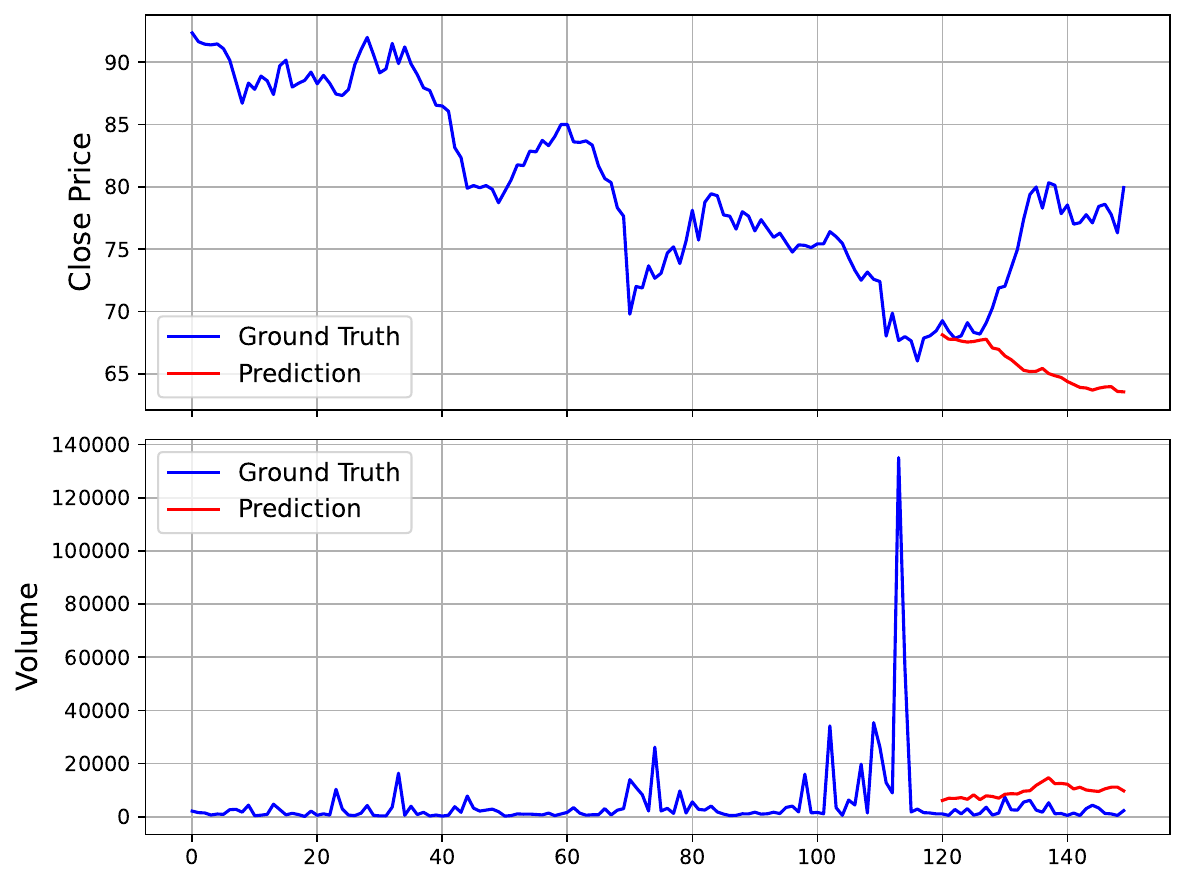}
        \caption{iTransformer}
    \end{subfigure}
    \hfill
    \begin{subfigure}[b]{0.33\textwidth}
        \centering
        \includegraphics[width=\linewidth]{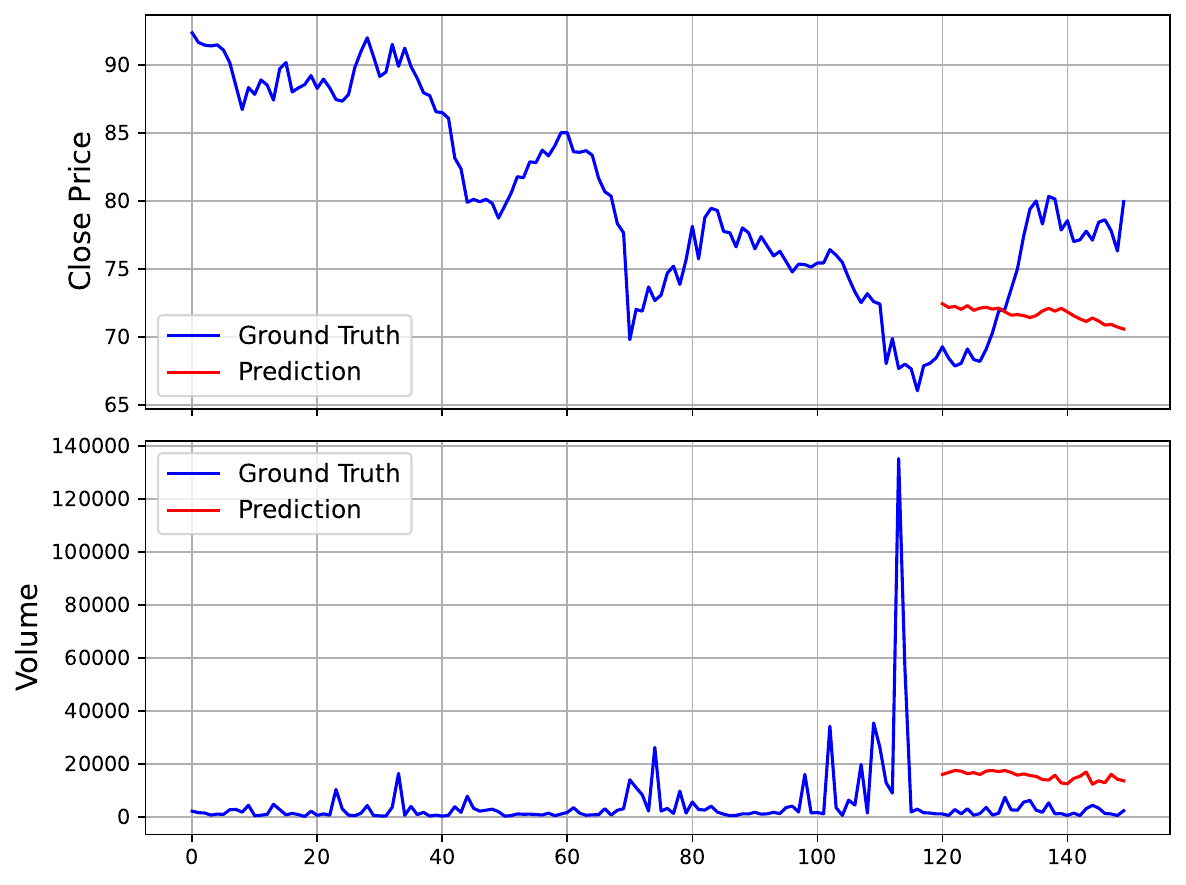}
        \caption{DLinear}
    \end{subfigure}
    \hfill
    \begin{subfigure}[b]{0.33\textwidth}
        \centering
        \includegraphics[width=\linewidth]{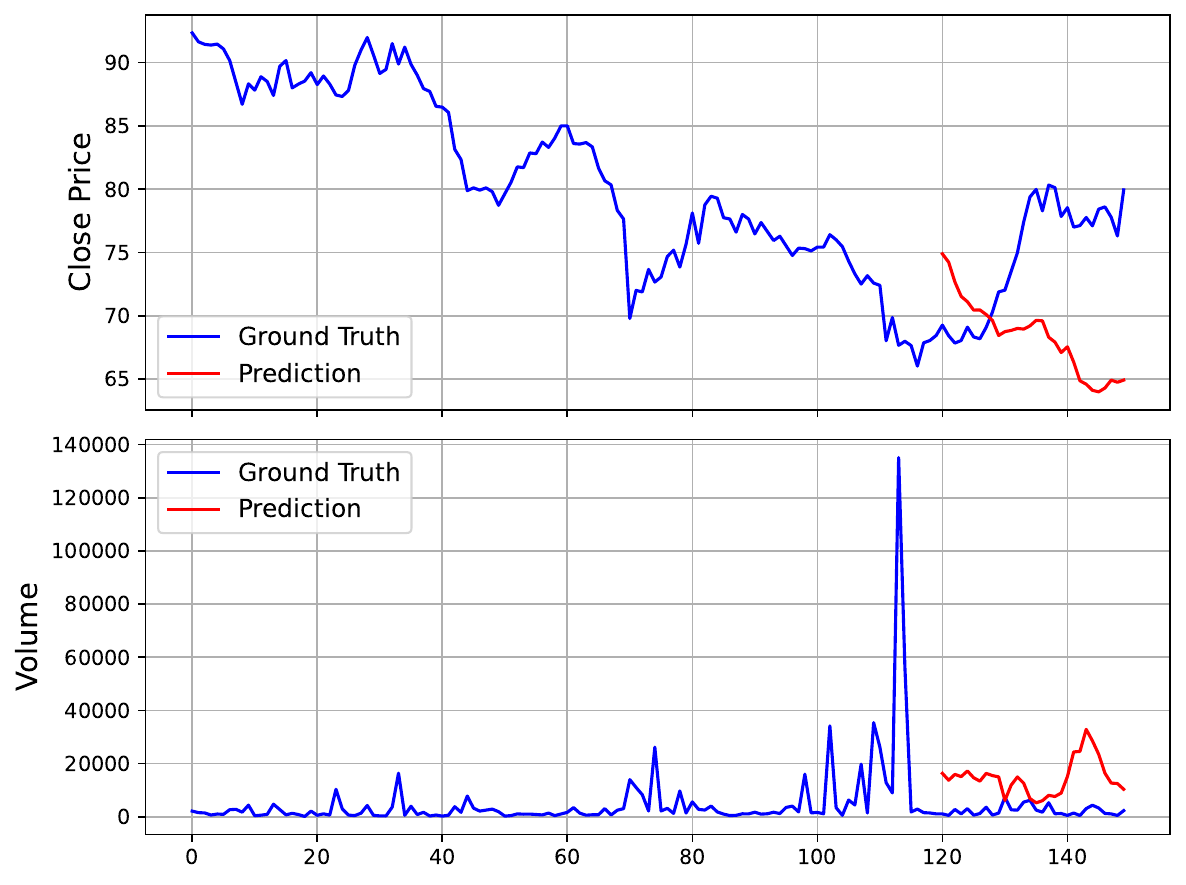}
        \caption{TimesNet}
    \end{subfigure}

    \caption{Forecasting results for the `Close Price' and `Volume' of BMW (FWB: BMW), based on daily K-line data. The model uses a 120-step look-back window to predict a 30-step horizon. \textcolor{blue}{Blue} lines represent the ground truths and \textcolor{red}{red} lines are the model's predictions.}
    \label{fig:pred_case_5}
\end{figure*}

Figures~\ref{fig:pred_case_1} to \ref{fig:pred_case_5} present the forecasting results of our proposed model, Kronos, against several baselines.
We select a few representative assets and showcase the predictions for two key features: closing price and trading volume.
As observed, the forecasts from Kronos not only achieve competitive predictive performance but also exhibit a strong qualitative resemblance to the ground-truth series.
Notably, our model adeptly captures the characteristic dynamics and patterns of the actual price and volume sequences, producing forecasts that are not only accurate but also visually plausible.

\section{Discussion}
\label{sec:discussion}

\subsection{Has K-line data embedded enough information to drive the price movement of capital market in short term? (Q1)}




In capital markets, the determinants of price dynamics are conventionally bifurcated into:

\begin{itemize}
\item \textbf{Long‐term driving factors}, which manifest as persistent trends and exert a lasting influence on intrinsic value;
\item \textbf{Short‐term driving factors}, which are typified by elevated volatility and immediate market impact.
  
\end{itemize}

Long‐term driving factors establish the market’s prevailing trajectory and valuation benchmarks, whereas short‐term ones introduce transient volatility and generate discrete trading opportunities.

Extensive empirical evidence demonstrates that kline data (\textbf{OHLCVA}, including \textbf{price} and \textbf{trading volume})~\cite{kim1991trading}, when analyzed in tandem, effectively encapsulate the informational content of short‐term driving factors—such as macroeconomic data releases ~\cite{flannery2002macroeconomic}, corporate event disclosures~\cite{kim1991trading}, and shifts in investor sentiment~\cite{baker2006investor, da2011search}.

The detail discussion about the above empirical evidences is beyond the scope of this paper.

\subsection{What makes Krono's tokenizer work? (Q2)}

The effectiveness of our vision-inspired quantization (BSQ) tokenizer can be analyzed from two key perspectives: its inherent noise suppression and its ability to create a structured, discrete state space suitable for sequence modeling.

\subsubsection{Noise Suppression and Stability}
Financial time-series data is often corrupted by noise and subject to extreme outliers, such as ``flash-crash'' events caused by anomalous trades. A primary challenge for regression-based models is that such outliers can lead to unbounded approximation errors, severely degrading model stability~\cite{brownlees2006financial}.

Our approach addresses this by transforming the representation learning into a more robust, classification-like framework. By quantizing continuous price-volume embeddings, we effectively cap the influence of any single data point. Specifically, BSQ's projection of embeddings onto a unit sphere prior to binarization guarantees that the expected distortion is strictly upper-bounded~\cite{zhao2024image}:
\[
  \mathbb{E}_u \,\big\|u - \widehat u\big\|\;<\;\sqrt{2 - 2/\sqrt{L}}\;<\sqrt{2}.
\]
This bound tightens as the codebook dimension \(L\) increases. In contrast, simpler methods like sign-based quantization without normalization (e.g., LFQ) lack such a guarantee, leaving them vulnerable to arbitrarily large errors from outlier inputs~\cite{zhao2024image}. This bounded error property is crucial for building reliable financial forecasting models.

\subsubsection{Learning in a Compact and Discrete State Space}
High-frequency financial data exists in a high-dimensional, continuous state space, posing significant challenges for sequence models. Our tokenizer maps these infinite states into a finite, discrete vocabulary of tokens. This discretization serves as a powerful form of regularization with two main benefits~\cite{rabanser2020effectiveness}:

\begin{itemize}
    \item \textbf{Improved Sample Efficiency and Generalization}: Instead of learning a complex function over a continuous space, a downstream model like a Transformer learns to predict transitions and patterns among a finite set of abstract states (tokens). This simplifies the learning task. Different but semantically similar input vectors can be mapped to the same token, effectively increasing the number of observations for each discrete state. This allows the model to learn robust patterns from fewer examples, which is particularly critical for modeling rare market phenomena like responses to liquidity shocks, where data is sparse.
    \item \textbf{Reduced Overfitting}: The quantization process inherently discards fine-grained, potentially noisy variations within each quantization cell. This prevents the model from fitting to spurious artifacts in the training data.
\end{itemize}

\begin{table}[h!]
\centering

\resizebox{\columnwidth}{!}{%
\begin{tabular*}{\columnwidth}{@{\extracolsep{\fill}} ccc}
\toprule 
\textbf{Codebook Type} & \textbf{Size} & \textbf{Usage} \\
\midrule 
Coarse-Level-Subtoken Codebook      & $2^{10}$ & 97.66\% \\
Fine-Level-Subtoken Codebook        & $2^{10}$ & 85.25\% \\
\bottomrule
\end{tabular*}}

\caption{Codebook usage for coarse-level subtoken and fine-level subtoken.}
\label{tab:codebook_usage}
\end{table}

\begin{table*}[t!]
\centering

\resizebox{\textwidth}{!}{%
\begin{tabular}{@{}l c c r r r r c@{}}
\toprule
\textbf{Setup} & \textbf{Splits ($n$)} & \textbf{Sub-Vocab ($2^{k/n}$)} & \textbf{Core Params (M)} & \textbf{Vocab Params (M)} & \textbf{Fusion Params (M)} & \textbf{Total Params (M)} & \textbf{Inference Steps per Token} \\
\midrule
No Split & 1 & 1,048,576 & 97.5 & 1744.8 & 0.0 & 1842.3 & 1$\times$ \\
\textbf{Ours} & \textbf{2} & \textbf{1,024} & \textbf{97.5} & \textbf{3.4} & \textbf{1.4} & \textbf{102.3} & \textbf{2$\times$} \\
More Splits & 4 & 32 & 97.5 & 0.2 & 2.8 & 100.5 & 4$\times$ \\
& 5 & 16 & 97.5 & 0.1 & 3.5 & 101.1 & 5$\times$ \\
\bottomrule
\end{tabular}%
}
\caption{
    Trade-off analysis for factorizing a $k=20$ bit token into $n$ subtokens, based on the $\text{Kronos}_{base}$ architecture. The model's core Transformer blocks have $\approx$97.5M parameters.
}
\label{tab:factorization_analysis_revised}
\end{table*}

The effectiveness of our tokenizer is further evidenced by its codebook utilization. As shown in Table \ref{tab:codebook_usage}, the code usage of BSQ reaches 97.66\% at the coarse level and 85.25\% at the fine level. Such high utilization indicates that our method creates an expressive vocabulary, effectively partitioning the feature space without suffering from codebook collapse (where many codes are left unused)~\cite{zhu2024addressing}. This expressiveness provides the rich foundation necessary for a model to capture the nuanced and diverse states of market microstructure.

\begin{figure*}[ht]
    \centering

    \begin{subfigure}[b]{\textwidth}
        \centering
        \includegraphics[width=0.32\textwidth]{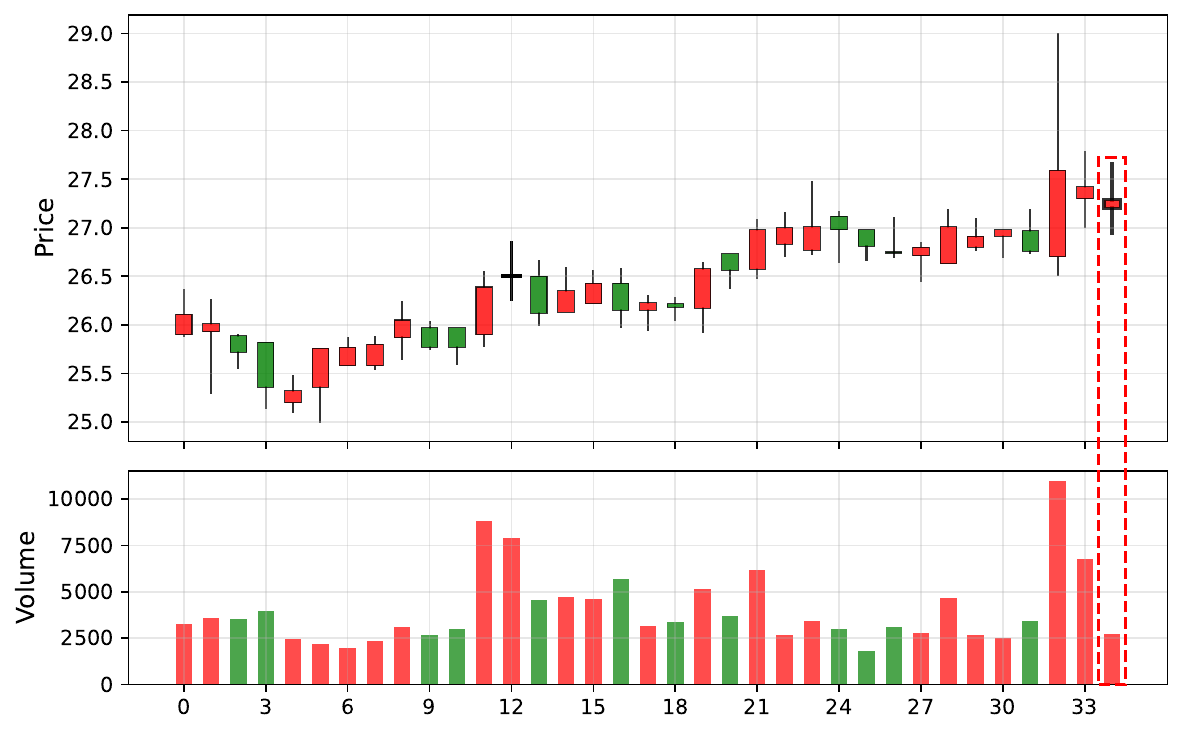}
        \includegraphics[width=0.32\textwidth]{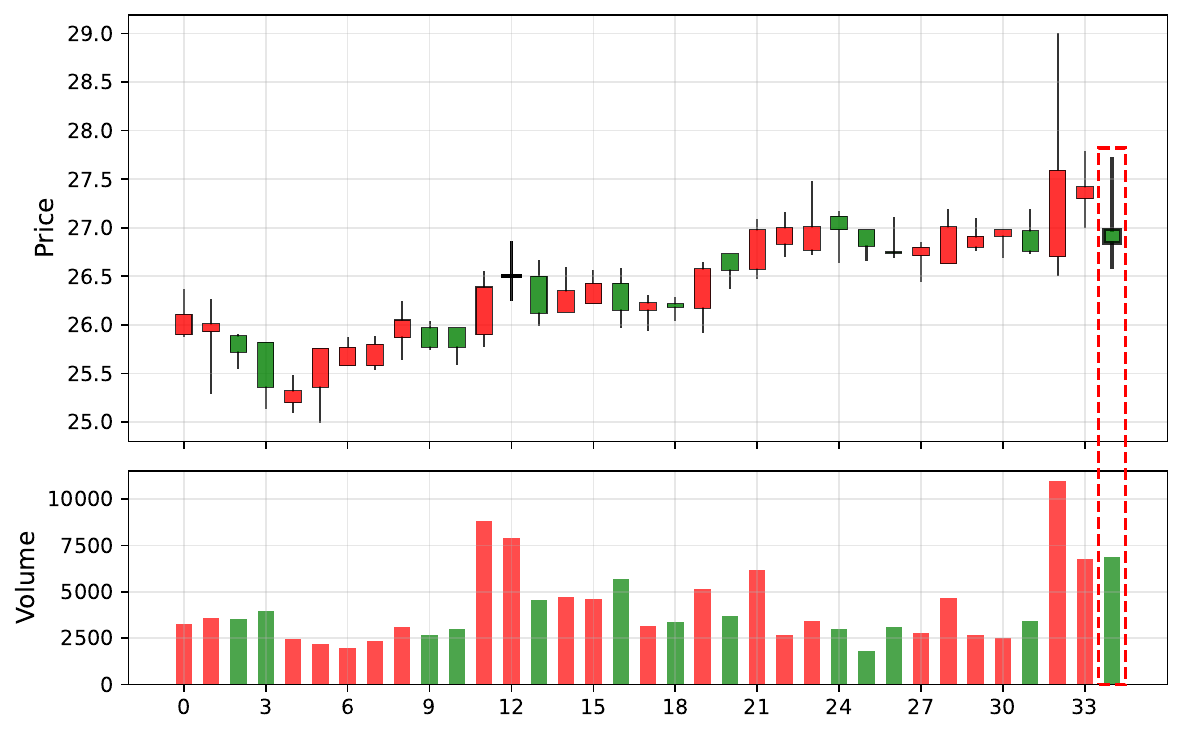}
        \includegraphics[width=0.32\textwidth]{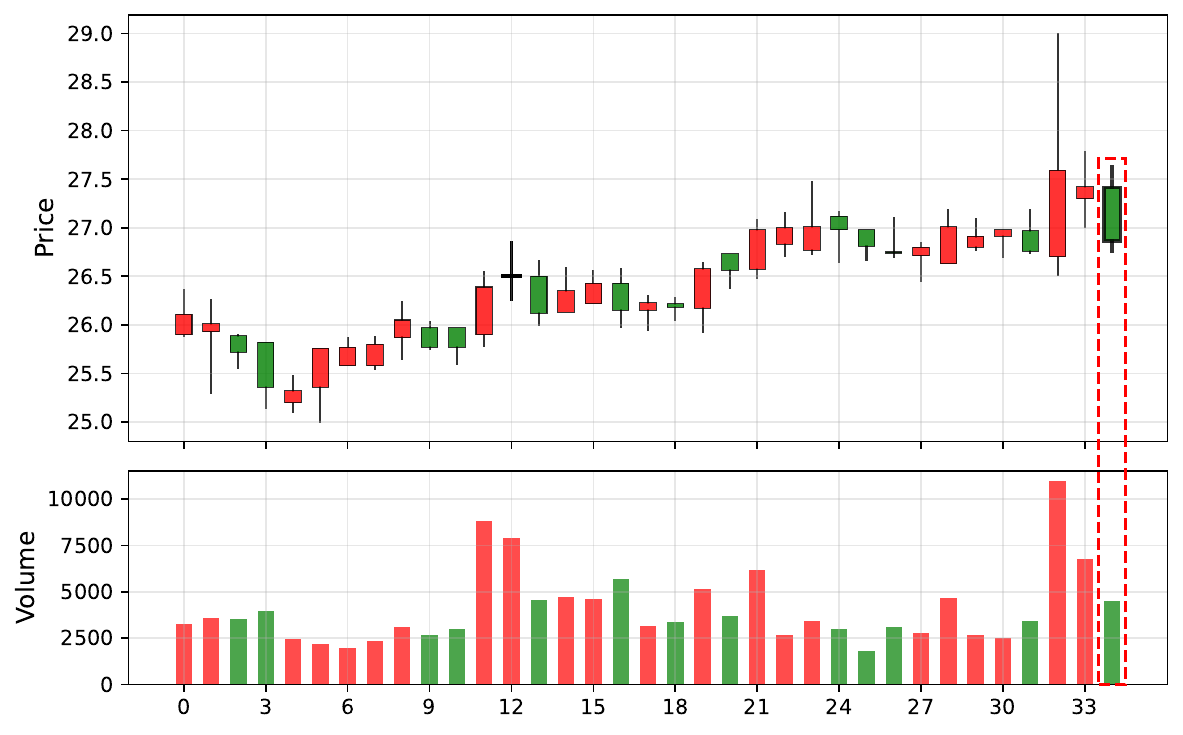}
        \caption{Examples of high-frequency tokens.}
    \end{subfigure}
    \vspace{0.1cm}

    \begin{subfigure}[b]{\textwidth}
        \centering
        \includegraphics[width=0.32\textwidth]{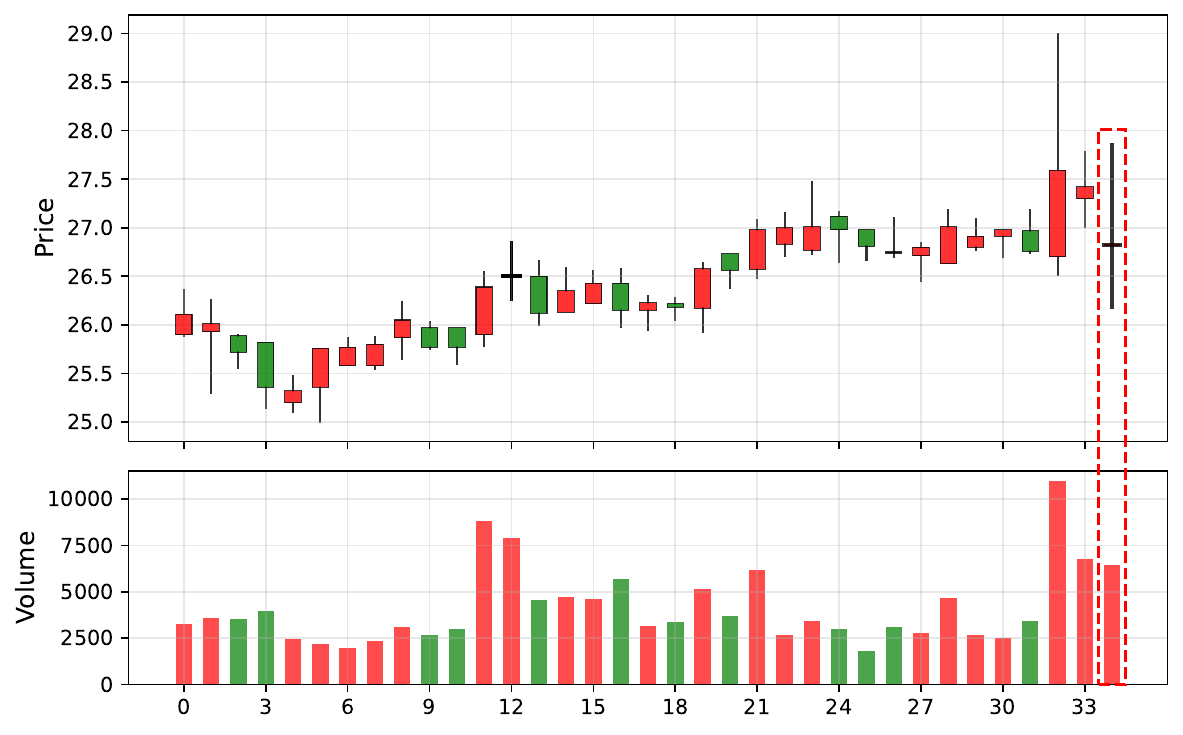}
        \includegraphics[width=0.32\textwidth]{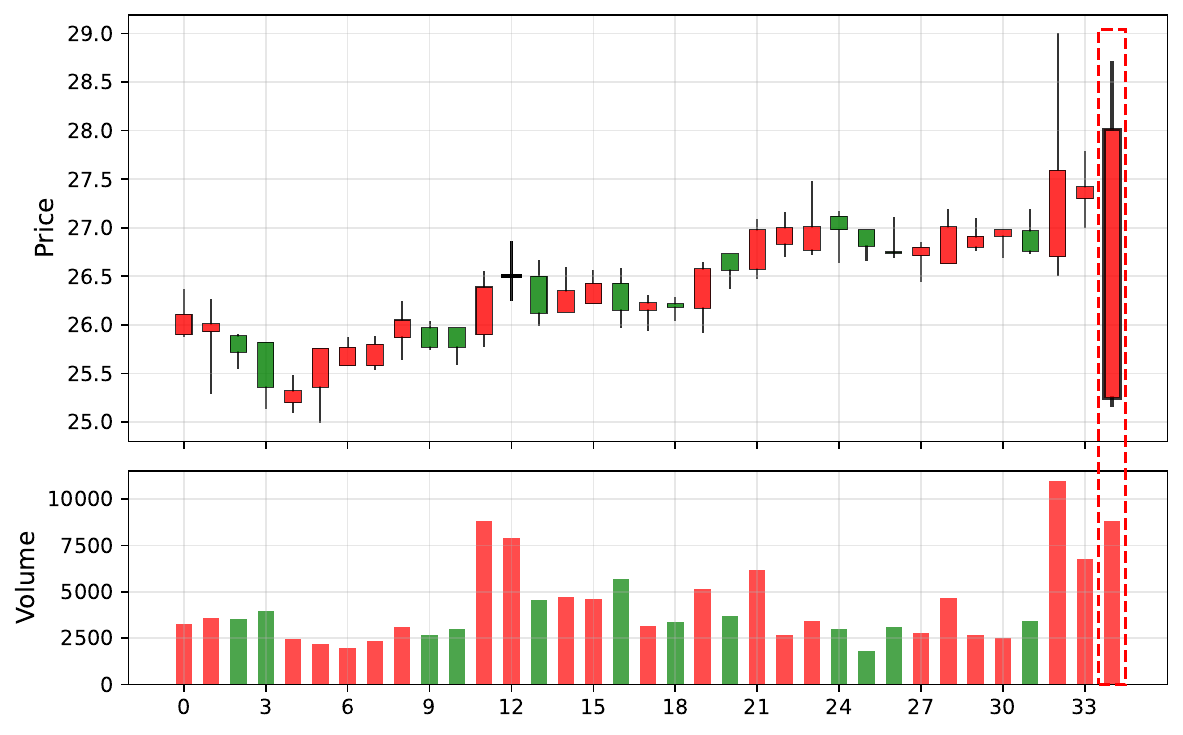}
        \includegraphics[width=0.32\textwidth]{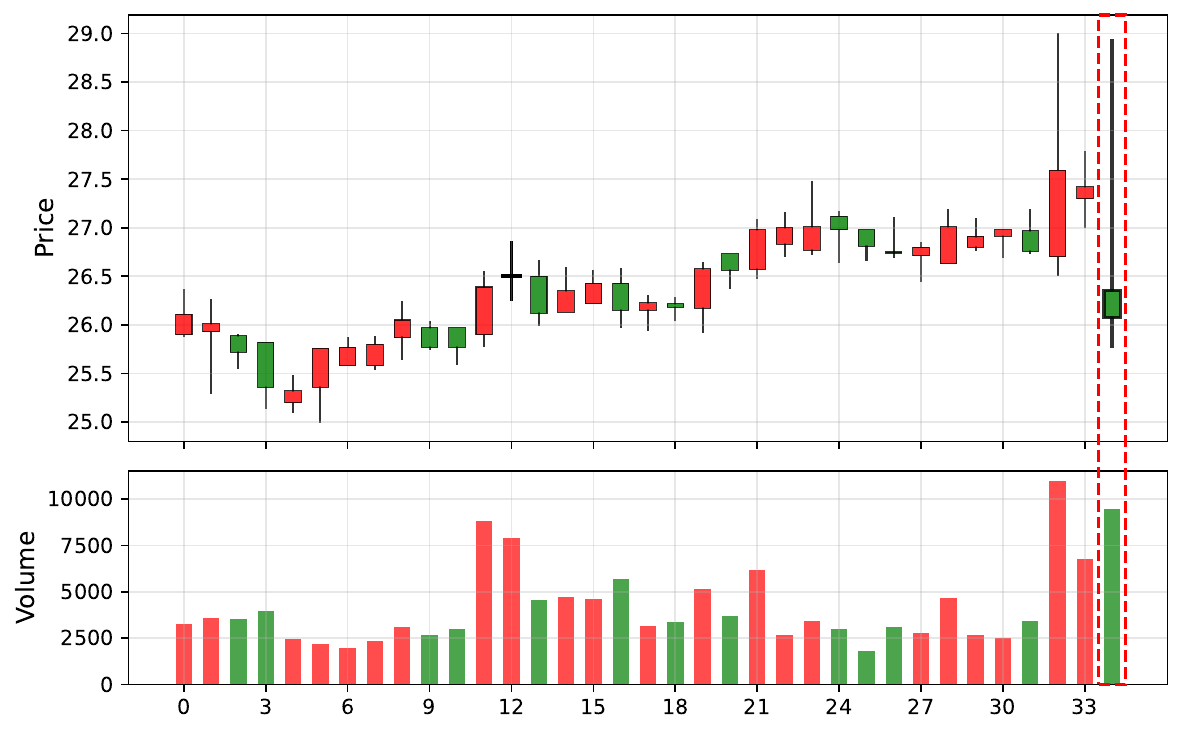}
        \caption{Examples of low-frequency tokens.}
    \end{subfigure}
    \vspace{0.1cm}

    \begin{subfigure}[b]{\textwidth}
        \centering
        \includegraphics[width=0.32\textwidth]{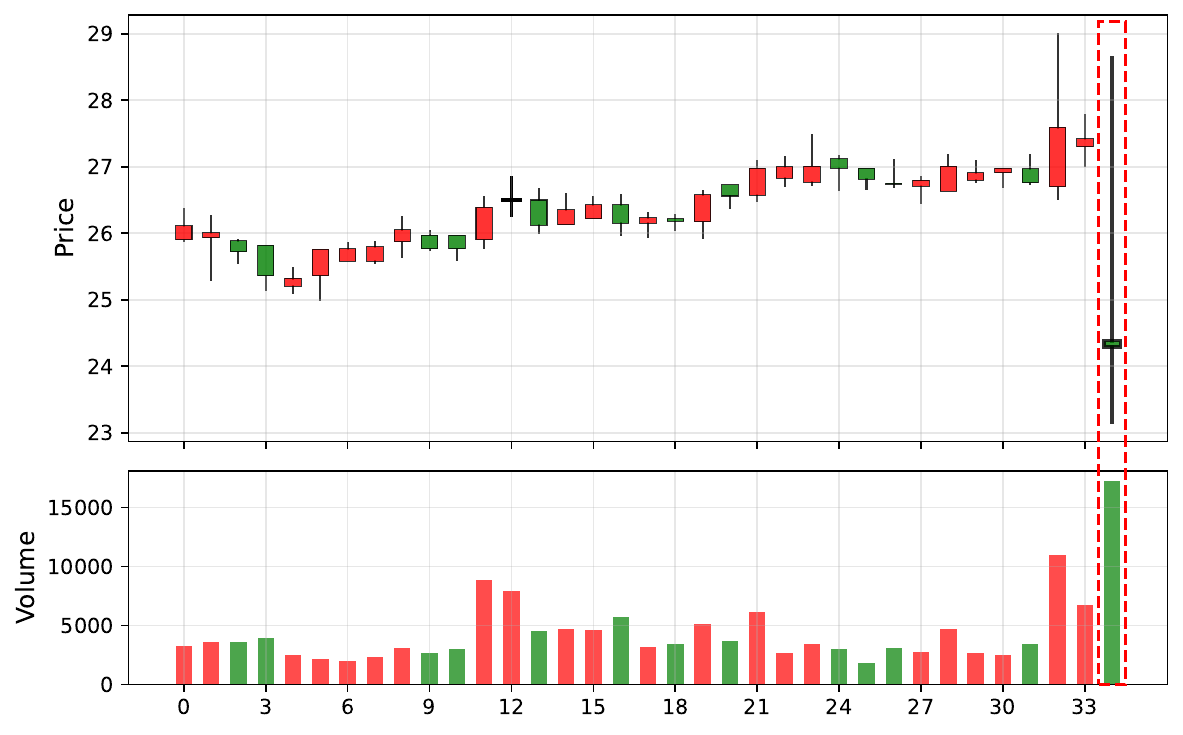}
        \includegraphics[width=0.32\textwidth]{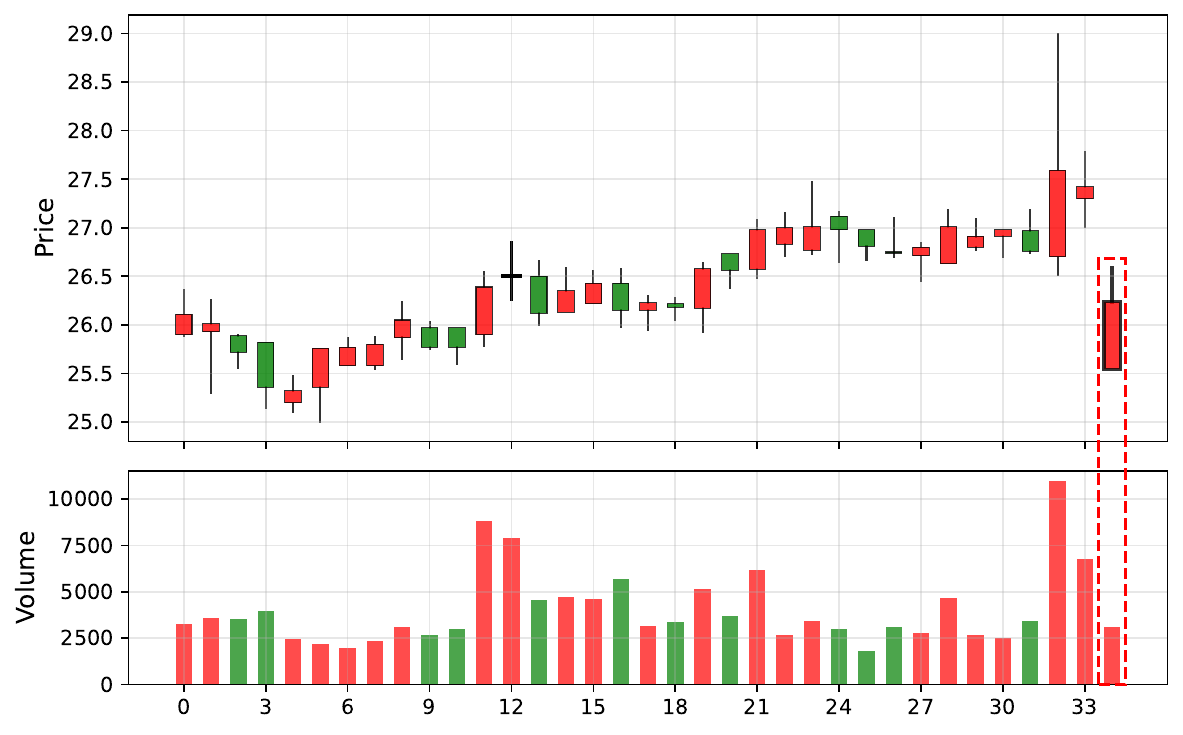}
        \includegraphics[width=0.32\textwidth]{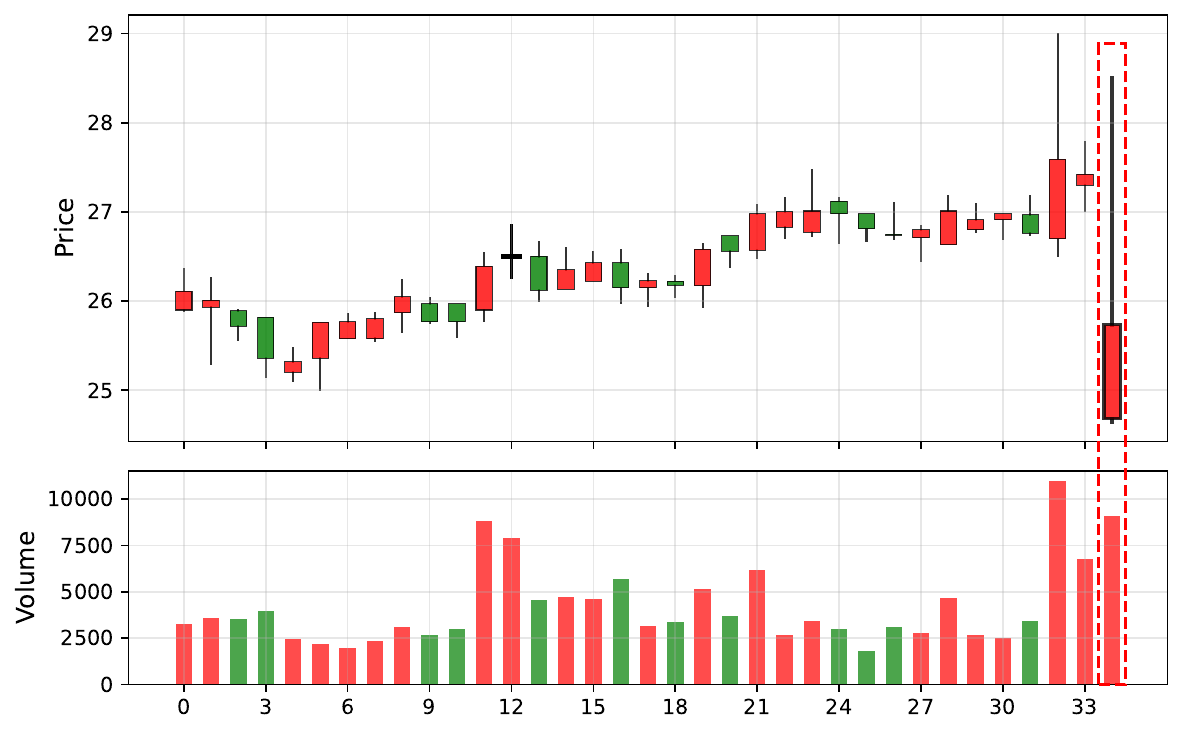}
        \caption{Examples of unused tokens from the vocabulary.}
    \end{subfigure}
    \vspace{0.1cm}

    \begin{subfigure}[b]{\textwidth}
        \centering
        \includegraphics[width=0.32\textwidth]{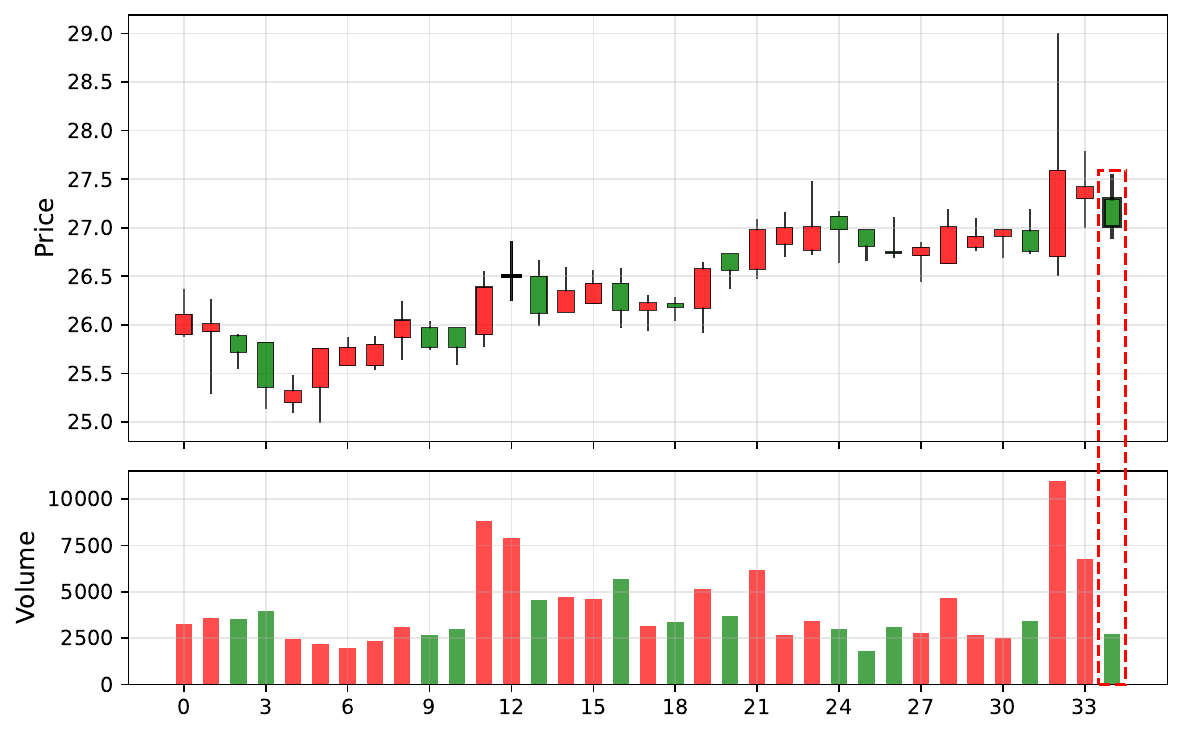}
        \caption{A sample of an original token sequence.}
    \end{subfigure}

    \caption{Visualization of token usage patterns. The figure illustrates token categories based on their occurrence frequency in the corpus: (a) high-frequency, (b) low-frequency, and (c) unused (zero-frequency) tokens. A sample from an original sequence (d) is shown for reference. The sequences in (a), (b), and (c) are constructed by replacing the last token of (d) with a randomly sampled token from the corresponding category. In the visualization, the candlesticks follow a ``red for up, green for down'' convention (where up/down is determined by the close price relative to the open price), and the volume bars are colored accordingly.}
    \label{fig:token_show_case}
\end{figure*}

Additionally, the vocabulary is stratified into three categories based on usage frequency: (a) high-frequency, (b) low-frequency, and (c) unused tokens. To investigate their representational characteristics, we conduct an analysis where we replace the final token of an encoded sequence with a token from each category and then decode it back to a K-line. Figure \ref{fig:token_show_case} presents the results of this procedure. We observe a clear correspondence between token frequency and pattern typicality. High-frequency tokens (a) map to common K-bar shapes, indicative of stable market conditions. Conversely, low-frequency (b) and unused (c) tokens generate more extreme and atypical K-bars, such as those with long bodies or wicks, signifying rare, high-volatility events. This suggests that the learned codebook captures a meaningful semantic hierarchy, effectively distinguishing between common and significant market patterns based on token frequency.

\subsubsection{Hyperspherical geometry for tail sensitivity}
In financial contexts, market returns and price changes often exhibit heavy tails (or fat tails)~\cite{mandelbrot1963variation}. The heavy-tail distribution of price changes is one of the key sources of trading profits in quantitative investment and cannot be ignored.

Unlike standard vector-quantization on the Euclidean sphere, BSQ’s binary encoding preserves angular information very efficiently, making it more sensitive to fat-tail data that manifest as sharp directional changes in feature space. This aligns well with how microstructure events often appear as abrupt shifts in the “direction” of the joint price-volume vector~\cite{podobnik2009cross}.

\begin{figure*}[htbp]
    \centering

    \begin{subfigure}{\textwidth}
        \centering
        \includegraphics[width=1.0\textwidth]{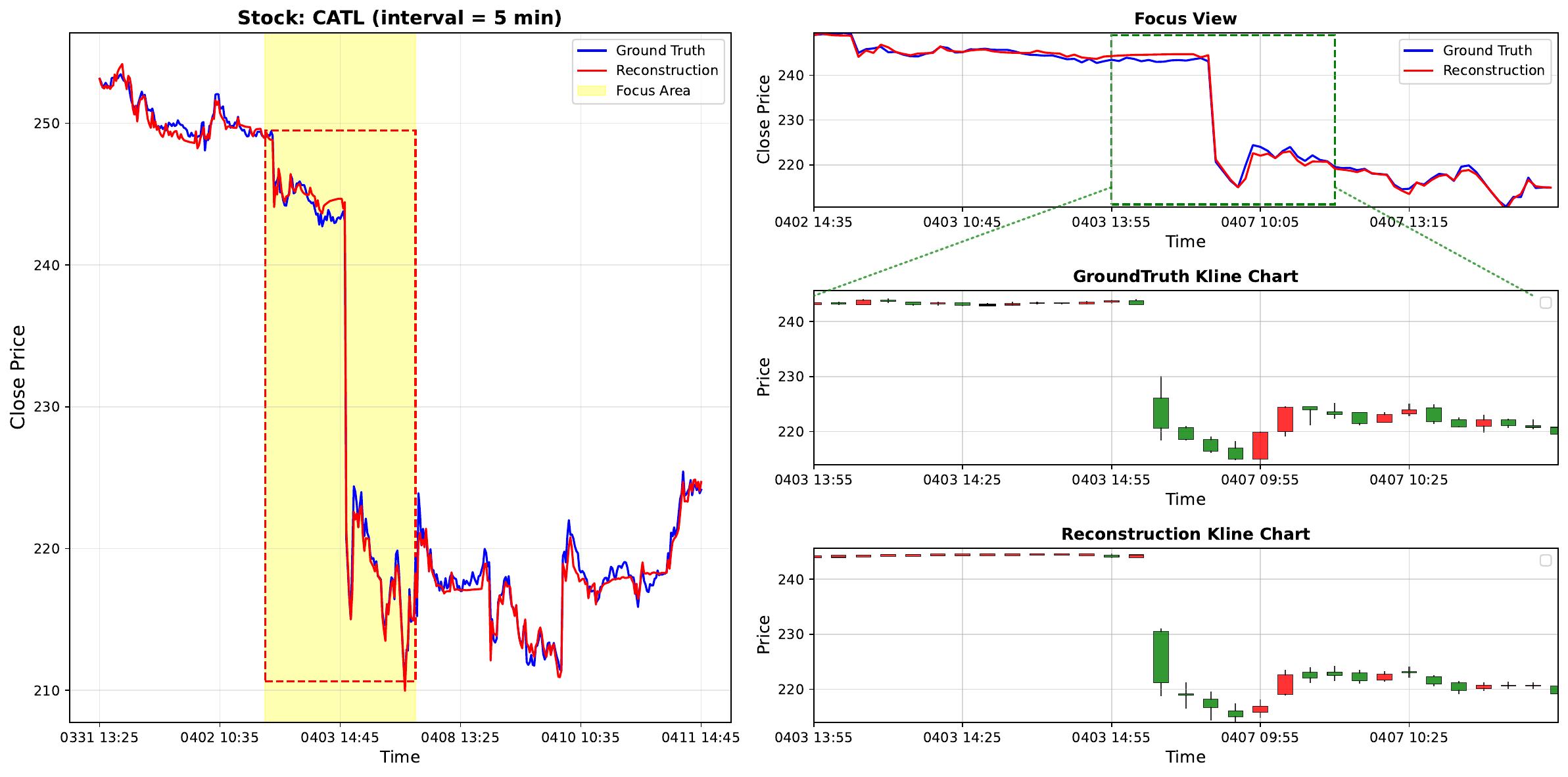}
        \caption{Stock K-line Reconstruction, Price Part }
    \end{subfigure}

    \begin{subfigure}{\textwidth}
        \centering
        \includegraphics[width=1.0\textwidth]{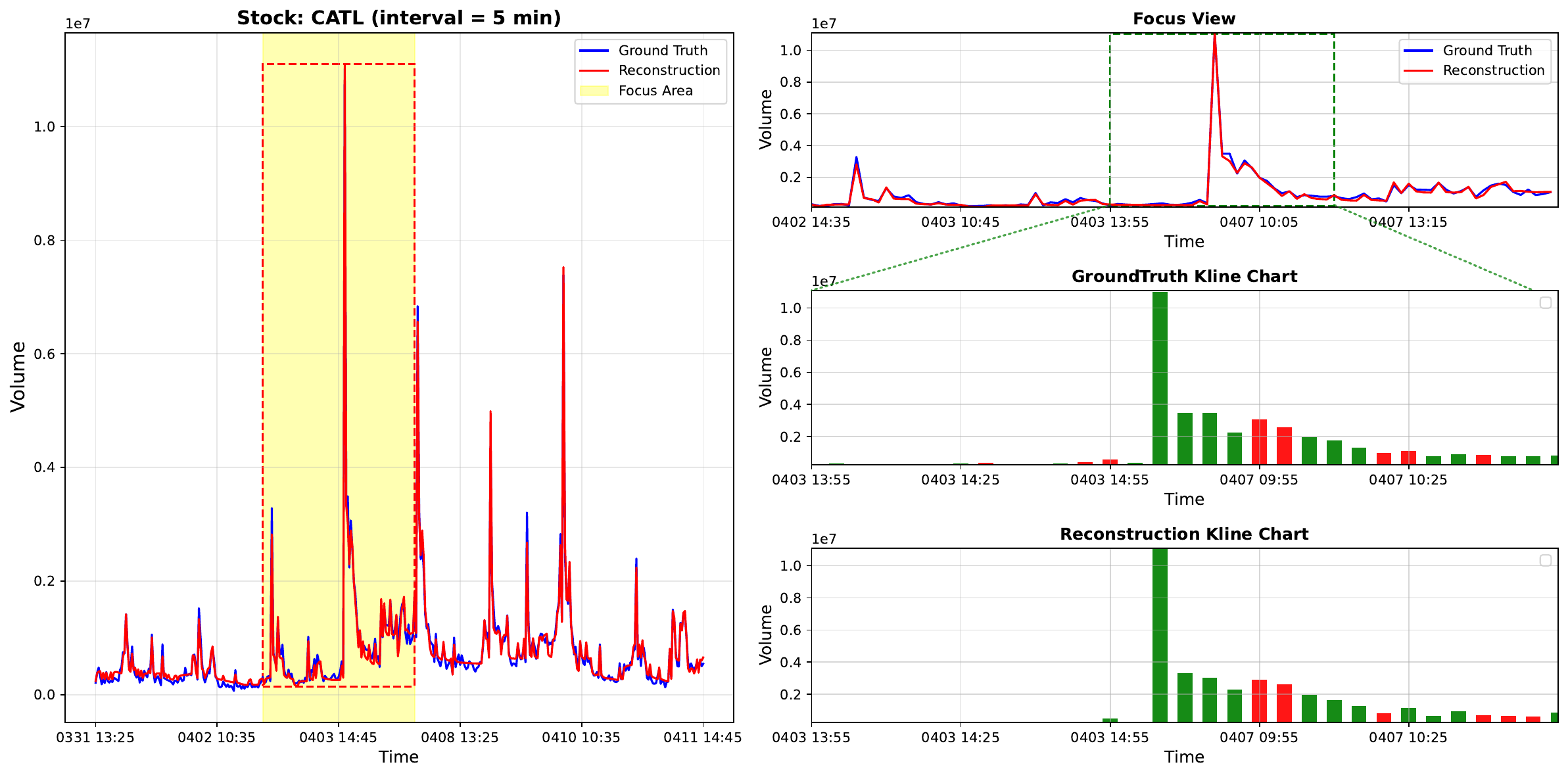}
        \caption{Stock K-line Reconstruction, Volume Part}
    \end{subfigure}

    \vspace{0.1cm}

    \caption{
    Illustration of the reconstruction performance of 5-minute K-line data for CATL (Contemporary Amperex Technology Co., Limited) on April 7th, 2025, in the economic context of Trump’s Trade War~\cite{mckibbin2025global}. In the visualization, the candlesticks follow a ``red for up, green for down'' convention (where up/down is determined by the close price relative to the open price), and the volume bars are colored accordingly.
    }
    \label{fig:kline_reconstruction_tradewar}
\end{figure*}

Figure~\ref{fig:kline_reconstruction_tradewar} illustrates the tokenizer's ability to capture and reconstruct the long-tailed market microstructure under short-term high volatility and during extreme gap events (in the economic context of Trump’s Trade War~\cite{mckibbin2025global}).

Above all, we summarize the concrete advantages of BSQ for K-line time series data, leveraging its ability to preserve angular information and capture sharp directional changes, which are crucial for modeling financial time series with heavy tails and abrupt shifts due to microstructure events.

\subsection{Analysis of Subtoken Factorization (Q3)}
\label{sec:subtoken_factorization}

Our methodology factorizes a $k$-bit token into $n$ subtokens to manage a large vocabulary size. A key design choice is the number of factors, $n$. While further factorization (e.g., $n>2$) could reduce sub-vocabulary sizes even more (e.g., from $2^{10}$ to $2^5$ for a $k=20$ token), we argue that $n=2$ offers the best trade-off between parameter efficiency and inference latency.

This factorization introduces a fundamental trade-off. On one hand, it significantly reduces the size of \textbf{vocabulary-dependent parameters} in the input embedding and output projection layers, replacing a single large table for a $2^k$ vocabulary with $n$ smaller tables for $2^{k/n}$ sub-vocabularies. On the other hand, it introduces two costs: (1) a new \textbf{fusion layer} ($W_{\text{fuse}}$ in Equation~\ref{eq:embedding}), whose parameters $(n \times \mathbf{d}_{\text{model}}) \times \mathbf{d}_{\text{model}}$ grow linearly with $n$, and (2) increased \textbf{inference latency}, as generating a full token requires $n$ sequential autoregressive steps.

Table~\ref{tab:factorization_analysis_revised} quantifies this trade-off for our $\text{Kronos}_{base}$ model. The most significant parameter reduction is achieved by moving from no factorization ($n=1$) to a 2-way split. This single step reduces vocabulary-dependent parameters by over 99.8\% (from $\approx$1.7B to 3.4M), shrinking the total model size by nearly 95\% and making a large effective vocabulary computationally feasible.

However, further factorization yields diminishing returns while incurring rising costs. Moving from $n=2$ to $n=4$ reduces vocabulary parameters by only 3.2M, a saving that is partially offset by a 1.4M increase in fusion layer parameters. This results in a marginal total parameter reduction of just $\approx$2\%. As $n$ increases to 5, the overhead from the fusion layer outweighs the savings from the smaller vocabularies, causing the total parameter count to increase. Crucially, these marginal or negative parameter benefits come at a direct and substantial latency cost: moving from $n=2$ to $n=4$ doubles the number of sequential generation steps required per token.

In summary, our choice of $n=2$ represents an effective balance. It captures the vast majority of the parameter-reduction benefits, making our large vocabulary practical, while avoiding the significant latency penalties and growing architectural overhead associated with finer-grained splits.

\onecolumn

\begin{footnotesize} 
\begin{center} 
\begin{longtable}{@{\extracolsep{\fill}} l l l r r l @{}} 
    
    \toprule
    \textbf{Exchange / Country} & \textbf{Asset Types} & \textbf{Timeframes} & \textbf{\# Assets} & \textbf{\# Observations} & \textbf{Start Date} \\
    \midrule
    \endfirsthead 
    \multicolumn{6}{c}%
    {{\bfseries\tablename\ \thetable{} -- Continued from previous page}} \\
    \toprule
    \textbf{Exchange / Country} & \textbf{Asset Types} & \textbf{Timeframes} & \textbf{\# Assets} & \textbf{\# Observations} & \textbf{Start Date} \\
    \midrule
    \endhead 
    \midrule
    \multicolumn{6}{r@{}}{\textit{Continued on next page}} \\ 
    \endfoot 

    \bottomrule \\
    \caption{Descriptive statistics of the multi-exchange, multi-asset K-line dataset. The timeframe abbreviations are: T (1-min), H (1-hour), D (1-day), W (1-week).}
    \label{tab:dataset_overview} \\ 
    \endlastfoot 

    Binance            & Crypto,Perpetual Swap             & T, 5T, 15T, 30T, H, D, W        & 997   & 1,237,002,843   & 2021/1/31 \\
        Athens Stock Exchange          & Stock, ETF             & D, W        & 180   & 226,315   & 2023/4/11 \\
        Beijing Stock Exchange         & Stock       & 5T, 15T, 30T, H, D, W & 272      & 10,197,628   & 2021/11/19 \\
        Brazil Stock Exchange          & Stock, ETF       & D, W & 2,058      & 1,315,290   & 2020/1/31 \\
        Moscow Exchange             & Stock, ETF       & D, W              & 514     &  567,351    & 2020/1/31 \\
        Euronext Amsterdam             & Stock, ETF           & D, W                  & 514    & 602,083    & 2020/1/31 \\
        Australian Securities Exchange     & Stock, ETF           & 5T, 15T, 30T, H, D, W              & 3,381    & 86,613,897    & 2020/1/31 \\
        Stock Exchange of Thailand            & Stock, ETF        & 5T, 15T, 30T, H, D, W                  & 1,664   & 49,590,394    & 2020/1/31 \\
        Bombay Stock Exchange & Stock, ETF       & 5T, 15T, 30T, H, D, W              & 5,491 & 284,428,211 & 2020/1/31 \\
        Euronext Brussels & Stock, ETF       & D, W              & 166 & 195,491 & 2020/1/31 \\
        Bucharest Stock Exchange & Stock, ETF       & D, W  & 247 & 176,080 & 2020/1/31 \\
        Budapest Stock Exchange & Stock, ETF       & D, W  & 50 & 57,586 & 2022/1/14 \\
        Buenos Aires Stock Exchange & Stock & D, W  & 183 & 225,352 & 2020/1/31 \\
        Colombo Stock Exchange & Stock & D, W  & 292 & 372,627 &  2020/1/31 \\
        Copenhagen Stock Exchange & Stock & D, W  & 825 & 617,464 & 2020/1/31 \\
        Frankfurt Stock Exchange & Stock, ETF & D, W  & 17,054 & 21,547,744 & 2020/1/31 \\
        Ghana Stock Exchange &  Stock & D, W  & 44 & 57,690 & 2020/1/31 \\
        Hong Kong Stock Exchange & Stock, ETF & 5T, 15T, 30T, H, D, W & 3,500 & 359,434,220  & 2020/1/31 \\
        Japan Exchange Group &  Stock, ETF & 5T, 15T, 30T, H, D, W & 4,467 & 280,601,980 &  2020/1/31 \\
        Indonesia Stock Exchange & Stock & 5T, 15T, 30T, H, D, W & 935 & 38,627,125 &  2020/1/31 \\
        Borsa Istanbul &  Stock & D, W  & 627 & 784,147 &  2020/1/31 \\
        Johannesburg Stock Exchange & Stock, ETF & D, W  & 562 & 681,587 & 2020/1/31 \\
        Pakistan Stock Exchange & Stock, ETF & D, W  & 660 & 595,505 & 2020/1/31 \\
        Kuala Lumpur Stock Exchange & Stock, ETF & 5T, 15T, 30T, H, D, W & 1,150 & 45,938,559 & 2020/1/31 \\
        Korea Exchange & Stock, ETF & 5T, 15T, 30T, H, D, W & 2,928 & 205,061,301 & 2020/1/31 \\
        Lima Stock Exchange & Stock, ETF & D, W  & 166 & 63,503 & 2020/1/31 \\
        Euronext Lisbon & Stock, ETF & D, W  & 60 & 65,753 & 2020/1/31 \\
        London Stock Exchange & Stock, ETF & 5T, 15T, 30T, H, D, W & 8,660 & 177,947,624 & 2020/1/31 \\
        Luxembourg Stock Exchange & Stock & D, W  & 5 & 7,598 & 2020/1/31 \\
        Madrid Stock Exchange & Stock, ETF & D, W  & 309 & 331,745 & 2020/1/31 \\
        Mexican Stock Exchange & Stock, ETF & D, W  & 775 & 937,637 & 2020/1/31 \\
        Nasdaq Stock Exchange & Stock, ETF & T, 5T, 15T, 30T, H, D, W & 8,725 & 2,478,662,459 & 2000/1/1 \\
        National Stock Exchange of India & Stock, ETF & 5T, 15T, 30T, H, D, W & 2,554 & 242,429,169 & 2020/1/31 \\
        New York Stock Exchange & Stock, ETF & T, 5T, 15T, 30T, H, D, W & 7,073 & 2,133,143,549 & 2000/1/1 \\
        Euronext Paris & Stock, ETF & D, W  & 1,781 & 1,981,059 & 2020/1/31 \\
        Philippine Stock Exchange & Stock, ETF & 5T, 15T, 30T, H, D, W & 351 & 4,388,378 & 2020/1/31 \\
        Prague Stock Exchange & Stock & D, W  & 50 & 62,666 & 2020/1/31 \\
        Santiago Stock Exchange & Stock & D, W  & 225 & 160,638 & 2020/1/31 \\
        Shenzhen Stock Exchange & Stock, ETF & T, 5T, 15T, 30T, H, D, W & 3,519 & 1,754,519,331 & 1990/12/19 \\
        Shenzhen Stock Exchange (B-shares) & Stock & 5T, 15T, 30T, H, D, W & 46 & 4,198,702 & 2020/2/3 \\
        Shanghai Stock Exchange & Stock, ETF & T, 5T, 15T, 30T, H, D, W & 3,064 & 1,967,996,343 & 1990/12/19 \\
        Shanghai Stock Exchange (B-shares) & Stock & 5T, 15T, 30T, H, D, W & 50 & 4,526,152 & 2020/2/3 \\
        Stockholm Stock Exchange & Stock, ETF & D, W  & 1,305 & 1,463,722 & 2020/1/31 \\
        SIX Swiss Exchange & Stock, ETF & D, W  & 1,981 & 2,451,675 & 2020/1/31 \\
        Taiwan Stock Exchange & Stock, ETF & 5T, 15T, 30T, H, D, W & 1,252 & 71,619,260 & 2020/1/31 \\
        Toronto Stock Exchange & Stock, ETF & D, W  & 3,035 & 3,356,561 & 2020/1/31 \\
        Vienna Stock Exchange & Stock & D, W  & 98 & 123,643 & 2020/1/31 \\
        China & Future & T, 5T, 15T, D & 75 & 63,318,960 & 2010/1/1 \\
        \textbackslash & Foreign Exchange & 5T, 15T, 30T, H, D, W & 1,023 & 462,434,562 & 2020/1/31 \\
        Australia & Stock Index & 5T, 15T, 30T, H, D, W & 40 & 183,158 & 2020/1/31 \\
        Belgium & Stock Index & D, W  & 5 & 8,109 & 2020/1/31 \\
        Brazil & Stock Index & D, W  & 3 & 4,766 & 2020/1/31 \\
        Canada & Stock Index & D, W  & 18 & 27,622 & 2020/1/31 \\
        China & Stock Index & 5T, 15T, 30T, H, D, W & 597 & 55,884,065 & 2020/2/3 \\
        Germany & Stock Index & D, W  & 18 & 28,622 & 2020/1/31 \\
        Spain & Stock Index & D, W  & 2 & 3,257 & 2020/1/31 \\
        France & Stock Index & D, W  & 38 & 55,945 & 2020/1/31 \\
        Britain & Stock Index & 5T, 15T, 30T, H, D, W   & 51 & 5,355,869  & 2020/1/31 \\
        Greece & Stock Index & D, W  & 1 & 1,589 & 2020/1/31 \\
        Hong Kong, China & Stock Index & 5T, 15T, 30T, H, D, W   & 4 & 453,016 & 2020/1/31 \\
        Hungary & Stock Index & D, W  & 1 & 1,602 & 2020/1/31 \\
        Indonesia & Stock Index & 5T, 15T, 30T, H, D, W  & 2 & 47,816 & 2020/1/31 \\
        India & Stock Index & 5T, 15T, 30T, H, D, W  & 113 & 3,189,450 & 2020/1/31 \\
        Japan & Stock Index & 5T, 15T, 30T, H, D, W  & 9 & 125,024 & 2020/1/31 \\
        Korea & Stock Index & 5T, 15T, 30T, H, D, W  & 5 & 274,292 & 2020/1/31 \\
        Mexico & Stock Index & D, W  & 1 & 1,619 & 2020/1/31 \\
        Malaysia & Stock Index & D, W  & 2 & 3,145 & 2020/1/31 \\
        Netherlands & Stock Index & D, W  & 4 & 6,475 & 2020/1/31 \\
        Pakistan & Stock Index & D, W  & 3 & 3,184  & 2020/1/31 \\
        Philippines & Stock Index & D, W  & 2 & 3,187  & 2020/1/31 \\
        Portugal & Stock Index & D, W  & 1 & 1,632 & 2020/1/31 \\
        Romania & Stock Index & D, W  & 5 & 7,726 & 2020/1/31 \\
        Russia & Stock Index & D, W  & 15 & 19,079 & 2020/1/31 \\
        Sweden & Stock Index & D, W  & 11 & 16,389 & 2020/1/31 \\
        Thailand & Stock Index & D, W  & 4 & 5,005 & 2020/1/31 \\
        Taiwan, China & Stock Index & 5T, 15T, 30T, H, D, W  & 1 & 85,318 & 2020/1/31 \\
        America & Stock Index & 5T, 15T, 30T, H, D, W  & 670 & 37,887,535 & 2020/1/31 \\

    \midrule
    \multicolumn{3}{@{}l}{\textbf{Approximate Totals}} & \textbf{96569} & \textbf{12.11B} & -- \\

\end{longtable}
\end{center}
\end{footnotesize}

\twocolumn

\twocolumn

\begin{table*}[t]
  \centering
  \footnotesize
  \setlength{\tabcolsep}{4pt}

  \adjustbox{max width=\textwidth}{
  \begin{tabular}{l|c|*{11}{S}}
    \toprule
    \multicolumn{2}{c}{\textbf{Models}} &
    \multicolumn{3}{c}{\textbf{\includegraphics[height=0.25cm]{img/logo_paper.png}\hspace{0.2em}Kronos (Ours)}} &
    \multicolumn{8}{c}{\textbf{Full-shot Time Series Models}}\\
    \cmidrule(lr){3-5} \cmidrule(lr){6-13}
    \multicolumn{2}{c}{\textbf{Metrics}} &
    \textbf{$\text{Kronos}_S$} & \textbf{$\text{Kronos}_B$} & \textbf{$\text{Kronos}_L$} &
    \textbf{TimeXer} & \textbf{TimeMixer} & \textbf{iTransformer} & \textbf{PatchTST} &
    \textbf{TimesNet} & \textbf{DLinear} & \textbf{FEDformer} & \textbf{NSTransformer} \\
    \midrule
        
        \multirow{2}{*}{XSHG} & IC     & \best{0.0549} & \second{0.0564} & 0.0546 & 0.0280 & 0.0291 & 0.0350 & 0.0450 & 0.0424 & 0.0405 & 0.0233 & 0.0433 \\
                              & RankIC & 0.0375 & \best{0.0390} & \second{0.0381} & 0.0053 & 0.0079 & 0.0128 & 0.0088 & 0.0175 & 0.0181 & 0.0107 & 0.0155 \\
        \midrule     
        \multirow{2}{*}{XNAS} & IC     & \second{0.0343} & 0.0322 & \best{0.0361} & 0.0132 & 0.0097 & 0.0204 & 0.0116 & 0.0174 & 0.0197 & 0.0165 & 0.0253 \\
                              & RankIC & 0.0155 & \second{0.0190} & \best{0.0191} & 0.0106 & 0.0048 & 0.0111 & 0.0083 & 0.0084 & 0.0084 & -0.0014 & 0.0016 \\
        \midrule   
        \multirow{2}{*}{XJPX} & IC     & 0.0314 & \second{0.0332} & \best{0.0360} & 0.0094 & 0.0017 & 0.0137 & 0.0053 & 0.0099 & 0.0118 & 0.0046 & 0.0281 \\
                              & RankIC & 0.0199 & 0.0209 & \best{0.0277} & 0.0159 & 0.0036 & \second{0.0271} & 0.0056 & 0.0127 & 0.0149 & 0.0024 & 0.0212 \\
        \midrule
        \multirow{2}{*}{XNSE} & IC     & \second{0.0634} & \best{0.0648} & \second{0.0634} & -0.0055 & 0.0094 & -0.0252 & 0.0082 & 0.0566 & 0.0024 & 0.0063 & 0.0514 \\
                              & RankIC & 0.0434 & \second{0.0464} & \best{0.0486} & -0.0371 & 0.0024 & -0.0248 & 0.0084 & 0.0379 & -0.0024 & 0.0003 & 0.0225 \\
        \midrule
        \multirow{2}{*}{XKRX} & IC     & 0.0550 & \best{0.0575} & \second{0.0567} & -0.0328 & 0.0036 & -0.0442 & 0.0248 & 0.0416 & 0.0001 & -0.0070 & 0.0416 \\
                              & RankIC & 0.0362 & \best{0.0393} & \second{0.0373} & -0.0160 & 0.0033 & -0.0284 & 0.0214 & 0.0285 & 0.0006 & -0.0049 & 0.0058 \\
        \midrule
        \multirow{2}{*}{XHKG} & IC     & \second{0.0435} & \best{0.0439} & 0.0428 & 0.0318 & 0.0322 & 0.0336 & 0.0401 & 0.0333 & 0.0392 & 0.0296 & 0.0366 \\
                              & RankIC & 0.0226 & \best{0.0236} & \second{0.0228} & -0.0051 & -0.0009 & -0.0021 & -0.0068 & -0.0040 & -0.0009 & -0.0078 & -0.0017 \\
        \midrule
        \multirow{2}{*}{XIDX} & IC     & \second{0.0551} & \second{0.0551} & \best{0.0573} & -0.0139 & 0.0116 & -0.0233 & 0.0194 & 0.0468 & 0.0158 & 0.0169 & 0.0381 \\
                              & RankIC & 0.0214 & \second{0.0216} & \best{0.0223} & 0.0025 & 0.0046 & 0.0011 & 0.0149 & 0.0171 & 0.0037 & 0.0084 & 0.0051 \\
        \midrule
        \multirow{2}{*}{XKLS} & IC     & \second{0.0411} & 0.0408 & \best{0.0466} & -0.0283 & 0.0079 & -0.0281 & -0.0037 & 0.0341 & 0.0306 & -0.0102 & 0.0101 \\
                              & RankIC & \best{0.0215} & 0.0149 & 0.0167 & 0.0051 & 0.0171 & 0.0024 & -0.0078 & -0.0025 & \second{0.0208} & -0.0169 & -0.0103 \\
        \midrule
        \multirow{2}{*}{XTAI} & IC     & 0.0424 & \second{0.0443} & \best{0.0448} & 0.0282 & 0.0197 & 0.0275 & 0.0328 & 0.0312 & 0.0394 & 0.0249 & 0.0334 \\
                              & RankIC & 0.0301 & \second{0.0320} & \best{0.0342} & -0.0042 & 0.0015 & 0.0111 & 0.0147 & 0.0095 & 0.0192 & 0.0059 & 0.0129 \\
        \midrule
        \multirow{2}{*}{Crypto} & IC     & \best{0.0247} & 0.0209 & \second{0.0211}  & 0.0105 & 0.0128 & 0.0155 & 0.0149 & 0.0192 & 0.0137 & 0.0081 & 0.0164 \\
                                & RankIC & 0.0138 & 0.0135 & 0.0129  & 0.0022 & 0.0038 & 0.0134 & \best{0.0192} & \second{0.0146} & 0.0040 & 0.0000 & 0.0096\\
        \midrule
        \multirow{2}{*}{Forex} & IC     & \second{0.0279} & \best{0.0292} & 0.0244 & 0.0124 & 0.0102 & 0.0142 & 0.0158 & 0.0167 & 0.0227 & 0.0153 & 0.0228 \\
                               & RankIC & \best{0.0177} & 0.0141 & 0.0137 & 0.0134 & 0.0128 & 0.0090 & 0.0085 & \second{0.0175} & 0.0168 & 0.0120 & 0.0079 \\
        \midrule

        \rowcolor{AvgGray}
          & IC &
        0.0431 & \second{0.0435} & \best{0.0440} & 0.0048 & 0.0134 & 0.0036 & 0.0195 & 0.0317 & 0.0214 & 0.0117 & 0.0316 \\
        \rowcolor{AvgGray} \multirow{2}{*}[2.5ex]{\textbf{Average}} & RankIC &
        0.0254 & \second{0.0258} & \best{0.0267} & -0.0007 & 0.0055 & 0.0030 & 0.0087 & 0.0143 & 0.0094 & 0.0008 & 0.0082 \\
         \midrule
    \rowcolor{CntGray}
    \multicolumn{2}{c}{\textbf{$\mathbf{1^{st}}$ Count}} &
\cnt{4} & \cnt{7} & \cnt{10} & \cnt{0} & \cnt{0} & \cnt{0} &
\cnt{1} & \cnt{0} & \cnt{0} & \cnt{0} & \cnt{0} \\
    \bottomrule
  \end{tabular}}
  \caption{Full results of price series forecasting experiments (Part 1): Our model (Kronos) and full-shot time series models. A higher IC or RankIC indicates a better prediction. Best and second best results are marked with \best{red underline} and \second{blue underline}, respectively.}
  \label{tab:full_price_forecasting_1}
\end{table*}

\begin{table*}[t]
    \centering
    \footnotesize  
    \setlength{\tabcolsep}{4pt}

    \adjustbox{max width=\textwidth}{
    \begin{tabular}{l|c|*{13}{S}}
        \toprule
        \multicolumn{2}{c}{\textbf{Models}} &
        \multicolumn{12}{c}{\textbf{Zero-shot Time Series Models}} \\
        \cmidrule(lr){3-14}
        \multicolumn{2}{c}{\textbf{Metrics}} &
        \textbf{$\text{Time\text{-}MOE}_{S}$} & \textbf{$\text{Time\text{-}MOE}_{B}$} &
        \textbf{$\text{Moirai}_{S}$} & \textbf{$\text{Moirai}_{B}$} & \textbf{$\text{Moirai}_{L}$} &
        \textbf{TimesFM} &
        \textbf{$\text{Moment}_{S}$} & \textbf{$\text{Moment}_{B}$} & \textbf{$\text{Moment}_{L}$} &
        \textbf{$\text{Chronos}_{S}$} & \textbf{$\text{Chronos}_{B}$} & \textbf{$\text{Chronos}_{L}$} \\
        \midrule
        \multirow{2}{*}{XSHG} & IC     & 0.0463 & 0.0493 & -0.0007 & -0.0005 & -0.0002 & 0.0174 & 0.0028 & -0.0032 & -0.0009 & 0.0147 & 0.0069 & 0.0195 \\
                              & RankIC & 0.0304 & 0.0317 & 0.0000 & -0.0012 & 0.0003 & 0.0020 & 0.0003 & -0.0037 & -0.0017 & -0.0026 & -0.0108 & 0.0025 \\
        \midrule
        \multirow{2}{*}{XNAS} & IC     & -0.0032 & -0.0045 & -0.0008 & -0.0005 & 0.0000 & 0.0076 & 0.0010 & -0.0023 & -0.0003 & -0.0025 & -0.0005 & 0.0020 \\
                              & RankIC & -0.0033 & -0.0042 & -0.0008 & 0.0013 & 0.0007 & 0.0112 & -0.0015 & -0.0027 & -0.0007 & -0.0001 & 0.0008 & 0.0030 \\
        \midrule
        \multirow{2}{*}{XJPX} & IC     & 0.0268 & 0.0280 & 0.0012 & 0.0004 & 0.0000 & 0.0076 & -0.0010 & -0.0003 & -0.0027 & 0.0117 & 0.0113 & 0.0067 \\
                              & RankIC & 0.0228 & 0.0230 & 0.0025 & 0.0019 & 0.0016 & 0.0073 & -0.0031 & 0.0010 & -0.0006 & 0.0110 & 0.0123 & 0.0070 \\
        \midrule
        \multirow{2}{*}{XNSE} & IC     & 0.0173 & 0.0190 & -0.0005 & -0.0008 & -0.0006 & 0.0025 & 0.0063 & -0.0129 & -0.0039 & -0.0012 & -0.0049 & 0.0014 \\
                              & RankIC & 0.0155 & 0.0169 & -0.0021 & -0.0021 & -0.0029 & 0.0009 & 0.0060 & -0.0104 & -0.0055 & -0.0041 & -0.0066 & -0.0005 \\
        \midrule
        \multirow{2}{*}{XKRX} & IC     & 0.0113 & 0.0141 & -0.0014 & 0.0006 & -0.0011 & -0.0105 & 0.0056 & -0.0083 & -0.0114 & -0.0009 & -0.0018 & 0.0061 \\
                              & RankIC & 0.0088 & 0.0118 & -0.0020 & 0.0006 & -0.0002 & -0.0097 & 0.0041 & -0.0082 & -0.0069 & -0.0006 & -0.0009 & 0.0072 \\
        \midrule
        \multirow{2}{*}{XHKG} & IC     & 0.0174 & 0.0189 & 0.0000 & -0.0001 & -0.0013 & 0.0117 & 0.0013 & -0.0050 & -0.0003 & 0.0159 & 0.0140 & 0.0166 \\
                              & RankIC & 0.0186 & 0.0201 & 0.0003 & 0.0031 & 0.0011 & 0.0058 & -0.0034 & -0.0013 & 0.0009 & 0.0190 & 0.0179 & 0.0192 \\
        \midrule
        \multirow{2}{*}{XIDX} & IC     & -0.0053 & -0.0052 & -0.0009 & -0.0009 & -0.0003 & 0.0026 & 0.0052 & -0.0094 & -0.0007 & 0.0021 & 0.0042 & 0.0080 \\
                              & RankIC & 0.0002 & 0.0000 & 0.0008 & 0.0012 & 0.0007 & 0.0042 & -0.0015 & -0.0029 & 0.0014 & 0.0087 & 0.0122 & 0.0153 \\
        \midrule
        \multirow{2}{*}{XKLS} & IC     & 0.0123 & 0.0125 & -0.0003 & -0.0028 & 0.0005 & 0.0106 & 0.0045 & -0.0093 & -0.0065 & -0.0080 & -0.0076 & -0.0077 \\
                              & RankIC & 0.0112 & 0.0135 & 0.0010 & 0.0027 & 0.0047 & -0.0052 & 0.0000 & 0.0017 & -0.0031 & 0.0113 & 0.0118 & 0.0114 \\
        \midrule
        \multirow{2}{*}{XTAI} & IC     & 0.0296 & 0.0292 & 0.0005 & 0.0001 & -0.0004 & -0.0002 & 0.0025 & -0.0047 & -0.0046 & 0.0028 & -0.0002 & 0.0080 \\
                              & RankIC & 0.0234 & 0.0224 & 0.0011 & 0.0013 & 0.0003 & -0.0028 & 0.0001 & -0.0023 & -0.0009 & 0.0088 & 0.0060 & 0.0125 \\
        \midrule
        \multirow{2}{*}{Crypto} & IC     & 0.0054 & 0.0037 & -0.0008 & -0.0006 & -0.0004 & -0.0009 & -0.0002 & -0.0004 & -0.0030 & -0.0114 & -0.0129 & -0.0096 \\
                                & RankIC & 0.0069 & 0.0050 & 0.0004 & 0.0011 & 0.0000 & 0.0014 & -0.0011 & -0.0061 & -0.0007 & -0.0051 & -0.0061 & -0.0045 \\
        \midrule
        \multirow{2}{*}{Forex} & IC     & 0.0265 & 0.0267 & -0.0011 & -0.0011 & 0.0000 & 0.0092 & -0.0007 & 0.0008 & 0.0024 & 0.0176 & 0.0143 & 0.0155 \\
                               & RankIC & 0.0115 & 0.0114 & -0.0010 & 0.0005 & -0.0003 & 0.0076 & -0.0014 & -0.0010 & 0.0022 & 0.0168 & 0.0147 & 0.0127 \\
        \midrule

        \rowcolor{AvgGray}
         & IC &
        0.0168 & 0.0174 & -0.0004 & -0.0006 & -0.0003 & 0.0052 & 0.0025 & -0.0050 & -0.0029 & 0.0037 & 0.0021 & 0.0060 \\
        \rowcolor{AvgGray} \multirow{2}{*}[2.5ex]{\textbf{Average}} & RankIC &
        0.0133 & 0.0138 & 0.0000 & 0.0009 & 0.0005 & 0.0021 & -0.0001 & -0.0033 & -0.0014 & 0.0057 & 0.0047 & 0.0078 \\
        \midrule
        \rowcolor{CntGray}
        \multicolumn{2}{c}{\textbf{$\mathbf{1^{st}}$ Count}} &
\cnt{0} & \cnt{0} & \cnt{0} & \cnt{0} & \cnt{0} & \cnt{0} &
\cnt{0} & \cnt{0} & \cnt{0} & \cnt{0} & \cnt{0} & \cnt{0} \\
        \bottomrule
    \end{tabular}}
\caption{Full results of price series forecasting experiments (Part 2): Zero-shot time series models. A higher IC or RankIC indicates a better prediction. Best and second best results are marked with \best{red underline} and \second{blue underline}, respectively.}
    \label{tab:full_price_forecasting_2}
\end{table*}

\begin{table*}[t]
  \centering
  \footnotesize
  \setlength{\tabcolsep}{4pt}

  \adjustbox{max width=\textwidth}{
  \begin{tabular}{l|c|*{11}{S}}
    \toprule
    \multicolumn{2}{c}{\textbf{Models}} &
    \multicolumn{3}{c}{\textbf{\includegraphics[height=0.25cm]{img/logo_paper.png}\hspace{0.2em}Kronos (Ours)}} &
    \multicolumn{8}{c}{\textbf{Full-shot Time Series Models}}\\
    \cmidrule(lr){3-5} \cmidrule(lr){6-13}
    \multicolumn{2}{c}{\textbf{Metrics}} &
    \textbf{$\text{Kronos}_S$} & \textbf{$\text{Kronos}_B$} & \textbf{$\text{Kronos}_L$} &
    \textbf{TimeXer} & \textbf{TimeMixer} & \textbf{iTransformer} & \textbf{PatchTST} &
    \textbf{TimesNet} & \textbf{DLinear} & \textbf{FEDformer} & \textbf{NSTransformer}\\
    \midrule
        
        \multirow{2}{*}{XSHG} & IC     & \second{0.0677} & 0.0652 & 0.0662 & 0.0456 & 0.0114 & 0.0371 & 0.0467 & 0.0563 & 0.0626 & 0.0589 & \best{0.0777} \\
                              & RankIC & 0.0617 & 0.0653 & 0.0642 & 0.0306 & -0.0072 & 0.0266 & 0.0437 & 0.0421 & 0.0461 & 0.0568 & 0.0595 \\
        \midrule     
        \multirow{2}{*}{XNAS} & IC     & 0.0563 & \second{0.0626} & \best{0.0639} & 0.0051 & 0.0270 & 0.0340 & 0.0569 & -0.0193 & 0.0144 & 0.0219 & 0.0377 \\
                              & RankIC & 0.0513 & \second{0.0544} & \best{0.0601} & 0.0061 & 0.0204 & 0.0251 & 0.0446 & 0.0352 & 0.0518 & 0.0254 & 0.0335 \\
        \midrule   
        \multirow{2}{*}{XJPX} & IC     & 0.0618 & \second{0.0667} & \best{0.0668} & 0.0309 & 0.0211 & 0.0439 & 0.0655 & 0.0656 & 0.0621 & 0.0409 & 0.0436 \\
                              & RankIC & 0.0583 & 0.0623 & 0.0687 & 0.0474 & 0.0145 & 0.0399 & 0.0446 & 0.0556 & 0.0253 & 0.0373 & 0.0428 \\
        \midrule
        \multirow{2}{*}{XNSE} & IC     & 0.0501 & \second{0.0523} & \best{0.0585} & -0.0021 & -0.0126 & 0.0117 & 0.0216 & 0.0238 & 0.0144 & 0.0238 & 0.0314 \\
                              & RankIC & 0.0541 & \second{0.0550} & \best{0.0639} & 0.0031 & 0.0044 & 0.0146 & 0.0238 & 0.0277 & 0.0442 & 0.0130 & 0.0312 \\
        \midrule
        \multirow{2}{*}{XKRX} & IC     & 0.0749 & 0.0778 & \second{0.0792} & 0.0389 & 0.0253 & 0.0309 & 0.0589 & \best{0.0844} & 0.0704 & 0.0726 & 0.0754 \\
                              & RankIC & 0.0707 & 0.0763 & 0.0790 & -0.0024 & -0.0071 & 0.0282 & 0.0422 & \best{0.0801} & 0.0439 & 0.0354 & \second{0.0792} \\
        \midrule
        \multirow{2}{*}{XHKG} & IC     & \best{0.0678} & 0.0661 & 0.0654 & \second{0.0666} & -0.0276 & 0.0106 & 0.0470 & 0.0276 & 0.0404 & 0.0496 & 0.0210 \\
                              & RankIC & 0.0671 & 0.0646 & \second{0.0703} & \best{0.0707} & -0.0063 & 0.0091 & 0.0631 & 0.0288 & 0.0558 & 0.0605 & 0.0264 \\
        \midrule
        \multirow{2}{*}{XIDX} & IC     & \second{0.0998} & 0.0990 & \best{0.1046} & 0.0039 & -0.0095 & 0.0393 & 0.0003 & 0.0301 & 0.0195 & -0.0007 & 0.0244 \\
                              & RankIC & \second{0.0943} & 0.0924 & \best{0.1007} & -0.0111 & -0.0018 & 0.0341 & 0.0280 & 0.0304 & 0.0184 & 0.0358 & 0.0610 \\
        \midrule
        \multirow{2}{*}{XKLS} & IC     & \second{0.1213} & 0.1153 & \best{0.1359} & 0.0144 & 0.0074 & 0.0252 & 0.0605 & 0.0941 & 0.0781 & -0.0016 & 0.1046 \\
                              & RankIC & \second{0.1047} & 0.1009 & \best{0.1145} & -0.0261 & 0.0097 & 0.0237 & 0.0685 & 0.0712 & 0.0800 & -0.0050 & 0.0851 \\
        \midrule
        \multirow{2}{*}{XTAI} & IC     & \best{0.0549} & \second{0.0524} & 0.0511 & 0.0382 & -0.0038 & 0.0313 & 0.0421 & 0.0216 & 0.0514 & 0.0489 & 0.0143 \\
                              & RankIC & \second{0.0597} & 0.0584 & \best{0.0609} & 0.0404 & -0.0027 & 0.0163 & 0.0363 & 0.0261 & 0.0431 & 0.0444 & 0.0159 \\
        \midrule
        \multirow{2}{*}{Crypto} & IC     & 0.0373 & \second{0.0376} & 0.0368  & 0.0286 & 0.0250 & 0.0372 & 0.0163 & 0.0348 & \best{0.0446} & 0.0065 & 0.0274 \\
                                & RankIC & 0.0332 & \best{0.0336} & \second{0.0333}  & 0.0154 & 0.0135 & 0.0151 & 0.0213 & 0.0272 & 0.0283 & 0.0027 & 0.0111\\
        \midrule
        \multirow{2}{*}{Forex} & IC     & 0.0398 & \best{0.0555} & \second{0.0441} & 0.0079 & 0.0203 & 0.0266 & 0.0124 & 0.0054 & 0.0254 & 0.0146 & 0.0122 \\
                               & RankIC & 0.0289 & \best{0.0343} & 0.0274 & 0.0275 & \second{0.0322} & 0.0152 & 0.0148 & 0.0037 & 0.0279 & 0.0148 & 0.0169 \\
        \midrule

        \rowcolor{AvgGray}
          & IC &
        0.0665 & \second{0.0682} & \best{0.0702} & 0.0253 & 0.0076 & 0.0298 & 0.0389 & 0.0386 & 0.0439 & 0.0305 & 0.0427\\
        \rowcolor{AvgGray} \multirow{2}{*}[2.5ex]{\textbf{Average}} & RankIC &
        0.0622 & \second{0.0634} & \best{0.0675} & 0.0183 & 0.0063 & 0.0225 & 0.0392 & 0.0389 & 0.0423 & 0.0292 & 0.0421\\
         \midrule
    \rowcolor{CntGray}
    \multicolumn{2}{c}{\textbf{$\mathbf{1^{st}}$ Count}} &
\cnt{2} & \cnt{3} & \cnt{10} & \cnt{1} & \cnt{0} & \cnt{0} &
\cnt{0} & \cnt{2} & \cnt{1} & \cnt{0} & \cnt{1} \\
    \bottomrule
  \end{tabular}}
 \caption{Full results of return forecasting experiments (Part 1): Our model (Kronos) and full-shot time series models. A higher IC or RankIC indicates a better prediction. Best and second best results are marked with \best{red underline} and \second{blue underline}, respectively.}
  \label{tab:full_return_forecasting_1}
\end{table*}

\begin{table*}[t]
    \centering
    \footnotesize   
    \setlength{\tabcolsep}{4pt}

    \adjustbox{max width=\textwidth}{
    \begin{tabular}{l|c|*{13}{S}}
        \toprule

        \multicolumn{2}{c}{\textbf{Models}} &
        \multicolumn{12}{c}{\textbf{Zero-shot Time Series Models}} \\
        \cmidrule(lr){3-14}
        \multicolumn{2}{c}{\textbf{Metrics}} &
        \textbf{$\text{Time\text{-}MOE}_{S}$} & \textbf{$\text{Time\text{-}MOE}_{B}$} &
        \textbf{$\text{Moirai}_{S}$} & \textbf{$\text{Moirai}_{B}$} & \textbf{$\text{Moirai}_{L}$} &
        \textbf{TimesFM} &
        \textbf{$\text{Moment}_{S}$} & \textbf{$\text{Moment}_{B}$} & \textbf{$\text{Moment}_{L}$} &
        \textbf{$\text{Chronos}_{S}$} & \textbf{$\text{Chronos}_{B}$} & \textbf{$\text{Chronos}_{L}$} \\
        \midrule

        \multirow{2}{*}{XSHG} & IC     & 0.0507 & 0.0501 & 0.0507 & 0.0579 & 0.0534 & 0.0322 & 0.0575 & 0.0579 & 0.0575 & -0.0152 & -0.0055 & -0.0019 \\
                              & RankIC & 0.0612 & 0.0621 & \second{0.0657} & 0.0647 & \best{0.0661} & 0.0445 & 0.0527 & 0.0530 & 0.0525 & -0.0277 & -0.0116 & -0.0048 \\
        \midrule
        \multirow{2}{*}{XNAS} & IC     & 0.0416 & 0.0399 & 0.0275 & 0.0281 & 0.0271 & 0.0226 & 0.0290 & 0.0288 & 0.0287 & 0.0545 & 0.0504 & 0.0572 \\
                              & RankIC & 0.0480 & 0.0457 & 0.0280 & 0.0290 & 0.0304 & 0.0271 & 0.0300 & 0.0297 & 0.0296 & 0.0448 & 0.0405 & 0.0461 \\
        \midrule
        \multirow{2}{*}{XJPX} & IC     & 0.0639 & 0.0642 & 0.0441 & 0.0417 & 0.0446 & 0.0498 & 0.0509 & 0.0508 & 0.0512 & 0.0326 & 0.0323 & 0.0276 \\
                              & RankIC & 0.0473 & 0.0487 & \second{0.0790} & \second{0.0790} & \best{0.0793} & 0.0579 & 0.0490 & 0.0491 & 0.0493 & 0.0175 & 0.0174 & 0.0126 \\
        \midrule
        \multirow{2}{*}{XNSE} & IC     & 0.0348 & 0.0343 & 0.0356 & 0.0357 & 0.0354 & 0.0068 & 0.0356 & 0.0357 & 0.0354 & 0.0190 & 0.0179 & 0.0168 \\
                              & RankIC & 0.0476 & 0.0483 & 0.0518 & 0.0518 & 0.0514 & 0.0180 & 0.0518 & 0.0518 & 0.0514 & 0.0116 & 0.0175 & 0.0161 \\
        \midrule
        \multirow{2}{*}{XKRX} & IC     & 0.0573 & 0.0566 & 0.0545 & 0.0546 & 0.0512 & 0.0392 & 0.0545 & 0.0546 & 0.0544 & 0.0523 & 0.0508 & 0.0532 \\
                              & RankIC & 0.0599 & 0.0592 & 0.0617 & 0.0619 & 0.0545 & 0.0465 & 0.0617 & 0.0619 & 0.0618 & 0.0348 & 0.0347 & 0.0394 \\
        \midrule
        \multirow{2}{*}{XHKG} & IC     & 0.0373 & 0.0385 & 0.0324 & 0.0314 & 0.0304 & 0.0281 & 0.0358 & 0.0357 & 0.0357 & 0.0271 & 0.0286 & 0.0297 \\
                              & RankIC & 0.0439 & 0.0431 & 0.0485 & 0.0487 & 0.0486 & 0.0369 & 0.0485 & 0.0487 & 0.0486 & 0.0315 & 0.0331 & 0.0328 \\
        \midrule
        \multirow{2}{*}{XIDX} & IC     & 0.0611 & 0.0565 & 0.0487 & 0.0475 & 0.0474 & 0.0555 & 0.0487 & 0.0488 & 0.0489 & 0.0514 & 0.0560 & 0.0615 \\
                              & RankIC & 0.0638 & 0.0597 & 0.0586 & 0.0586 & 0.0587 & 0.0582 & 0.0586 & 0.0586 & 0.0587 & 0.0404 & 0.0486 & 0.0522 \\
        \midrule
        \multirow{2}{*}{XKLS} & IC     & 0.0971 & 0.0963 & 0.0815 & 0.0782 & 0.0852 & 0.0585 & 0.0856 & 0.0854 & 0.0854 & 0.0804 & 0.0788 & 0.0772 \\
                              & RankIC & 0.0954 & 0.0952 & 0.1004 & 0.1001 & 0.0999 & 0.0710 & 0.0803 & 0.0800 & 0.0799 & 0.0723 & 0.0698 & 0.0697 \\
        \midrule
        \multirow{2}{*}{XTAI} & IC     & 0.0386 & 0.0369 & 0.0418 & 0.0414 & 0.0412 & 0.0332 & 0.0418 & 0.0414 & 0.0412 & 0.0361 & 0.0359 & 0.0338 \\
                              & RankIC & 0.0238 & 0.0202 & 0.0494 & 0.0488 & 0.0487 & 0.0505 & 0.0394 & 0.0388 & 0.0387 & 0.0264 & 0.0326 & 0.0312 \\
        \midrule
        \multirow{2}{*}{Crypto} & IC     & 0.0291 & 0.0293 & -0.0051 & -0.0081 & -0.0046 & -0.0042 & -0.0042 & -0.0039 & -0.0043 & 0.0041 & 0.0067 & 0.0107 \\
                                & RankIC & 0.0122 & 0.0112 & 0.0157 & 0.0172 & 0.0159 & 0.0105 & 0.0058 & 0.0071 & 0.0059 & -0.0069 & -0.0064 & 0.0009 \\
        \midrule
        \multirow{2}{*}{Forex} & IC     & 0.0334 & 0.0336 & 0.0355 & 0.0357 & 0.0347 & 0.0353 & 0.0155 & 0.0157 & 0.0157 & 0.0289 & 0.0255 & 0.0274 \\
                               & RankIC & 0.0217 & 0.0215 & 0.0262 & 0.0264 & 0.0264 & 0.0276 & 0.0162 & 0.0164 & 0.0164 & 0.0194 & 0.0218 & 0.0184 \\
        \midrule

        \rowcolor{AvgGray}
         & IC &
        0.0495 & 0.0487 & 0.0407 & 0.0404 & 0.0405 & 0.0325 & 0.0410 & 0.0410 & 0.0409 & 0.0337 & 0.0343 & 0.0357 \\
        \rowcolor{AvgGray} \multirow{2}{*}[2.5ex]{\textbf{Average}} & RankIC &
        0.0477 & 0.0468 & 0.0532 & 0.0533 & 0.0527 & 0.0408 & 0.0449 & 0.0450 & 0.0448 & 0.0240 & 0.0271 & 0.0286 \\
        \midrule
        \rowcolor{CntGray}
        \multicolumn{2}{c}{\textbf{$\mathbf{1^{st}}$ Count}} &
\cnt{0} & \cnt{0} & \cnt{0} & \cnt{0} & \cnt{2} & \cnt{0} &
\cnt{0} & \cnt{0} & \cnt{0} & \cnt{0} & \cnt{0} & \cnt{0} \\
        \bottomrule
    \end{tabular}}
\caption{Full results of return forecasting experiments (Part 2): Zero-shot time series models. A higher IC or RankIC indicates a better prediction. Best and second best results are marked with \best{red underline} and \second{blue underline}, respectively.}
    \label{tab:full_return_forecasting_2}
\end{table*}

\begin{table*}[t]
  \centering
  \footnotesize
  \setlength{\tabcolsep}{4pt}

  \adjustbox{max width=\textwidth}{
  \begin{tabular}{l|c|*{13}{S}}
    \toprule
    \multicolumn{2}{c}{\textbf{Models}} &
    \multicolumn{3}{c}{\textbf{\includegraphics[height=0.25cm]{img/logo_paper.png}\hspace{0.2em}Kronos (Ours)}} &
    \multicolumn{8}{c}{\textbf{Full-shot Time Series Models}} & \multicolumn{2}{c}{\textbf{Eco. Volatility Models}}\\
    \cmidrule(lr){3-5} \cmidrule(lr){6-13} \cmidrule(lr){14-15}
    \multicolumn{2}{c}{\textbf{Metrics}} &
    \textbf{$\text{Kronos}_S$} & \textbf{$\text{Kronos}_B$} & \textbf{$\text{Kronos}_L$} &
    \textbf{TimeXer} & \textbf{TimeMixer} & \textbf{iTransformer} & \textbf{PatchTST} &
    \textbf{TimesNet} & \textbf{DLinear} & \textbf{FEDformer} & \textbf{NSTransformer} & \textbf{ARCH} & \textbf{GARCH} \\
    \midrule
        
        \multirow{2}{*}{XSHG} & MAE     & \best{0.0199} & 0.0205 & \second{0.0203} & 0.0510 & 0.0349 & 0.0593 & 0.0356 & 0.0348 & 0.0398 & 0.0231 & 0.0348 & 0.0247 & 0.0219 \\
                              & $R^2$ & 0.2597 & \second{0.2630} & \best{0.2809} & 0.1500 & 0.1585 & 0.2191 & 0.2401 & 0.1429 & 0.2400 & 0.2301 & 0.1232 & 0.1969 & 0.1986 \\
        \midrule     
        \multirow{2}{*}{XNAS} & MAE     & 0.1540 & 0.1407 & 0.1503 & 0.3323 & 0.3473 & 0.3223 & 0.2926 & 0.2492 & 0.2416 & 0.2223 & 0.2168 & 0.1472 & 0.1259 \\
                              & $R^2$ & 0.1169 & 0.0961 & 0.0978 & 0.0819 & 0.0071 & 0.0876 & 0.1036 & 0.0452 & 0.1192 & 0.0512 & 0.0963 & \second{0.2174} & \best{0.2271} \\
        \midrule   
        \multirow{2}{*}{XJPX} & MAE     & \second{0.0198} & \second{0.0198} & \best{0.0196} & 0.1309 & 0.1324 & 0.0425 & 0.0842 & 0.0365 & 0.1527 & 0.0316 & 0.0353 & 0.0320 & 0.0271 \\
                              & $R^2$ & 0.1626 & 0.1912 & 0.1996 & 0.1818 & 0.0229 & 0.1245 & 0.0383 & 0.1277 & 0.0133 & 0.0467 & 0.1531 & \second{0.2421} & \best{0.2434} \\
        \midrule
        \multirow{2}{*}{XNSE} & MAE     & \best{0.0264} & 0.0269 & \second{0.0267} & 0.0667 & 0.0347 & 0.0502 & 0.0784 & 0.0555 & 0.1272 & 0.0614 & 0.0497 & 0.0269 & 0.0271 \\
                              & $R^2$ & \second{0.1803} & 0.1445 & \best{0.1815} & 0.1184 & 0.0708 & 0.1140 & 0.0153 & 0.0486 & 0.0152 & 0.0286 & 0.0365 & 0.1424 & 0.1548 \\
        \midrule
        \multirow{2}{*}{XKRX} & MAE     & 0.0271 & \second{0.0255} & \best{0.0246} & 0.0332 & 0.0424 & 0.0408 & 0.0449 & 0.0537 & 0.0608 & 0.0715 & 0.0552 & 0.0347 & 0.0316 \\
                              & $R^2$ & 0.5936 & \best{0.6190} & \second{0.6156} & 0.1966 & 0.0175 & 0.1967 & 0.1792 & 0.2695 & 0.0795 & 0.0842 & 0.2223 & 0.4617 & 0.4641 \\
        \midrule
        \multirow{2}{*}{XHKG} & MAE     & \second{0.0352} & 0.0402 & \best{0.0349} & 0.0435 & 0.0746 & 0.0679 & 0.0547 & 0.0608 & 0.0529 & 0.0702 & 0.0499 & 0.0464 & 0.0402 \\
                              & $R^2$ & 0.1935 & 0.1875 & 0.1824 & 0.1423 & 0.0515 & 0.0394 & 0.0408 & 0.0396 & 0.0482 & 0.0176 & 0.0051 & \second{0.3294} & \best{0.3295} \\
        \midrule
        \multirow{2}{*}{XIDX} & MAE     & 0.0566 & \second{0.0544} & \best{0.0501} & 0.1412 & 0.2504 & 0.0925 & 0.0728 & 0.0827 & 0.1263 & 0.0987 & 0.0836 & 0.0647 & 0.0592 \\
                              & $R^2$ & 0.1275 & 0.1884 & 0.1467 & 0.1443 & 0.0163 & 0.1730 & 0.0433 & 0.1053 & 0.0322 & 0.0391 & 0.1065 & \best{0.2209} & \second{0.2092} \\
        \midrule
        \multirow{2}{*}{XKLS} & MAE     & \second{0.0370} & \best{0.0367} & 0.0376 & 0.1570 & 0.0823 & 0.0456 & 0.1355 & 0.0759 & 0.0533 & 0.0787 & 0.0827 & 0.0397 & 0.0406 \\
                              & $R^2$ & \best{0.5369} & 0.4781 & \second{0.4967} & 0.1867 & 0.1378 & 0.2245 & 0.1201 & 0.1409 & 0.0529 & 0.0540 & 0.1172 & 0.2148 & 0.2247 \\
        \midrule
        \multirow{2}{*}{XTAI} & MAE     & \second{0.0217} & 0.0220 & \best{0.0213} & 0.0230 & 0.0254 & 0.0267 & 0.0229 & 0.0318 & 0.0262 & 0.0223 & 0.0271 & 0.0263 & 0.0240 \\
                              & $R^2$ & \second{0.2607} & 0.2074 & \best{0.2915} & 0.1755 & 0.1797 & 0.1740 & 0.2171 & 0.1591 & 0.2592 & 0.1853 & 0.1783 & 0.2021 & 0.2320 \\
        \midrule
        \multirow{2}{*}{Crypto} & MAE     & \second{0.0147} & 0.0148 & \best{0.0145}  & 0.1438 & 0.0705 & 0.0346 & 0.0926 & 0.0289 & 0.0446 & 0.0642 & 0.0375 & 0.0286 & 0.0292 \\
                                & $R^2$ & 0.1772 & 0.2179 & \best{0.2658}  & 0.0468 & 0.0711 & 0.1212 & 0.1475 & \second{0.2372} & 0.0547 & 0.0286 & 0.1095 & 0.1642 & 0.1575\\
        \midrule
        \multirow{2}{*}{Forex} & MAE     & 0.0097 & \second{0.0074} & \best{0.0069} & 0.0277 & 0.0277 & 0.0205 & 0.0300 & 0.0187 & 0.0212 & 0.0171 & 0.0176 & 0.0219 & 0.0185 \\
                               & $R^2$ & \best{0.1301} & 0.1235 & \second{0.1277} & 0.0002 & 0.0302 & 0.0290 & 0.0270 & 0.0029 & 0.0901 & 0.0382 & 0.0034 & 0.1169 & 0.1141 \\
        \midrule

        \rowcolor{AvgGray}
          & MAE &
        0.0384 & \second{0.0372} & \best{0.0370} & 0.1046 & 0.1021 & 0.0730 & 0.0858 & 0.0662 & 0.0861 & 0.0692 & 0.0627 & 0.0448 & 0.0405 \\
        \rowcolor{AvgGray} \multirow{2}{*}[2.5ex]{\textbf{Average}} & $R^2$ &
        \second{0.2490} & 0.2470 & \best{0.2624} & 0.1295 & 0.0694 & 0.1366 & 0.1066 & 0.1199 & 0.0913 & 0.0731 & 0.1047 & 0.2281 & 0.2323 \\
         \midrule
    \rowcolor{CntGray}
    \multicolumn{2}{c}{\textbf{$\mathbf{1^{st}}$ Count}} &
\cnt{4} & \cnt{2} & \cnt{11} & \cnt{0} & \cnt{0} & \cnt{0} &
\cnt{0} & \cnt{0} & \cnt{0} & \cnt{0} & \cnt{0} & \cnt{1} & \cnt{3} \\
    \bottomrule
  \end{tabular}}
  \caption{Full results of realized volatility forecasting experiments (Part 1): Our model (Kronos) and full-shot time series models. A lower MAE or higher $R^2$ indicates a better prediction. Best and second best results are marked with \best{red underline} and \second{blue underline}, respectively.}
  \label{tab:full_volatility_forecasting_1}
\end{table*}

\begin{table*}[t]
    \centering
    \footnotesize        
    \setlength{\tabcolsep}{4pt}        

    \adjustbox{max width=\textwidth}{
    \begin{tabular}{l|c|*{13}{S}}
        \toprule
        \multicolumn{2}{c}{\textbf{Models}} &
        \multicolumn{12}{c}{\textbf{Zero-shot Time Series Models}} \\
        \cmidrule(lr){3-14}
        \multicolumn{2}{c}{\textbf{Metrics}} &
        \textbf{$\text{Time\text{-}MOE}_{S}$} & \textbf{$\text{Time\text{-}MOE}_{B}$} &
        \textbf{$\text{Moirai}_{S}$} & \textbf{$\text{Moirai}_{B}$} & \textbf{$\text{Moirai}_{L}$} &
        \textbf{TimesFM} &
        \textbf{$\text{Moment}_{S}$} & \textbf{$\text{Moment}_{B}$} & \textbf{$\text{Moment}_{L}$} &
        \textbf{$\text{Chronos}_{S}$} & \textbf{$\text{Chronos}_{B}$} & \textbf{$\text{Chronos}_{L}$} \\
        \midrule
        \multirow{2}{*}{XSHG} & MAE     & 0.0462 & 0.0471 & 0.1158 & 0.0994 & 0.1048 & 0.0408 & 0.0357 & 0.0343 & 0.0366 & 0.0386 & 0.0384 & 0.0382 \\
                              & $R^2$ & 0.2423 & 0.2417 & 0.2118 & 0.2233 & 0.2191 & 0.0995 & 0.2479 & 0.2461 & 0.2336 & 0.1946 & 0.1922 & 0.1663 \\
        \midrule
        \multirow{2}{*}{XNAS} & MAE     & 0.2713 & 0.2498 & 0.3537 & 0.1927 & 0.2502 & 0.1902 & \second{0.1034} & \best{0.1020} & 0.1168 & 0.1896 & 0.1863 & 0.1881 \\
                              & $R^2$ & 0.1255 & 0.0901 & 0.1782 & 0.1228 & 0.1306 & 0.0740 & 0.0872 & 0.0882 & 0.0804 & 0.0811 & 0.0340 & 0.0982 \\
        \midrule
        \multirow{2}{*}{XJPX} & MAE     & 0.0372 & 0.0367 & 0.1065 & 0.0829 & 0.0878 & 0.0345 & 0.0291 & 0.0278 & 0.0306 & 0.0331 & 0.0331 & 0.0329 \\
                              & $R^2$ & 0.1392 & 0.1374 & 0.1150 & 0.1541 & 0.1493 & 0.1213 & 0.1489 & 0.1450 & 0.01375 & 0.1812 & 0.1794 & 0.1769 \\
        \midrule
        \multirow{2}{*}{XNSE} & MAE     & 0.0420 & 0.0415 & 0.1029 & 0.0873 & 0.0924 & 0.0437 & 0.0364 & 0.0358 & 0.0397 & 0.0414 & 0.0413 & 0.0411 \\
                              & $R^2$ & 0.0411 & 0.0457 & 0.0455 & 0.0588 & 0.0554 & 0.0394 & 0.0483 & 0.0468 & 0.0422 & 0.0454 & 0.0563 & 0.0439 \\
        \midrule
        \multirow{2}{*}{XKRX} & MAE     & 0.0452 & 0.0447 & 0.1109 & 0.0909 & 0.0982 & 0.0508 & 0.0418 & 0.0413 & 0.0461 & 0.0485 & 0.0484 & 0.0482 \\
                              & $R^2$ & 0.2248 & 0.2321 & 0.2235 & 0.2576 & 0.2229 & 0.1249 & 0.2914 & 0.2811 & 0.2588 & 0.3357 & 0.3371 & 0.3132 \\
        \midrule
        \multirow{2}{*}{XHKG} & MAE     & 0.0701 & 0.0671 & 0.1824 & 0.1367 & 0.1499 & 0.0551 & 0.0500 & 0.0475 & 0.0499 & 0.0526 & 0.0523 & 0.0521 \\
                              & $R^2$ & 0.1757 & 0.1475 & 0.0900 & 0.1838 & 0.1576 & 0.1862 & 0.1537 & 0.1502 & 0.1432 & 0.1064 & 0.1018 & 0.1090 \\
        \midrule
        \multirow{2}{*}{XIDX} & MAE     & 0.0725 & 0.0718 & 0.2321 & 0.1687 & 0.1876 & 0.0766 & 0.0652 & 0.0695 & 0.0663 & 0.0744 & 0.0735 & 0.0732 \\
                              & $R^2$ & 0.1558 & 0.1572 & 0.1228 & 0.1118 & 0.1144 & 0.0952 & 0.1607 & 0.1093 & 0.1471 & 0.1445 & 0.1820 & 0.1692 \\
        \midrule
        \multirow{2}{*}{XKLS} & MAE     & 0.0572 & 0.0553 & 0.1142 & 0.0914 & 0.1037 & 0.0733 & 0.0571 & 0.0597 & 0.0699 & 0.0706 & 0.0705 & 0.0703 \\
                              & $R^2$ & 0.0828 & 0.1021 & 0.1451 & 0.1559 & 0.1669 & 0.0541 & 0.1714 & 0.1725 & 0.1393 & 0.1673 & 0.1745 & 0.1645 \\
        \midrule
        \multirow{2}{*}{XTAI} & MAE     & 0.0387 & 0.0384 & 0.1047 & 0.0900 & 0.0954 & 0.0386 & 0.0335 & 0.0319 & 0.0341 & 0.0371 & 0.0369 & 0.0366 \\
                              & $R^2$ & 0.1901 & 0.1913 & 0.1611 & 0.1704 & 0.1729 & 0.0789 & 0.1885 & 0.1850 & 0.1672 & 0.1868 & 0.1804 & 0.1588 \\
        \midrule
        \multirow{2}{*}{Crypto} & MAE     & 0.0374 & 0.0373 & 0.0570 & 0.0574 & 0.0572 & 0.0352 & 0.0209 & 0.0236 & 0.0327 & 0.0341 & 0.0340 & 0.0339 \\
                                & $R^2$ & 0.1416 & 0.1387 & 0.1061 & 0.1004 & 0.2016 & 0.0881 & 0.1685 & 0.1310 & 0.1758 & 0.1566 & 0.1608 & 0.1584 \\
        \midrule
        \multirow{2}{*}{Forex} & MAE     & 0.0225 & 0.0110 & 0.0119 & 0.0151 & 0.0120 & 0.0171 & 0.0151 & 0.0155 & 0.0158 & 0.0102 & 0.0124 & 0.0218 \\
                               & $R^2$ & 0.1173 & 0.0286 & 0.0145 & 0.0306 & 0.0504 & 0.0141 & 0.0744 & 0.0592 & 0.0717 & 0.0391 & 0.0245 & 0.0453 \\
        \midrule

        \rowcolor{AvgGray}
         & MAE &
        0.0673 & 0.0637 & 0.1356 & 0.1011 & 0.1127 & 0.0596 & 0.0444 & 0.0444 & 0.0490 & 0.0573 & 0.0570 & 0.0579 \\
        \rowcolor{AvgGray} \multirow{2}{*}[2.5ex]{\textbf{Average}} & $R^2$ &
        0.1487 & 0.1375 & 0.1285 & 0.1427 & 0.1492 & 0.0887 & 0.1380 & 0.1468 & 0.1339 & 0.1490 & 0.1475 & 0.1458 \\
        \midrule
        \rowcolor{CntGray}
        \multicolumn{2}{c}{\textbf{$\mathbf{1^{st}}$ Count}} &
\cnt{0} & \cnt{0} & \cnt{0} & \cnt{0} & \cnt{0} & \cnt{0} &
\cnt{0} & \cnt{1} & \cnt{0} & \cnt{0} & \cnt{0} & \cnt{0} \\
        \bottomrule
    \end{tabular}}
    \caption{Full results of realized volatility forecasting experiments (Part 2): Zero-shot time series models. A lower MAE or higher $R^2$ indicates a better prediction. Best and second best results are marked with \best{red underline} and \second{blue underline}, respectively.}
    \label{tab:full_volatility_forecasting_2}
\end{table*}

\begin{table*}[t]
  \centering
  \footnotesize
  \setlength{\tabcolsep}{4pt}

  \adjustbox{max width=\textwidth}{
  \begin{tabular}{l|c|*{6}{c}}
    \toprule
    \multicolumn{2}{c|}{\textbf{Models}} &
    \multicolumn{3}{c}{\textbf{\includegraphics[height=0.23cm]{img/logo_paper.png}\hspace{0.2em}Kronos (Ours)}} &
    \multicolumn{3}{c}{\textbf{Time-series Generative Models}} \\ 
    \cmidrule(lr){3-5} \cmidrule(lr){6-8}
    \multicolumn{2}{c|}{\textbf{Metrics}} & 
    \textbf{$\text{Kronos}_{small}$} & \textbf{$\text{Kronos}_{base}$} & \textbf{$\text{Kronos}_{large}$} &
    \textbf{DiffusionTS} & \textbf{TimeVAE} & \textbf{TimeGAN} \\
    \midrule

    \multirow{2}{*}{XSHG} & 15min & \second{0.2313} & 0.2317 & \best{0.2393} & 0.0885 & 0.0015 & 0.2241 \\ 
                          & daily & 0.1865 & \second{0.2227} & 0.2105 & \best{0.2532} & 0.0142 & 0.1193 \\
    \midrule
    \multirow{2}{*}{XTAI} & 15min & 0.1733 & 0.1478 & \second{0.1788} & 0.1420 & 0.0387 & \best{0.2689} \\ 
                          & daily & \second{0.2088} & 0.2023 & \best{0.2235} & 0.1712 & 0.0097 & 0.0622 \\
    \midrule
    \multirow{2}{*}{Crypto} & 15min & 0.4100 & \second{0.4185} & \best{0.4187} & 0.3005 & 0.0637 & 0.0680 \\ 
                            & daily & 0.2792 & 0.2575 & \second{0.2835} & \best{0.3188} & 0.0402 & 0.2114 \\
    \midrule
    \multirow{2}{*}{Forex} & 15min & \second{0.4783} & 0.\best{4903} & 0.4688 & 0.4112 & 0.0492 & 0.4015 \\ 
                           & daily & 0.3337 & \best{0.4363} & \second{0.4152} & 0.3177 & 0.0295 & 0.2387 \\
    \midrule

    \rowcolor{AvgGray}
    \multicolumn{2}{c|}{\textbf{Average}} &
      0.2876 & \second{0.3009} & \best{0.3048} & 0.2504 & 0.0308 & 0.1993 \\
    \midrule

    \rowcolor{CntGray}
    \multicolumn{2}{c|}{\textbf{$\mathbf{1^{st}}$ Count}} &
      \cnt{0} & \cnt{2} & \cnt{4} & \cnt{2} & \cnt{0} & \cnt{1} \\
    \bottomrule
  \end{tabular}}
  \caption{Full discriminative score results for synthetic K-line generation experiments. A higher score indicates a better generation quality. Best and second best results are marked with \best{red underline} and \second{blue underline}, respectively.}
  \label{tab:generate_discriminative}
\end{table*}

\begin{table*}[t]
  \centering
  \footnotesize
  \setlength{\tabcolsep}{4pt}

  \adjustbox{max width=\textwidth}{
  \begin{tabular}{l|c|c|*{6}{S}}
    \toprule
    \multicolumn{3}{c|}{\textbf{Models}} &
    \multicolumn{3}{c}{\textbf{\includegraphics[height=0.23cm]{img/logo_paper.png}\hspace{0.2em}Kronos (Ours)}} &
    \multicolumn{3}{c}{\textbf{Time-series Generative Models}} \\ 
    \cmidrule(lr){4-6} \cmidrule(lr){7-9}
    \multicolumn{3}{c|}{\textbf{Metrics}} & 
    \textbf{$\text{Kronos}_{small}$} & \textbf{$\text{Kronos}_{base}$} & \textbf{$\text{Kronos}_{large}$} &
    \textbf{DiffusionTS} & \textbf{TimeVAE} & \textbf{TimeGAN} \\
    \midrule

    \multirow{4}{*}{XSHG} & \multirow{2}{*}{15min} & IC      & 0.0223 & \second{0.0231} & \best{0.0236} & 0.0103 & 0.0098 & 0.0102 \\
                          &                        & RankIC  & 0.0144 & \second{0.0147} & \best{0.0151} & 0.0087 & 0.0134 & 0.0081 \\
                          \cmidrule(lr){2-9}
                          & \multirow{2}{*}{daily} & IC      & \best{0.0918} & \second{0.0902} & 0.0845 & 0.0760 & -0.0789 & 0.0108 \\
                          &                        & RankIC  & \best{0.0854} & \second{0.0839} & 0.0796 & 0.0684 & -0.0720 & 0.0150 \\
    \midrule
    \multirow{4}{*}{XTAI} & \multirow{2}{*}{15min} & IC      & 0.0230 & \second{0.0274} & \best{0.0281} & 0.0074 & -0.0118 & 0.0045 \\
                          &                        & RankIC  & 0.0226 & \second{0.0276} & \best{0.0299} & 0.0037 & -0.0092 & -0.0003 \\
                          \cmidrule(lr){2-9}
                          & \multirow{2}{*}{daily} & IC      & \second{0.0460} & 0.0437 & \best{0.0560} & 0.0013 & -0.0213 & 0.0118 \\
                          &                        & RankIC  & \second{0.0445} & 0.0431 & \best{0.0551} & -0.0001 & -0.0193 & 0.0118 \\
    \midrule
    \multirow{4}{*}{Crypto} & \multirow{2}{*}{15min} & IC      & \second{0.0237} & \best{0.0243} & \second{0.0237} & -0.0016 & -0.0012 & 0.0096 \\
                            &                        & RankIC  & \second{0.0222} & \best{0.0231} & \best{0.0231} & -0.0026 & -0.0016 & 0.0079 \\
                            \cmidrule(lr){2-9}
                            & \multirow{2}{*}{daily} & IC      & 0.0027 & \best{0.0051} & \second{0.0037} & -0.0085 & -0.0130 & -0.0330 \\
                            &                        & RankIC  & 0.0028 & \best{0.0049} & \second{0.0031} & -0.0111 & -0.0100 & -0.0301 \\
    \midrule
    \multirow{4}{*}{Forex} & \multirow{2}{*}{15min} & IC      & \best{0.0202} & \second{0.0172} & 0.0171 & 0.0156 & -0.0150 & 0.0095 \\
                           &                        & RankIC  & \best{0.0183} & \second{0.0158} & 0.0150 & 0.0142 & -0.0140 & 0.0094 \\
                           \cmidrule(lr){2-9}
                           & \multirow{2}{*}{daily} & IC      & 0.0044 & \second{0.0069} & 0.0042 & 0.0016 & \best{0.0140} & -0.0044 \\
                           &                        & RankIC  & 0.0042 & \second{0.0066} & 0.0045 & 0.0007 & \best{0.0160} & -0.0058 \\
    \midrule

    \rowcolor{AvgGray}
      &  & IC     & 0.0293 & \second{0.0297} & \best{0.0301} & 0.0128 & -0.0147 & 0.0024 \\
    \rowcolor{AvgGray}
    \multirow{-2}{*}{\textbf{Average}} &
      & RankIC & 0.0268 & \second{0.0275} & \best{0.0282} & 0.0102 & -0.0121 & 0.0020 \\
    \midrule
    
    \rowcolor{CntGray}
    \multicolumn{3}{c|}{\textbf{$\mathbf{1^{st}}$ Count}} &
      \cnt{4} & \cnt{4} & \cnt{9} & \cnt{0} & \cnt{2} & \cnt{0} \\
    \bottomrule
  \end{tabular}}
  \caption{Full results of predictive usefulness (IC and RankIC) for synthetic K-line generation experiments. Higher IC and RankIC scores suggest the generated data is more useful for building predictive financial models. Best and second best results are marked with \best{red underline} and \second{blue underline}, respectively.}
  \label{tab:generate_usefulness}
\end{table*}

\end{document}